# VECTOR SOLITON FIBER LASERS

**ZHANG HAN**

**School of Electrical & Electronic Engineering**

A thesis submitted to the Nanyang Technological University

in partial fulfillment of the requirement for the degree of

Doctor of Philosophy

**2010**

## __Statement of Originality__

I hereby certify that the work embodied in this thesis is the result of original research and has not been submitted for a higher degree to any other university or institution.

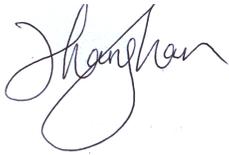

July 17, 2010

Zhang Han                                        Date

# Summary


Solitons, as stable localized wave packets that can propagate long distance in dispersive media without changing their shapes, are ubiquitous in nonlinear physical systems. Since the first experimental realization of optical bright solitons in the anomalous dispersion single mode fibers (SMF) by Mollenauer *et al*. in 1980 and optical dark solitons in the normal dispersion SMFs by P. Emplit *et al*. in 1987, optical solitons in SMFs had been extensively investigated. In reality a SMF always supports two orthogonal polarization modes. Taking fiber birefringence into account, it was later theoretically predicted that various types of vector solitons, including the bright-bright vector solitons, dark-dark vector solitons and dark-bright vector solitons, could be formed in SMFs. However, except the bright-bright type of vector solitons, other types of vector solitons are so far lack of clear experimental evidence.

Optical solitons have been observed not only in the SMFs but also in mode locked fiber lasers. It has been shown that the passively mode-locked erbium-doped fiber lasers offer a promising experimental platform for studying the scalar optical solitons. Vector solitons can also be formed in mode locked fiber lasers. In this dissertation, the author presents results of a series of theoretical and experimental investigations on the vector solitons in fiber lasers.




First of all, passively mode-locked erbium-doped fiber lasers with the semiconductor saturable absorber mirror (SESAM) as a mode locker were designed and constructed. Formation of various vector bright-bright solitons in the fiber lasers was demonstrated, and whose features and dynamics were investigated. There are either the coherently or incoherently coupled vector solitons in the fiber lasers. Under coherent coupling between the two orthogonal polarization components of the fiber lasers, a type of polarization-locked high-order vector soliton as well as the polarization-rotating vector solitons was observed. Worth of mentioning is the special features of the high-order vector soliton: besides that its two orthogonal polarization components are phase locked, the two polarization components also have different soliton profiles. While the stronger polarization component is a single hump pulse, the weaker component has a double-humped structure with 180° phase difference between the humps. The features of the experimentally observed high-order vector soliton well match those of the theoretical predictions. Moreover, we show that in the normal dispersion cavity fiber lasers dissipative vector solutions could be formed, and under stronger cavity birefringence multi-wavelength dissipative soliton operation of the fiber lasers is possible.

In addition of the experimental studies, numerical simulations on the vector soliton operation of the fiber lasers were also carried out. To closely simulate the vector soliton evolution in a fiber laser, a model that is based on the coupled extended Ginzburg-Landau equations that take into account not only the fiber dispersion and nonlinearity, laser gain and cavity losses, saturable absorber effect,



but also the cavity feedback and boundary condition was built up. The coupled extended GLEs were solved numerically using the standard split-step method under the experimental laser cavity conditions. It was found that all the experimentally observed vector soliton features could be numerically reproduced.

Special attention was paid on the soliton formation in all normal dispersion cavity fiber lasers. It was firstly revealed that scalar dark solitons described by the nonlinear Schrodinger equation could be formed in an all dispersion cavity fiber laser with a polarizer inserted in the cavity. If the polarizer was replaced with a graphene based saturable absorber, vector dark-dark soliton and trapping of vector dark-dark solitons were further experimentally obtained. Numerical simulations have also well confirmed the experimental observations of dark soliton formation in the fiber lasers.

Independent on the laser cavity dispersion, another novel type of optical solitons known as the optical domain wall solitons was also firstly experimentally identified in the experiments. Domain wall soliton is also a kind of vector soliton. However, different from the vector solitons mentioned above, its formation is a result of the nonlinear coupling between two coexisting eign states in an optical system. Either the coherently coupled optical domain wall solitons, represented as a phase-locked dark-bright pulse pair or dark-dark pulse pair, or the incoherently coupled optical domain wall solitons, represented as vector dark solitons that separating two stable optical domains were first experimentally demonstrated and numerically confirmed.



Lastly, vector soliton operation of graphene mode locked fiber lasers was demonstrated. Graphene is a newly discovered 2D single atomic layer nano-material with unique electronic properties, whose applications in the nano-electronics have attracted enormous attention. Graphene also possesses a number of special optical properties including the wavelength independent ultrafast saturable absorption. The author has first experimentally demonstrated the saturable absorber of graphene, and further exploited the saturable absorption of graphene for mode locking fiber lasers. It was shown that using graphene as a saturable absorber to mode lock the erbium-doped fiber lasers, large energy mode locked pulses with pulse energy up to 7.3 nJ could be achieved. Moreover, wide-band tuning of the mode locked pulses and various types of the previously discussed vector solitons could also be obtained in the graphene mode locked fiber lasers. Comparing with other types of the conventionally used mode lockers, e.g. SESAM, graphene as a mode locker has the following merits: polarization independent broadband saturable absorption, ultrafast saturation recovery time, tunable saturation modulation deepth, lower non-saturable loss, cheaper fabrication cost, easy integration in a laser system. It is envisaged that other applications of graphene in photonics and optoelectronics could be further identified.



# Acknowledgements

I am grateful to all the people who had given me support whilst researching and accomplishing this dissertation.

First and foremost, I wish to express my sincere gratitude to my supervisor, Prof. Tang Dingyuan, for introducing me to the wonderful world of nonlinear optics. His wide knowledge, logical way of thinking, passion and enthusiasm for research, has been lifelong wealth for me. I appreciate all his contribution of efforts, ideas, time and fund to make my Ph. D experience fruitful and stimulating. His understanding, encouraging and personal guidance have provided a good basis for the present thesis.

I would like to thank Prof. Pan Chunxu in Wuhan University of China for recommending me as a Ph. D candidate to NTU and offering me continuous supports. I wish to thank Prof. Randall Knize in United States Air Force Academy, who has given me many exciting ideas. I warmly thank Dr. Zhao Luming, who taught me a lot of experimental skills. I am thankful to Dr. Bao Qiaoliang, with who I could extend my research from laser to graphene optoelectronics. Many thanks also go to Ms. Wu Xuan for the productive discussions, and Dr. Xie Guoqiang, Mr. Tan Wei De, Mr. Lin Bo for helping me a lot.

I would like to acknowledge Prof. Shum Ping, director of NTRC of NTU, who provided first-class experimental facilities for fiber laser research. I also wish



to thank all my colleagues and friends in Photonics Research Center and NTRC of NTU. Without their friendship and kindly assistance it would have been much tougher to finally complete this dissertation. In particular, I would like to show my appreciation to the technicians in the two labs for their kindly assistance with the technical problems.

This dissertation is dedicated to my family. My dad Zhang Guoxin, my mum Chen Zhiling and my young sister Zhang Jiayi, I am forever indebted to you for your great love, considerate understanding, eternal patience and encouragement all the way along. Deeply Appreciated!

The financial support of NTU is gratefully acknowledged!



# Table of Contents









# List of Figures















# List of Acronyms

**Abbreviations**                     **Full Expressions**

CW                                    Continuous wave

DBVS                                  Dark-bright vector soliton

DDVS                                  Dark-dark vector soliton

DSs                                   Dissipative soliton

DVSs                                  Dissipative vector solitons

DWS                                   Domain wall soliton

EDF                                   Erbium-doped fiber

EDFAs                                 Erbium-doped fiber amplifiers

EDFLs                                 Erbium-doped fiber lasers

FWHM                                  Full width at half maximum

FWM                                   Four-wave-mixing

GLE                                   Ginzburg-Landau equation

GHz                                   Gigahertz

GVD                                   Group velocity dispersion

GVLVSs                                Group velocity locked vector solitons

HOVS                                  High-order vector soliton

LCPDB                                 Linear cavity phase delay bias

MHz                                   Megahertz

NPR                                   Nonlinear Schrödinger equation



| | |
|---|---|
| OSA | Optical Spectrum Analyzer |
| PBS | Polarization beam splitter |
| PC | Polarization controller |
| PDWSs | Polarization domain-wall solitons |
| PEF | Polarization evolution frequency |
| PMI | Polarization modulation instability |
| RF | Radio-frequency |
| SESAM | Semiconductor saturable saturaber |
| SMF | Single mode fiber |
| SPM | Self-phase modulation |
| SWCNT | Single wall carbon nanotube |
| PEF | Polarization evolution frequency |
| PLVSs | Polarization-locked vector solitons |
| PMF | Polarization-maintaining fibers |
| WDM | Wavelength-division-multiplexing |
| XPM | Cross-phase modulation |





# Chapter 1.  Introduction

## 1.1    Background and motivation

Solitons, as particle-like nonlinear localized waves due to an interior self-reinforcing against dispersion through nonlinearity, universally existed in a large amount of nonlinear physics scenarios from fluids and plasma physics to optics, biological and atmospheric systems [1-9]. Optical solitons have attracted everlasting interest in the last three decades thanks to their theoretical values and attractive practical applications in the optical communication and signal processing systems. In the context of optical fibers, conventional solitons are ultra-fast, high-intensity optical pulses that do not dissipate and temporally retain their shapes through the balance between the dispersion and nonlinearity. One of the most promising applications of the soliton theory could be credited to the field of optical fiber communications [1-7]. Hasegawa and Tappert firstly predicted the existence of solitons in optical fibers and proposed the potential application of solitons for optic communications in 1973 [1]. Seven years later, Mollenauer *et al*. at the Bell Laboratories experimentally demonstrated the propagation of optical bright solitons in anomalous dispersion fibers [2], and subsequently in 1987, P. Emplit *et al*. from the Universities of Brussels and Limoges, made the first experimental observation of the propagation of a dark soliton in normal dispersion fibers, both of which heralded a new era of soliton and whipped up a storm of studying on optical





solitons that continue to amaze [8]. Optical bright solitons were characterized by temporally localized intensity peaks while the optical dark solitons were featured by temporally localized dips on a continuous background. Decades of study on optical solitons show that their dynamics could be well understood by the nonlinear Schrödinger equation (NLSE). However, in reality, an optical single mode fiber (SMF) is not rigorously single "mode" but always endorses two orthogonal polarization modes. Considering the fiber birefringence, it was later theoretically found that depending on the sign of fiber dispersion and the strength of fiber birefringence, three types of vector solitons: bright-bright, dark-dark, and dark-bright [9], could be yielded in SMFs. Although theoretical investigations on optical solitons had explosively progressed, experimental studies on the vector solitons seriously lagged.

A short pulse laser source is one of the prerequisites of optical soliton formation. Usually, "mode lock" is employed to generate ultra-short pulses in lasers [10-11]. In the case of mode-locked fiber lasers, apart from the cavity components that are necessary for achieving mode locking, laser cavity is mainly made up of optical fibers. If the fiber laser cavity is free of any polarization dependent elements, particularly, passive mode lockers purely contributed by material based saturable absorbers other than nonlinear polarization rotation technique (NPR); it is natural to anticipate that under certain conditions optical vector solitons can also propagate along the laser cavity without polarizer limitations. Lately, research interests have been shifted from the anomalous to normal dispersion regime with the purpose of large energy formation. In normal dispersion domain, dissipative solitons are





formed as a result of the mutual nonlinear interaction among the normal cavity dispersion, cavity fiber nonlinear Kerr effect, laser gain saturation and gain bandwidth filtering [12-27]. Since dissipative solitons manifested entirely different features from conventional solitons, it is meaningful to experimentally and theoretically study the kinetics of vector sides of dissipative solitons.

Except the soliton formation induced by mode locking, NLSEs automatically permit the dark soliton in normal dispersion regime and the gain competition among the two orthogonal polarizations allows the formation of domain wall pulse. Besides mode locking operation, NLSE- or domain wall- type dark solitons could be experimentally realized under particular cavity parameters. Thus, we could naturally conclude that the fiber laser is indeed a well-controlled platform to reveal the dynamics of vector solitons from vector bright soliton, vector dark soliton to domain wall soliton.

This chapter is intended to provide a basic overview on soliton theory and mode-locked fiber soliton lasers. Section 1.1 gives a brief review on the development of mode-locked fiber soliton lasers. The basic theories and development of mode locking and soliton generation are provided in Section 1.2. Section 1.3 presents the motivation and objectives of the research. Section 1.4 discusses the main contribution in this thesis. An overview of the dissertation is summarized in Section 1.5.

## 1.2    Development of mode-locked soliton fiber lasers

Since the low-loss silica fiber was available in 1970s, fiber lasers have gained incessant worldwide attention due to their flexible, simplicity, durability, high





efficiency, compact size, modest energy and cost. Many different rare-earth ions, such as erbium ($Er^{3+}$), neodymium ($Nd^{3+}$), and ytterbium ($Yb^{3+}$) have been used to dope the normal fibers. These rare-earth-doped fibers then can be used as the gain medium in fiber lasers or amplifiers depending on the required operating wavelength range.

With the rapid development of optical communications in the last three decades, the demand for stable ultrashort pulse laser sources has become the central goal of research. Mode-locked fiber lasers are capable of producing pulses with widths in a very wide range of from tens of fs to nanosecond and a wide repetition rate range of from less than 1 MHz to 1 THz. From the technical point of view, mode-locked, erbium-doped fiber lasers (EDFLs) stand out in that they are capable of generating ultra-short optical pulses in the spectral range of 1.5 µm, which allows use of readily available telecommunications components and can be easily frequency doubled to 760 nm to 820 nm, offering a promising alternative to expensive Ti:sapphire laser. In addition, in contrast with other wavelength, both single mode anomalous and normal dispersion fibers are effortlessly available in EDFLs, providing an attractive experimental setup to investigate solitons from anomalous-, zero- to normal- dispersion cavity. This is a two-way street; the developed soliton theory in EDFL also supports the advancement of laser operations at other wavelengths.

### 1.2.1   Fundamentals of Mode locking

Laser is essentially an optical oscillator requiring the two basic constituents of any oscillator, namely amplification and feedback. The stimulated emission in a gain





medium provided the amplification while the laser cavity, composed by sets of mirrors reflecting light, supplied the feedback. The word "mode locking" firstly appeared in the paper of Hargroves in 1964 [28]. The requirement that the electromagnetic field be unchanged after one round trip in the laser makes lasing only occurring for discrete frequencies such that the cavity length is an integer number of wavelengths. The "cavity" of a laser ensures that light is emitted at well-defined wavelengths, known as modes. By introducing a relatively weak modulation synchronous with the round trip time of the laser, the coherence between the phases of different modes could be realized, and pulsed radiation could be produced. The history of laser mode locking is a progression of new and better ways to generate shorter and shorter pulses, and of improvements in the understanding of mode-locking processes. Over the last three decades, mode locking has been used in all kinds of lasers and even ultra-short pulses as short as 47 fs were achieved [29]. In the temporal domain, mode locking actually produces a pulse train, where the time interval between neighboring pulses equals the cavity round trip time. It should be mentioned that there are others types of lasers that generate pulses, the most common being called a Q-switched laser. Furthermore, the pulse evolution in those lasers is unrelated to soliton dynamics. Non-pulsed lasers are defined as "continuous wave" (CW).

The theory of mode locking is a little complicated. Here, we only give the basic physics of mode locking. A large number of longitude modes can be stimulated simultaneously in the gain bandwidth of a laser provided that the pump is strong. The frequency spacing among the modes is given by





$$\Delta \nu = {}^{c}\!\!\big/\!\!{}_{L_{opt}} \qquad (1.1.1)$$

where $L_{opt}$ is the optical length during one round trip inside the cavity. Therefore, the total optical field can be written as:

$$E(t) = \sum_{m=-M}^{M} E_m \exp[i(\phi_m - \omega_m t)] \qquad (1.1.2)$$

where $E_m$, $\phi_m$ and $\omega_m$ are the amplitude, phase and frequency of the specific modes among $(2M + 1)$ modes permitted by the laser gain bandwidth. If all these modes operate independently of each other without definite phase relationship between them, the interference terms in the total intensity $|E(t)|^2$ averages out to zero. Under this situation, the laser works in a multimode CW state.

Mode locking occurs when the phases of various longitude modes are synchronized such that the phase difference between any two neighboring modes is locked to a constant value, viz. $\phi_m = \phi_0 + m\phi$. Since the mode frequency $\omega_m = \omega_0 + 2m\pi\Delta\nu$, if we assume for simplicity that all modes have the same amplitude $E_0$, the total intensity can be analytically calculated and presented as:

$$|E(t)|^2 = \frac{\sin^2\left[(2M+1)\pi\Delta\nu t + \phi/2\right]}{\sin^2\left(\pi\Delta\nu t + \phi/2\right)} E_0^2 \qquad (1.1.3)$$

The intensity shows as a periodic function with period $\tau_r = 1/\Delta\nu$, which is just the cavity round trip time. Under mode locking, the laser output is in the form of a pulse train with a repetition rate equals to $\Delta\nu$.

Under the modulation in the cavity, the initiated pulse is shortened every time it passes through the resonator. This shortening process continues until the pulse





becomes so short and its spectrum so wide that the pulse lengthening or spectrum narrowing mechanisms, such as the finite bandwidth of the gain, spring to action. The pulse width is estimated from Eq. (1.1.3) $\tau_p = \left[ (2M+1)\Delta\nu \right]^{-1}$. Since the $\left[ (2M+1)\Delta\nu \right]$ represents the total bandwidth of all mode-locked modes, the pulse width is inversely related to the spectral bandwidth over which phases of various modes can be synchronized. In practice, the exact relationship between the pulse width and the gain bandwidth depends on the nature of the gain broadening. In rare-earth doped fibers, the fiber characters such as the birefringence also affect the pulse width. In general, there are two sorts of mode locking: active mode locking, and passive mode locking. Both methods have been used in fiber lasers to achieve ultra-short optical pulses.

### 1.2.2 Conservative/dissipative soliton operation

Conservative soliton operation is an intrinsic feature of mode-locked fiber lasers with anomalous dispersion and has been intensively studied [30-42]. In general, a pulse propagating in optical fibers is affected by both the group velocity dispersion (GVD) and the nonlinear optical Kerr effect. The GVD broadens the pulse in time domain, while the nonlinear self-phase modulation (SPM) broadens the pulse in frequency domain, which corresponds to narrowing the pulse in time domain. If these two effects on a pulse totally compensate with each other, it will maintain its pulse shape and pulse width during the propagation in the fiber i.e. forming a soliton. The formation of solitons in optical fibers is described by the NLSE [42].

In a fiber laser, after mode locking, an ultra-short pulse is firstly formed in the laser.





If the peak power of the mode-locked pulse is strong, due to the nonlinear optical Kerr effect in the cavity, SPM occurs, which narrows the pulse width. If the strength of the SPM is strong enough that it can balance the pulse width broadening caused by the cavity dispersion, the pulse will propagate in the fiber without changing its pulse width. In fact, since the SPM is the pulse intensity dependent, as a pulse propagates in anomalous-dispersion fiber, it can adjust its intensity so that the effects induced by the optical Kerr effect and the anomalous GVD are automatically balanced. Namely, optical solitons is actually a generic property of mode-locked fiber lasers with anomalous cavity dispersion. Due to the existence of gain and loss, the soliton formation in a fiber laser is also governed by the Ginzburg-Landau equation (GLE) [12].

Current soliton theory considers the nonlinear systems as a conservative system or takes the energy import-export dynamics as a small perturbation to the conservative system, which is different from the intrinsic dynamics of dissipative systems. Therefore, more appropriate description of the nonlinear systems is compulsory. Dissipative solitons (DSs) are stable solitary localized structures that arise in nonlinear spatio-temporal dissipative systems due to mechanisms of self-organization [43]. They can be considered as an extension of the classical soliton concept of conservative systems. In fact, DSs are the real product of nonlinear systems. Therefore, detailed studies for DSs are desirable as it can help us to get insight into the concrete dynamics in nonlinear systems without approximation.

In addition, we point out that the balance between gain and loss is certainly a necessary condition for the dissipative soliton formation in a fiber laser, but the





balance between the linear gain and linear loss is not a sufficient condition. It is inappropriate to employ the linear gain-loss balance as the sole condition to judge that all solitons formed in a laser are dissipative solitons. For a dissipative soliton in the anomalous dispersion fiber lasers, the spectral filtering effect, which is actually frequency dependent loss, must be strong enough (like the case of soliton formation in the normal dispersion fiber lasers), but in the practice so far in all anomalous dispersion fiber lasers, due to the pulse peak clamping effect this effect did not appear, consequently it did not contribute to the pulse shaping, indicating that dissipative soliton was hardly observed in anomalous dispersion fiber lasers. The situation is completely different in the normal dispersion regime, where no NLSE pulse shaping exists and the spectral filtering could become a significant effect. Therefore, dissipative soliton is always formed.

Soliton operation can be achieved in both actively and passively mode-locked fiber lasers through the soliton shaping of the mode-locked pulses. Once the soliton operation is achieved in a laser, the pulse characteristics are no longer determined by the mode locking mechanism but by the soliton shaping. Compared with a conventional mode-locked pulse, solitons have narrower pulse width and therefore higher peak power.

### 1.2.3   Active laser mode locking

A typical arrangement for an integrated actively mode-locked fiber laser is shown in **Figure 1.1**. An active modulator is necessary in actively mode-locked lasers to modulate either the amplitude or the phase of the intracavity optical field at a





frequency that equals to integer multiples of the cavity longitude mode spacing [44-47]. As far as the modulation frequency is matched to the cavity length, active mode locking of fiber lasers can be easily obtained. With this mode locking technique, high repetition rate pulses with good noise performance can be generated. For instance, optical pulses of less than 6 ps duration generated at repetition rate to 40 GHz have been reported in an actively mode-locked erbium-doped fiber ring laser [48].

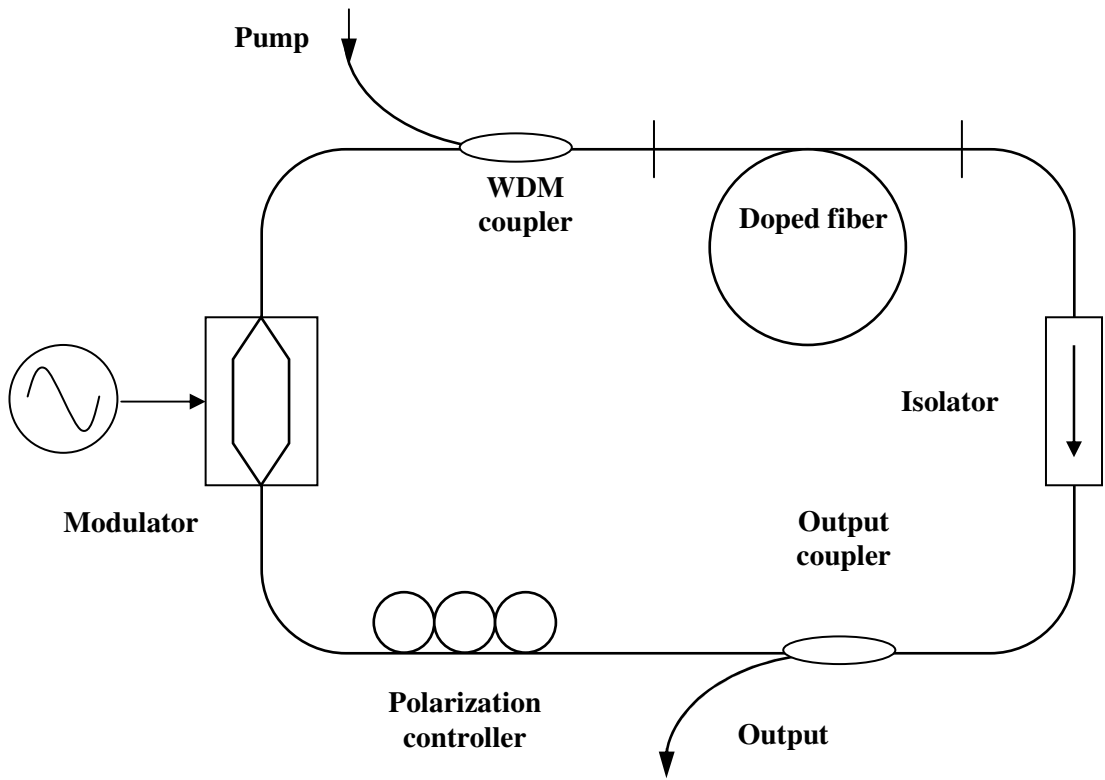

**Figure 1.1:** Typical experimental setup of an actively mode-locked fiber laser

Without soliton shaping an actively mode-locked fiber laser produces mode-locked pulses of Gaussian pulse profile, whose width is determined by the laser gain





bandwidth, and the modulation frequency and strength [44]. For the EDFLs this is normally in the range of several tens of picoseconds. If the peak power of the pulses is strong enough, soliton shaping happens and shortens the mode-locked pulse as well as changes the pulse profile from a Gaussian shape to a Sech$^2$ shape [49, 50]. However, comparing with the passively mode-locked fiber lasers, soliton operation cannot be easily achieved in actively mode-locked fiber lasers. The main reason for the difficulty is that harmonic mode locking is normally implemented. In a mode locked state too many mode-locked pulses are generated in the laser cavity, and they share the laser cavity energy. The energy of each mode locked pulse is therefore weak. And determined by the mode-locking technique, the mode-locked pulse width is initially broad, leading to that the peak power of the pulse achievable is also low. Therefore, only very weak nonlinear SPM could be actually generated by the pulses, which limits the strength of their soliton shaping. To obtain ultra narrow soliton pulses in the laser either strong pumping and/or low repetition rate operation are necessary. Actively mode-locked fiber lasers also suffer from the pulse drop-out problem [51], which could also be traced back to the gain competition between the pulses. In short, although actively mode-locked fiber lasers have the advantages such as high repetition rates, narrow line-width, they also have the drawbacks of broad pulse width, low peak power, and expensive as a modulator is required to be inserted in the cavity. Moreover, as actively mode locked pulses have only weak nonlinearity, they are impossible to be shaped into optical solitons that possess the born preponderance: good stability, low time





jittering, short pulse width and high peak power. Fortunately, passive mode locking might overcome those shortcomings.

### 1.2.4    Passive laser mode locking

The distinction between the actively and passively mode locking technique is that active mode-lockers are based on externally modulated media or device while passive mode-lockers are using an optical effect in a material without any time varying intervention. In this dissertation, we mainly discuss the passive elements because they are more fundamental. Passive mode lockers could be divided into two categories: artificial saturable absorber based on nonlinear light interference and real saturable absorber based on material's nonlinear optical absorption property. Basically, nonlinear polarization rotation (NPR), semiconductor saturable absorber mirror (SESAM), carbon nano-tube and graphene based saturable absorbers are the mainly recognized passive mode lockers. Their operation principles can be generalized: making use of a nonlinear device whose response to the entering optical pulse is intensity dependant so that the optical pulse is shortened every time passing through it. Comparing with actively mode-locked fiber lasers, passively mode-locked fiber lasers can produce much shorter and more intense optical pulses as well as keep a relatively low component count simultaneously. As a result, soliton operation can be easily obtained after mode locking in a passively mode-locked fiber laser.

### Type 1.    Nonlinear polarization rotation mode-locked fiber laser

The NPR mode locking technique exploits the nonlinear birefringence of the single





mode optical fibers for the generation of an artificial saturable absorber effect in the laser cavity. The same effect of optical fibers was used previously for the polarization switching and optical pulse shaping [52, 53]. Hoffer *et al*. were the first who used the effect as a self-sustaining mechanism for passive mode locking of fiber lasers [54]. However, owing to the short length and relatively high birefringence of the fiber used, mode locking of their fiber laser could not self-start. The first successful demonstration of the effect for a self-started mode locking was shown by Matsas *et al* [55]. The technique was then widely used for the self-started mode locking of the passively mode-locked ultra-short pulse fiber lasers. We use a configuration to illustrate the mode locking principle of the technique, as shown in **Figure 1.2a**. A piece of linearly birefringent optical fiber is placed between two linear polarizers. Light of arbitrary polarization incident to the setup is transferred into a linearly polarized light by the polarizer before the fiber. When the light propagates in the fiber, it splits into two components along the two polarization axes of the fiber, respectively. After passing through the fiber, generally the polarization of the light becomes elliptically polarized. If the light intensity is weak, then the ellipticity and azimuth of the light polarization are fully determined by the linear birefringence of the fiber and the orientation of the polarizer. However, if the intensity of the light is strong, the nonlinear effects of the fiber must be considered. Due to the nonlinear optical Kerr effect of the fiber, the polarization of the light after passing through the fiber will depend on the light intensity. Furthermore, the transmission of the light through the analyzer will also depend on the light intensity. Through appropriately selecting the orientations of the polarizer and the analyzer, a





situation could be achieved with the setup that the stronger the light, the larger is the light intensity transmission through the analyzer. Physically, such a result is equivalent to that generated by a saturable absorber. Saturable absorber mode locking is a well-known technique that has been extensively investigated and widely used [56]. Therefore, through incorporating the setup in a fiber laser, an artificial saturable absorber effect is automatically generated, which results in mode locking of the laser. As the artificial saturable absorption is generated based on the optical Kerr effect, which has a recovery time in the order of several femto-seconds, from the laser physics point of view, ultra-short mode-locked pulses can be generated by the technique. **Figure 1.2** shows the cavity configuration of a typical soliton fiber laser mode locked by the NPR technique. To achieve the self-started mode locking, a ring cavity configuration is normally used. A polarization independent isolator is inserted in the cavity to force the unidirectional operation of the ring. A polarizer is put in the cavity to set the polarization of light at the cavity position. As the cavity is a ring, the polarizer plays both the roles of the polarizer and the analyzer shown in **Figure 1.2**. To provide gain for the laser operation, a segment of erbium-doped fiber is incorporated in the cavity, which can be directly pumped either by a 980 nm or a 1480 nm laser diode. The pump light is coupled into the cavity by a wavelength-division-multiplexing (WDM) coupler, and the laser output at the 1550 nm is coupled out of the cavity by a fiber coupler. With above basic cavity components, if the cavity length is appropriately selected, mode locking of the laser could always self start. However, in the practice one or two polarization controllers are inserted in the cavity to fine-tune the linear phase delay





between the two polarization components. For the purpose of accurate control on the linear cavity phase delay, frequently a set of two quarter-wave-plates is used. An advantage of using the set of quarter-wave plates is that changing the relative orientations between the quarter-wave plates could not only generate a linear phase delay with a value between 0 and $2\pi$, but also introduces no influence on the other parts of the cavity.

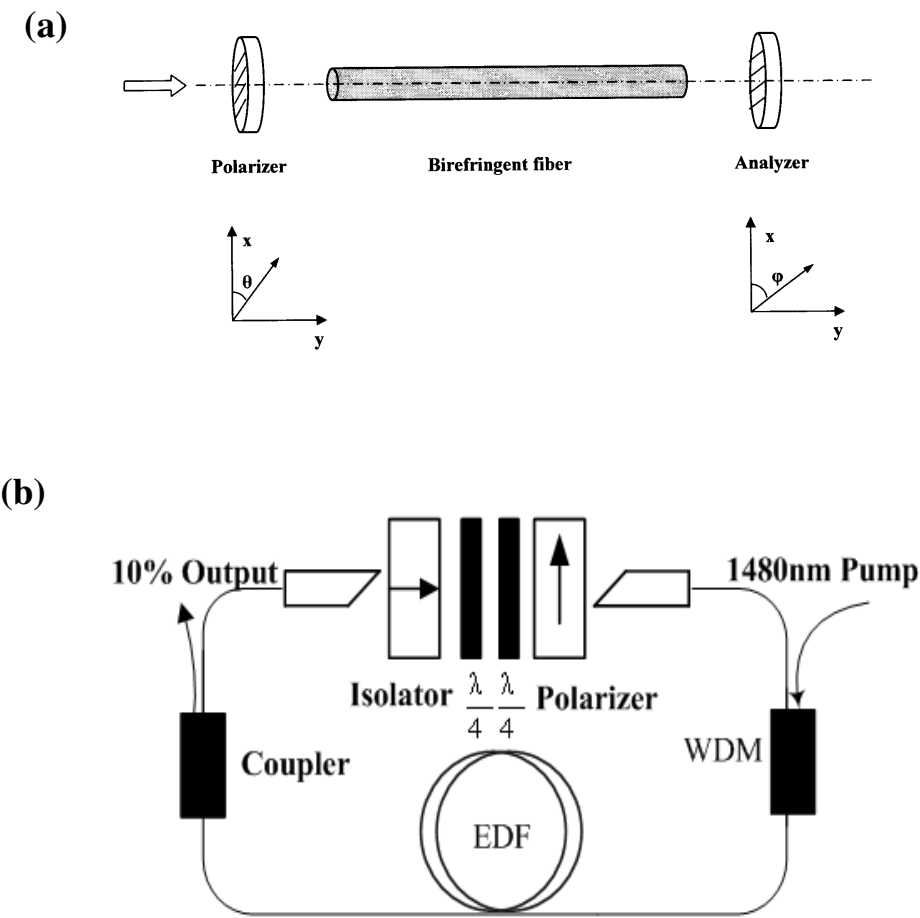

**Figure 1.2**: (a) schematic of NPR mode locking principle; (b) configuration of a typical NPR mode locking fiber laser.

Experimentally, erbium-doped fiber lasers of similar cavity configuration but with





different cavity lengths, fiber birefringence and fiber dispersion properties have been setup and investigated [57, 58]. As far as the linear cavity phase delay is appropriately selected, and the pump power is set beyond the mode-locking threshold, self-started mode locking has always been obtained. Depending on the concrete laser parameters and operation condition, various soliton features have been experimentally observed. However, free-space elements must be incorporated into the laser cavity, which additionally introduces a considerable amount of loss, breaks the all-fiber integrated format and also makes the lasers environmentally unstable. In order to sidestep this drawback and fullfill the all-fiber requirement, a real passive mode locker should be used.

**Type 2.   Semiconductor saturable absorber mode-locked fiber laser**

Semiconductor saturable absorbers are also widely used for passively mode locking lasers. The mode locking technique bases on the mutual interactions of light with the laser gain medium and the saturable absorber. Therefore, properties of the saturable absorber such as the recovery time, saturable absorption strength play an important role on the mode locking quality. For mode-locked fiber lasers using this technique, the most suitable saturable absorbers are the specially designed semiconductor structures. One typical structure of SESAM was grown on n-GaAs (1 0 0) substrates by means of solid-source molecular beam epitaxy (MBE), as schematically shown in **Figure 1.3a**. The nominal structure has 25 pairs of GaAs (77 nm)/AlAs (91 nm) distributed Bragg reflector grown on the GaAs substrate, followed by GaAs (90 nm) and $GaAs_{0.43}P_{0.57}$ (10 nm) spacer layers, and five 6 nm $Ga_{0.69}In_{0.31}As$ QWs. In between the QWs were four 22 nm $GaAs_{0.43}P_{0.57}$ barrier





layers. On top of the last QW was a cap formed by 10 nm GaAs$_{0.43}$P$_{0.57}$ and 90 nm GaAs layers. The growth temperatures were 600$^0$C for GaAs/AlAs DBRs, 510$^0$C for GaInAs/GaAs(P) QWs/barriers, and 590$^0$C for GaAs buffers and caps.

Saturable absorption is a property of materials where the absorption of light decreases with increasing light intensity. Essential parameters of the SESAM are the recovery time, the modulation depth, the bandwidth, the saturation intensity and the non-saturable losses. Usually, the Bragg stack layer can be chosen to be either anti-resonant or resonant. The only difference is: SESAMs based on resonant Bragg stacks can have quite large modulation depths, but with the limited bandwidth of the resonant structure while anti-resonant SESAMs can have quite large bandwidths (e.g. 100 nm) but at the expense of a smaller modulation depth. A larger modulation depth can be obtained from an anti-resonant design at the expense of higher non-saturable losses. In solid state lasers where the single pass gain is low, the non-saturable losses of the SESAM are very crucial and must be as low as possible, but in fiber lasers where the single pass gain is much higher, non-saturable losses are less significant.





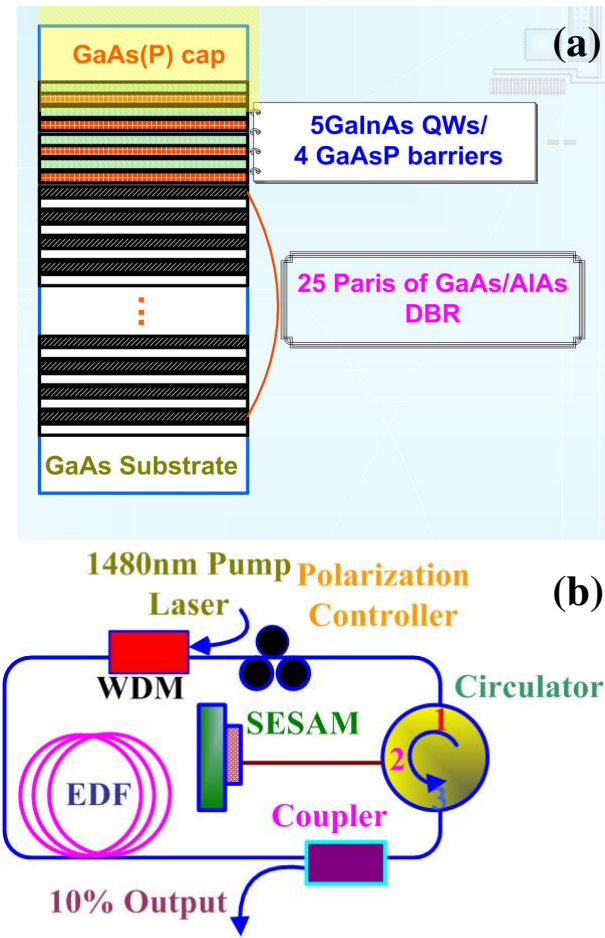

**Figure 1.3**: (a) Schematic structure of the SESAM; (b) Schematic configuration of a fiber laser mode locked using a SESAM.

Ideally, smaller recovery time is always desirable provided that mode locking operation is still ensured. If the pulse is chirped, pulse will develop asymmetric spectra at the impact of SESAMs when their recovery times are the same orders of magnitude as the pulse duration, indicating that recovery times of SESAM could strongly affect the pulse dynamics inside the cavity. Even larger recovery times can limit the obtainable pulse duration. Since the relaxation time due to the spontaneous photon emission in a semiconductor is at nanosecond scale, some





precautions must be taken to shorten the relaxation time drastically. There are two technologies adopted to introduce lattice defects in the absorber layer for fast non-irradiative relaxation of the carriers: low-temperature molecular beam epitaxial growth and ion implantation, which could reduce the relaxation time from nano-second scale down to a few picoseconds.

SESAMs are well known to show a bi-temporal recovery time with the shortest time in the picosecond or sub-picosecond range. This bi-temporal recovery time is suitable for mode-locked lasers in that the short recovery time enables ultrafast pulses generation while the longer recovery time initiates mode-locking. For fast saturable absorbers with recovery times much faster than the pulse duration, the reflection can be described by:

$$q(t) = \frac{q_0}{1 + \frac{\left|A(t)\right|^2}{P_{SA}}}$$

where $q_0$ is the non-saturated but saturable loss, $P_{SA} = \frac{E_{SA}}{\tau_{SA}}$ the saturation power, $E_{SA}$ the saturation energy and $\tau_{SA}$ the recovery time. To more accurately describe the saturable absorption, the following rate equation was usually used:

$$\frac{\partial}{\partial t} q(t) = -\frac{q - q_0}{\tau_{SA}} - q \frac{\left|A(t)\right|^2}{E_{SA}}$$

This differential equation can be numerically integrated to give $q(t)$, and from $q(t)$ the reflection from the SESAM can be determined as:





$$R(t) = 1 - q(t) - l_0$$

Here, $l_0$ is the linear non-saturable loss. The saturation energy can be calculated as the product of the saturation fluency and the effective area on the SESAM.

**Figure 1.3**b shows a typical laser configuration of an erbium-doped fiber soliton laser mode locked by using a SESAM. SESAM mode locked lasers have been extensively studied by a lot of researchers [59-67]. As the relaxation times of semiconductor saturable absorbers are generally quite long, limited by the saturable absorber recovery time, the conventional mode-locked pulses have broad pulse width. However, with the soliton shaping, pulse width is no longer determined by the recovery time of the absorber, but by the soliton effect in the laser. In this case the pulse width could be significantly narrower than the absorber recovery time. After a soliton is formed in the cavity, the function of the saturable absorber turns to stabilize the soliton through suppressing the background noise. Like the soliton operation of other fiber lasers, here again the saturable absorber initiates the mode locking in the lasers.

Apart from mode locking, semiconductor saturable absorbers were also used for the timing stabilization of harmonically mode-locked fiber lasers [68]. It was suggested that the phase effects in semiconductor saturable absorber could lead to pulse repulsion, which provides self-organization of the pulse repetition rate. However, for the applications, the recovery time and saturation energy density of the saturable absorbers must be carefully designed.

Although appreciable success has been achieved with SESAMs, yet their





fabrication involves MBE growth and treatment such as low temperature growth or post-growth ion-implantation is needed to reduce the relaxation time. Problems such as lattice-mismatch and poor thermal properties have seriously limited the quality of SESAMs. The operation bandwidth of SESAMs was very narrow, restricting the potential applications of SESAMs as wideband saturable absorbers operating at other wavelengths. Consequently, there is always strong demand of seeking new materials to replace SESAMs, achieving the goal of perfect saturable absorbers with wider operation range, faster response time and lower cost.

**Type 3.   Carbon material based saturable absorber mode-locked fiber laser**

Nowadays, the ability to extensively control/tune photonic property of material through different physical, chemical and nano-technological approaches is now at the center of modern photonics. Researchers gradually diverted their research focuses from the previous III-V semiconductors to IV materials in that silicon is silicon is the second most abundant element after oxygen in Earth's crust while carbon is the *materia prima* for life and the basis of all organic chemistry. Due to the flexibility of its bonding, carbon-based materials show an unlimited amount of different structures with an equally large variety of optical properties. These optical properties are, in great part, the result of the dimensionality of these structures. Among systems with only carbon atoms, graphene—a two-dimensional allotrope of carbon—plays a central role since it is the basis for the understanding of the electronic/optical properties in other allotropes. Graphene is made out of carbon atoms arranged on a honeycomb structure made out of hexagons, also well known as chicken-wire array, as shown in **Figure 1.4**, and can be thought of as composed





of benzene rings stripped out from their hydrogen atoms. From the physical point of view, fullerenes are molecules where graphene layer are arranged spherically and are zero-dimensional with discrete energy levels; carbon nano-tubes are produced by rolling graphene along a given direction and reconnecting the carbon bonds and hence carbon nano-tubes have only hexagons and can be thought of as one- dimensional object; graphite is generated by stacking graphene layers that are weakly coupled by van der Waals forces and can be recognized as three-dimensional allotrope of carbon. Although graphene is the parent for all these different allotropes, it was only recently isolated by A. K. Geim and co-workers at the University of Manchester through the simple "scotch tape technique" in 2004 [70,71].





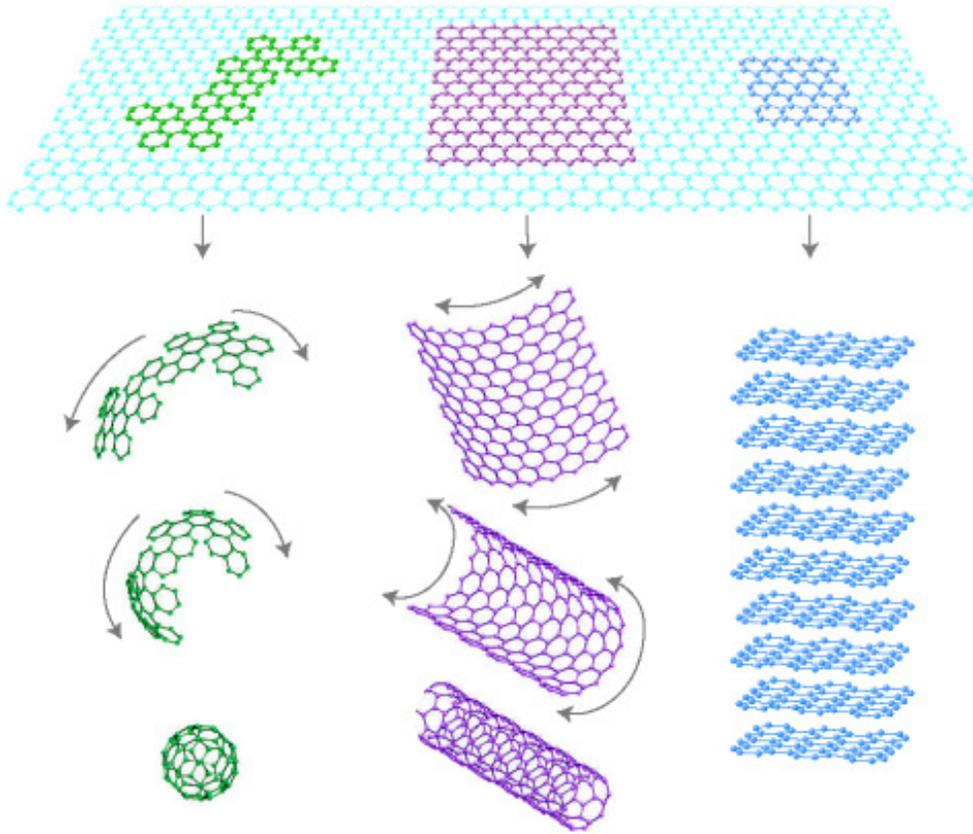

**Figure 1.4**: graphene, the parent of all graphitic forms [70, 71].

Since fullerene, graphene, carbon nano-tubes and graphite have different dimensionalities from 0-D to 3-D, they show distinctive electronic band structures, respectively, leading to unique optical property. Briefly, fullerene was experimentally verified to have optical limiting property (or reversed saturable absorption) while graphite completely absorbs any light with trivial features, indicating that both fullerene and graphite might have little potential applications as saturable absorbers to mode lock lasers. A single wall carbon nanotube (SWCNT) is a hexagonal network of carbon atoms rolled into a cylinder with each end capped with half of a fullerene molecule [72, 73]. The diameter of SWCNTs is normally





close to one nanometre while their lengths can be up to orders of centimetres. The chirality of SWNTs is defined by a single vector called chiral vector. $C_h = na_1 + ma_2$, where, $a_1$ and $a_2$ are unit vectors of the hexagonal lattice, as shown in **Figure 1.5**. This vector connects two crystallographically equivalent sites on a two-dimensional graphene sheet and different chiral vectors are differentiated using two integers $(n, m)$. Depending on the value of the chiral vector, SWCNTs are classified as armchair, zigzag and chiral. Armchair type of SWCNTs corresponds to the case where $n = m$ while zigzag type satisfy either $n = 0$ or $m = 0$. All other cases correspond to chiral type. A SWCNT may be semiconducting or metallic depending on its chirality. Metallic nano–tube arises when $n–m$ is an integer multiple of three while for all other arrangements of $(n, m)$ the corresponding nano-tubes exhibit semiconductor properties. Because current synthesis methods for SWCNTs cannot yield accurately controlled and desired diameter or chirality, it is a usually mixer of both semiconductor and metallic tubes. Statistically, 1/3 of them are metallic nano-tubes and the others are semiconducting nano-tubes.





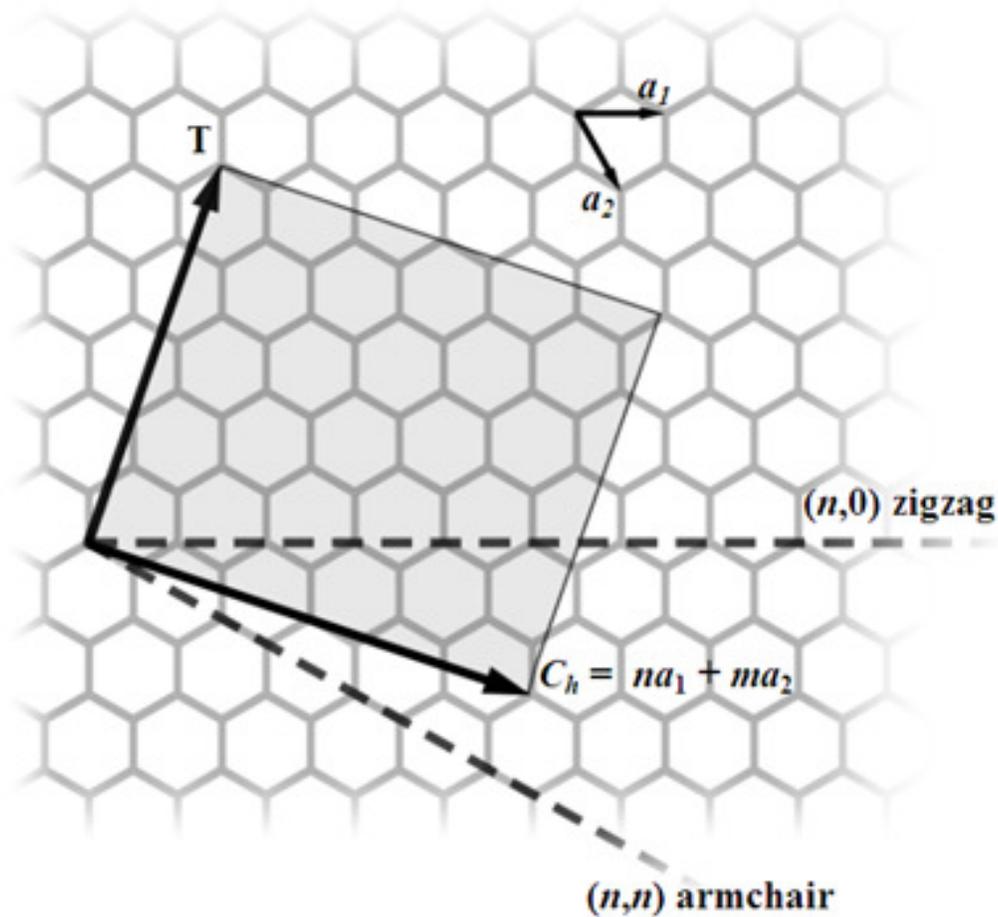

**Figure 1.5**: illustration of chiral vector. The ($n$, $m$) nanotube naming scheme can be thought of as a vector ($C_h$) in an infinite graphene sheet that describes how to "roll up" the graphene sheet to make the nanotube, $T$ denotes the tube axis, and $a_1$ and $a_2$ are the unit vectors of graphene in real space [72].

Semiconducting SWCNTs are direct band-gap materials having series of van-Hove singularities in the density of states (**Figure 1.6**) [72]. The band gap of SWCNTs depends on the tube diameter [72], by growing of SWCNTs with a proper diameter distribution, it is possible to set the absorption peak positions in a broad spectral range between visible and near infrared [74]. **Figure 1.6** illustrated that SWCNTs





have different absorption peaks. It should be noted that, this spectral tuning capability for SWNTs growing is very useful for the operation of saturable absorber devices.

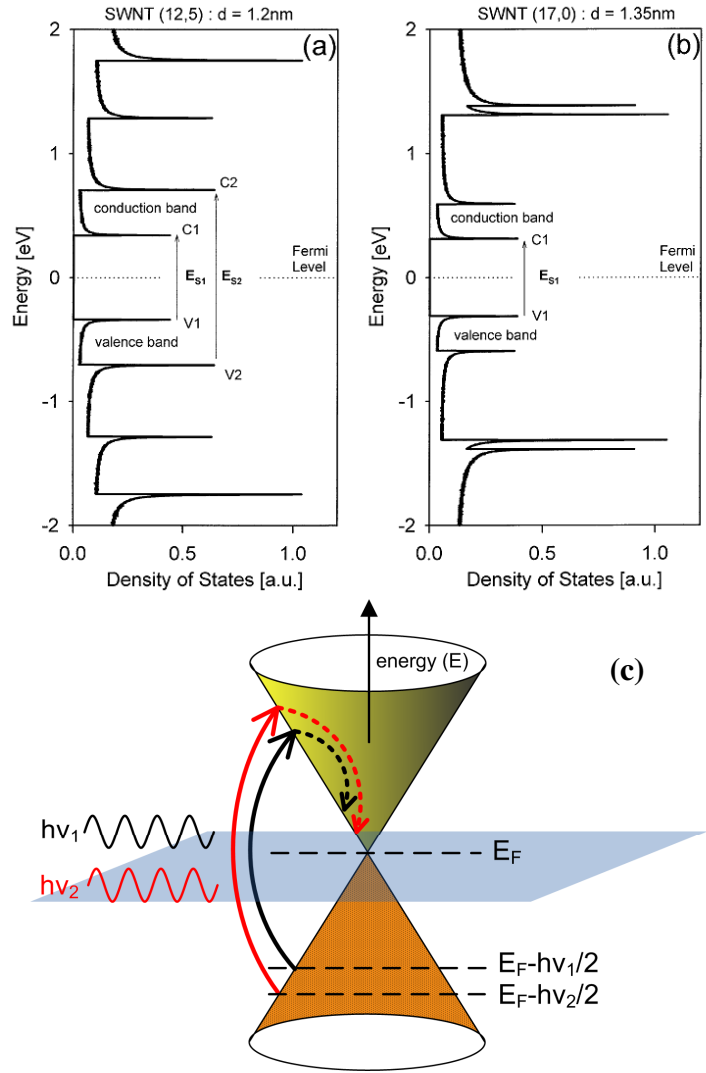

**Figure 1.6**: (a) & (b) schematic illustration of energy band structure of single wall carbon nano-tube with different diameters and chiralities [74]; (c) graphene's energy band structureand photon absorption.

By setting the absorption peak of SWNTs-based saturable absorbers coincident at





the lasers' operating wavelengths, passive mode-locking operation has been achieved for a wide range of wavelengths between 1 μm and 1.55 μm for both solid state lasers and fiber lasers [75-95]. It has been shown that SWCNT mode lockers have the advantages such as intrinsically ultrafast recovery time, large saturable absorption, easy to fabricate, and low cost. In particular, as SWCNTs are direct-bandgap materials with a gap that depends on the nano-tubes' diameter and chirality, through mixing SWCNTs with different diameters, a broadband saturable absorption mode locker could be made. A wideband wavelength tunable erbium-doped fiber laser mode locked with SWCNTs was experimentally demonstrated [96]. However, the broadband SWCNT mode locker suffers intrinsic drawbacks: SWCNTs with a certain diameter only contribute to the saturable absorption of a particular wavelength of light, and SWCNTs tend to form bundles that finish up as scattering sites. Therefore, coexistence of SWCNTs with different diameters introduces extra linear losses to the mode locker, making mode locking of a laser difficult to achieve.

Quite recently, we experimentally found that these drawbacks could be circumvented if graphene is used as a broadband saturable absorber. Unlike the conventional semiconductor saturable absorbers, the energy band diagram of graphene has zero band-gap and a linear dispersion relation. These unique energy band properties combined with the Pauli blocking principle renders graphene a full band ultrafast saturable absorber [97-103]. Because of the Pauli Exclusion Principle, when pumping of electrons in the excited state is quicker than the rate at which they relax, the absorption saturates. **Figure 1.6**c shows a schematic





illustration of the energy band structure and photon absorption of graphene. Li *et al.* had experimentally shown that atomic layer graphene could absorb a considerable amount of infrared light without any bandwidth limitation [104]. Due to interactions between electrons and lattice vibration modes of the carbon atoms, or mutual interactions among electrons, graphene could absorb a considerable amount of infrared light without any bandwidth limitation. Through the absorption of photon with energy $hv$ and energy transference from photon towards electron, electron at the valence band with energy $hv/2$ lower than Fermi energy can be excited to its corresponding conduction band with energy $hv/2$ higher than Fermi energy. But such absorption is photon number dependent. When the photons number is low, photons could be continuously depleted through the excitation of the electrons from valence band to conduction band. When the photons number is large enough, owing to the Pauli blocking principle, the newly generated carriers fill the valence bands, preventing further excitation of electrons at valence band and thus allowing photons transmitted without absorption, interpreting graphene's wide-band saturable absorption. By the virtue of graphene's broadband nonlinear optical property, graphene mode locked fiber lasers at 1 μm or 1.55 μm with widely tunable central wavelength were successfully achieved.

## 1.3    Motivation and objectives

Due to their intrinsic stability in propagation, optical solitons have been proposed as information carriers for the long-distance optic fiber communications. Self-starting passively mode-locked erbium-doped fiber lasers, as attractive sources for ultrashort soliton pulses for their simplicity, tunability and ultrashort pulse





operation, have been intensively studied. Soliton operation is a natural consequence of mode-locked fiber lasers in the anomalous cavity dispersion regime under strong pumping. Due to the balance effect between the GVD and SPM, soliton operation is ready to be obtained after the mode locking of the fiber laser. However, as in a fiber laser, besides the optical fibers, there also exist other components e.g. the gain medium and the output coupler, whose existence further affects the dynamics of the formed solitons. Therefore, compared with the solitons formed in single-mode fibers, the solitons obtained in a fiber laser are of new characteristics, which required further investigations. However, a real long-distance soliton fiber optic communication system also involves in the soliton losses and periodic amplifications. Therefore, the dynamics of solitons in the system is no longer described by the NLSE, which is integral and only describes the conservative systems, but by the GLE, where soliton is formed not only as a result of the balanced interaction between the fiber Kerr nonlinearity and dispersion, but also as a result of the balance between fiber losses and gain generated by the fiber amplifiers. As a soliton propagating in the system periodically experiences loss and amplification, dispersive waves with discrete spectra are generated, which resonantly draw energy from the soliton. A fiber laser is also periodical gain-and-loss system, in which, an optical pulse periodically experience the amplification of the fiber amplifier and the output loss. In this sense, a passively mode-locked soliton fiber ring laser can be regarded as a miniature of a soliton optic fiber communication system. Study on the soliton propagation and interaction in the laser cavity therefore gives a direct insight into the soliton interactions in the long-





haul optic communication systems. So far, soliton operation in passively mode-locked fiber lasers has been extensively investigated [16-42].

Generally, the soliton formation in optical fibers is governed by the GLE, and the soliton pulses have a single-peak bell-shape intensity profile. Although the fiber birefringence always existed, due to the polarization limitation from NPR mode locking technique, the formed solitons had fixed polarization state once they passed the polarizer. Consequently, the polarization states of solitons were unable to evolve under the actions of fiber birefringence and the formed solitons could only be termed as scalar solitons. Extensive theoretical and experimental investigations on scalar solitons in NPR mode-locked fiber lasers have been carried out including soliton bunching, stable randomly spaced soliton distribution, bound scalar solitons, long range or direct interaction of scalar solitons, gain-guided or dissipative scalar solitons, modulation instability of scalar solitons, period-doubling scalar solitons and multi-pulse formation mechanism of scalar solitons [16-39]. Although the basic principles of the operation of a scalar soliton fiber laser are generally understood, many features of the vector soliton laser still remain uninvestigated.

To establish the vector soliton emission in a fiber laser, any polarization dependent element must be removed so that the polarization states of the formed soliton could freely evolve. In contrast with scalar soliton, due to additional polarization freedom, vector soliton manifested richer and more fascinating features than scalar soliton. How do the two-orthogonal polarizations of the vector soliton interact together and bound together? How does the cavity birefringence influence the vector soliton dynamics? What are the cavity parameters for the generation of polarization





rotating or locked vector soliton? In normal dispersion cavity, could the dissipative nature affect the polarization state of dissipative vector soliton? How about the interaction forces among the conservative/dissipative vector soliton? Could the two-orthogonal polarizations of the vector soliton show completely different soliton features? For example, one polarization was fundamental soliton while the other polarization was high-order (bound) soliton.

In the field of optical soliton, although many soliton theoretists had predicted the existence of dark soliton family including vector dark-dark soliton, vector bright-dark soliton and domain wall soliton, those patterns widely existing in other physics systems, no direct experimental evidences were provided. So, it should be fundamentally interesting to study the dynamics of those novel dark solitons. All these questions have so far not been clearly addressed. To clarify them is not only important for understanding the vector soliton in fiber lasers, but also potentially useful for the future application of those novel vector solitons in the ultra-high-bit-rate optical communication systems.

Another interesting topic is whether it is achievable to extend the family of saturable absorbers for passively mode locking. Although SESAMs were widely used, there are many drawbacks inherited by SESAMs. Specifically, SESAMs require complex and costly clean-room-based fabrication systems , an additional substrate removal process is needed in some cases; high-energy heavy-ion implantation required to introduce defect sites in order to reduce the device recovery time (typically a few nanoseconds) to the picosecond regime, the reflection properties of SESAM limted the structure of fiber lasers, and low optical





damage threshold. Quite recently, by the virtue of the carbon material based saturable absorption property, researchers have successfully produced a new type of effective saturable absorber, leading to the demonstration of mode-locked pico- or subpicosecond erbium-doped fiber (EDF) lasers. SWCNT films onto flat glass substrates, mirror substrates, or end facets of optical fibers were fabricated as the new saturable absorber devices [75-95]. But the non-uniform chiral properties of SWNTs present inherent problems for precise control of the properties of the saturable absorber. Furthermore, the emergence of bundled and entangled SWNTs, catalyst particles, and the formation of bubbles induced high nonsaturable losses in the cavity. Under large energy ultrashort pulses multi-photon effect induced oxidation occurs, which degrades the long term stability of the saturable absorbers. In order to overcome the above disadvantages, graphene, a single two-dimensional atomic layer of carbon atom arranged in a hexagonal lattice, was successfully veried as another effective saturable absorbers.

The main objective of this research is to investigate the properties of vector solitons in fiber lasers. In experiment, an erbium-doped fiber ring laser will be intentionally designed for experimental investigation. In theory, the pulse propagation and vector soliton formation in a fiber laser will be investigated. Numerical simulations on the laser system will also be carried out to get a better understanding on the operation of the laser. In particular, various types of vector solitons including: high-order vector bright-bright soliton, vector dark-dark soliton, vector bright-dark soliton and domain wall soliton were firstly experimentally discovered in our fiber lasers. Moreover, to extend the family of saturable





absorbers, we successfully fabricated graphene based saturable absorber which has the widest operation range, fastest response time and polarization insensitive optical properties, paving the way to studying the vector soliton in graphene mode-locked fiber lasers.

## 1.4 Main contributions of the research

First of all, fiber lasers passively mode-locked by SESAMs have been built up; various operation features of the laser have been examined. Different operation regimes of the lasers, from purely anomalous to normal dispersion, have been obtained and investigated. Vector soliton operations, as an intrinsic feature of the fiber laser have been achieved and comprehensively studied.

In an all-anomalous dispersion cavity with weak birefringence, one novel type of spectral sideband generation on the soliton spectra of the phase locked vector solitons in a passively mode-locked fiber ring laser was experimentally and numerically identified. The polarization resolved study on the soliton spectrum demonstrated that coherence energy exchange between the two orthogonal polarization components of the vector solitons accounted for the generation of such new spectral sidebands. When the cavity birefringence was strong enough, we experimentally and numerically found that the induced temporal solitons were formed by the XPM between the two orthogonal polarization components of the birefringence laser, and the induced solitons could either have the same or different soliton frequency to the inducing soliton. As the induced solitons always have the same group velocity as that of the inducing soliton, they form vector solitons in the laser. To our knowledge, this is the first experimental observation of temporal





induced solitons. Moreover, we also study the effect of XPM on the vector solitons including the induced vector solitons, trapping of vector solitons both in purely negative dispersion cavity and dispersion managed cavity with net positive dispersion. Under relatively higher pumping and sufficiently weak cavity birefringence, a new type of high order phase locked vector soliton in a passively mode-locked fiber laser was experimentally observed and numerically verified. The high order vector soliton is characterized by that its two orthogonal polarization components are phase locked, while the stronger polarization component is a single hump pulse, the weaker component has a double-humped structure with 180° phase difference between the humps. Our experimental result firstly confirmed the theoretical prediction on the high order phase locked vector solitons in birefringent dispersive media. In normal dispersion cavity, dissipative vector solitons (DVSs) have been experimentally demonstrated in a dispersion-managed fiber laser passively mode locked by a SESAM. It was found that despite of their large frequency chirp of the gain-guided solitons, polarization rotating and polarization locked DVSs could still be formed. In addition, formation of multiple DVSs with identical soliton parameters and stable harmonic DVSs mode-locking are also experimentally obtained. Except the stable existence of vector solitons, group interactions of vector solitons have been both experimentally and numerically observed; our experimental observation shows that two groups of vector soliton traveling at different group velocities because their polarization states are orthogonal to each other. Correspondingly, they collide with each other endlessly. Polarization rotating and locking of DVSs have been both





experimentally and numerically obtained in either a dispersion-managed or purely normal dispersion fiber laser cavity. The period of DVSs polarization rotation could be still locked to integer multiple of the cavity roundtrip. We have also experimentally shown than despite of the existence of the laser gain competition, the angle of polarization ellipse orientation between two sets of DVSs can be varied from $0^0$ to $90^0$ because larger chirp and the broader pulse separation could partially counteract the influence of laser gain competition. Numerical simulations have confirmed such polarization rotation. If the cavity birefringence was further increased and therefore an artificial birefringence filter was deliberately introduced into the cavity, multi-wavelength dissipative solitons in an all normal dispersion fiber laser passively mode-locked with SESAM were firstly experimentally observed. Depending on the strength of the cavity birefringence, stable single-, dual- and triple-wavelength DVSs can be formed in the laser. The multi-wavelength soliton operation of the laser was experimentally investigated, and the formation mechanisms of the multi-wavelength DVSs are discussed.

Secondly, in the non-mode-locking regime where SESAM is replaced with a polarization dependent isolator, scalar dark soliton emission in a fiber laser was firstly observed. We anticipated that within a narrow operation regime, the mode locking behavior could be suppressed and the fiber laser could be operated in the non-mode-locking regime. Correspondingly, dark pulse rather than bright pulse was established. Through nonlinearity accumulation and continuous pulse shaping, eventually, dark solitons could be formed. Moreover, if a weak saturable absorber instead of the polarization dependent isolator was used, we could observe the





vector dark-dark soliton and even the trapping of vector dark-dark soliton under moderate cavity birefringence. Numerical simulions confirmed the above experimental observations, and further proved that the observed dark soliton could be a genetic feature of NLSE. In both the normal and anomalous dispersion cavity, optical domain walls, characterized by topological structures separating different components (different polarization or wavelength), were experimentally observed in a fiber laser. It is noticed that the optical domain walls are irrelative to the cavity dispersion but dependent on the cavity birefringence. Particularly, similar to the bright-bright vector soliton, when the cavity is weakly birefringent, the two polarization components of the optical domain wall solitons are coherently coupled and phased locked. Moreover, they could be interpreted as dark-bright vector soliton, which is a special case of domain wall solitons. When the cavity is largely birefringent, the two components of the optical domain wall solitons are incoherently coupled and the wall widths are strongly related to the cavity birefringence. The numerical simulation could well reproduce the above experimental observations.

Lastly, we proposed and achieved another novel saturable absorber made of graphene. Although it is a zero bandgap semiconductor, like the SWCNTs, graphene also holds the feature of saturable absorption due to Pauli blocking effect. As a single 2D atomic layer of carbon atom arranged in a hexagonal lattice, graphene based saturable absorbers exibit superior advantages including: (i) Controllable saturable absorption strength through controlling the number of graphene layers or chemical functionalization; (ii) Super broadband saturable





absorption; (iii) Ultrafast saturation recovery time; (iv) Easy to be fabricated. Correspondingly, graphene was fabricated as a wideband saturable absorber for ultrafast mode-locking fiber lasers. The polarization indepdence of graphene's saturable absorption property also guaranteed graphene as a polarization insenstive saturable absorber to generate vector soliton, like SESAMs. The ultrafast recovery time of graphene also facilitates ultrashort pulse generation. The optical modulation depth can be tuned in a wide range by using single to multilayer graphene or doping/intercalating with other materials. We believe those features of the graphene can also find other interesting photonic applications.

## 1.5    Overview of the dissertation

The dissertation is organized as follows:

Chapter 1 serves as an introduction of the research.

Chapter 2 presents the experimental and theoretical background of the ultra-short in optical fibers. Major fiber characteristics that affect the pulse evolution in an optical fiber including the GVD, fiber nonlinearity, and fiber birefringence are discussed. As an important factor, the gain and gain dispersion in erbium-doped fibers are also introduced. Equations that describe the pulse evolution in a real gain and loss fiber system, e.g. a fiber laser, are derived out step by step in Chapter 2.

Chapter 3 demonstrates the experimental and numerical investigations on the mutual interaction forces among the two orthogonal polarizations of a vector soliton. Specifically, how the Four-wave mixing (also called as coherent energy exchange) and cross polarization coupling influence the dynamics of vector soliton.





We found that a new kind of spectral sideband could be generated due to the coherent energy exchange between these two polarizations of vector solitons and a weak soliton in one polarization of a vector soliton could be induced by its perpendicular strong component owing to the cross polarization coupling.

Chapter 4 introduces one novel vector soliton: high order vector soliton. The high order vector soliton is characterized by that its two orthogonal polarization components are phase locked, and while the stronger polarization component is a single hump pulse, the weaker component has a double-humped structure with 180° phase difference between the humps. Our experimental result firstly confirmed the theoretical predictions on the high order phase locked vector solitons in birefringent dispersive media.

Chapter 5 addresses the generation of dissipative vector soliton in all-normal-dispersion cavity or a dispersion managed cavity with net normal dispersion. Different from the conventional vector solitons which only survives in the anomalous dispersion regime, the prerequisite of dissipative vector soliton generation is a natural balance among the normal cavity dispersion, cavity fiber nonlinear Kerr effect, laser gain saturation and gain bandwidth filtering. Both polarization-locked and rotating dissipative vector solitons have been obtained. In the end, multi-wavelength dissipative soliton is also generated due to the existence of an artificial birefringent filter when cavity birefringence is very strong.

Chapter 6 provides the observation of dark soliton in a fiber laser. Both scalar and vector dark soliton have been experimentally and numerically obtained. Dark





solitons in all normal dispersion and dispersion managed cavity possess different soliton dynamics, which will be experimentally and numerically studied. Trapping of vector dark solitons will also be studied under very strong cavity birefringence.

Chapter 7 discusses the experimental and numerical observation of two types of domain wall solitons: polarization domain wall and dual wavelength domain wall. When cavity birefringence was small enough, dark-bright and dark-dark type polarization domain wall soliton could be formed; while in large cavity birefringence, incoherent interaction of two wavelengths with the same polarizations could lead to the generation of dual wavelength domain wall.

Chapter 8 demonstrates a novel type of saturable absorber: graphene and its application for ultra-fast mode locked lasers. Graphene as an ideal two-dimension nano-material has unique optical and photonic properties. We show that graphene possesses wavelength independent ultrafast saturable absorption due to Pauli blocking, which can be exploited as a super broadband saturable absorber for passive mode locking of lasers. Various graphene-based saturable absorbers for ultrafast lasers have summarized in Chapter 8.

Chapter 9 concludes the dissertation, summarizes the achievements of the research and gives recommendations for future work.





# Chapter 2.  Theory of vector soliton propagation in a fiber laser

The envelope of a light wave in a single-mode optical fiber is deformed by the dispersion (variation of the group velocity as a function of the frequency) and nonlinearity (variation of the phase velocity as a function of the wave intensity) of the optical fibers. The dispersive property of the light wave envelope is decided by the GVD while the nonlinear properties of the light wave envelope are characterized by a combination of Kerr effect (an effect of the increase in the refractive index $n$ in proportion to the light intensity). In a fiber laser, the gain and gain dispersion introduced by the erbium-doped fiber amplifier will also complicate the pulse propagation. However, "single" mode optical fiber actually supports two orthogonal polarization modes while the term "single" mode refers only to the transverse profile. The polarization modes could only be completely degenerate provided that optical fiber was perfectly isotropic. In reality, manufacturing imperfections, externally applied stress or bending easily breaks down the degeneracy between the modes. Thus fiber supports two orthogonally-polarized modes with differing propagation constants; i.e., fiber is birefringent [40]. The difference in phase velocities of the two modes causes the polarization state of a pulse to evolve as it propagates. Fortunately, although being exposed to mediate fiber birefringence, theoretical analysis showed that the orthogonally-polarized components of the soliton could stick together and propagate as one non-dissipative entity through shift their center frequency slightly, forming the state of vector soliton [40, 109].

The fiber birefringence induced polarization evolution is very crucial aspect





that must be taken into account when analyzing the vector soliton propagation in practical fiber systems. For a good understanding of the vector soliton operation in fiber lasers, it is necessary to firstly consider the theory of wave propagation in fibers in the presence of these effects.

The objective of Chapter 2 is to derive the basic equations that govern the vector soliton pulse propagation in a fiber laser cavity. Section 2.1 introduces the group velocity dispersion in fibers and its effect on the pulse propagation. Section 2.2 provides the theory of nonlinear pulse propagation in optical fibers. In Section 2.2, the NLSE is derived and the fundamental soliton solution of this equation is given. Section 2.3 discusses the gain property introduced by the erbium-doped fibers. By including the gain and gain dispersion, the pulse propagation is then described by the Ginzburg-Landau equation. Section 2.4 considered the polarization characters and derivation of the coupled Ginzburg-Landau equations, paradigm equation for vector soliton generation. Finally, various vector solitons including vector conservative/dissipative bright-bright soliton, vector dark-dark soliton, vector dark-bright soliton and domain wall solitons will be summarized in the Section 2.5.

## 2.1   Linear pulse propagation in optical fibers

When an optical pulse propagates within an optical fiber, it becomes distorted due to intramodal (i.e. chromatic) dispersion and intermodal delay effects [40]. The distortion effects can be explained by analyzing the group velocities of the guided modes. Group velocity is the speed at which energy in a particular mode travels along the fiber. For an optical signal that normally contains a finite number of frequency modes, a phenomenon known as the group velocity





dispersion (GVD) occurs. GVD also referred to as the chromatic dispersion, exhibit as pulse spreading that occurs within a single-mode fiber (SMF), which arises due to the finite spectral width of the optical signal. This phenomenon is called GVD since the dispersion is a result of the group velocity being a function of the wavelength.

Since the refractive index of the medium is frequency dependent, which can be written as $n(\omega)$, chromatic dispersion always exists with the propagation of light in fibers. For CW light or pulse with big pulsewidth (usually $T_0 > 100$ ps), this dispersion can be ignored. But in the case of short pulses, especially for the ultrashort pulses as those presented in this dissertation, fiber dispersion plays a critical role because different spectral components associated with the pulse travel at different speeds given by $c/n(\omega)$, where $c$ is the speed of light in vaccum. Because of dispersion, a pulse will broaden with the propagation in fibers, which can be detrimental for optical communication systems.

On a fundamental level, the origin of dispersion is related to the characteristic resonance frequencies at which the medium absorbs the electromagnetic radiation through oscillations of bound electrons. The refractive index can be well approximated by the Sellmeier equation at frequencies far from the material resonances:

$$n^2(\omega) = 1 + \sum_{j=1}^{m} \frac{B_j \omega_j^2}{\omega_j^2 - \omega^2} \quad , \qquad (2.1.1)$$

where $\omega_j$ is the resonance frequency and $B_j$ is the strength of $j^{\text{th}}$ resonance. The sum in Eq. (2.1.1) extends over all material resonances that contribute to the





frequency range of interest. For silica, a three-term Sellmeier equation is typically used, accounting for the resonances in the ultraviolet and infrared.

$$n^2(\omega) = 1 + \frac{B_1 \omega_1^{\;2}}{\omega_1^{\;2} - \omega^2} + \frac{B_2 \omega_2^{\;2}}{\omega_2^{\;2} - \omega^2} + \frac{B_3 \omega_3^{\;2}}{\omega_3^2 - \omega^2} \qquad (2.1.2)$$

Mathematically, the effects of fiber dispersion are accounted for by expending the mode-propagation constant $\beta$ in a Taylor series about the frequency $\omega_0$ at which the pulse spectrum is centered:

$$\beta(\omega) = n(\omega)\frac{\omega}{c} = \beta_0 + \beta_1(\omega - \omega_0) + \frac{1}{2}\beta_2(\omega - \omega_0)^2 + \cdots \qquad (2.1.3)$$

$$\beta_m = \left(\frac{d^m \beta}{d\omega^m}\right)_{\omega = \omega_0} \qquad (m = 0,1,2,\ldots) \;. \qquad (2.1.4)$$

The parameters $\beta_1$ and $\beta_2$ are related to the refractive index $n$ and its derivatives through the relations

$$\beta_1 = \frac{1}{v_g} = \frac{n_g}{c} = \frac{1}{c}\left(n + \omega \frac{dn}{d\omega}\right), \qquad (2.1.5)$$

$$\beta_2 = \frac{1}{c}\left(2\frac{dn}{d\omega} + \omega \frac{d^2 n}{d\omega^2}\right), \qquad (2.1.6)$$

where $n_g$ is the group index and $v_g$ is the group velocity. Physically speaking, the envelope of an optical pulse moves at the group velocity while parameter $\beta_2$ represents dispersion of the group velocity and is responsible for pulse broadening. $\beta_2$ is referred to as the group velocity dispersion (GVD) parameter. **Figure 2.1** shows how $\beta_2$ vary with wavelength $\lambda$ in fused silica [40].





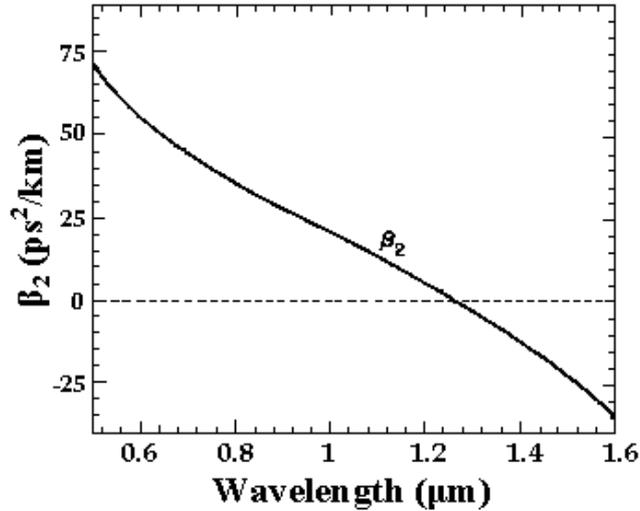

**Figure 2.1**: Group velocity dispersion $\beta_2$ for fused silica as a function of

wavelength (After Ref. [40]).

In the figure, $\beta_2$ vanishes at the wavelength of about 1.27 µm, which is referred

to as the zero-dispersion wavelength and is denoted as $\lambda_D$. For wavelengths

such that $\lambda < \lambda_D$, the fiber is said to exhibit normal dispersion as $\beta_2 > 0$. In the

normal dispersion regime, the higher frequency (blue shifted) components of an

optical pulse travel slower than lower frequency (red-shifted) components. By

contrast, the fiber shows anomalous dispersion as $\beta_2 < 0$ when the light

wavelength exceeds the zero-dispersion wavelength ($\lambda > \lambda_D$). And in this

regime, the blue-shifted components of an optical pulse travel faster than red-

shifted components. The anomalous-dispersion regime is of considerable

interest for the study of solitons. Because it is in this regime that optical fibers

support solitons through a balance between the dispersive and nonlinear effects.

## 2.2 Pulse propagation in dispersive media

Basically, like all electromagnetic phenomena, the propagation of optical fields

in the media is governed by Maxwell's equations. Then deriving from the





Maxwell's equations, one can get the fundamental wave equation

$$\nabla \times \nabla \times \vec{E} = -\frac{1}{c^2}\frac{\partial^2 \vec{E}}{\partial t^2} - \mu_0 \frac{\partial^2 \vec{P}}{\partial t^2}, \tag{2.1.7}$$

where $\vec{E}$ and $\vec{P}$ are electric field vector and the induced electric polarization, respectively. And $c$ is the speed of light in vacuum and has a relation $1/c^2 = \mu_0 \varepsilon_0$, where $\mu_0$ is the vacuum permeability and $\varepsilon_0$ is the vacuum permittivity.

Before solving the wave equation, two simplifications are made. First, by just considering linear polarized light propagation in the media, the vectors in Eq. (2.1.7) can be replaced by scalars. Second, by ignoring the nonlinear effects temporarily in this section, Eq. (2.1.7) is linear and can be easily converted to the frequency domain.

Using the two simplifications and also adopting the operation

$$\nabla \times \nabla \times \vec{E} = \nabla(\nabla \cdot \vec{E}) - \nabla^2 \vec{E}, \tag{2.1.8}$$

One can easily get the wave propagating equation in frequency domain:

$$\nabla^2 E(z, \omega) + n^2(\omega)\frac{\omega^2}{c^2}E(z, \omega) = 0 \tag{2.1.9}$$

where, $E(z, \omega)$ is the Fourier transform of $E(z, t)$ defined as:

$$E(z, \omega) = \int_{-\infty}^{\infty} E(z, t)\exp(i\omega t)dt \tag{2.1.10}$$

In addition, the paraxial approximation is also implied in Eq. (2.1.9), which is applicable and reasonable in laser optics by replace $\vec{r}$ with $z$ along which the light propagates.





Next, another function $F(z,\omega)$ is brought in to further simplify the wave equation. $F(z,\omega)$ is assumed to be a slowly varying function of $z$.

$$F(z,\omega) = E(z,\omega)\exp[-i\beta(\omega)z] \qquad (2.1.11)$$

Therefore, the wave equation (2.1.9) can be further simplified as:

$$\frac{\partial F(z,\omega)}{\partial z} = -i\beta(\omega)F(z,\omega) \qquad (2.1.12)$$

where, $\beta(\omega) = n(\omega)\dfrac{\omega}{c}$ is the propagation constant varying with frequency $\omega$. In obtaining Eq. (2.1.12), the second derivative $\partial^2 F/\partial z^2$ is neglected since $F(z,\omega)$ is assumed to be a slowly varying function of $z$.

Now let us go on to study the propagation equation of an optical pulse. The electric field of an optical pulse can be generally written as:

$$f(z,t) = E(z,t)\exp\{i[\omega_0 t - \beta(\omega_0)z]\} \quad , \qquad (2.1.13)$$

$E$ $(z, t)$ is the pulse envelop, which contains both the amplitude and phase information. $\omega_0$ is the central frequency of the pulse and $\beta(\omega_0)$ is the propagation constant at the frequency of $\omega_0$. Correspondingly, in the frequency domain, an optical pulse can be described by the Fourier transform of $f(z,t)$ and written as:

$$F(z,\omega) = E(z,\omega-\omega_0)\exp[i\beta(\omega_0)z] \qquad (2.1.14)$$

The function of $F(z,\omega)$ obeys the conventional transmission relation of Eq. (2.1.12) If the pulse width is big enough so that the dispersion effect can be ignored, each spectral component of the pulse at $\omega$ can be well approximated





to propagate with the same phase constant $\beta(\omega_0)$. But when the width of an optical pulse $\tau_p$ is so short that the spectral bandwidth $\delta\omega \propto \dfrac{1}{\tau_p}$ is a significant fraction of $\omega_0$, one must account for the fact that each spectral component of $F(z,\omega)$ has its own phase constant $\beta(\omega)$. In this condition, by substituting Eq. (2.1.14) and Eq. (2.1.3) into Eq. (2.1.12), one can finally obtain an equation describing the scalar envelops function $E(z,\omega-\omega_0)$:

$$\frac{\partial E(z,\omega-\omega_0)}{\partial z} = -i\delta\beta E - i\beta_1 E - i\frac{\beta_2}{2}(\omega-\omega_0)^2 E \qquad (2.1.15)$$

Since a perturbation induced by the amplitude of the field itself usually exists (i.e. $n = n\ (\omega) + \Delta n$), the factor $\delta\beta$ is imported to denote the perturbation it induced on $\beta$ (i.e. $\beta = \beta\ (\omega) + \delta\beta$).

At this point, one can go back to the time domain by taking inverse Fourier transform of Eq. (2.1.15), and then obtain the propagation equation for $E(z,t)$:

$$\frac{\partial E(z,t)}{\partial z} + \beta_1\frac{\partial E(z,t)}{\partial t} = i\frac{\beta_2}{2}\frac{\partial^2 E(z,t)}{\partial t^2} - i\delta\beta E(z,t) \quad . \qquad (2.1.16)$$

In practice, a synchronous coordinate system moving at the group velocity $v_g$ is always used to make Eq. (2.1.16) even simpler in form. The transformation between the original and synchronous coordinate systems is as:

$$\begin{cases} \tau = t - \dfrac{z}{v_g} \\ z' = z \end{cases} \qquad (2.1.17)$$

Finally, the propagation equation of an optical pulse can be reformed as:





$$\frac{\partial E(z',\tau)}{\partial z'} = i\frac{\beta_2}{2}\frac{\partial^2 E(z',\tau)}{\partial \tau^2} - i\delta\beta E(z',\tau) \qquad (2.1.18)$$

Eq. (2.1.18) describes the propagation of a short pulse in dispersive media. It will be further used to acquire pulse propagation in single-mode fibers in the next section of Chapter 2.

Based on Eq. (2.1.18), one can get some features of pulse propagation in dispersive media. If the GVD parameter $\beta_2 = 0$, i.e. there is no group velocity dispersion, then

$$\frac{\partial E}{\partial z'} = 0 \qquad (2.1.19)$$

Therefore, any envelope function can propagate any distance without distortion. While unfortunately, the only medium where $\beta_2 = 0$ is a vacuum.

As $\beta_2 \neq 0$, then the pulse shapes become distorted as they propagate in the media. In practice, GVD always plays an important role in pulse propagation, especially in ultrashort pulse propagation.

## 2.3    Nonlinear pulse propagation in optical fibers

### 2.3.1    Nonlinear effects in optical fibers

Besides the frequency dependence, the refractive index of the optical media is always intensity dependent. Hence when the peak power of a pulse is very strong, an interesting manifestation of that occurs through self–phase modulation (SPM) arising from the Kerr nonlinearity. This phenomenon leads to the spectral broadening of an optical pulse [40].





On a fundamental level, the origin of nonlinearity is related to the nonlinear response of the electric dipolar polarization $\vec{P}$ under the influence of an applied electric field $\vec{E}$:

$$\vec{P} = \varepsilon_0 \left( \chi^{(1)} \cdot \vec{E} + \chi^{(2)} : \vec{E}\vec{E} + \chi^{(3)} : \vec{E}\vec{E}\vec{E} + \ldots \right), \qquad (2.2.1)$$

where, $\varepsilon_0$ is the vacuum permittivity, and $\chi^{(j)}$ ($j$ = 1, 2...) is the $j^{\text{th}}$ order susceptibility. The linear susceptibility $\chi^{(1)}$ represents the dominant contribution to $P$, whose effects are included through the refractive index $n$. The second-order susceptibility $\chi^{(2)}$ is caused by the asymmetry in molecular structure, which is responsible for nonlinear effects such as the second-harmonic generation and sum-frequency generation. Since $SiO_2$ has a symmetric molecular structure, $\chi^{(2)}$ vanishes for silica glasses. Therefore, in the case of a silica fiber, the lowest-order nonlinear effects originate from the third-order susceptibility $\chi^{(3)}$. These third-order nonlinear effects include the third-order harmonic generation, four-wave mixing, and nonlinear refraction. However, as the third-harmonic generation and the four-wave mixing both needs strict phase matching condition, they are not efficient in single-mode fibers. Most of the nonlinear effects in optical fibers such as the optical Kerr effect, stimulated Raman and Brillouin scattering, originate from nonlinear refraction, which can automatically occur provided that the light intensity reach some thresholds. In its simplest form, the nonlinear refractive index can be written as:

$$n\left(\omega, |E|^2\right) = n(\omega) + n_2 |E|^2 \quad , \qquad (2.2.2)$$

where, $n(\omega)$ is the linear part given by Eq. (2.1.1), $|E|^2$ is the optical intensity inside the fiber, and $n_2$ is the nonlinear-index coefficient determined by $\chi^{(3)}$.





Normally, $n_2$ of optical fibers has a very small value, which is in the range of $2.2 \sim 3.4 \times 10^{-20}$ m$^2$W$^{-1}$ [40]. However, because the core of SMFs is very small, which is normally < 10 μm in diameter, and the light is confined to propagate in the fiber core with almost ignorable loss, a very long distance of light-matter interaction is thus possible, which in consequence makes the efficiency of the nonlinear optical process very high in optical fibers. The intensity dependence of the refractive index leads to a large number of interesting nonlinear effects. The two most widely studied are self-phase modulation (SPM) and cross-phase modulation (XPM). Self-phase modulation refers to the self-induced phase shift experienced by an optical field during its propagation in optical fibers. Its magnitude can be obtained by noting that the phase of an optical field changes by

$$\phi = nk_0L = \left(n + n_2|E|^2\right)k_0L \quad , \tag{2.2.3}$$

where $k_0 = \dfrac{2\pi}{\lambda}$ , and $L$ is the fiber length. The intensity-dependent nonlinear phase shift due to SPM is $\phi_{NL} = n_2|E|^2 k_0L$. SPM is responsible for the spectral broadening of ultrashort pulses and the formation of optical solitons in the anomalous-dispersion regime of fibers.

Cross-phase modulation refers to the nonlinear phase shift of an optical field induced by another field having a different wavelength, direction, or polarization state. Its origin can be understood by noting for example that the total electrical field $E$ is given by





$$E = \frac{1}{2}\hat{x}\left[E_1\exp\left(-i\omega_1 t\right) + E_2\exp\left(-i\omega_2 t\right) + c.c.\right] \qquad (2.2.4)$$

When two optical fields at frequencies $\omega_1$ and $\omega_2$, polarized along the same axis, propagate simultaneously inside the fiber. The nonlinear phase shift introduced by the field $E_2$ to the field $E_1$ is $\phi_{nl} = n_2 k_0 L \cdot 2|E_2|^2$. Therefore in this case the total nonlinear phase shift for the field $E_1$ would be

$$\phi_{nl} = n_2 k_0 L\left(|E_1|^2 + 2|E_2|^2\right) , \qquad (2.2.5)$$

The two terms on the right side are due to SPM and XPM, respectively. Among other things, XPM is responsible for the asymmetric spectral broadening of the co-propagating optical pulses. XPM also gives an anticipation of soliton interaction and the existence of paired solitons in fiber soliton lasers.

### 2.3.2 Nonlinear Schrödinger equation

By including the nonlinear part of refractive index into $\delta\beta$ of Eq. (2.1.16) using $\beta(\omega) = n(\omega)\dfrac{\omega}{c}$, one can get:

$$\frac{\partial E}{\partial z} + \beta_1 \frac{\partial E}{\partial t} = i\frac{\beta_2}{2}\frac{\partial^2 E}{\partial t^2} - i\gamma|E|^2 E , \qquad (2.2.5)$$

$$\gamma = \frac{\pi n_2}{\lambda} . \qquad (2.2.6)$$

When introducing the normalized time $\tau$, distance $\xi$ and wave envelope $U$ by:





$$\tau = \frac{t - z\big/ v_g}{\tau_p}$$

$$\xi = \frac{1}{\tau_p^2} |\beta_2| z \qquad\qquad (2.2.7)$$

$$U = \tau_p \left( \frac{\alpha}{\beta_2} \right)^{\!\!{}^{1\!/\!2}} E$$

where $\tau_p$ is the pulse width, Eq. (2.2.5) is then reduced to the standard NLSE:

$$i \frac{\partial U}{\partial \xi} + \frac{1}{2} \frac{\partial^2 U}{\partial \tau^2} + |U|^2 U = 0 \qquad\qquad (2.2.8)$$

Eq. (2.2.8) gives a simple description over the propagation of an optical pulse in nonlinear dispersive media, which is affected by both the GVD and nonlinear effects denoted respectively by the second and third term in the equation.

### 2.3.3 Fundamental Soliton

From Eq. (2.2.5) or (2.2.8), it is easy to understand that the propagation of an optical pulse is the combinative effect of both GVD and nonlinear effects. The nonlinear optical Kerr effect always generates a positive frequency chirp in the center part of a pulse, while the frequency chirp induced by the GVD can be either positive or negative depending on the sign of the GVD parameter. Since in anomalous dispersion regimes ($\beta_2 < 0$), the fiber generates a negative linear frequency chirp, under the simultaneous action of the nonlinear Kerr effect, the frequency chirp on a pulse may be eliminated and no pulse distortion will happen as it propagating in the fiber. In fact, as the chirp induced by the nonlinear Kerr effect is the pulse intensity dependent, as a pulse with strong intensity propagates in anomalous-dispersion fiber, it can adjust its intensity so that the chirp induced by the two effects is automatically balanced. Namely,





optical solitons is actually a generic property of fibers in the anomalous dispersion regime.

The fundamental soliton solution can be derived theoretically from NLSE by assuming the balance between the GVD and nonlinear terms. By solving the NSLE using the inverse scattering method [40], one can get the solution of fundamental solitons written as:

$$E = \left( \frac{\beta_2}{\alpha} \right)^{\!\!1/2} \frac{1}{\tau_p} \operatorname{sec} h \left( \frac{t - z/v_g}{\tau_p} \right) \exp\left( -i\frac{\pi z}{4 z_0} \right) \tag{2.2.9}$$

where $\alpha$ is the fiber loss. $z_0$ represents the soliton period that is defined as:

$$z_0 = \frac{\pi}{2} \cdot \frac{\tau_p^{\,2}}{|\beta_2|}. \tag{2.2.10}$$

The wave packet in Eq. (2.2.9) is referred to as the fundamental soliton because its shape does not change with propagation.

## 2.4    Pulse propagation in erbium-doped fiber

### 2.4.1    Gain profile of erbium-doped fibers

Rare-earth doped fibers manufactured by doping the rare-earth ions in SMF core have been widely used for amplifiers in optic fiber communication systems. Various doping ions are chosen to provide gain for light in different wavelength regions. Among them, erbium-doped fiber amplifiers (EDFAs) have attracted especial attention, as they have a broadband gain in the 1.55 μm optical fiber communication window.

By pumping the ions from their ground state to the exited states, population inversion could be achieved between the two energy levels of the ions. Then the





stimulated emission happens and amplifies the light incident to the fiber whose frequency is within the gain profile. The optical spectrum of a fiber amplifier is mainly determined by the property of the dopants in the fiber. Based on the rate equations, the gain coefficient of an optical amplifier can be calculated by $g = \sigma(N_2 - N_1)$, where $\sigma$ is the transition cross-section, and $N_1$ and $N_2$ are the atomic densities in the two energy levels of the ions. In practice, there are many factors that affect the gain coefficient of doped fiber amplifiers. Different rare-earth doped fiber amplifier or even the fiber amplifiers doped with the same rare-earth ion but with different doping densities and fiber properties could have different gain profiles. The pumping scheme of the $Er^{3+}$ ions can be regarded as a three-level system, which is schematically illustrated as **Figure 2.2**. Efficient pumping is possible using semiconductor lasers operating near 980 nm and 1480 nm wavelengths.

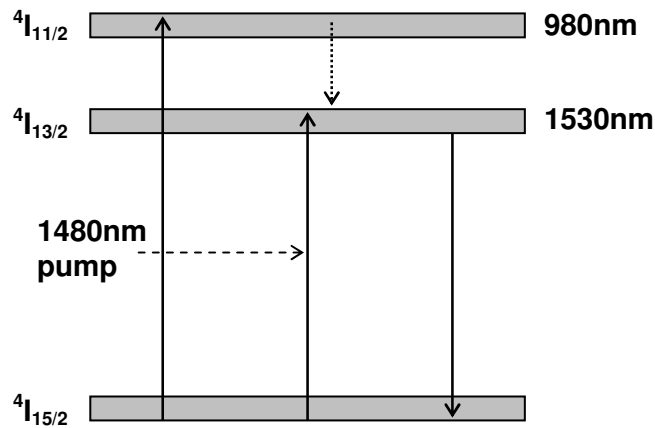

**Figure 2.2**: Energy levels of $Er^{3+}$ ions.

Generally, an amplifier is characterized by its small signal gain, gain bandwidth, gain saturation power and the noise figure. For a homogeneously broadened gain medium as $Er^{3+}$ ions, the gain coefficient can be written as [40]





$$g(\omega) = \frac{g_0}{1 + (\omega - \omega_0)^2 T_2^2 + P/P_S} \quad, \tag{2.3.1}$$

$g_0$ is the small signal gain, $\omega$ is the frequency of the incident light, $\omega_0$ is the resonant frequency of the dopants, $T_2$ is the dipole relaxation time of the dopants, $P$ is the power of the incident light, and $P_S$ is the saturation power of the gain medium. The saturation power depends on the stimulated emission cross-section and the upper energy level lifetime of the dopants. In the case of three-level pumping scheme, it is also the pump strength dependent. In practice, the gain saturation can be negligible for most doped fiber amplifier for single pulse amplification, i.e. $P/P_S << 1$. By neglecting the $P/P_S$ term, the gain coefficient becomes:

$$g(\omega) = \frac{g_0}{1 + (\omega - \omega_0)^2 T_2^2} \approx g_0 \left[ 1 - \frac{(\omega - \omega_0)^2}{\Omega_g^2} \right]. \tag{2.3.2}$$

This equation shows that the gain is governed by a Lorentzian profile with the maximum value happening when the signal frequency coincides with the atomic transition frequency $\omega_0$. The gain bandwidth $\Omega_g$ is defined as the full width at half maximum (FWHM) of the gain spectrum $g(\omega)$. For a Lorentzian profile, $\Omega_g$ is inversely related to the dipole relaxation time $T_2$ of the dopants. **Figure 2.3** shows a typical absorption and emission spectrum of erbium-doped fiber amplifiers [105]. The shapes of the absorption and emission spectrum of erbium-doped fiber amplifiers can be associated with the energy-level transitions of $Er^{3+}$ ions shown in **Figure 2.2**. Details of the spectrum also depend on the fiber composition and the co-dopants. Worthy of mentioning here, the gain of isolated $Er^{3+}$ ions is homogenously broadened, which exhibit a Lorentzian profile. However, as $Er^{3+}$ ions are doped into silica fiber, due to the





interaction of the ions with the silica host and other co-dopants such as germania and alumina within the fiber core, their energy levels are no longer discrete, but consist of a number of closely spaced levels. As a result, the gain and absorption spectrum is broadened with a double peak structure as shown in **Figure 2.3**. For erbium-doped fiber, the gain bandwidth could reach approximately 40 nm, which can correspondingly support as short to 60 fs pulse. But in practice, not the whole part of the gain can be efficiently used due to some effects such as the nonuniform gain profile, the fiber birefringence, and also the cavity effect in fiber lasers.

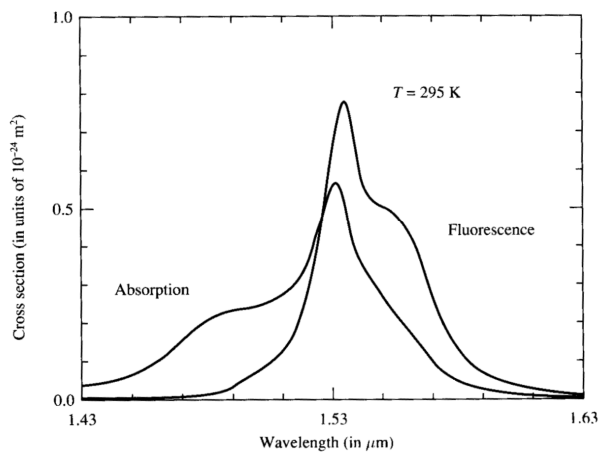

**Figure 2.3**: Typical absorption and emission spectra of the erbium-doped fibers (After Ref. [106]).

### 2.4.2 Ginzburg-Landau equation

When an optical pulse propagates in the erbium-doped fibers, the effect of the light amplification must be considered. Generally, the interaction of the doped ions with the light is governed by the Maxwell-Bloch equations [40]. However, in the case of that the pulse widths is larger comparing with the dipole





relaxation time $T_2$ of the $Er^{3+}$ ions ($T_2$ < 0.1 ps), the rate equation approximation can be made and consequently the gain term can be simply added into the NLSE.

As the gain dispersion is to reduce the gain for spectral components away from the gain peak $\omega_0$, therefore, the frequency dependence of the gain can be approximated by:

$$g(\omega) = g(\omega_0) + \frac{\partial g}{\partial \omega}\Big|_{\omega=\omega_0}(\omega - \omega_0) + \frac{1}{2}\frac{\partial^2 g}{\partial \omega^2}\Big|_{\omega=\omega_0}(\omega - \omega_0)^2 + ... \qquad (2.3.3)$$

Without loss of the generality, assuming that the pulse carrier frequency coincides with the gain peak, the equation describing optical pulse propagation in the doped fiber becomes:

$$\frac{\partial E}{\partial z} = -\beta_1\frac{\partial E}{\partial t} - \frac{i}{2}\beta_2\frac{\partial^2 E}{\partial t^2} + i\gamma|E|^2 E + \frac{g}{2}E + \frac{g}{2\Omega_g^2}\frac{\partial^2 E}{\partial t^2} \qquad (2.3.4)$$

where $g$ is the peak gain coefficient, $\Omega_g$ is the gain bandwidth. The time dependence of the peak gain coefficient varies with time due to the gain saturation and is determined by [40]

$$\frac{\partial g}{\partial t} = \frac{g_0 - g}{T_1} - \frac{g|E|^2}{E_s} \qquad , \qquad (2.3.5)$$

where $g_0$ is the small signal gain, $T_1$ is the population decay time, and $E_s$ is the saturation energy. As for the erbium doped ions $T_1$ is in the time scale of 10 ms, if the pulse width is far narrower than it, which is normally the case, the $T_1$ term is negligible during pulse amplification, and $g$ $(t)$ becomes:

$$g(t) = g_0 \exp\left[-\frac{1}{E_s}\int_{-\infty}^{t}|E|^2 dt\right] \qquad (2.3.6)$$





Typical values of $E_s$ for erbium-doped fibers are about 10 μJ. As the pulse energies are normally much smaller than the saturation energy $E_s$, gain saturation is negligible over the duration of a single pulse. However, in the case of a pulsed fiber laser, the pulse circulating in the laser cavity, the average power of the light may still saturate the gain and determines the saturated gain value.

In reality, one may also need to take into account the fiber loss (~0.2 dB/km) for long distance light propagation in fibers.

$$\frac{\partial E}{\partial z} = -\beta_1 \frac{\partial E}{\partial t} - \frac{i}{2} \beta_2 \frac{\partial^2 E}{\partial t^2} + i\gamma |E|^2 E + \frac{g - \alpha_f}{2} E + \frac{g}{2\Omega_g^2} \frac{\partial^2 E}{\partial t^2} \quad (2.3.7)$$

where $\alpha_f$ represents the fiber loss. When $\alpha_f = 0$ and $g = 0$, the equation reduces back to NLSE.

Eq. (2.3.7) can be also written in a domainless form by making the transformation of Eq. (2.2.7):

$$i\frac{\partial U}{\partial \xi} + \frac{1}{2}(1 - id)\frac{\partial^2 U}{\partial \tau^2} + N^2 |U|^2 U = \frac{i}{2} \mu U \quad (2.3.8)$$

$$d = \frac{gT_2^2}{n_0 |\beta_2|}$$
$$\mu = \left( \frac{g}{n_0} - \alpha_f \right) L_D \quad (2.3.9)$$

$$N = \left( \frac{\gamma P_0 T_0^2}{|\beta_2|} \right)^{1/2} \quad (2.3.10)$$

### 2.4.3   Soliton solution

Different to the NLSE, which is integrable and has an exact soliton solution, the





GLEQ is non-integrable. However, it is found that in the case of anomalous fiber dispersion, the equation also has solitary wave solutions in the sense of optical pulses whose shape does not change during propagation. When $N = 1$, fundamental soliton solution can be obtained in the case of a single eigen-value by using a trial-and-error method [40]:

$$U(\xi, \tau) = \mathrm{sech}(p\tau)\exp\left[iq\ln\cosh(p\tau) + i\Gamma\xi\right] \qquad (2.3.11)$$

where $p$, $q$ and $\Gamma$ are constants, comparing with the case in un-doped fiber, due to the influence of the fiber gain, the solitons in the doped fibers become frequency chirped. And for a SMF of GVD $\beta_2$, the peak power required to form a fundamental soliton is

$$P_0 = \frac{|\beta_2|}{\gamma T_0^2} \approx \frac{3.11|\beta_2|}{\gamma T_{FWHM}^2} \qquad (2.3.12)$$

$N > 1$ corresponds to high-order solitons. However, although theoretically predicted, by now no high-order soliton has been ever experimentally observed to our knowledge due to its intrinsic instability.

## 2.5    Pulse propagation in linearly birefringent fibers

### 2.5.1    Fiber birefringence

In all the above processes, we have ignored the polarization characteristic of light in the fiber. Actually, even a SMF is not truly single mode because it can support two degenerate modes that polarized in two orthogonal directions. Under ideal conditions, the two polarization states would not couple to each other. But in practice, all fibers exhibit some modal birefringence because of small departures from cylindrical symmetry due to random variations in core





shape and/or stress-induced anisotropy along the fiber length. Mathematically, the strength of modal birefringence is defined as:

$$B = \delta\beta / k_0 = \left| n_x - n_y \right| \tag{2.4.1}$$

where $\delta\beta = \left| \beta_x - \beta_y \right|$ is the difference in the propagation constant, and $n_x$ and $n_y$ are the respective modal refractive indices for the two orthogonally polarized components. Here it is assumed the fiber is linearly birefringent, i.e. the fiber has two principle axes along which it is capable of maintaining the state of linear polarization of the incident light in the absence of nonlinear effects. The axis along which the mode index is smaller, namely the group velocity is larger for light propagating in this direction, is called the fast axis. And similarly, the axis with the larger mode index is called the slow axis. This assumption is ideally the case for polarization-maintaining fibers (PMF), where the built-in birefringence is made much larger than random changes occurring due to stress and core-shape variations.

For a given value of $B$, after propagating a length of $L$, the two polarization components of the incident light will gain a phase difference of

$$\Delta\phi = \frac{2\pi\left(n_x - n_y\right)L}{\lambda} = \frac{2\pi BL}{\lambda} = 2\pi \cdot \frac{L}{L_B} \tag{2.4.2}$$

where $L_B = 2\pi / \left| \beta_x - \beta_y \right| = \lambda / B$ is called the beat length.

From Eq. (2.4.2), it can be seen that the light varies its polarization periodically along the fiber. At a multiple distance of $L_B$, a multiple phase difference of $2\pi$ will be generated between the two polarization components, which mean that the light returns to its initial polarization state. From a viewpoint of physics, the





two polarization modes exchange their powers as propagating in the fiber with a period of the beat length $L_B$. Different single-mode fibers could have very different beat lengths. For a standard SMF, $L_B$ is in the range of several meters, while for the PMF, as a strong birefringence is deliberately built in the fibers, they could have a beat length of several millimeters.

When the nonlinear effects in optical fibers become dominant, a sufficiently intense optical field can induce nonlinear birefringence whose magnitude is intensity dependent. Consequently, the refractive index is described as

$$n_j = n_j^L + \Delta n_j \qquad (j = x, y) \qquad (2.4.3)$$

where $n_j^L$ is the linear part of the refractive index. And the nonlinear contributions $\Delta n_x$ and $\Delta n_y$ are given by

$$\Delta n_x = n_2 \left( \left| E_x \right|^2 + \frac{2}{3} \left| E_y \right|^2 \right), \Delta n_y = n_2 \left( \left| E_y \right|^2 + \frac{2}{3} \left| E_x \right|^2 \right), \qquad (2.4.4)$$

Normally one takes the value of $n_2$ the same for the two polarizations, which is also practically just the case. In Eq. (2.4.4), the first term is responsible for SPM. The second term results in XPM since the nonlinear phase shift acquired by one polarization component dependent on the intensity of the other polarization component. The presence of this term induces a nonlinear coupling between the two polarization eigen-modes $E_x$ and $E_y$.

### 2.5.2 Coupled Ginzburg-Landau equations

The NLSE remains universal due to two reasons. The first reason is it predicts the behavior of the observed pulses and secondly it can be manually derived [40]





$$\frac{\partial A'}{\partial z} + \beta_1 \frac{\partial A'}{\partial t} + i \frac{\beta_2}{2} \frac{\partial^2 A'}{\partial t^2} - \frac{\beta_3}{6} \frac{\partial^3 A'}{\partial t^3} + \frac{\alpha}{2} A' = i\gamma |A'|^2 A' \qquad (2.4.5)$$

where, the nonlinear coefficient $\gamma$ is defined as:

$$\gamma = \frac{n_2 \omega_0}{c A_{eff}} \qquad (2.4.6)$$

The NLSE above only describe the propagation in one spatial dimension. Pulse propagation in a fibre laser needs to consider coupling between the two orthogonally polarized components. Therefore, a coupled version of the NLSE is required.

$$\frac{\partial A_x}{\partial z} + \beta_{1x} \frac{\partial A_x}{\partial t} + \beta_2 \frac{i}{2} \frac{\partial^2 A_x}{\partial t^2} - \frac{\beta_3}{6} \frac{\partial^3 A_x}{\partial t^3} + \frac{\alpha}{2} A_x = i\gamma \left( |A_x|^2 + \frac{2}{3} |A_y|^2 \right) A_x + i \frac{\gamma}{3} A_x^* A_y^2 \exp[-j2\Delta\beta z] \quad (2.4.7)$$

$$\frac{\partial A_y}{\partial z} + \beta_{1y} \frac{\partial A_y}{\partial t} + \beta_2 \frac{i}{2} \frac{\partial^2 A_y}{\partial t^2} - \frac{\beta_3}{6} \frac{\partial^3 A_y}{\partial t^3} + \frac{\alpha}{2} A_y = i\gamma \left( |A_y|^2 + \frac{2}{3} |A_x|^2 \right) A_y + i \frac{\gamma}{3} A_y^* A_x^2 \exp[+j2\Delta\beta z] \quad (2.4.8)$$

where $A_x$ and $A_y$ are the two normalized slowly varying pulse envelopes along the slow and the fast axes, $A_x^*$ and $A_y^*$ represent their conjugates and $\Delta\beta$ is the wave-number difference. The NLSE can be further developed into the coupled GLEs, which describes the pulse propagation within the whole cavity including the amplification effect of the erbium doped fiber gain.

$$\begin{cases} \dfrac{\partial u}{\partial z} = i\beta u - \delta \dfrac{\partial u}{\partial t} - \dfrac{ik''}{2} \dfrac{\partial^2 u}{\partial t^2} + \dfrac{k'''}{6} \dfrac{\partial^3 u}{\partial t^3} + i\gamma \left( |u|^2 + \dfrac{2}{3} |v|^2 \right) u + \dfrac{i\gamma}{3} v^2 u^* + \dfrac{g}{2} u + \dfrac{g}{2\Omega_g^2} \dfrac{\partial^2 u}{\partial t^2} \\ \dfrac{\partial v}{\partial z} = -i\beta v + \delta \dfrac{\partial v}{\partial t} - \dfrac{ik''}{2} \dfrac{\partial^2 v}{\partial t^2} + \dfrac{k'''}{6} \dfrac{\partial^3 v}{\partial t^3} + i\gamma \left( |v|^2 + \dfrac{2}{3} |u|^2 \right) v + \dfrac{i\gamma}{3} u^2 v^* + \dfrac{g}{2} v + \dfrac{g}{2\Omega_g^2} \dfrac{\partial^2 v}{\partial t^2} \end{cases} \quad (2.4.9)$$

where, $u$ and $v$ are the normalized envelopes of the optical pulses along the two orthogonal polarized modes of the optical fiber. $k''$ is the second order dispersion coefficient; $k'''$ is the third order dispersion coefficient and $\gamma$ represents the nonlinearity of the fiber. $g$ is the saturable gain coefficient of the fiber and $\Omega_g$ represent the bandwidth of the laser gain.





In this thesis, a numerical approach is applied to understand the concept of the nonlinear pulse propagation in the soliton fiber laser. To this end the Coupled Ginzburg-Landau equations are numerically solved.

## 2.6    Procedure of numerically solve the Coupled Ginzburg-Landau equations

Nonlinear pulse propagation in optical fibers can be represented by the coupled NLSE as shown below:

$$\frac{\partial A_x}{\partial z} + \beta_{1x}\frac{\partial A_x}{\partial t} + \beta_2\frac{i}{2}\frac{\partial^2 A_x}{\partial t^2} - \frac{\beta_3}{6}\frac{\partial^3 A_x}{\partial t^3} + \frac{\alpha}{2}A_x = i\gamma\left(|A_x|^2 + \frac{2}{3}|A_y|^2\right)A_x + i\frac{\gamma}{3}A_x{}^*A_y{}^2\exp[-j2\Delta\beta z] \quad (2.4.10)$$

$$\frac{\partial A_y}{\partial z} + \beta_{1y}\frac{\partial A_y}{\partial t} + \beta_2\frac{i}{2}\frac{\partial^2 A_y}{\partial t^2} - \frac{\beta_3}{6}\frac{\partial^3 A_y}{\partial t^3} + \frac{\alpha}{2}A_y = i\gamma\left(|A_y|^2 + \frac{2}{3}|A_x|^2\right)A_y + i\frac{\gamma}{3}A_y{}^*A_x{}^2\exp[+j2\Delta\beta z] \quad (2.4.11)$$

In the above equations:

$$\beta_x(\omega) = \beta_{0x} + \beta_{1x}(\omega - \omega_0) + \frac{1}{2}\beta_{2x}(\omega - \omega_0)^2 + \frac{1}{6}\beta_{3x}(\omega - \omega_0)^3 \ldots$$

$$\beta_y(\omega) = \beta_{0y} + \beta_{1y}(\omega - \omega_0) + \frac{1}{2}\beta_{2y}(\omega - \omega_0)^2 + \frac{1}{6}\beta_{3y}(\omega - \omega_0)^3 \ldots$$

The term $\Delta\beta = \beta_{0x} - \beta_{0y} = \frac{2\pi}{\lambda}B_m = \frac{2\pi}{L_B}$ is related to birefringence of the fiber. The additional term in $\gamma\left(|A_y|^2 + \frac{2}{3}|A_x|^2\right)A_y$ consider the XPM effect and the term $i\frac{\gamma}{3}A_y{}^*A_x{}^2\exp[+i2\Delta\beta z]$ is due to FWM. The term $|A_y|^2$ represents optical power.

### Step 1–Transform into a retarded reference frame

In practice, a coordinate system (so called "retarded frame") moving at the mean group- velocity $\overline{v_g}$ (the average of the group-velocities of the two polarization components, which may differ from the mutual group-velocity





taken by the two polarization modes) is employed. The transformation/mapping between the original coordinate system and the retarded frame is defined as:

$$\begin{cases} T = t - \dfrac{z}{v_g} = t - \overline{\beta_1} \times z \\ Z = z \end{cases}$$

$$\overline{\beta_1} = \frac{\beta_{1x} + \beta_{1y}}{2}$$ . Thus, we have:

$$\frac{\partial A_x}{\partial t} = \frac{\partial A_x}{\partial T} \cdot \frac{\partial T}{\partial t} + \frac{\partial A_x}{\partial Z} \cdot \frac{\partial Z}{\partial t} = \frac{\partial A_x}{\partial T}$$

$$\frac{\partial A_x}{\partial z} = \frac{\partial A_x}{\partial T} \cdot \frac{\partial T}{\partial z} + \frac{\partial A_x}{\partial Z} \cdot \frac{\partial Z}{\partial z} = \frac{\partial A_x}{\partial T} = -\overline{\beta_1} \frac{\partial A_x}{\partial T} + \frac{\partial A_x}{\partial Z}$$

$$\frac{\partial^2 A_x}{\partial t^2} = \frac{\partial \left( \dfrac{\partial A_x}{\partial T} \right)}{\partial t} = \frac{\partial \left( \dfrac{\partial A_x}{\partial T} \right)}{\partial T} \cdot \frac{\partial T}{\partial t} + \frac{\partial \left( \dfrac{\partial A_x}{\partial T} \right)}{\partial Z} \cdot \frac{\partial Z}{\partial t} = \frac{\partial^2 A_x}{\partial T^2} \cdot \frac{\partial T}{\partial t} + \frac{\partial \left( \dfrac{\partial A_x}{\partial T} \right)}{\partial Z} \cdot 0 = \frac{\partial^2 A_x}{\partial T^2}$$

$$\frac{\partial^3 A_x}{\partial t^3} = \frac{\partial \left( \dfrac{\partial^2 A_x}{\partial T^2} \right)}{\partial t} = \frac{\partial \left( \dfrac{\partial^2 A_x}{\partial T^2} \right)}{\partial T} \cdot \frac{\partial T}{\partial t} + \frac{\partial \left( \dfrac{\partial^2 A_x}{\partial T^2} \right)}{\partial Z} \cdot \frac{\partial Z}{\partial t} = \frac{\partial^3 A_x}{\partial T^3} \cdot \frac{\partial T}{\partial t} + \frac{\partial \left( \dfrac{\partial^2 A_x}{\partial T^2} \right)}{\partial Z} \cdot 0 = \frac{\partial^3 A_x}{\partial T^3}$$

$$\frac{\partial A_x}{\partial t} = \frac{\partial A_x}{\partial T} \tag{2.4.12}$$

$$\frac{\partial A_x}{\partial z} = -\overline{\beta_1} \frac{\partial A_x}{\partial T} + \frac{\partial A_x}{\partial Z} \tag{2.4.13}$$

$$\frac{\partial^2 A_x}{\partial t^2} = \frac{\partial^2 A_x}{\partial T^2} \tag{2.4.14}$$

$$\frac{\partial^3 A_x}{\partial t^3} = \frac{\partial^3 A_x}{\partial T^3} \tag{2.4.15}$$

Substituting $\overline{\beta_1} = \dfrac{\beta_{1x} + \beta_{1y}}{2}$ and equations (2.4.12), (2.4.13), (2.4.14) and (2.4.15) into equations (2.4.7) and (2.4.8) while ignoring fiber attenuation ($\alpha$ = 0) gives:





$$\frac{\partial A_x}{\partial Z} + \delta \frac{\partial A_x}{\partial T} + \beta_2 \frac{i}{2} \frac{\partial^2 A_x}{\partial T^2} - \frac{\beta_3}{6} \frac{\partial^3 A_x}{\partial T^3} = i\gamma \left( |A_x|^2 + \frac{2}{3} |A_y|^2 \right) A_x + i \frac{\gamma}{3} A_x^* A_y^2 \exp[-i2\Delta\beta z] \quad (2.4.16)$$

$$\frac{\partial A_y}{\partial Z} - \delta \frac{\partial A_y}{\partial T} + \beta_2 \frac{i}{2} \frac{\partial^2 A_y}{\partial T^2} - \frac{\beta_3}{6} \frac{\partial^3 A_y}{\partial T^3} = i\gamma \left( |A_y|^2 + \frac{2}{3} |A_x|^2 \right) A_y + i \frac{\gamma}{3} A_y^* A_x^2 \exp[+i2\Delta\beta z] \quad (2.4.17)$$

where $\delta = \dfrac{\beta_{1x} - \beta_{1y}}{2}$ .

**Step 2–Eliminate the exponential components in the FWM terms**

Define: $A_x = F_h \exp\left( \dfrac{-j\Delta\beta z}{2} \right)$ and $A_y = F_v \exp\left( \dfrac{+j\Delta\beta z}{2} \right)$. After some simple

algebra, we obtain:

$$\frac{\partial F_h}{\partial Z} - i \frac{\Delta\beta}{2} F_h + \delta \frac{\partial F_h}{\partial T} + \beta_2 \frac{i}{2} \frac{\partial^2 F_h}{\partial T^2} - \frac{\beta_3}{6} \frac{\partial^3 F_h}{\partial T^3} = i\gamma \left( |F_h|^2 + \frac{2}{3} |F_v|^2 \right) F_h + i \frac{\gamma}{3} F_h^* F_v^2 \quad (2.4.18)$$

$$\frac{\partial F_v}{\partial Z} + i \frac{\Delta\beta}{2} F_v - \delta \frac{\partial F_v}{\partial T} + \beta_2 \frac{i}{2} \frac{\partial^2 F_v}{\partial T^2} - \frac{\beta_3}{6} \frac{\partial^3 F_v}{\partial T^3} = i\gamma \left( |F_v|^2 + \frac{2}{3} |F_h|^2 \right) F_v + i \frac{\gamma}{3} F_v^* F_h^2 \quad (2.4.19)$$

**Step 3–Take into account the amplification by the EDF**

The gain provided by the Erbium-doped fiber (EDF) and its gain dispersion
need to be considered as gain medium is an indispensable part of a laser. Thus,
the coupled NLSE can now be written as:

$$\frac{\partial F_h}{\partial Z} = i \frac{\Delta\beta}{2} F_h - \delta \frac{\partial F_h}{\partial T} - \beta_2 \frac{i}{2} \frac{\partial^2 F_h}{\partial T^2} + \frac{\beta_3}{6} \frac{\partial^3 F_h}{\partial T^3} + i\gamma \left( |F_h|^2 + \frac{2}{3} |F_v|^2 \right) F_h + i \frac{\gamma}{3} F_h^* F_v^2$$
$$+ \frac{g_p(T)}{2} F_h + \frac{g_p(T)}{2.\Omega_g^2} \cdot \frac{\partial^2 F_h}{\partial T^2} \quad (2.4.20)$$

$$\frac{\partial F_v}{\partial Z} = -i \frac{\Delta\beta}{2} F_v + \delta \frac{\partial F_v}{\partial T} - \beta_2 \frac{i}{2} \frac{\partial^2 F_v}{\partial T^2} + \frac{\beta_3}{6} \frac{\partial^3 F_v}{\partial T^3} + i\gamma \left( |F_v|^2 + \frac{2}{3} |F_h|^2 \right) F_v + i \frac{\gamma}{3} F_v^* F_h^2$$
$$+ \frac{g_p(T)}{2} F_v + \frac{g_p(T)}{2.\Omega_g^2} \cdot \frac{\partial^2 F_v}{\partial T^2} \quad (2.4.21)$$

The new term $\dfrac{g_p(T)}{2} F_h$ in equation (2.4.20) and $\dfrac{g_p(T)}{2} F_v$ in equation

(2.4.21) account for the gain of the EDF. Whereas, the last term in equation





(2.4.21) is due to the gain dispersion of the EDF. $g_p(t)$ in equation is defined as :

$$g_p(t) = g_o \exp\left[-\frac{1}{E_s}\int_{-\infty}^{t}\left(|u|^2 + |v|^2\right)dt'\right]$$

where $g_0$ refers to the small signal gain and $E_s$ is the saturation energy, which has a typical value of 1 μJ. Equations (2.4.20) and (2.4.21) symbolize the coupled Ginzburg-Landau equation (GLE). [40]

**Step 4–Symmetrized split-step Fourier method**

The above coupled Ginzburg-Landau equations (GLE) can be solved using the symmetrized Fourier split-step method. The procedures of transforming the coupled GLE can be summarized as follows:

1. Ignore the nonlinear terms. The coupled GLE then becomes:

$$\frac{\partial F_h}{\partial Z} = i\frac{\Delta\beta}{2}F_h - \delta\frac{\partial F_h}{\partial T} - \beta_2\frac{i}{2}\frac{\partial^2 F_h}{\partial T^2} + \frac{\beta_3}{6}\frac{\partial^3 F_h}{\partial T^3} + \frac{g_p(T)}{2.\Omega_g^2}.\frac{\partial^2 F_h}{\partial T^2} \qquad (2.4.22)$$

$$\frac{\partial F_v}{\partial Z} = -i\frac{\Delta\beta}{2}F_v + \delta\frac{\partial F_v}{\partial T} - \beta_2\frac{i}{2}\frac{\partial^2 F_v}{\partial T^2} + \frac{\beta_3}{6}\frac{\partial^3 F_v}{\partial T^3} + \frac{g_p(T)}{2.\Omega_g^2}.\frac{\partial^2 F_v}{\partial T^2} \qquad (2.4.23)$$

Since Fourier Transform can be defined as $\widetilde{F}(\omega) = \int_{-\infty}^{+\infty} F(T)\exp(-i\omega T)dT$ , the Fourier Transform of $\partial F(T)\big/\partial T$ is $j\omega\widetilde{F}(\omega)$ .

Perform Fourier Transform on Equations (2.4.22) and (2.4.23), we obtain:

$$\frac{\partial\widetilde{F}_h}{\partial Z} = i\frac{\Delta\beta}{2}\widetilde{F}_h - i\omega\delta\widetilde{F}_h + \frac{i}{2}\beta_2\omega^2\widetilde{F}_h - i\frac{\beta_3}{6}\omega^3\widetilde{F}_h - \frac{g_p(T)}{2.\Omega_g^2}\omega^2\widetilde{F}_h \quad (2.4.24)$$

$$\frac{\partial\widetilde{F}_v}{\partial Z} = -i\frac{\Delta\beta}{2}\widetilde{F}_v + i\omega\delta\widetilde{F}_v + \frac{i}{2}\beta_2\omega^2\widetilde{F}_v - i\frac{\beta_3}{6}\omega^3\widetilde{F}_v - \frac{g_p(T)}{2.\Omega_g^2}\omega^2\widetilde{F}_v \quad (2.4.25)$$





For an increment step of $\dfrac{\partial z}{2}$, $\tilde{F}_h(z + \dfrac{\partial z}{2}, T)$ and $\tilde{F}_v(z + \dfrac{\partial z}{2}, T)$ are given as:

$$\tilde{F}_h\left(z + \frac{\partial z}{2}, \omega\right) = \tilde{F}_h(z, \omega) \times \exp\left[\left(-\frac{g_p(T)}{2.\Omega_g^2}\omega^2 + i\left(\frac{\beta_2}{2}\omega^2 - \frac{\beta_3}{6}\omega^3 + \frac{\Delta\beta}{2} - \omega\delta\right)\right)\frac{\partial z}{2}\right] \quad (2.4.26)$$

$$\tilde{F}_v\left(z + \frac{\partial z}{2}, \omega\right) = \tilde{F}_v(z, \omega) \times \exp\left[\left(-\frac{g_p(T)}{2.\Omega_g^2}\omega^2 + i\left(\frac{\beta_2}{2}\omega^2 - \frac{\beta_3}{6}\omega^3 - \frac{\Delta\beta}{2} + \omega\delta\right)\right)\frac{\partial z}{2}\right] \quad (2.4.27)$$

$F_h\left(z + \dfrac{\partial z}{2}, T\right)$ and $F_v\left(z + \dfrac{\partial z}{2}, T\right)$ can then be obtained by performing Inverse Fourier Transform on equations (2.4.26) and (2.4.27).

2. Ignore the dispersion terms. The coupled GLE then becomes:

$$\frac{\partial F_h}{\partial Z} = i\gamma\left(|F_h|^2 + \frac{2}{3}|F_v|^2\right)F_h + i\frac{\gamma}{3}F_h^{*}F_v^2 + \frac{g_p(T)}{2}F_h \quad (2.4.28)$$

$$\frac{\partial F_v}{\partial Z} = i\gamma\left(|F_v|^2 + \frac{2}{3}|F_h|^2\right)F_v + i\frac{\gamma}{3}F_v^{*}F_h^2 + \frac{g_p(T)}{2}F_v \quad (2.4.29)$$

In order to eliminate the FWM terms $i\dfrac{\gamma}{3}F_v^{*}F_h^2$ and $i\dfrac{\gamma}{3}F_h^{*}F_v^2$ in the coupled GLE, a so-called Circular Transformation is employed.

Define: $F_r = \dfrac{F_h + jF_v}{\sqrt{2}}$ and $F_l = \dfrac{F_h - jF_v}{\sqrt{2}}$, from which we can obtain:

$$F_h = \frac{F_r + F_l}{\sqrt{2}} \text{ and } F_v = \frac{F_r - F_l}{\sqrt{2}\,j}$$

Substitute them into equations (2.4.28) and (2.4.29), we obtain:





$$\frac{\partial F_r}{\partial Z} + \frac{\partial F_l}{\partial Z} = \frac{g_p(T)}{2}F_r + \frac{g_p(T)}{2}F_l + i\frac{\gamma}{6}\left(\begin{array}{c} 5|F_r|^2 F_r + 6|F_r|^2 F_l + F_r^2 F_l^* \\ + 6F_r|F_l|^2 + F_r^* F_l^2 + 5|F_l|^2 F_l \end{array}\right)$$

$$-i\frac{\gamma}{6}\left(\begin{array}{c} |F_r|^2 F_r - 2|F_r|^2 F_l + F_r^2 F_l^* \\ - 2F_r|F_l|^2 + F_r^* F_l^2 + |F_l|^2 F_l \end{array}\right) \tag{2.4.30}$$

$$\frac{\partial F_r}{\partial Z} - \frac{\partial F_l}{\partial Z} = \frac{g_p(T)}{2}F_r - \frac{g_p(T)}{2}F_l + i\frac{\gamma}{6}\left(\begin{array}{c} 5|F_r|^2 F_r - 6|F_r|^2 F_l - F_r^2 F_l^* \\ + 6F_r|F_l|^2 + F_r^* F_l^2 - 5|F_l|^2 F_l \end{array}\right)$$

$$-i\frac{\gamma}{6}\left(\begin{array}{c} |F_r|^2 F_r + 2|F_r|^2 F_l - F_r^2 F_l^* \\ - 2F_r|F_l|^2 + F_r^* F_l^2 - |F_l|^2 F_l \end{array}\right) \tag{2.4.31}$$

(2.4.30) + (2.4.31) and divide the sum by 2, we obtain:

$$\frac{\partial F_r}{\partial Z} = \frac{g_p(T)}{2}F_r + i\frac{2\gamma}{3}\left(|F_r|^2 + 2|F_l|^2\right)F_r \tag{2.4.32}$$

(2.4.30) – (2.4.31) and divide the difference by 2, we obtain:

$$\frac{\partial F_l}{\partial Z} = \frac{g_p(T)}{2}F_l + i\frac{2\gamma}{3}\left(|F_l|^2 + 2|F_r|^2\right)F_l \tag{2.4.33}$$

Hence for an increment step of $\partial z$, $F_r(z + \partial z, T)$ and $F_l(z + \partial z, T)$ are given as:

$$F_r(z + \partial z, T) = F_r(z, T) \times \exp\left[\left(\frac{g_p(T)}{2} + i\frac{2\gamma}{3}\left(|F_r|^2 + 2|F_l|^2\right)\right)\partial z\right] \tag{2.4.34}$$

$$F_l(z + \partial z, T) = F_l(z, T) \times \exp\left[\left(\frac{g_p(T)}{2} + i\frac{2\gamma}{3}\left(|F_l|^2 + 2|F_r|^2\right)\right)\partial z\right] \tag{2.4.35}$$

$F_h(z + \partial z, T)$ and $F_v(z + \partial z, T)$ can then be obtained by performing inverse Circular Transform on equations (2.4.34) and (2.4.35).





# Chapter 3.  Coherent and Incoherent Interaction of Vector Solitons

Among all the features of vector soliton phenomena, probably, interactions of vector solitons are the most fascinating ones. Generally, the interaction forces could be divided into two types: coherent and incoherent interaction. Coherent interaction occurs when the nonlinear medium can respond to interference effects that take place when solitons overlap [108]. This interaction depends on the relative phases of the interaction fields. To maximize the strength of coherent interaction, phase matching condition must be fulfilled. Corresponding, it only occurs when the nonlinear medium is weakly anisotropic or weakly birefringent.

However, incoherent interactions, on the other hand, occur when the relative phase between the soliton varies much faster than relaxation time of the nonlinear medium. It arises from the third order susceptibility: propagation refractive index of one light beam could be modified/modulated by another light beam. This process is phase insensitive and only determined by the relative strength of individual light beam.

I presented, in Chapter 3, the investigation of these two kinds of interactions among the two polarizations of a vector soliton. Specifically, the Four-wave-mixing effect (also known as coherent energy exchange) would be discussed in the Section 3.1. To obviously observe this effect, the cavity birefringence must be kept as weak as possible in order to fulfill the coherent interaction requirement. The Section 3.2 addressed the incoherent interaction, termed as





cross phase modulation (XPM) in fiber laser and reported the first experimental observation of induced vector soliton in a fiber laser.

## 3.1   Coherent Interactions

Passive mode-locking of erbium-doped fiber lasers with SESAM has been extensively investigated [108]. In contrast to the NPR mode-locking, mode-locking incorporating a SESAM does not require any polarization element inside the laser cavity, thereby under suitable condition of the cavity birefringence, vector solitons could be formed in the lasers [109-114]. Recently, it was reported that even the polarization-locked vector solitons (PLVSs) could be formed in the mode-locked fiber lasers [114]. Formation of a PLVS requires not only that the group velocities of the two orthogonal polarization components of a vector soliton are locked but also that their phase velocities are also locked. It is well known that through the SPM and XPM between the two polarization-modes of a fiber, the group velocity locked vector solitons (GVLVSs) could be formed [109-114]. Although it was also pointed out that the four-wave-mixing (also called coherent energy exchange) coupling between the polarization components of a vector soliton could have contributed to the formation of the PLVSs [114], so far no experimental evidence on the soliton internal FWM has been shown.

In this section, we report on the experimental observation of coherent interaction between the two orthogonal polarization components of a vector soliton formed in a fiber laser passively mode locked with a SESAM. A new type of spectral sidebands was first experimentally observed on the polarization





resolved soliton spectra of the PLVSs of the fiber lasers. The new spectral sidebands are characterized by that their positions on the soliton spectrum vary with the strength of the linear cavity birefringence, and while on one vector soliton polarization component the sideband appears as a spectral peak, then on the orthogonal polarization component it is a spectral dip, indicating the energy exchange between the two orthogonal polarization components of the vector solitons. Numerically we confirmed that the formation of the new type of spectral sidebands was formed by the FWM between the two polarization components of the vector solitons.

The fiber laser is illustrated in **Figure 3.1**. It has a ring cavity consisting of a piece of 4.6 m EDF with GVD parameter of 10 ps/km/nm and a total length of 5.4 m standard SMF with GVD parameter of 18 ps/km/nm. The cavity has a length of $4.6_{EDF} + 5.4_{SMF} = 10$ m. Note that within one cavity round-trip the signal propagates twice in the SMF between the circulator and the SESAM. A circulator is used to force the unidirectional operation of the ring and simultaneously to incorporate the SESAM in the cavity. An intra cavity polarization controller is used to change the cavity's linear birefringence. The laser is pumped by a high power Fiber Raman Laser source (BWC-FL-1480-1) of wavelength 1480 nm. A 10% fiber coupler is used to output the signals. The laser operation is monitored with an optical spectrum analyzer (Ando AQ-6315B), a 26.5 GHz RF spectrum analyzer (Agilent E4407BESA-E SERIES) and a 350 MHz oscilloscope (Agilent 54641A) together with a 5 GHz photodetector. A commercial autocorrelator (Femtochrome FR-103MN) is used to measure the pulse width of the soliton pulses. The SESAM used is made





based on a GaInNAs quantum well and has a saturable absorption modulation depth of 5%, a saturation fluence of 180 µJ/cm$^2$ and 10 ps relaxation time. The central absorption wavelength of the SESAM is at 1550 nm.

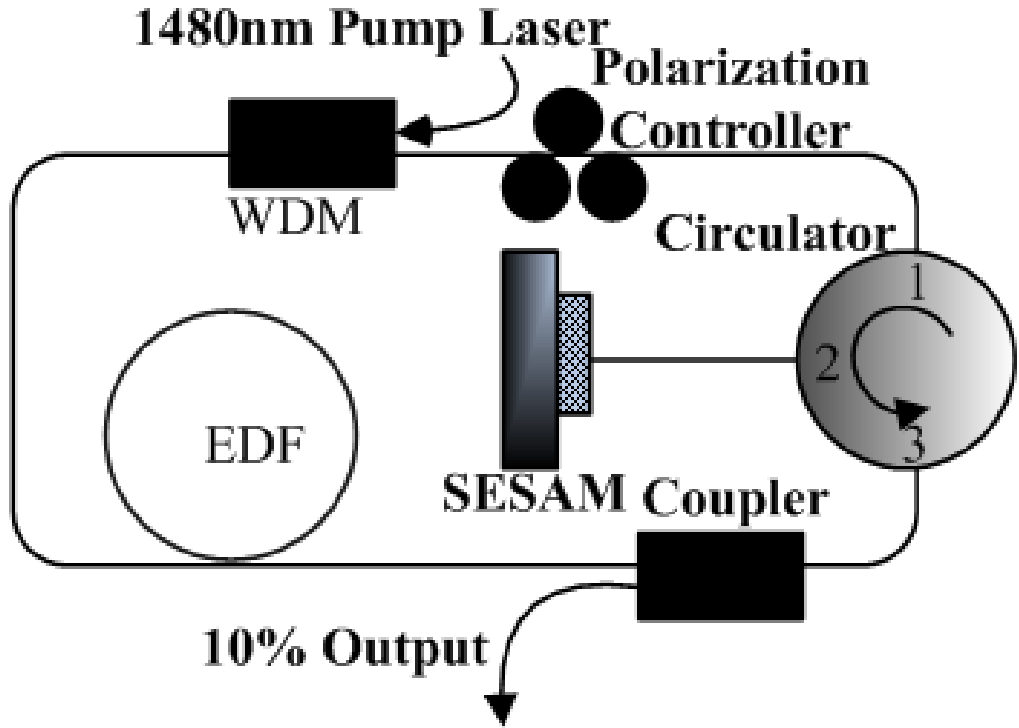

**Figure 3.1**: Schematic of the SESAM mode locked fiber laser.

Experimentally, it was noticed that after mode-locking multiple mode locked pulses were always initially formed in the cavity. Depending on the net cavity birefringence, they were either the GVLVSs, characterized by the rotation of soliton polarization state along the cavity, or the PLVSs, characterized by the fixed polarization at the laser output. With multiple vector solitons in cavity, as a result of mutual soliton interaction complicated relative soliton movement or vector soliton bunches with random fixed soliton separations were observed. To exclude the complications caused by soliton interactions, we have always reduced the number of solitons in cavity through carefully decreasing pump





power so that only one or a few widely separated solitons exist in cavity.

**Figure 3.2** shows typical measured optical spectra of the PLVSs of the laser. The soliton feature of the mode-locked pulses is confirmed by the existence of soliton sidebands. However, apart from the existence of the conventional Kelly soliton sidebands, on the vector soliton spectrum there are also extra sets of spectral sidebands. Experimentally it was noticed that different from the Kelly sidebands whose positions are almost independent of the laser operation conditions, such as the pump strength and polarization controller orientation change (linear cavity birefringence change), the positions of the new spectral sidebands varied sensitively with the linear cavity birefringence. We note that Cundiff *et al.* have reported a new type of spectral sidebands on the GVLVS spectrum and interpreted their formation as caused by the vector soliton polarization evolution in the cavity [114]. However, in our experiment the sidebands were observed on the PLVSs, whose polarization remains unchanged as they propagate along the laser cavity.

To determine the physical origin of the extra sideband formation, we then conducted polarization resolved measurement of the vector soliton spectrum. To this end the laser output was first passed through a rotatable external cavity linear polarizer. To separate the two orthogonal polarization components of a vector soliton, we always first located the orientation of the polarizer to the maximum soliton transmission, which sets the long axis of an elliptically polarized vector soliton; we then rotated the polarizer by 180 degree to determine the soliton polarization component along the short axis of the polarization ellipse. Through separating the two orthogonal polarization





components of the vector solitons, it turned out that the formation of the extra spectral sidebands was due to the coherent energy exchange between the two soliton polarization components. As can be clearly seen from the polarization resolved spectra, at the positions of extra spectral sidebands, while the spectral intensity of one soliton polarization component has a spectral peak, the orthogonal polarization component then has a spectral dip, indicting coherent energy exchange between them. We note that the energy flow between the two polarization components is not necessarily from the strong one to the weak one. Energy flow from the weak component to the strong component was also observed. In addition, it is to see from the polarization resolved spectra that the extra sidebands are symmetric with respect to the soliton peak frequency and at different wavelength positions the peak-dip can also alternate, suggesting that the energy exchange is the relative phase of the coupled components dependent. **Figure 3.2**a and **Figure 3.2**b were obtained from the same laser but under different intra cavity polarization controller orientations. Obviously, the positions of the extra sidebands are the cavity birefringence dependent.





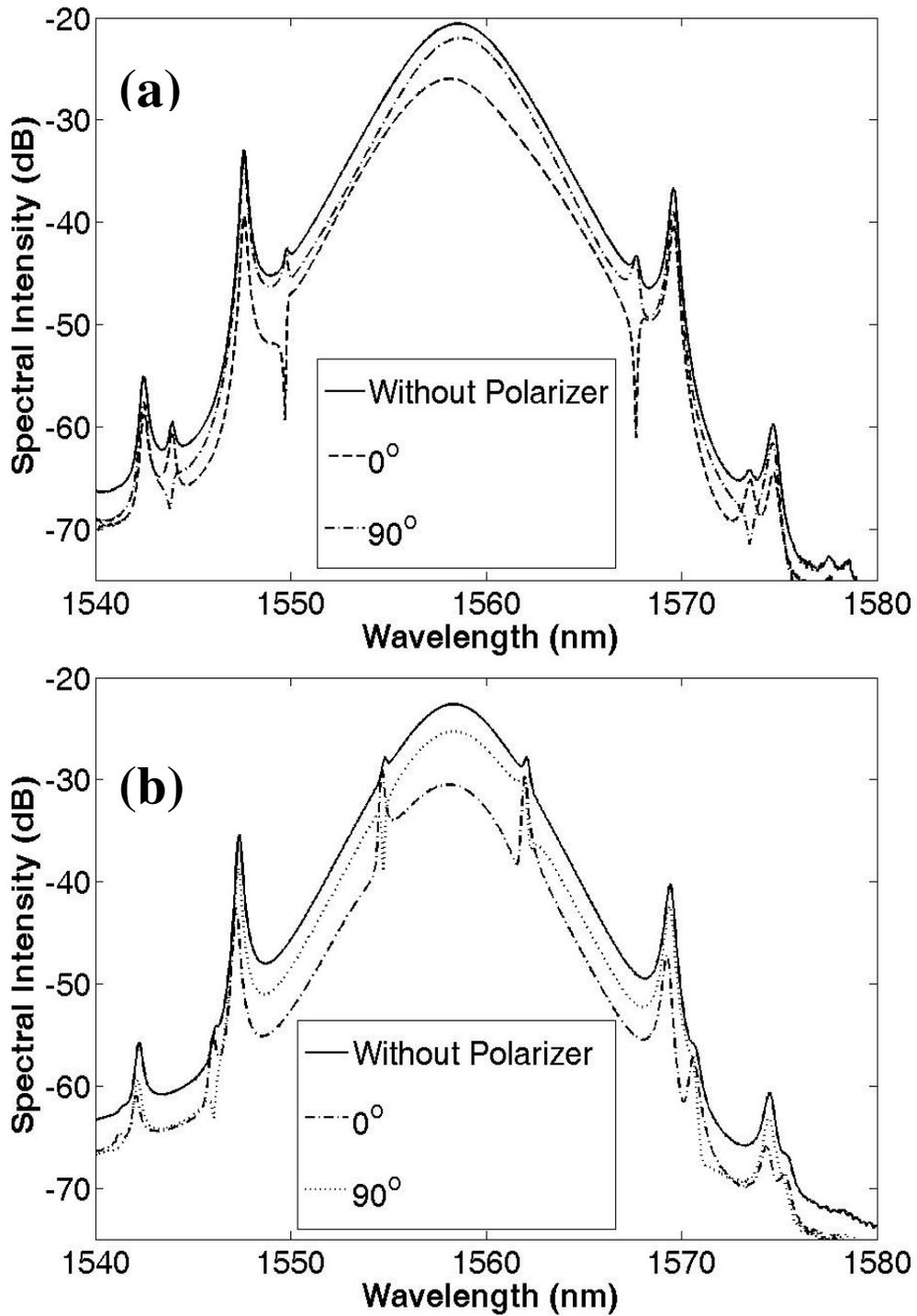

**Figure 3.2**: Optical spectra of the phase locked vector solitons of the laser measured without passing and passing through a polarizer: (a) and (b) were measured under different linear cavity birefringence.

To verify our experimental observations and determine the extra sideband





formation mechanism, we further numerically simulated the effects of the FWM between the two polarization-components of a vector soliton formed in the laser. We used a round-trip model as described in [115] for the simulations. Briefly, we used the following coupled extended Ginzburg-Landau equations to describe the pulse propagation in the weakly birefringent fibers:

$$\frac{\partial u}{\partial z} = i\beta u - \delta \frac{\partial u}{\partial t} - \frac{ik''}{2}\frac{\partial^2 u}{\partial t^2} + \frac{k'''}{6}\frac{\partial^3 u}{\partial t^3} + i\gamma\left(|u|^2 + \frac{2}{3}|v|^2\right)u + \frac{i\gamma}{3}v^2 u^* + \frac{g}{2}u + \frac{g}{2\Omega_g^2}\frac{\partial^2 u}{\partial t^2}$$

$$\frac{\partial v}{\partial z} = -i\beta v + \delta \frac{\partial v}{\partial t} - \frac{ik''}{2}\frac{\partial^2 v}{\partial t^2} + \frac{k'''}{6}\frac{\partial^3 v}{\partial t^3} + i\gamma\left(|v|^2 + \frac{2}{3}|u|^2\right)v + \frac{i\gamma}{3}u^2 v^* + \frac{g}{2}v + \frac{g}{2\Omega_g^2}\frac{\partial^2 v}{\partial t^2}$$
(1)

*u* and *v* are the normalized envelopes of the optical pulses along the two orthogonal polarized modes of the optical fiber. $2\beta = 2\pi\Delta n/\lambda$ is the wave number difference between the two modes and $L_b = \lambda/\Delta n$ is the beat length. $2\delta = 2\beta\lambda/2\pi c$ is the inverse group velocity difference. $k''$ is the second order dispersion coefficient; $k'''$ is the third order dispersion coefficient and $\gamma$ represents the nonlinearity of the fiber. *g* is the saturable gain coefficient of the fiber and $\Omega_g$ is the bandwidth of the laser gain. For undoped fibers $g = 0$; for erbium doped fiber, we considered its gain saturation as

$$g = G\exp\left[-\frac{\int(|u|^2 + |v|^2)dt}{P_{sat}}\right]$$
(2)

where *G* is the small signal gain coefficient and $P_{sat}$ is the normalized saturation energy. The saturable absorption of the SESAM is described by the rate equation [116]:

$$\frac{\partial l_s}{\partial t} = -\frac{l_s - l_0}{T_{rec}} - \frac{|u|^2 + |v|^2}{E_{sat}}l_s$$
(3)





where $T_{rec}$ is the absorption recovery time, $l_0$ is the initial absorption of the absorber, and $E_{sat}$ is the absorber saturation energy. To make the simulation possibly close to the experimental situation, we used the following parameters: $\gamma = 3$ W$^{-1}$km$^{-1}$, $\Omega_g = 24$ nm, $P_{sat} = 100$ pJ, $k''_{SMF} = -23$ ps$^2$/km, $k''_{EDF} = -13$ ps$^2$/km, $k''' = -0.13$ ps$^3$/km, $E_{sat} = 1$ pJ, $l_0 = 0.15$, and $T_{rec} = 6$ ps, Cavity length $L = 10$ m.

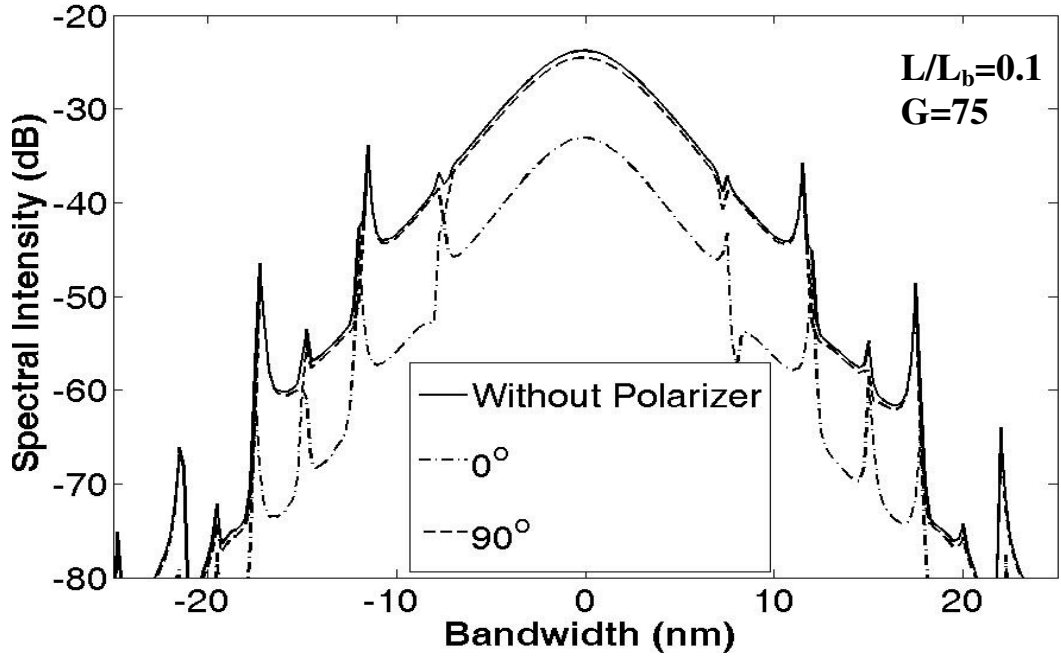

**Figure 3.3**: Numerically calculated optical spectra of the vector solitons formed in fiber ring lasers.

**Figure 3.3** shows the results obtained. Extra spectral sidebands appeared clearly on the vector soliton spectrum of the laser. In particular, the extra sidebands of the orthogonal soliton polarization components exhibited out-of-phase intensity variations. To verify that the extra spectral sidebands were caused by the FWM between the orthogonal soliton polarization components, we deliberately removed the coherent coupling terms in our simulations.





Without the FWM terms no extra spectral sidebands were observed. Numerically, it was also noticed that the appearance of the extra sidebands is related to the small linear cavity birefringence. When the linear cavity birefringence is set zero, although strong energy exchange exists between the two polarization components, no extra sidebands were observed, instead the overall soliton spectrum exhibits "peak-dip" alternation as the soliton propagates in cavity. Moreover, numerical simulations have also exhibited the dependence of extra sideband positions with the linear cavity birefringence.

The numerical simulations well reproduced the extra spectral sidebands and confirmed that their appearance is indeed caused by the FWM between the orthogonal soliton components. Based on the numerical model we further investigated the FWM interaction between the soliton polarization components and its impact on the vector soliton. Numerically it was observed that as a result of the weak linear cavity birefringence, FWM between the two vector soliton components occurred. The FWM caused an antiphase type of periodic pulse intensity variation between the two orthogonally polarized soliton components, and the stronger the linear cavity birefringence the weaker the periodic pulse intensity variation. However, independent of the cavity birefringence the pulse intensity of the vector soliton always remained constant. The observed features of the vector solitons could be easily understood. Due to small linear cavity birefringence, coherent coupling between the two polarization components of a vector soliton can no longer be neglected. Its existence causes coherent energy exchange between the two orthogonal soliton polarization components. Nevertheless, as far as the linear cavity birefringence is not zero, energy





exchange does not occur at whole soliton spectrum, but only at certain wavelengths where the phase matching condition is fulfilled under the aid of the laser cavity, which then leads to the formation of the discrete extra spectral sidebands. However, as the FWM is a parametric process and occurs between the internal components of a vector soliton, its appearance only causes an antiphase periodic intensity variation between the coupled soliton components but not the intensity of the vector soliton.

In conclusion, we have experimentally observed the evidence of coherent interaction among the two orthogonal polarizations of a vector soliton. A novel type of spectral sideband generation on the soliton spectra of the phase locked vector solitons in a passively mode-locked fiber ring laser has been generated due to such coherent interaction. Further polarization resolved study on the soliton spectrum revealed that the new sidebands were caused by the coherence energy exchange between the two orthogonal polarization components of the vector solitons. Numerical simulations have confirmed our experimental observation.

## 3.2    Incoherent Interaction

Optical solitons were first experimentally observed in SMF by Maulenauer *et al.* in 1980 [2]. The formation of the solitons was a result of the balanced interaction between the effects of anomalous fiber dispersion and the pulse SPM. To observe the solitons it is necessary that the intensity of a pulse be above a threshold where the nonlinear length of the pulse becomes comparable with the dispersion length. Apart from SPM, theoretical studies have also





shown that the incoherent interaction, or specifically XPM, could lead to soliton formation. Soliton formation through XPM was known as the induced soliton formation. An important potential application of the effect is the light controlling light. Various cases of soliton formation in SMF caused by XPM were predicted, these include the formation of a bright soliton in the normal fiber dispersion regime supported by a dark soliton in the anomalous dispersion regime [117,118], and bright solitons formation in anomalous fiber dispersion regime supported by each other through XPM [119,120]. Spatial soliton formation through XPM has been experimentally observed [121]. However, to the best of our knowledge, no induced temporal soliton formation experiments have been reported. In Chapter 3, we report on the experimental observation of induced solitons in a passively mode-locked fiber laser. Using a birefringence cavity fiber laser, we observed that due to the cross coupling between the two orthogonal polarization components, if a strong soliton is formed along one principal polarization axis, a weak soliton can always be induced along the orthogonal polarization axis. Especially, the intensity of the weak soliton could be so weak that it alone cannot form a soliton by the SPM. Numerical simulations have well supported the experimental observations.

Our fiber laser has nearly the same experimental parameters with the **Figure 3.1**. However, the cavity birefringence is no longer weak but at a moderation value through adjusting the polarization controllers. As no polarizer was used in the cavity, due to the weak birefringence of the fibers the cavity exhibited obvious birefringence features, e. g. varying the linear cavity birefringence we could observe the various types of vector solitons in the laser [114]. In order to





identify features of the vector solitons, we explicitly investigated their polarization resolved spectra under various experimental conditions. To measure their polarization resolved spectra, we let the laser output first pass through a rotatable external cavity polarizer, based on the measured soliton intensity change with the orientation of the polarizer we then identify the long and short polarization ellipse axes of the vector solitons. In our measurements we found that apart from vector solitons with comparable coupled orthogonal polarization components, vector solitons with very asymmetric component intensity also exist. **Figure 3.4** shows for example two cases experimentally observed. **Figure 3.4**a shows a case that was measured under laser operation with a relatively large cavity birefringence. In this case the spectral intensity difference between the two orthogonal polarization directions at the center soliton wavelength is more than 30 dB. The soliton nature of the strong polarization component is obvious as characterized by the existence of the Kelly sidebands. Kelly sidebands have also appeared on the weak polarization component. We emphasize the different locations of the Kelly sidebands along different polarization directions. It excludes the possibility that the sidebands on the spectrum of the weak component were produced due to an experimental artifact. The first order Kelly sidebands of the strong component are located at 1547.4 nm and 1569.5 nm, respectively; those of the weak component are at 1546.8 nm and 1568.9 nm. The separations of both sets of sidebands are the same. The appearance of Kelly sidebands on spectrum of the weak component suggests that it is also a soliton. In particular, due to the large cavity birefringence the solitons formed along the two orthogonal polarization





directions have different center wavelengths. Therefore, their Kelly sidebands have different locations.





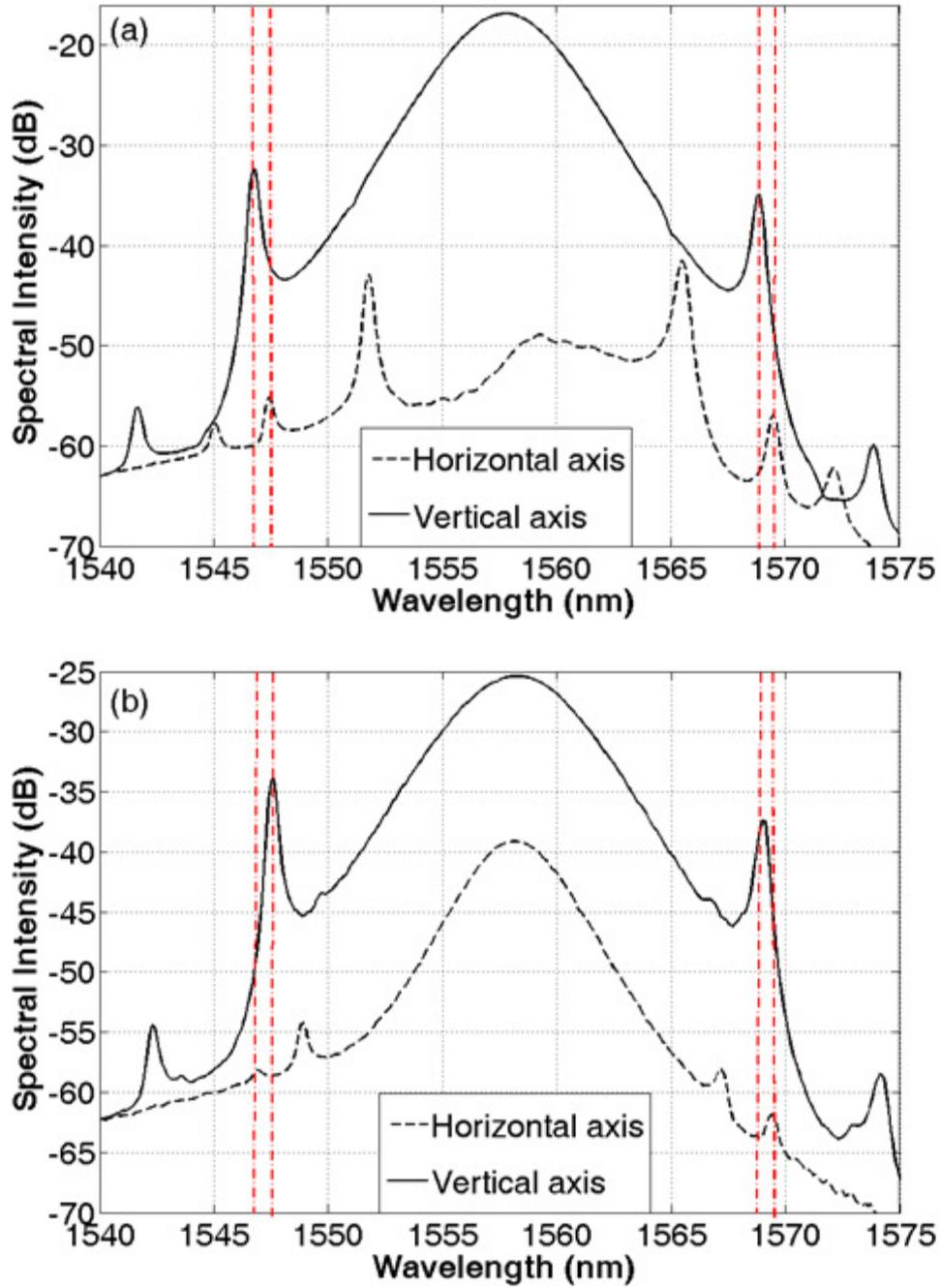

**Figure 3.4**: Polarization resolved optical spectra of the vector solitons experimentally observed. (a): Obtained under large cavity birefringence. (b) Obtained under relatively weak cavity birefringence.





Using a commercial auto-correlator we measured the soliton pulse width. Assuming a Sech$^2$ pulse profile it is about 1 ps. If only the SPM is considered, we estimate that the peak power of the fundamental solitons in our laser is about 24 W. This is well in agreement of the experimentally measured strong component soliton peak power of about 25 W. The experimentally measured weak component soliton peak power is only 0.5 W. Obviously with the intensity of the weak pulse it is impossible to form a soliton. The weak soliton should be an induced soliton.

Through adjusting the intra cavity polarization controller the net cavity birefringence could be changed. Another situation as shown in **Figure 3.5**b where both solitons have the same center wavelength was also obtained. Even in the case the two solitons have different Kelly sidebands. In particular, the first order Kelly sidebands of the weak soliton have slightly larger separation than that of the strong soliton, indicating that the induced soliton has a narrower pulse width than that of the strong soliton [122, 123].

To confirm the experimental observation, we numerically simulated the operation of the laser. We used the coupled Ginzburg-Landau equations to describe the pulse propagation in the weakly birefringent fibers. To make the simulation possibly close to the experimental situation, we used the following parameters $\gamma = 3$ W$^{-1}$km$^{-1}$, $\Omega_g = 24$ nm, $P_{sat} = 100$ pJ, $k''_{SMF} = -23$ ps$^2$/km, $k''_{EDF} = -13$ ps$^2$/km, $k''' = -0.13$ ps$^3$/km, $E_{sat} = 0.6$ nJ, $l_0 = 0.15$, and $T_{rec} = 6$ ps.

**Figure 3.5** shows the simulations obtained under different cavity birefringence. **Figure 3.5**a shows the case of the laser with a beat length $L_b = 0.1$ m. In this





case the strong soliton can either be formed along the slow or the fast axis of the cavity. Associated with the strong soliton there is always a weak soliton induced in the orthogonal polarization direction. The induced soliton has different central wavelengths, which causes that the Kelly sidebands of them have different locations. However, the wavelength shifts of their sidebands to the central soliton wavelength are the same, just like the experimental observation. **Figure 3.5**b shows a case of the cavity with a beat length of $L_b$ = 10 m. Due to the small cavity birefringence, the induced soliton has always exactly the same central wavelength as the strong soliton. Numerically we found that as $L_b$ changed from 10 m to 100 m, the strong soliton swapped from the slow axis to the fast axis of the cavity as a result of the polarization instability [124]. Nevertheless, in both cases the first order Kelly sidebands of the weak soliton have slightly larger separation than that of the strong soliton.

To verify that the weak soliton is induced by the strong soliton through XPM, the XPM terms were deliberately removed from the simulations. It was found that in this case the weak component kept continuously fading away, no stable soliton pulse could be formed. We note that apart from the Kelly sidebands, in **Figure 3.5**, there are some other discrete sharp spectral peaks. We have also numerically identified their formation as caused by the four-wave-mixing (FWM) between the two polarization-components [125]. By removing the FWM terms from Equation (1), these spectral sidebands completely disappeared from the numerically calculated soliton spectra. Like the experimental observations, the FWM sidebands are more pronounced on the spectra of the induced solitons due to their weak intensity. Especially, the first





order of the FWM sidebands appeared between the first order Kelly sidebands and the central soliton wavelength, and their exact positions varied with the cavity linear birefringence.

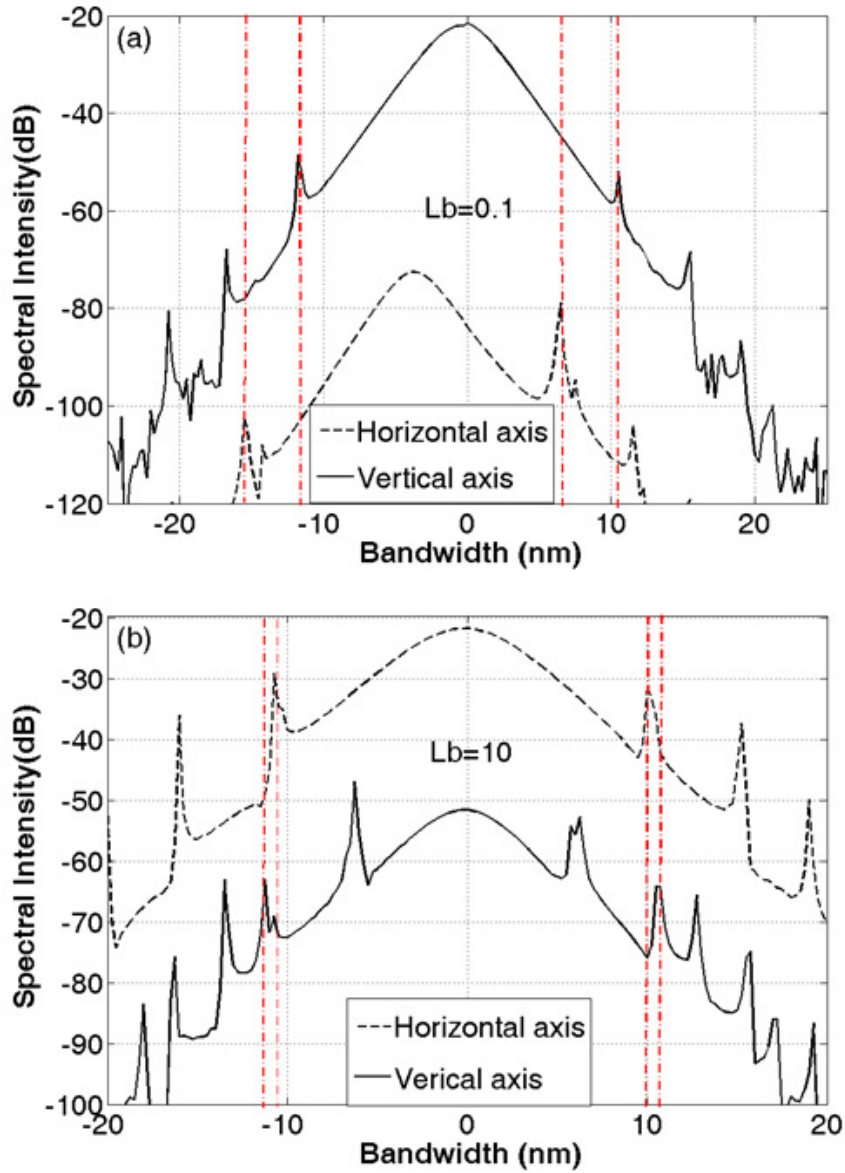

**Figure 3.5**: Numerically calculated optical spectra of the vector solitons. (a): Cavity beat length $L_b$ = 0.1 m. (b) Cavity beat length $L_b$ = 10 m. Pump strength $G_0$ = 80.





In conclusion, formation of induced temporal solitons has been experimentally observed in a passively mode-locked fiber laser with birefringence cavity. It was found that the induced solitons were formed by the XPM between the two orthogonal polarization components of the birefringence laser, and the induced solitons could either have the same or different soliton frequency to the inducing soliton. As the induced solitons always have the same group velocity as that of the inducing soliton, they form vector solitons in the laser. To our knowledge, this is the first experimental observation of temporal induced solitons.





# Chapter 4. High-order polarization-locked vector solitons

Actually, vector solitons discussed in Chapter 3 belong to the family of fundamental order vector soliton, i.e., each polarization component intensity distribution having the shape of Sech$^2$ profile with single hump. Previous studies on coupled NLSEs and the quintic complex Ginzburg-Landau equation have predicted that higher order soliton (also called as bound solitons) could be formed as a result of direct soliton interaction [126-131]. In contrast with the fundamental order soliton, the intensity profile high order soliton has more than one hump. Although the dynamics of high order scalar soliton in a fiber laser has been well-known, whether high order vector solitons exist or how they evolve was a question. In Chapter 4, a novel form of high order vector soliton is experimentally investigated and numerically confirmed. Moreover, similar to the fundamental order vector soliton, its polarization could be still locked under suitable condition. Finally, the concept of high order vector soliton could be extended to the dissipative/non-Hamiltonian system where the gain/loss effect dominates over the dispersion/nonlinearity.

Soliton as a stable localized nonlinear wave has been observed in various physical systems and been extensively studied [40]. Optical solitons were first experimentally observed in SMF by Mollenauer *et al*. in 1980 [2]. It was shown that dynamics of the solitons could be well described by the NLSE, a paradigm equation governing optical pulse propagation in ideal SMFs. However, in reality a SMF always supports two orthogonal polarization modes. Taking fiber





birefringence into account, it was later found that depending on the strength of fiber birefringence, different types of vector solitons, such as the group velocity locked vector solitons, the rotating polarization vector solitons, and the phase locked vector solitons [114], could also be formed in SMFs.

Optical solitons were also observed in mode-locked fiber lasers. Pulse propagation in a fiber laser cavity is different from that in a SMF. Apart from propagating in the fibers that form the laser cavity, a pulse propagating in a laser also subjects to actions of the laser gain and other cavity components. Dynamics of solitons formed in a fiber laser is governed by the Ginzburg-Landau equation, which takes account of not only the fiber dispersion and Kerr nonlinearity, but also the laser gain and losses. However, it was shown that under suitable conditions solitons formed in fiber lasers have analogous features to those of solitons formed in SMFs. Furthermore, vector solitons were also predicted in mode-locked fiber lasers and confirmed experimentally recently [114].

Among the various vector solitons formed in mode-locked fiber lasers or SMFs, the phase locked one has attracted considerable attention. Back to 1988 Christodoulides and Joseph first theoretically predicted a novel form of phase locked vector soliton in birefringent dispersive media [109], which is now known as a high order phase locked vector soliton in SMFs. The fundamental form of the phase locked temporal vector solitons was recently experimentally observed [114]. However, to the best of our knowledge, no high order temporal vector solitons have been demonstrated. Numerical studies have shown that the high order phase locked vector solitons are unstable in SMFs [124]. In Chapter





4, we report on the experimental observation of a stable phase-locked high order vector soliton in a mode-locked fiber laser. Multiple high order vector solitons with identical soliton parameters coexisting in laser cavity and harmonic mode-locking of the high order vector solitons were also observed. Moreover, based on a coupled Ginzburg-Landau equation model we show numerically that phase locked high order vector solitons are stable in mode-locked fiber lasers.

The experimental setup is shown in **Figure 4.1**. The fiber laser has a ring cavity consisting of a piece of 4.6 m EDF with GVD parameter 10 ps/km/nm and a total length of 5.4 m standard SMF with GVD parameter 18 ps/km/nm. Mode-locking of the laser is achieved with SESAM. Note that within one cavity round-trip the pulse propagates twice in the SMF between the circulator and the SESAM. A polarization independent circulator was used to force the unidirectional operation of the ring and simultaneously to incorporate the SESAM in the cavity. The laser was pumped by a high power Fiber Raman Laser source (BWC-FL-1480-1) of wavelength 1480 nm. A 10% fiber coupler was used to output the signals. The SESAM used is made based on GaInNAs quantum wells. It has a saturable absorption modulation depth of 5%, a saturation fluence of 180 µJ/cm$^2$ and a recovery time of 10 ps. The central absorption wavelength of the SESAM is at 1550 nm.





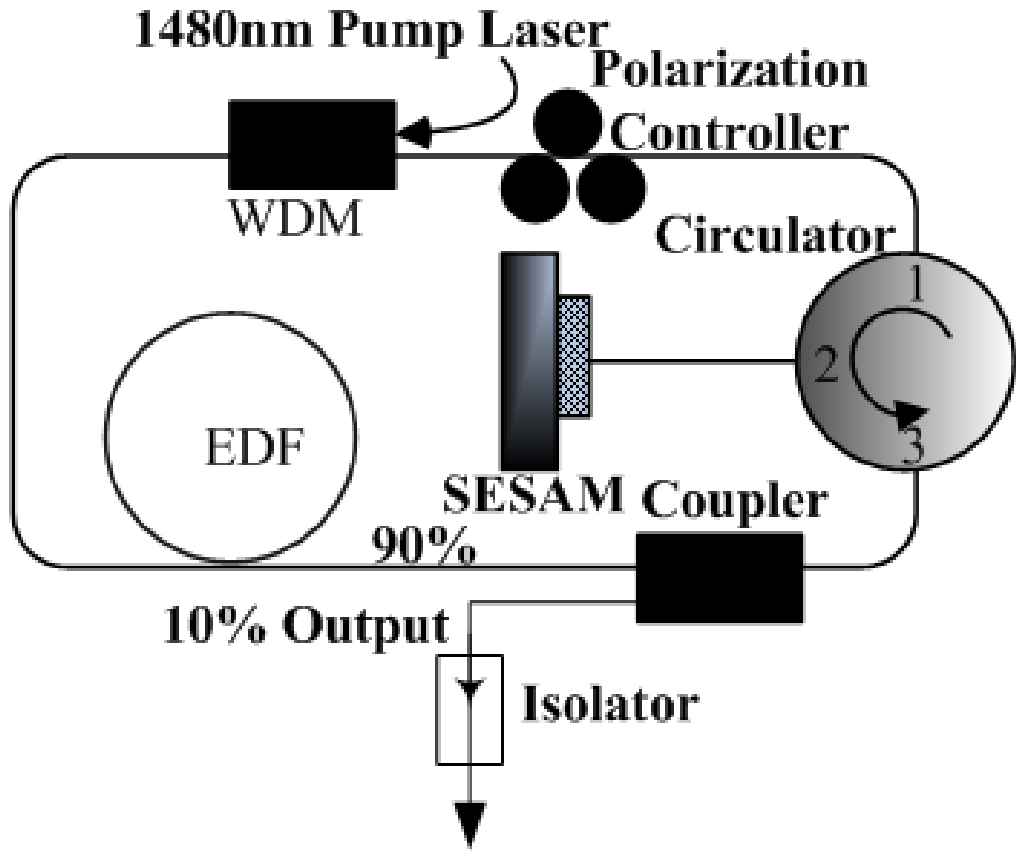

**Figure 4.1:** Schematic of the vector soliton fiber laser.

As no polarizer was used in the cavity, depending on the net cavity linear birefringence, various types of vector solitons such as the group velocity locked vector solitons, the polarization rotating vector solitons, and the fundamental phase locked vector solitons were obtained in the laser. Especially, we found that the experimentally observed features of these vector solitons could be well described by an extended coupled Ginzburg-Landau equation model, which also considered effects of the saturable absorber and the laser cavity [125]. Encouraged by the results we had further searched for the high order phase locked vector solitons theoretically predicted. Through splicing a fiber pigtailed





optical isolator between the output port and the external cavity measurement apparatus, which serves as suppressing the influence of spurious back reflection on the laser operation, we could indeed obtain one of such vector solitons. **Figure 4.2**a shows for example the optical spectra and autocorrelation traces of the soliton. Polarization locking of the soliton is identified by measuring the polarization evolution frequency (PEF) of the soliton pulse train [115]. No PEF could be detected. As the vector soliton has a stationary elliptic polarization, we could use an external polarizer to separate its two orthogonal polarization components. The optical spectra of the components are shown in **Figure 4.2**a. The spectra have the same central wavelength and about 10 dB peak spectral intensity differences. Both spectra display soliton sidebands. It shows that both of the components are optical soliton. In addition, coherent energy exchange between the two soliton components, represented by the appearance of spectral peak-dip sidebands [125], is also visible on the spectra. Different from the polarization resolved spectra of the fundamental phase locked vector solitons, there is a strong spectral dip at the center of the soliton spectrum of the weak component, while no such dip on the spectrum of the strong soliton component. To identify the formation mechanism of the spectral dip, we further measured the autocorrelation traces of each of the soliton components. It turned out that the weak component of the vector soliton had a double-humped intensity profile as shown in **Figure 4.2b**. The pulse width of the humps is about 719 fs if the Sech$^2$ profile is assumed, and the separation between the humps is about 1.5 ps. The strong component of the vector soliton is a single-hump soliton. It has a pulse width of about 1088 fs if the Sech$^2$ profile is assumed. The components of





the vector soliton have the pulse intensity profiles exactly as those predicted by Akhmediev *et al*. [132] and Christodoulides [109] for a high order phase locked vector soliton. Furthermore, the spectral dip at the center of the spectrum indicates that the two humps have 180° phase difference, which is also in agreement with the theoretical prediction.

Once the laser operation conditions were appropriately selected, the high order phase locked vector soliton operation was always obtained in the laser. Experimentally, multiple such vector solitons with identical soliton parameters were also obtained. Through carefully changing the pump strength one could even control the number of the vector solitons in cavity, and it did not change the structure of the vector solitons. Like the scalar solitons observed in the conventional soliton fiber lasers, harmonic mode locking of the high order phase locked vector solitons was also observed, as shown in **Figure 4.3**, where 8 such vector solitons were equally spaced in the cavity. All our experimental results show that formation of the high order phase locked vector solitons is an intrinsic feature of the fiber laser.





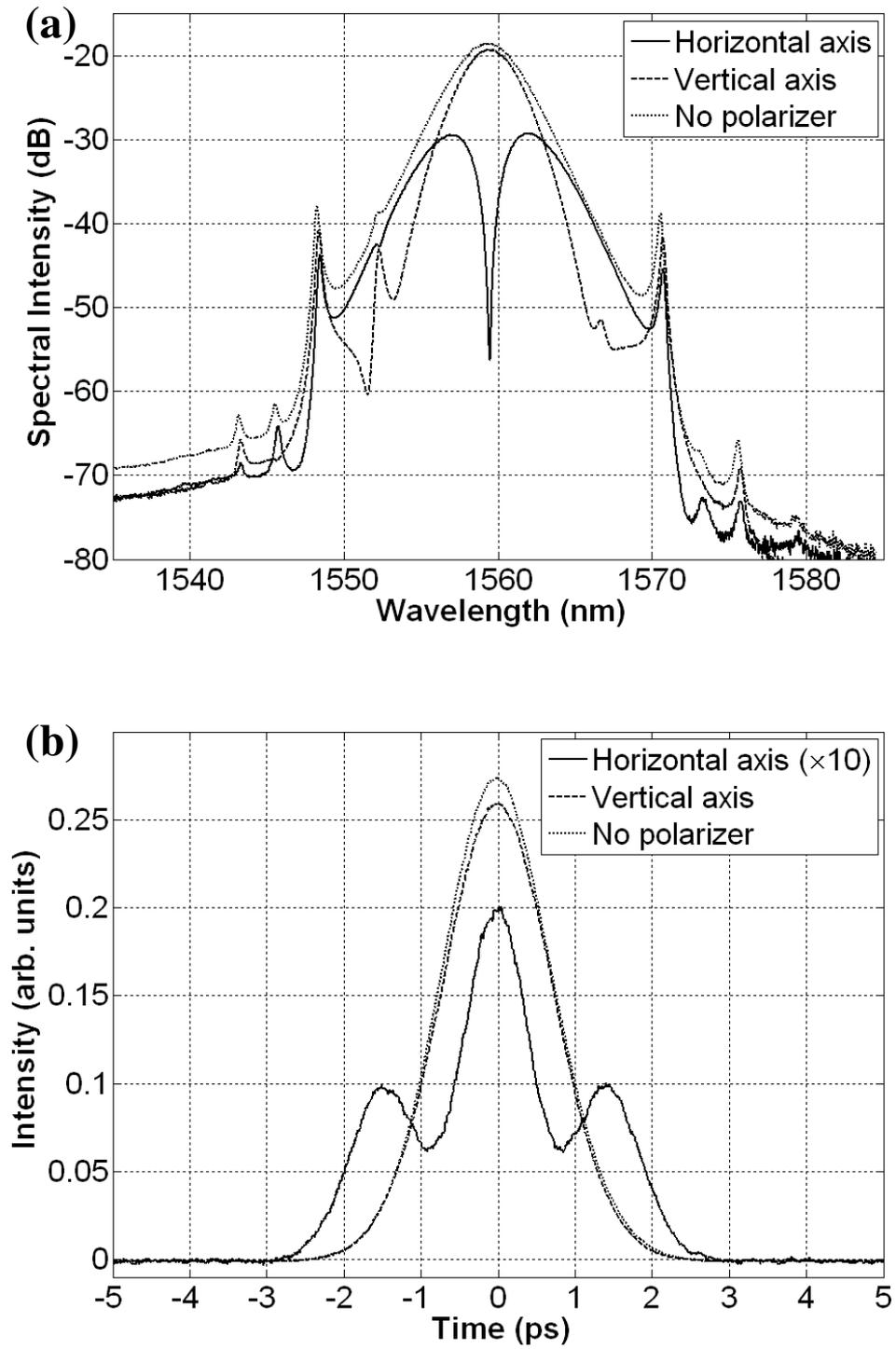

**Figure 4.2**: Polarization resolved soliton spectra and autocorrelation traces of the vector soliton observed. (a) Soliton spectra. (b) Autocorrelation traces.





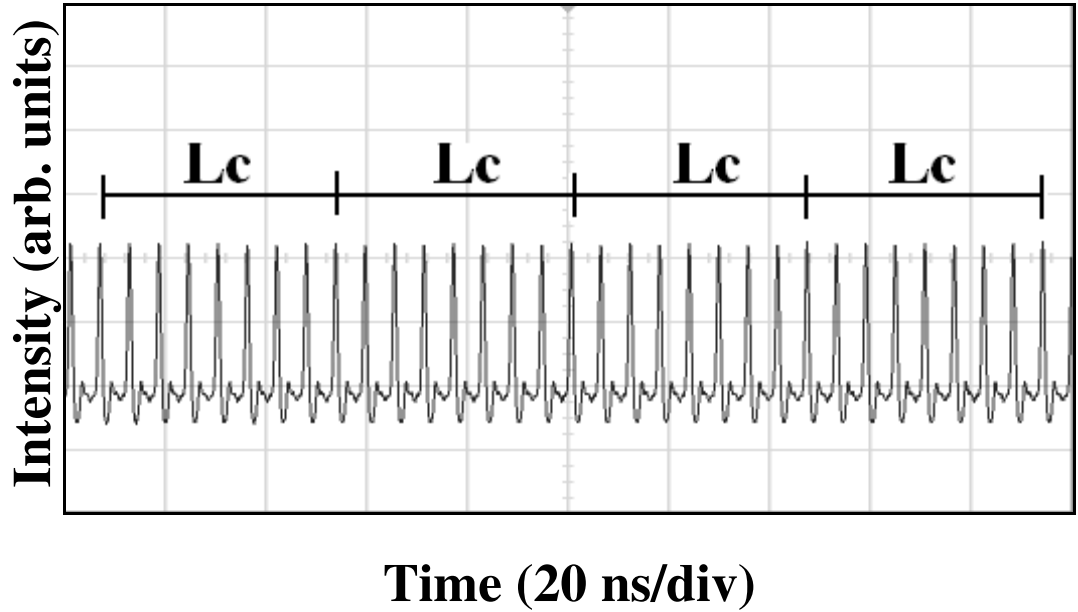

**Figure 4.3:** Oscilloscope trace of a harmonically mode-locked high order phase locked vector soliton state. Lc: cavity roundtrip time. 8 vector solitons coexist in cavity.

To confirm our experimental observations, we also numerically simulated the operation of the laser with the model used before [115]. In order to more accurately reflect the observation, we used the following parameters for our simulations for possibly matching the experimental conditions: $\gamma = 3$ W$^{-1}$km$^{-1}$, $\Omega_g = 24$ nm, $P_{sat} = 50$ pJ, $k''_{SMF} = -23$ ps$^2$/km, $k''_{EDF} = -3$ ps$^2$/km, $k''' = -0.13$ ps$^3$/km, $E_{sat} = 10$ pJ, $l_0 = 0.3$, and $T_{rec} = 2$ ps, cavity length $L = 10$ m.

We used the standard split-step Fourier technique to solve the equations and a so-called pulse tracing method to model the effects of laser oscillation [110]. We have always started our simulations with an arbitrary weak light input. **Figure 4.4** shows one of the typical results obtained. With a cavity linear





birefringence of $L_b = 3L$, a stable high order phase locked vector soliton state was obtained. The weak polarization component of the vector soliton consists of two bound solitons with pulse separation of about 1 ps, while the strong polarization component of the vector soliton is a single-hump soliton. It is interesting to see that the pulse of the strong component is only temporally overlapped with one of the two pulses of the weak component. Due to the strong cross-phase coupling between the temporally overlapped pulses, the two pulses of the weak components have different pulse widths and intensities. Propagating along the cavity, obvious coherent energy exchange between the two temporally overlapped solitons is visible. **Figure 4.4**b further gives the calculated spectra of the vector soliton components, which also show that the phase difference between the two bound solitons of the weak component is 180°.

Depending on the laser parameter selections, other high order phase locked vector soltions, such as the one with both soliton components having a double-humped structure, were also numerically obtained. We note that similar high order vector solitons were also predicted for pulse propagation in weakly birefringent fibers, but they are unstable. However, we found that all the numerically obtained high order phase locked vector solitons were stable in the laser. We believe that the different stability feature of the high order phase locked vector solitons in fiber and in fiber lasers could be traced back to their different soliton nature. While the soliton formed in a SMF is essentially a Hamiltonian soliton, the one formed in a fiber laser is a dissipative soliton, which is in fact a strong attractor of the laser system. The formation of multiple





identical high-order vector solitons in the fiber laser clearly shows the dissipative nature of the formed vector solitons.





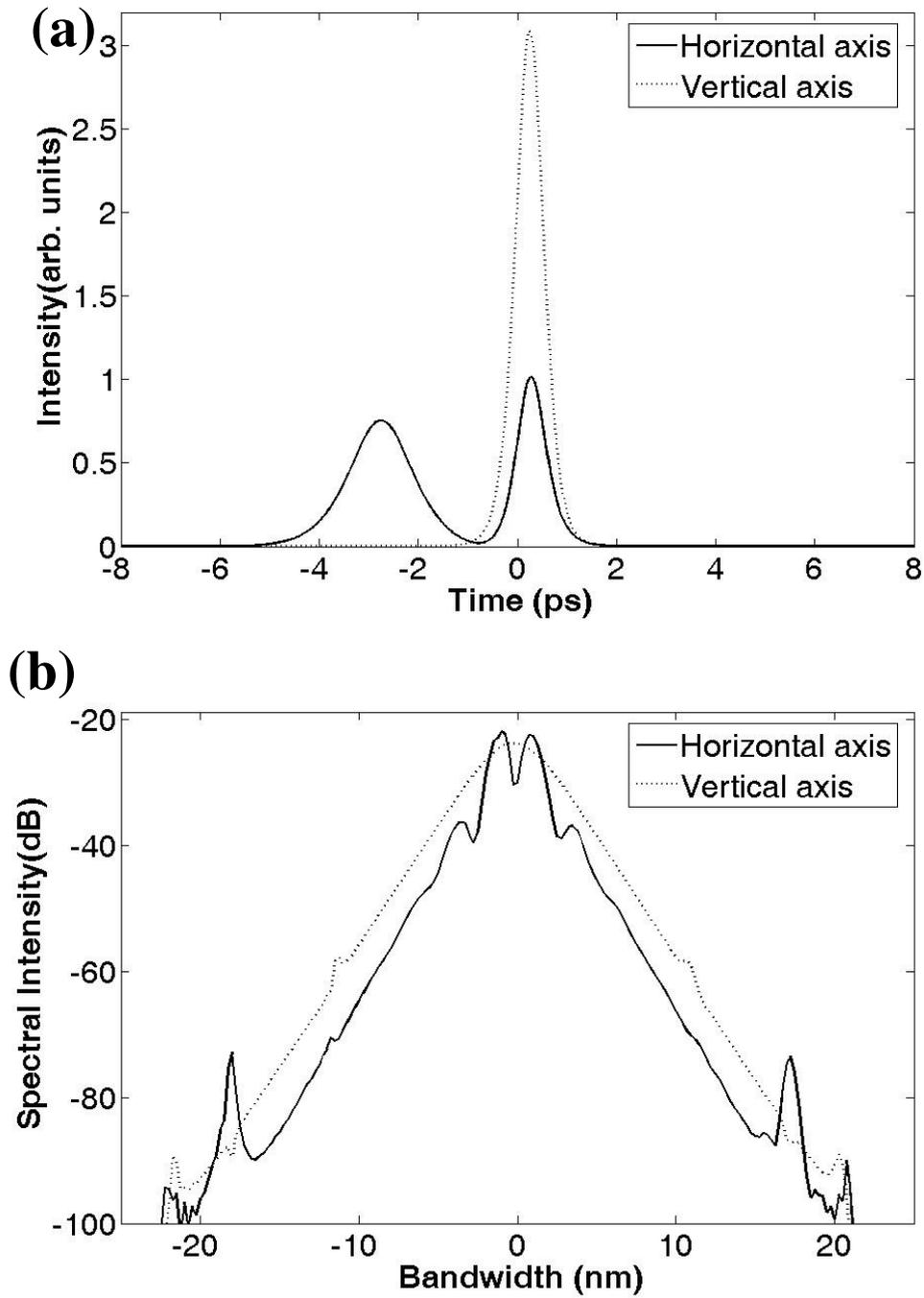

**Figure 4.4**: A stable high order phase locked vector soliton state numerically calculated. (a) Soliton intensity profiles of the two orthogonally polarized components. (b) The corresponding optical spectra of (a).





In conclusion, we have first experimentally observed a novel type of high order phase locked vector soliton in a passively mode-locked fiber laser. The high order vector soliton is characterized by that its two orthogonal polarization components are phase locked, and while the stronger polarization component is a single hump pulse, the weaker component has a double-humped structure with 180° phase difference between the humps. Our experimental result firstly confirmed the theoretical predictions on the high order phase locked vector solitons in birefringent dispersive media.





# Chapter 5.  Dissipative Vector Solitons

Although vector solitons investigated in Chapter 3 and Chapter 4 have different soliton profile, they have one common feature: nearly transform limited in that the fiber laser cavities are made of purely anomalous dispersion fibers and belonging to conservative NLSE solitons are formed. Their generations only require the balance between the anomalous dispersion and nonlinearity. However, in the normal dispersion regime, the formed solitons are much different in that they are largely chirped and more importantly, their formations need the mutual balance between gain, loss, normal dispersion and nonlinearity.

Dissipative solitons (DSs) are stable solitary localized structures that arise in nonlinear spatially extended dissipative systems due to mechanisms of self-organization. They exist for an extended period of time, even though parts of the structure experience gain and loss of energy and/or mass. Normal dispersion fiber laser is a natural dissipative system as the gain/loss effect must be taken into consideration.

Recently, formation of DSs in pure normal-dispersion-cavity fiber lasers passively mode locked by the NPR technique was reported [16-20]. It was shown that formation of the solitons is a result of the mutual nonlinear interaction among the normal cavity dispersion, cavity fiber nonlinear Kerr effect, laser gain saturation and gain bandwidth filtering. Although the NPR mode-locking process of the lasers unavoidably affected the detailed soliton features, it was found that its effect on the soliton shaping was minor [133]. In





addition, both the experimental and theoretical studies have shown that the dissipative solitons formed in the large net positive cavity GVD regime have very different characteristics from those formed in the net negative cavity GVD fiber lasers, and the differences can be traced back to the different soliton shaping mechanisms in the two cavity dispersion regimes. Mode-locking of a fiber laser can also be achieved with other techniques, e.g. using SESAM [61, 62]. Different from the NPR mode locking, mode-locking with a SESAM requires no polarizer in the laser cavity, which potentially allows the formation of a dissipative vector soliton (DVS) in the laser cavity.

As discussed in Chapter 3 and Chapter 4, the dynamics of vector soliton is strongly related with the cavity birefringence. Under different birefringence regime, the polarization effect and coupling strength might be much different. Thus, various types of vector solitons might appear. Chapter 5 is organized as follows. When the cavity birefringence is weak, both polarizations locked and rotating DVS could be experimentally observed, which is discussed firstly in Section 5.1. When the cavity birefringence is in moderate value, dual-wavelength dissipative solitons could be obtained; while the cavity birefringence is further strong, triple- wavelength dissipative solitons would appear. The generation mechanism of multi-wavelength dissipative soliton would be elaborately discussed in Section 5.2.

## 5.1 Polarization locked and rotating DVS

Formation of both the frequency locked and phase locked DVSs in the negative cavity GVD regime were theoretically predicted by Akhmediev *et al*. [134], and





experimentally observed in a fiber laser mode-locked with a SESAM [112-114]. Different from a dissipative soliton formed in the net negative cavity GVD regime, a dissipative soliton formed in the net large positive cavity GVD regime has different soliton shaping mechanism, and furthermore is strongly frequency chirped. It would be interesting to find out whether a phase-locked DVS could be formed in a fiber laser with large net positive cavity dispersion or not. In this section we will address this question. We show experimentally that either the polarization rotating or the phase-locked DVSs can be formed in the fiber lasers. In addition, we also show that multiple vector solitons with identical soliton parameters and harmonic mode locking of the vector solitons can be formed in the fiber lasers. Numerical simulations are found in agreement of the experimental observations, which confirm the DVS formation in the fiber lasers.

Our experimental setup is shown in **Figure 5.1**. The fiber laser has a ring cavity consisting of a piece of 1.5 m EDF with a GVD of about 40.8 $ps^2$/km, a total length of 3.5 m standard single mode fiber SMF with GVD of about –23 $ps^2$/km and 18.2 m DCF with GVD of about 2.55 $ps^2$/km. The total cavity length is 23.2 m. Mode-locking of the laser is achieved with a SESAM. A polarization independent circulator was used to force the unidirectional operation of the ring and simultaneously incorporate the SESAM in the cavity. Note that within one cavity round-trip the pulse propagates twice in the SMF between the circulator and the SESAM. The laser was pumped by a high power Fiber Raman Laser source (BWC-FL-1480-1) of wavelength 1480 nm. A 10% fiber coupler was used to output the signals. The SESAM used is made based on GaInNAs quantum wells. It has a saturable absorption modulation depth of 5%, a





saturation fluence of 90 µJ/cm$^2$ and a recovery time of 10 ps. The central absorption wavelength of the SESAM is at 1550 nm.

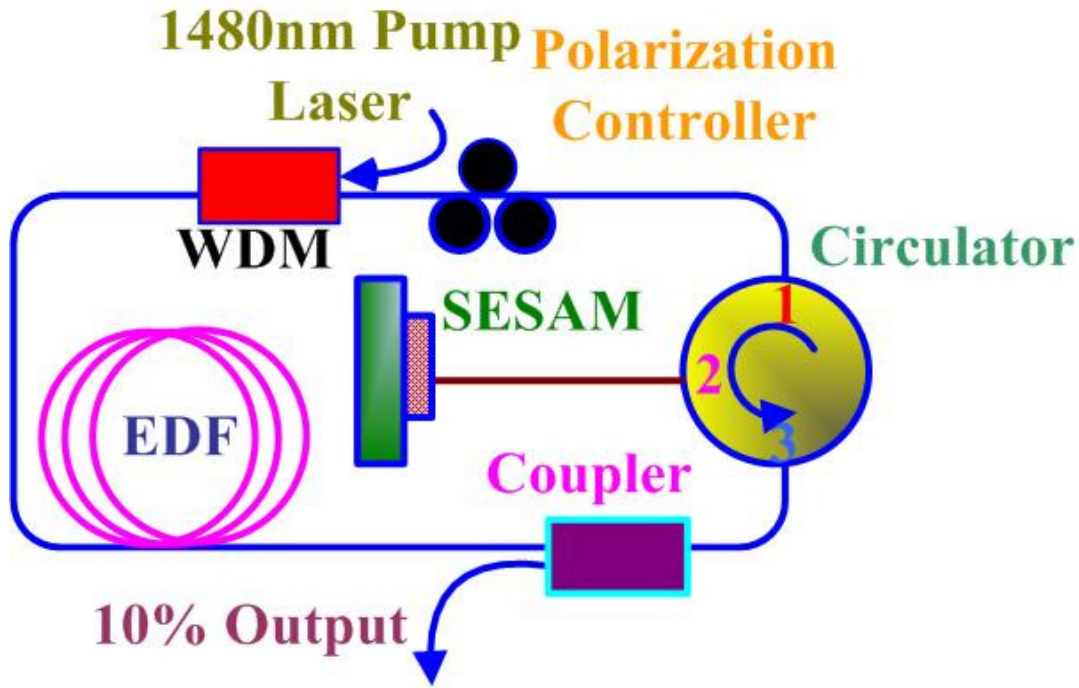

**Figure 5.1:** Schematic of the fiber laser.

This fiber laser has a typical dispersion-managed cavity with net normal cavity GVD of about 0.027 ps$^2$. To control the net cavity GVD, DCF with different lengths were inserted between the circulator and the SESAM. When the net cavity GVD was anomalous, it was observed that the spectrum of the mode-locked pulses had the typical features of those of the DVS reported in [136]. Increasing the length of the DCF, the net cavity GVD shifted to large positive values, correspondingly, the soliton operation of the laser shifted to a new dissipative soliton regime. Experimentally, the self-started mode locking of the laser could still be easily achieved at a large positive cavity GVD.





**Figure 5.2** shows a typical optical spectrum of the dissipative solitons of the laser obtained at a large net positive cavity GVD. The soliton spectrum has characteristic steep spectral edges. However, different from the dissipative solitons formed in the fiber lasers mode-locked with the NPR technique, the soliton consists of two orthogonal polarization components. To highlight the vector nature of the soliton, we let the soliton pulse train pass through a rotatable external cavity polarizer and compared the features of the pulse train before and after passing through the polarizer, either with a high speed oscilloscope or an RF-spectrum analyzer. We found that the soliton, shown in **Figure 5.2**a, was a polarization rotating vector soliton. Polarization rotation of the soliton could be easily identified e.g. by the oscilloscope trace measurement. Without passing through the external polarizer, the soliton pulse had identical pulse intensity on the oscilloscope trace for each cavity roundtrip, while after passing through the polarizer it became varying with the cavity roundtrips as shown in **Figure 5.2**b. It indicates that the polarization of the soliton rotated along the cavity. **Figure 5.2**a also shows the autocorrelation trace of the vector soliton. It has a width (FWHM) of 40 ps. If a $Sech^2$ pulse shape is assumed, the soliton pulse width is 24 ps. The 3 dB spectrum bandwidth of the soliton is 6.5 nm, which gives a time-bandwidth product about 18.9, indicating that the soliton is strongly chirped.





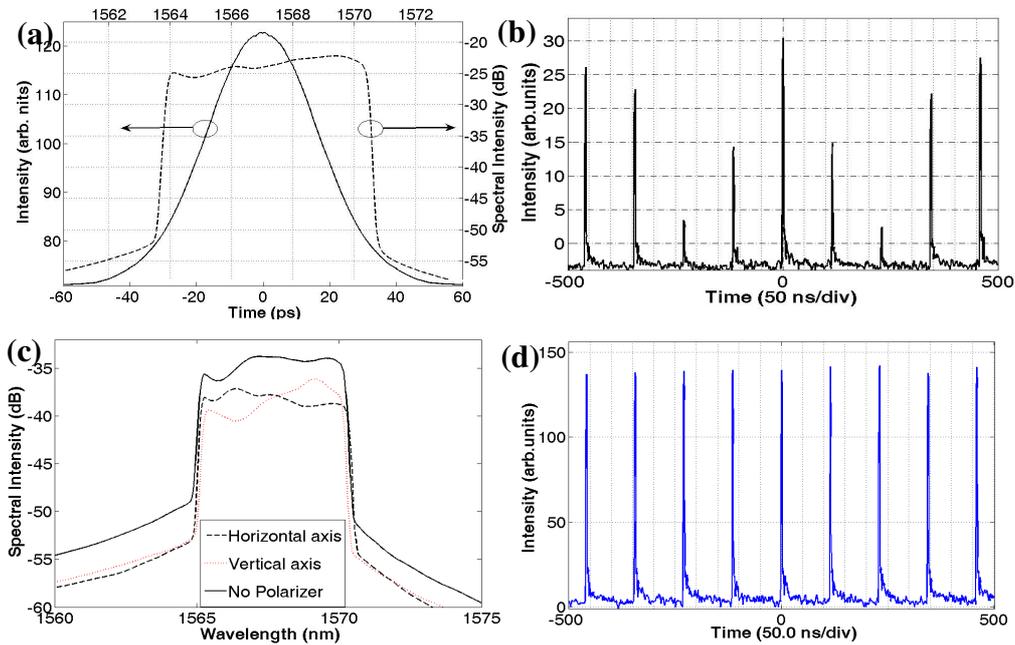

**Figure 5.2:** (a) Spectrum and corresponding autocorrelation trace of a polarization rotating DVS emission state of the laser; (b) Oscilloscope trace of (a) after passing through a polarizer; (c) Polarization resolved optical spectra of a phase locked DVS emission state of the laser; (d) Oscilloscope trace of (c) after passing through a polarizer.

Controlling the linear cavity birefringence through an intra-cavity polarization controller, polarization locked vector solitons were obtained. A polarization locked vector soliton has the characteristic that it has a fixed polarization during circulation in the laser cavity. Correspondingly, after passing through an external cavity polarizer the pulse height of such vector solitons on the oscilloscope trace would have identical value as shown in **Figure 5.2**d. For a phase locked vector soliton we could also measure its optical spectra along the long and the short polarization ellipse axes, as shown in **Figure 5.2**c. Different





from the polarization resolved dissipative vector solitons formed in the net negative cavity GVD fiber lasers, no four-wave-mixing spectral sidebands could be identified. We believe their absence could be traced back to the large frequency chirp of the soliton. The two orthogonal polarization components shown in **Figure 5.2**c have comparable spectral intensity and the same center wavelength, but clearly different spectral distributions. Experimentally, the phase locking between the two orthogonal polarizations was further confirmed by the polarization evolution frequency measurement as described in [125]. Apart from the above near circularly polarized polarization locked DVS, polarization locked DVSs with significant spectral intensity difference between the two orthogonal polarization components were also observed. In one case the spectral intensity difference at the center soliton wavelength was as large as 20 dB. We experimentally measured the peak power of the weak soliton component in the case. It was about 0.1 W. With the pulse peak power it is impossible to form a soliton in the laser. Therefore, we believe it could be an induced soliton formed by the cross coupling with the strong soliton component [136].

Adjusting the polarization controller, the central wavelength of the DVSs could be altered in a wide range from 1560 nm to 1575 nm. However, as the central soliton wavelength is increased, the formed DVS becomes less stable. This could be due to that the central absorption wavelength of the SESAM is about 1550 nm, which favors the formation of the DVS centered near 1550 nm. Multiple DVSs with identical soliton parameters were also obtained under strong pumping. In this case, harmonic mode locking was frequently obtained.





**Figure 5.3** illustrates a case where 8 DVSs were equally spaced in the cavity with a separation of about 14.5 ns. Other harmonic mode locking states of the DVS were also obtained by simply varying the pumping strength.

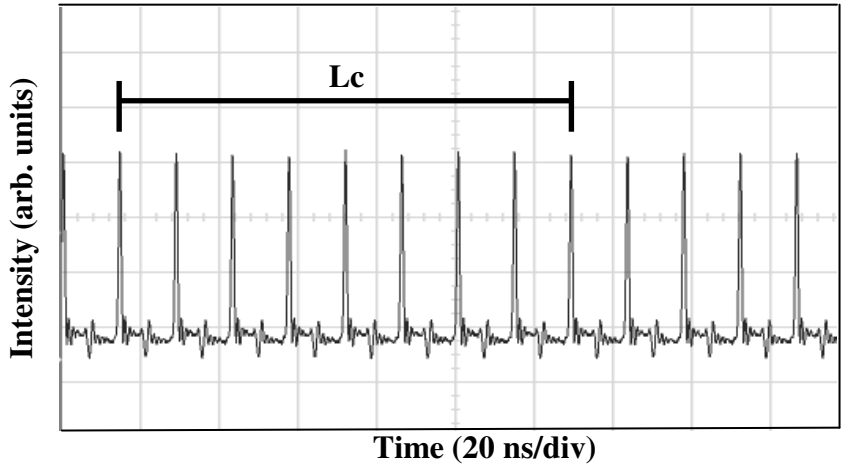

**Figure 5.3:** Oscilloscope trace of a harmonically mode-locked gain-guided vector soliton state. *Lc*: cavity roundtrip time. 8 DVS coexist in cavity.

To gain an insight into the DVS formation, we also numerically simulated the operation of the laser. We used a round-trip model to include the laser cavity effects as well as the saturable absorber effect in our simulations. To make the simulation possibly close to the experimental situation, we used the following parameters: $\gamma = 3$ W$^{-1}$km$^{-1}$, $\Omega_g = 16$ nm, $P_{sat} = 50$ pJ, $k''_{SMF} = -23$ ps$^2$/km, $k''_{EDF} = 41$ ps$^2$/km, $k''_{DCF} = 2.6$ ps$^2$/km, $k''' = -0.13$ ps$^3$/km, $E_{sat} = 35$ nJ, $l_0 = 0.2$, $T_{rec} = 2$ ps, and cavity length $L = 23.2$ m.

Numerically, it was found that under the current saturable absorber parameter selection, the formation of DVS in the laser is strongly the gain bandwidth dependent. When the gain bandwidth is large, i.e. 24 nm, DVS is unable to form. In this case the spectrum of the mode-locked pulses has a Gaussian profile.





Stable DVS is formed when the gain bandwidth is narrower than 16 nm. The narrower the gain bandwidth used for simulation, the smaller the spectrum bandwidth of the obtained DVS. **Figure 5.4**a shows a typical case of the calculated DVS evolution with the cavity roundtrips. Numerically we found that the total pulse intensity is unchanged with the cavity roundtrips, but intensity of the horizontal and the vertical component exhibits out-of-phase intensity variation, indicating that coherent energy exchange between them still exists [136]. After removing the four-wave-mixing terms from Equation (1), such an out-of-phase intensity variation between the vector soliton components then disappeared. The calculated pulse width is about 20 ps and its spectrum bandwidth is about 7 nm, which indicates that the formed DVS is strongly frequency chirped.

The result shown in **Figure 5.4** was obtained for a laser cavity with a beat length of $L_b$ = 100 m. The DVSs formed have two orthogonal polarization components with comparable intensity and coincident central wavelength. Numerically, as $L_b$ changed from 100 m to 0.1 m, i.e., increasing the cavity birefringence, intensity difference between the two orthogonal polarization components becomes larger, in accord with our experimental observations. We have also numerically investigated polarization evolutions of the formed DVSs by using the method reported in [137]. Depending on the cavity birefringence, either the polarization locked or polarization rotating DVSs were numerically obtained. Calculated based on the peak point of the DVSs, a polarization locked DVS has its polarization ellipse orientation fixed as it propagates along the cavity, while the polarization ellipse of a rotating DVS changes its orientation





along the cavity. The relative phase between the components of the polarization locked DVS at the pulse peak was always fixed at $\pi/2$. However, at the other points of the pulse it deviated from $\pi/2$, demonstrating the effect of frequency chirp. To determine how the frequency chirp of the solitons affects their polarization, we also numerically calculated the polarization states at different points of a pulse, e.g. at the pulse's peak and FWHM points. For the polarization locked DVSs it was found that the polarization ellipses calculated at different points had slightly different orientations, nevertheless, the difference remained the same along the propagation. This numerical result suggests that despite the strong frequency chirp of the DVSs, the temporal variation of their two orthogonally polarized components is always phase locked.

In conclusion, DVSs have been experimentally demonstrated in a dispersion-managed fiber laser with large net normal cavity GVD. It was found that despite of the large frequency chirp of the dissipative solitons formed in the lasers, the polarization rotating and locked DVSs could still be formed. In addition, formation of multiple DVSs with identical soliton parameters and stable harmonic DVS mode-locking are also experimentally obtained.





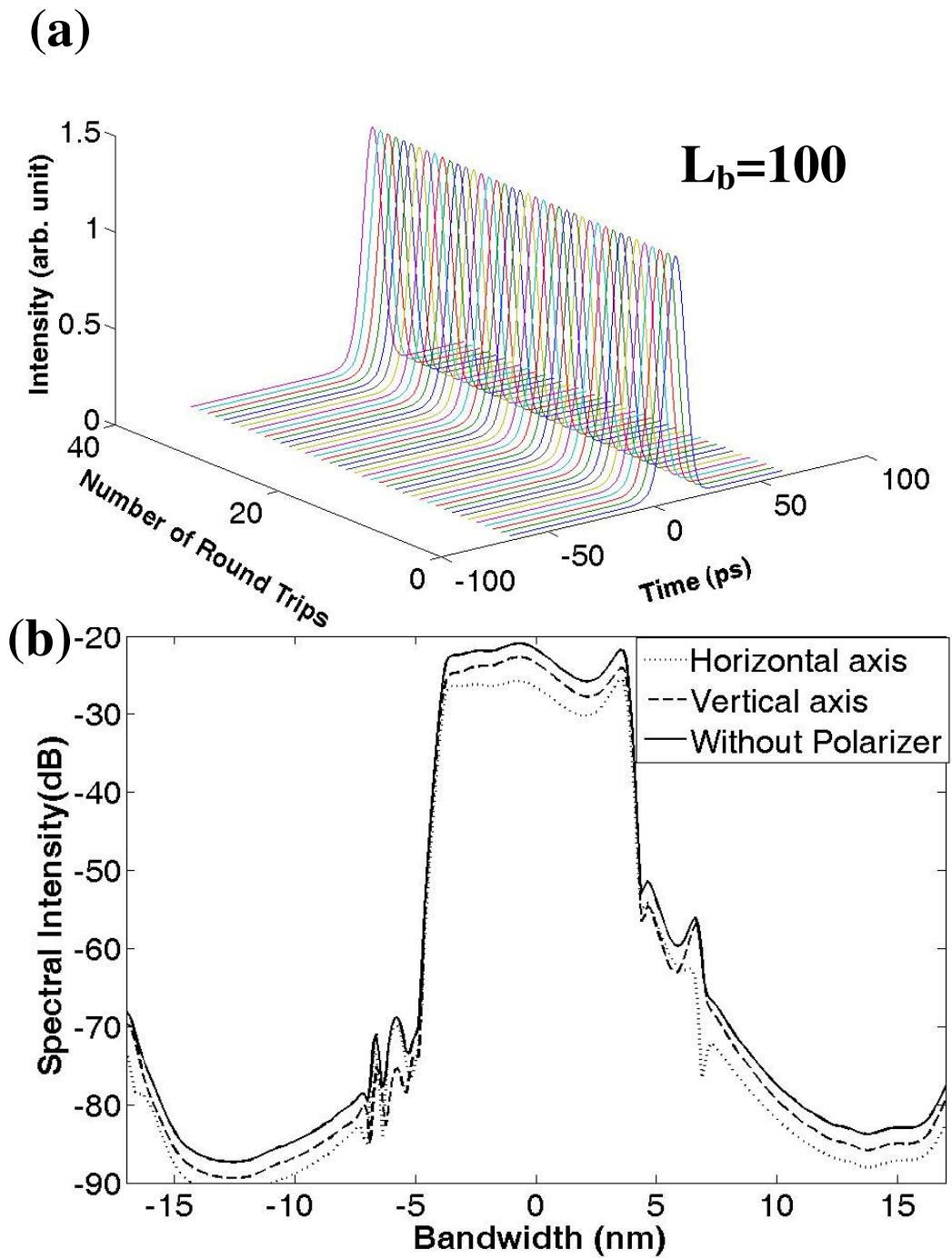

**(a)**

$L_b=100$

**(b)**

**Figure 5.4**: (a) Combined pulse intensity evolution; (b) corresponding optical spectra numerically calculated.





## 5.2    Multi-wavelength dissipative soliton

Multi-wavelength mode-locked fiber lasers have versatile applications, including fiber optic sensing, photonic component characterization and WDM optical communications. Several methods for achieving multi-wavelength mode locking have been studied. Li *et al*. reported the generation of triple-wavelength picosecond mode locked pulses using a self-seeded Fabry–Perot laser diode with fiber Bragg gratings [138]. Multiwavelength actively mode-locked fiber lasers incorporating either a single sampled fiber Bragg grating [139] or a biased semiconductor optical amplifier [140] in cavity was shown by Yao *et al*. By virtue of the NPR effect, simultaneous dual- and five-wavelength actively mode-locked erbium-doped fiber lasers at 10 GHz were demonstrated by Pan *et al*. [141]. Although multiwavelength actively mode-locked fiber lasers have the advantages such as high repetition rates, narrow linewidth, they also have the drawbacks of broad pulse width, low peak power, and expensive as a modulator is required to be inserted in the cavity. Moreover, as actively mode locked multiwavelength pulses have only weak nonlinearity, they are impossible to be shaped into optical solitons that possess the born preponderance: good stability, low time jittering, short pulse width and high peak power.

Back in 1992, Matsas *et al*. reported the experimental observation of dual-wavelength soliton emission in a fiber laser exploiting the NPR technique for mode locking [142]. A long laser cavity made of anomalous dispersion fibers was adopted in their experiment. However, no explanation on the formation mechanism of dual-wavelength solitons was given. In this section, we report on the experimental observation of multiple wavelength dissipative soliton





operation of an erbium-doped fiber laser. Formation of dissipative solitons in normal dispersion fiber lasers has recently attracted considerable attention [20-25, 144, 145]. Although in an all-normal dispersion fiber laser there is no natural balance between the actions of the fiber dispersion and fiber nonlinear optical Kerr effect, therefore, no natural NLSE soliton is formed. It was shown that as a result of the mutual interactions among the normal cavity dispersion, cavity fiber nonlinear Kerr effect, and the effective laser gain bandwidth, an optical soliton can still be formed in the laser. The formed solitons were known as the dissipative solitons [12]. Occasionally they were also called as the gain-guided solitons to distinguish from those solitons formed in the anomalous dispersion fiber lasers [40]. Single wavelength dissipative solitons have been observed in various fiber lasers [144, 145]. In a previous paper we have also reported the observation of dissipative vector solitons in a dispersion managed cavity fiber laser [20]. However, to the best of our knowledge, no multi-wavelength dissipative solitons operation of a fiber laser has so far been reported.

Our fiber laser is schematically shown in **Figure 5.5**a. It has a ring cavity made of pure normal dispersion fibers. A piece of 5.0 m EDF with GVD of –32 (ps/nm)/km was used as the gain medium, and all the other fibers are the DCF with GVD of –4 (ps/nm)/km. The cavity has a length of 13.5 m. Mode-locking of the laser is achieved with a SESAM. A polarization independent circulator was used to force the unidirectional operation of the ring and simultaneously incorporate the SESAM in the cavity. Note that within one cavity round-trip the pulse propagates twice in the DCF between the circulator and the SESAM. A





50% fiber coupler was used to output the signal, and the laser was pumped by a high power Fiber Raman Laser source (KPS-BT2-RFL-1480-60-FA) of wavelength 1480 nm. The maximum pump power can be as high as 5 W. All the passive components used (WDM, Coupler, and Circulator) were made of the DCF. An optical spectrum analyzer (Ando AQ-6315B) and a 350 MHz oscilloscope (Agilen 54641A) together with a 2 GHz photo-detector were used to simultaneously monitor the spectra and the mode locked pulse train, respectively. The SESAM used was made based on GaInNAs quantum wells. It has a saturable absorption modulation depth of 30%, a saturation fluence of 90 $\mu$J/cm$^2$ and a recovery time of 10 ps. The SESAM was pigtailed with 0.5 m DCF.

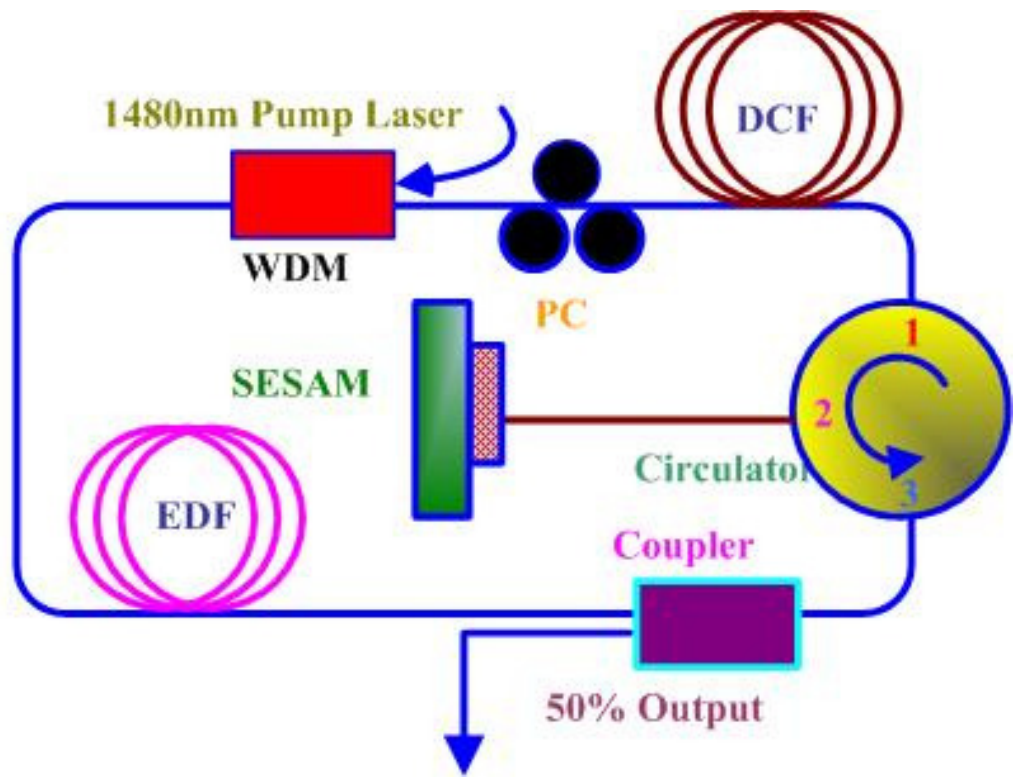

**Figure 5.5**: Schematic of the experimental setup.





**(a)**

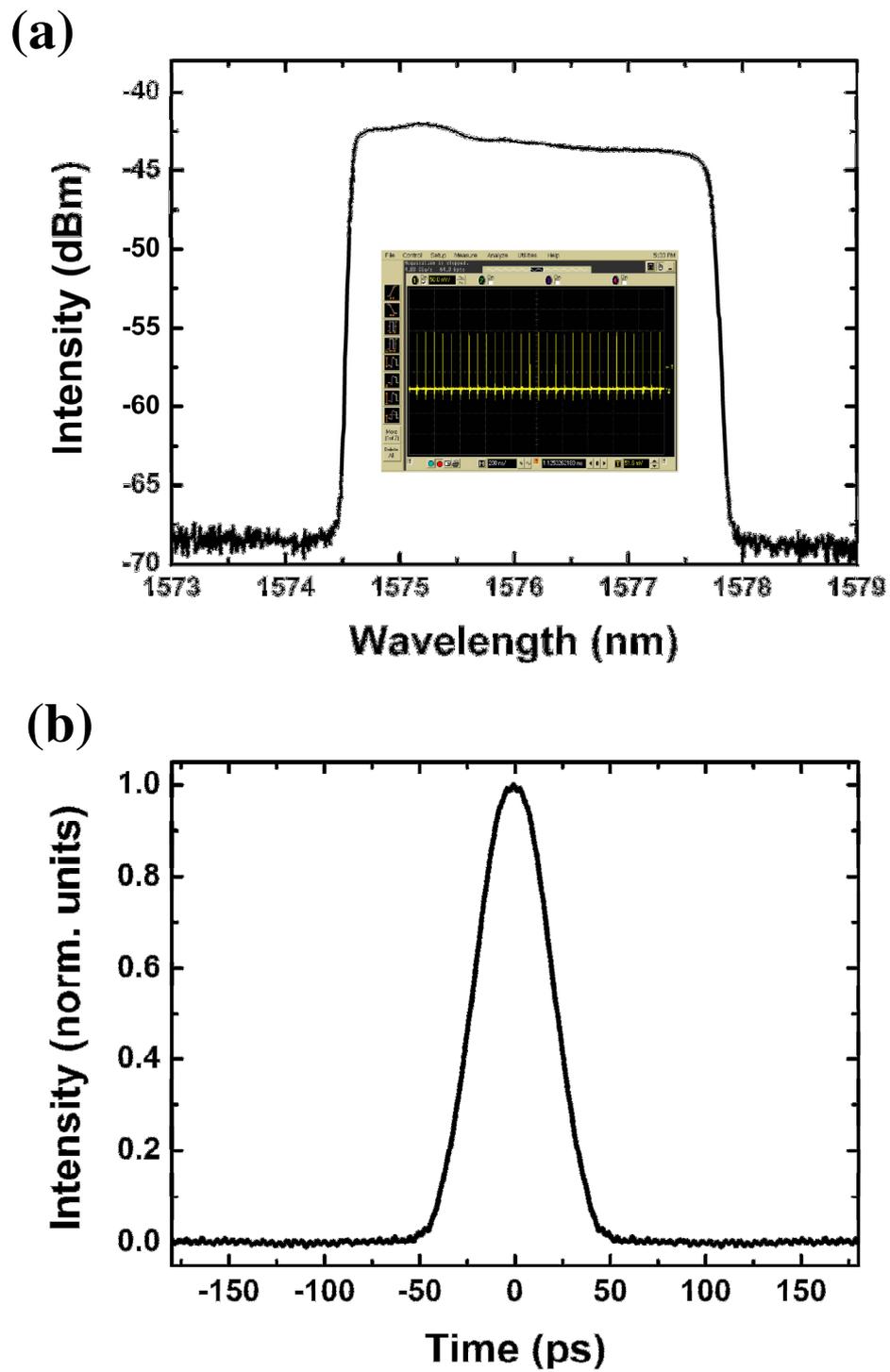

**(b)**

**Figure 5.6**: (a) Optical spectra of single wavelength dissipative soliton. Insert:

the oscilloscope trace. (b) The corresponding autocorrelation trace

Mode locking of the laser self-started as the pump power was increased above





the mode locking threshold. Immediately after the mode locking, multiple soliton pulses were generally formed in the cavity. However, through carefully decreasing the pump power, the number of soliton pulses could be reduced and eventually a single soliton operation state could be obtained. **Figure 5.6** shows a typical single soliton operation state of the laser. The optical spectrum of the pulse has the characteristic steep spectral edges, which shows that it is a dissipative soliton [20-25, 144, 145]. Based on the measured autocorrelation trace, the soliton pulse width is estimated 28.8 ps if a Sech$^2$-shape pulse is assumed. Experimentally we confirmed that the soliton pulse was linearly polarized. The center wavelength of the dissipative soliton shown in **Figure 5.6** is located at 1576.2 nm. Experimentally, adjusting the orientations of the paddles of the PC, which corresponds to changing the linear birefringence of the cavity, the central wavelength of the soliton could be varied in a wide range from 1570 nm to 1590 nm. Nevertheless, solitons with central wavelengths close to 1570 nm or 1590 nm were less stable.

Under stronger pumping, a new soliton was formed in the cavity. It was found that the features of the new soliton sensitively depended on the cavity birefringence. **Figure 5.7** shows a soliton operation state experimentally observed. **Figure 5.7**a is the optical spectrum of the laser emission. Compared with the spectrum shown in **Figure 5.6**, another steep-edge shaped spectrum with different central wavelength also appeared. **Figure 5.7**b shows the oscilloscope trace of the laser emission. There are two solitons in the cavity. Each soliton has different pulse energies as represented by the two different pulse heights in the oscilloscope trace. The two solitons also propagated with





different group velocities in the cavity. Therefore, when triggered with one soliton the other soliton then moved randomly on the oscilloscope screen, indicating that the two solitons have different group velocities in the cavity. We had further experimentally studied the polarization features of the pulses using an external cavity polarizer. Through rotating the external cavity polarizer it was identified that one soliton could be completely suppressed while the other still remained on the oscilloscope trace, associated with the suppression of one soliton pulse on the oscilloscope one squared-shaped spectrum also disappeared on the optical spectrum. **Figure 5.7** actually shows a state of the dual wavelength dissipative soliton operation of the fiber laser. Specifically, the dissipative soliton with the large pulse energy has a center wavelength of 1576.2 nm, and the one with the weak energy has a center wavelength of 1579.6 nm, and the two dissipative solitons have orthogonal polarizations.





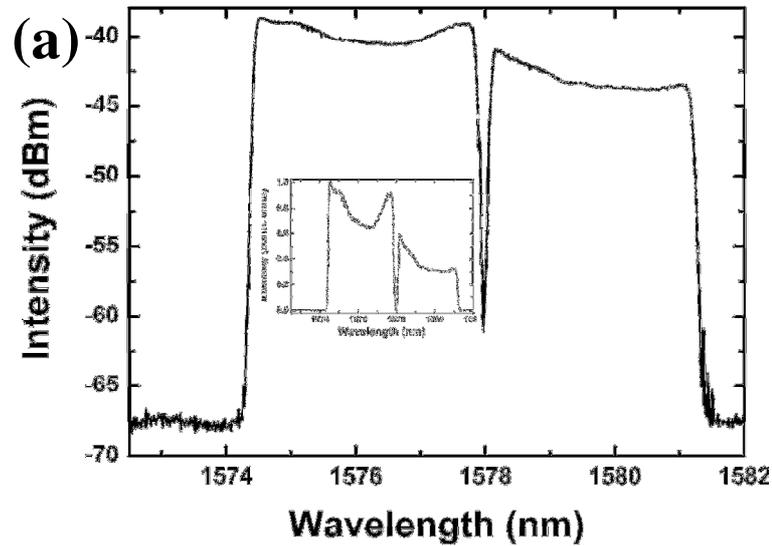

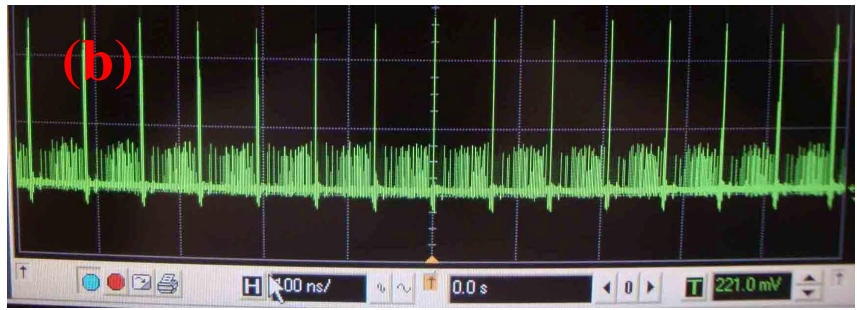

**Figure 5.7**: (a) Optical spectrum of dual wavelength dissipative solitons. Insert: the normalized optical spectrum; (b) Oscilloscope trace of dual wavelength dissipative solitons.

The relative strength of the solitons varied with the cavity birefringence. Slightly tuning the orientation of the PC, one could continuously change the relative soliton pulse energy. At a certain soliton intensity relation, it was found that two solitons could even have the same group velocity despite of the fact that they have different central wavelengths. **Figure 5.8** shows the oscilloscope of such a case, where the two solitons have fixed soliton separation as they





circulated in the cavity. However, such a state was unstable, after a short time the solitons lost their synchronization and moved independently again.

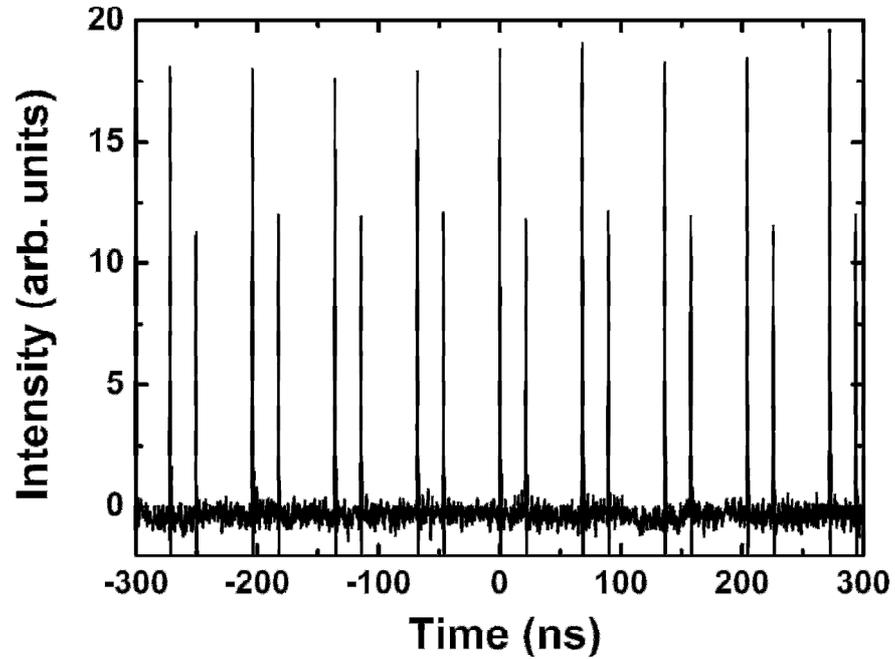

**Figure 5.8**: Oscilloscope traces of synchronized dual wavelength dissipative solitons.

Through Careful control of the cavity birefringence, experimentally we found that the dual wavelength solitons could also be transformed into a single wavelength vector dissipative soliton, as shown in **Figure 5.9**. To obtain the result we have significantly changed the orientation of the PC paddles while the pump strength was kept fixed. On the oscilloscope trace the previous two solitons now merged together, consequently only one pulse could be observed in the cavity. Checked with a high speed oscilloscope (50 GHz) combined with a commercial autocorrelator (FR-103MN), no fine structures were detected within the pulse, confirming it is vector soliton. Note that the central





wavelength of the vector dissipative soliton has now shifted to a value between those of the linearly polarized solitons shown in **Figure 5.7**a. The polarization resolved study on the state further confirmed that the soliton pulse was elliptically polarized, and its two orthogonal components have comparable spectral intensity and the same central wavelength [114].





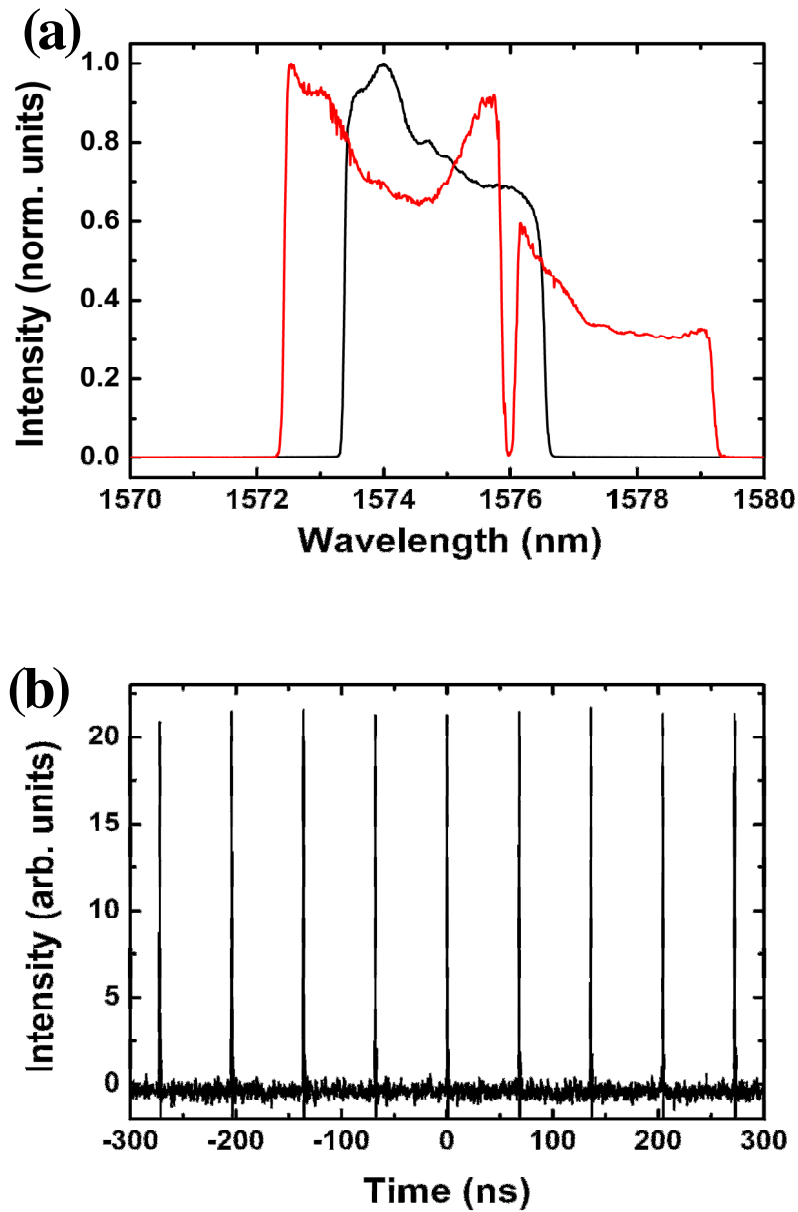

**Figure 5.9**: (a) Optical spectra of polarization locked gain guided vector soliton and dual wavelength spectrum obtained through rotating PCs but kept the pump strength fixed: normalized unit. (b) Oscilloscope trace of polarization locked gain guided vector soliton after passing through a polarizer.

Besides the above dual-wavelength dissipative soliton operation,





experimentally under certain laser operation conditions, triple-wavelength dissipative solitons as shown in **Figure 5.10** were also observed. Experimental studies turned out that all the solitons were linearly polarized, and the two solitons with the shortest and longest wavelength, respectively, have the same polarization, which is orthogonal to the soliton that has the middle value of the central wavelength.

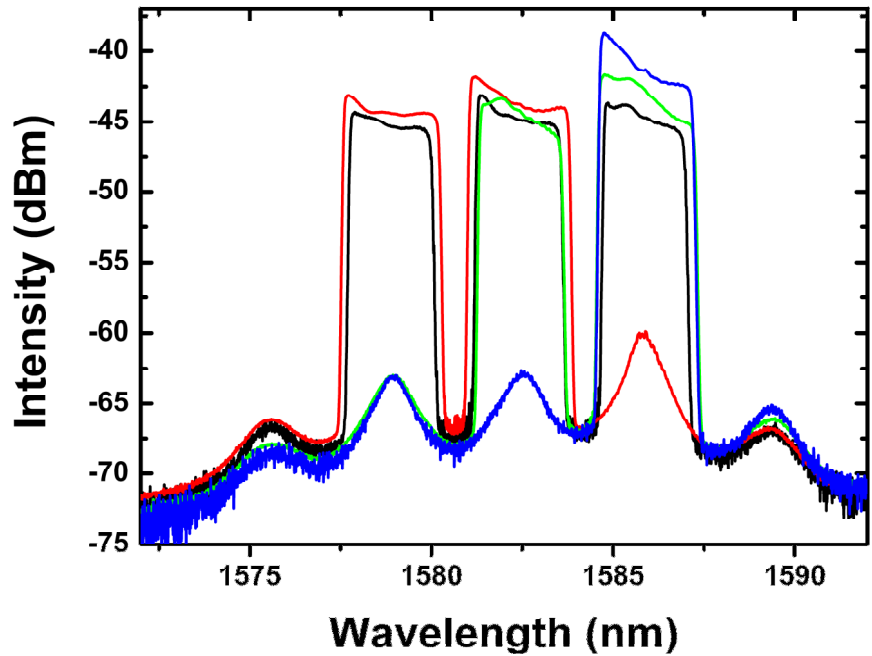

**Figure 5.10**: Single/dual/triple wavelength spectra obtained through rotating PCs but kept the pump strength fixed.

We have experimentally investigated the formation mechanism of the multiple wavelength solitons in our fiber laser. Starting from a state of dual-wavelength soliton operation as shown in **Figure 5.6**, slowly reducing the pump strength, it was found that only the spectral intensity became weaker, while the changes on other soliton parameters, such as the spectral bandwidth, central wavelength





and pulse duration were not obvious. As the pump power became about 110 mw, one of the soliton disappeared and the laser emission jumped to a single wavelength soliton operation state. If the pump power is further reduced till the mode locking is destroyed, continuous laser emission either at the 1576 nm or the 1580 nm, or simultaneously at both wavelengths could be observed. The CW laser emissions have the same polarization as those of the corresponding solitons. This experimental result shows that the dual-wavelength soliton emission of the laser should be related to the birefringence of the cavity. Similar feature has also been observed on the triple–wavelength soliton emission as have been shown in **Figure 5.10**. Decreasing pump power until mode locking was destroyed, on the spontaneous emission spectrum of the laser one could clearly see the existence of a periodic spectral intensity modulation, indicating the existence of a spectral filter in the cavity, which could also be traced back to the cavity birefringence effect.

Based on these experimental evidences and features of the multi-wavelength solitons, we explain the formation mechanism of the dual- and triple-wavelength solitons as the following: despite of the fact that no polarizing components were used in the cavity, the slight residual polarization asymmetry of the components used, such as the SESAM and the circulator, could still cause the formation of a linear artificial birefringent filter in the cavity. It is known that the wavelength spacing between the transmission maxima of a birefringence filter depends on the cavity birefringence, which is given by $\Delta\lambda = \lambda^2/(LB)$, where $\lambda$ is the central wavelength, $L$ is the cavity length and $B$ is the strength of birefringence [40]. The stronger the cavity birefringence, the smaller





is the spacing. The existence of the artificial birefringence filter results in the multiple wavelengths lasing of the laser, and the saturable absorption effect of the SESAM further causes the self-started mode locking of the laser. In the case of strong cavity birefringence, the artificial filter has narrow bandwidth, within the effective laser gain bandwidth range, three wavelength lasing and mode locking is possible. However, due to the polarization gain competition among the solitons, solitons with wavelength separation smaller than the homogeneous gain bandwidth cannot have the same polarization but orthogonal polarizations, while solitons with wavelength separation larger than the homogeneous gain bandwidth could simultaneously exist. With smaller cavity birefringence the filter bandwidth also becomes broader. Within the effective gain bandwidth, eventually only solitons with two different central wavelengths could be excited. As a result of the polarization gain competition the two soliton could not have the same polarization. Therefore, dual-wavelength soliton with orthogonal polarizations were observed. When the birefringence of the cavity is further decreased, the bandwidth of the filter became so broad that only a single wavelength soliton could be formed in the laser, in this case the cross-coupling between the two polarization-components of the cavity also became strong. As a result of the strong cross-phase modulation vector dissipative soliton could be formed in the laser. Based on this explanation we found the results of our experimental observations could be well understood.

In conclusion, we have experimentally observed multi-wavelength dissipative soliton operation of an erbium-doped fiber laser mode locked with a SESAM. It was shown that depending on the strength of the linear cavity birefringence





either dual- or triple-wavelength soliton operations could be obtained. We found that the multi-wavelength soliton operation of the laser could be well explained as caused by the existence of an artificial birefringence filter in the laser cavity. The multiple wavelength passively mode locked fiber laser may find applications in the fiber sensors or optical signal processing systems.





# Chapter 6.  Dynamics of dark soliton in fiber lasers

Strictly speaking, solitons mentioned/discussed in Chapter 3, Chapter 4 and Chapter 5 are bright solitons, because both polarization components have the shape of bell-shaped profile and vanish if t→±∞. In a passively mode locking fiber laser, bright pulse/soliton is a direct consequence of the mode locking operation because higher intensity optical wave will encounter much less attenuation that the lower intensity optical wave due to the saturable absorption behavior from the mode locking component (both NPR and SESAM). Correspondingly, the mode locker favors the formation of bright pulse. However, in a normal dispersion cavity fiber laser, within a very narrow operation regime which can be termed as non-mode-locking regime, the mode locking operation could be suppressed. Then bright pulse became unstable while its counterpart: dark pulse emission became dominant. Through further pulse shaping in normal dispersion fibers, dark pulses could evolve into dark solitons. In Chapter 6, the so-called special non-mode-locking regime was our target to study and dark solitons in a fiber lasers were presented.

The objective of Chapter 6 is to discuss the experimental and numerical observation of dark soliton in a fiber laser. Section 6.1 shows that the formation of scalar dark soliton is indeed a generic feature of an all normal dispersion fiber laser under strong CW operation. Section 6.2 further discusses that dark soliton emission could even be established in a dispersion-managed (DM) erbium-doped fiber laser with net normal dispersion, which could be





numerically confirmed as well. Section 6.3 address the problem of incoherently coupled dark vector soliton trapping in a fiber laser.

## 6.1    Scalar dark soliton

Since the first experimental observation of optical bright soliton in SMF by Mollenauer *et al*. in 1980 [2], soliton formation in optical fibers has been extensively investigated [40]. It is now well-known that dynamics of the solitons is governed by the NLSE, and bright solitons can be formed in the negative dispersion SMFs, whereas in the positive dispersion SMFs dark solitons, characterized as a localized intensity dip on a continuous wave (CW) background, are allowed to be formed [40]. Researchers had widely investigated the property of dark solitons in [146-155].

Bright scalar solitons have been observed in the mode locked fiber lasers with anomalous cavity dispersion. Although pulse propagation in a fiber laser is not exactly the same as that in the SMFs, it has been shown that the main features of the formed bright solitons are still governed by the NLSE [40]. A fiber laser can also be constructed with all-normal dispersion fibers. Recently, it has been shown that dissipative solitons could be formed in the fiber lasers as a combined result of the nonlinear pulse propagation in the fibers and the effective laser gain filtering. However, to our knowledge, no NLSE type of dark solitons has been observed in the lasers, despite of the fact that dark solitons could be naturally formed in the normal dispersion SMFs. We note that T. Sylvestre *et al*. have experimentally demonstrated dark-pulse train generation in a normal-dispersion fiber laser [155]. In their lasers because the mode locking





was based on a dissipative four-wave mixing process, no single dark soliton could be obtained. Strictly speaking, the dark solitons they reported are not fully governed by the NLSE.

Our fiber laser is schematically shown in **Figure 6.1**. Its cavity is made of allnormal dispersion fibers. A piece of 5.0 m EDF with GVD of –32 (ps/nm)/km was used as the gain medium. The rest of fibers used have altogether a length of 157.6 m, and they are all DCF with GVD of –4 (ps/nm)/km. All the passive components (WDM, Coupler, and Isolator) were made of the DCF. A polarization dependent isolator was used in the cavity to force the unidirectional operation of the ring, and an in-line polarization controller was inserted in the cavity to fine-tune the linear birefringence of the cavity. A 50% fiber coupler was used to output the signal. The laser was pumped by a high power Fiber Raman Laser source (KPS-BT2-RFL-1480-60-FA) of wavelength 1480 nm, and the maximum pump power can be as high as 5 W. An optical spectrum analyzer (Ando AQ-6315B) and a 350 MHz oscilloscope (Agilen 54641A) together with a 2 GHz photo-detector were simultaneously used to monitor the laser emission.

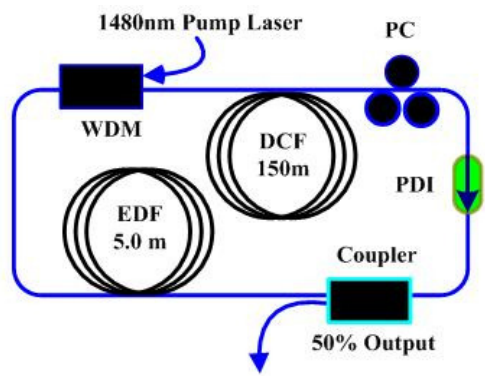

**Figure 6.1**: Schematic of the fiber laser.





The laser cavity has a typical configuration as that of a fiber laser that uses the nonlinear polarization rotation (NPR) technique for mode locking [156]. Indeed, under appropriate linear cavity phase delay bias (LCPDB) setting, self-started mode-locking could be achieved in the laser. Consequently either the gain-guided solitons or the flat-top dissipative solitons could be obtained [156]. Apart from the bright soliton operations, experimentally we have further identified another regime of laser operation, where a novel form of dark pulse emission as shown in **Figure 6.2**a was firstly revealed. **Figure 6.2**a (upper) shows a case of the laser emission where a single dark pulse was circulating in the cavity with the fundamental cavity repetition rate of 1.23 MHz (inset of **Figure 6.2**b). On the oscilloscope trace the dark pulse is represented as a narrow intensity dip in the strong CW laser emission background. The full width at the half minimum of the dark pulse is narrower than 500 ps, which is limited by the resolution of our detection system. Unfortunately, due to the low repetition rate of the dark pulses, their exact pulse width cannot be measured with the conventional autocorrelation technique, but a cross-correlation measurement is required. In **Figure 6.2**b we have shown the optical spectrum of the dark pulses. For the purpose of comparison, the CW spectrum of the laser emission is also shown in the same figure. Obviously, under the dark pulse emission the spectral bandwidth of the laser emission became broader. Based on the 3 dB spectral bandwidth and assuming that the dark pulses have a transform-limited hyperbolic-tangent profile, we estimate that their pulse width is about 8 ps.





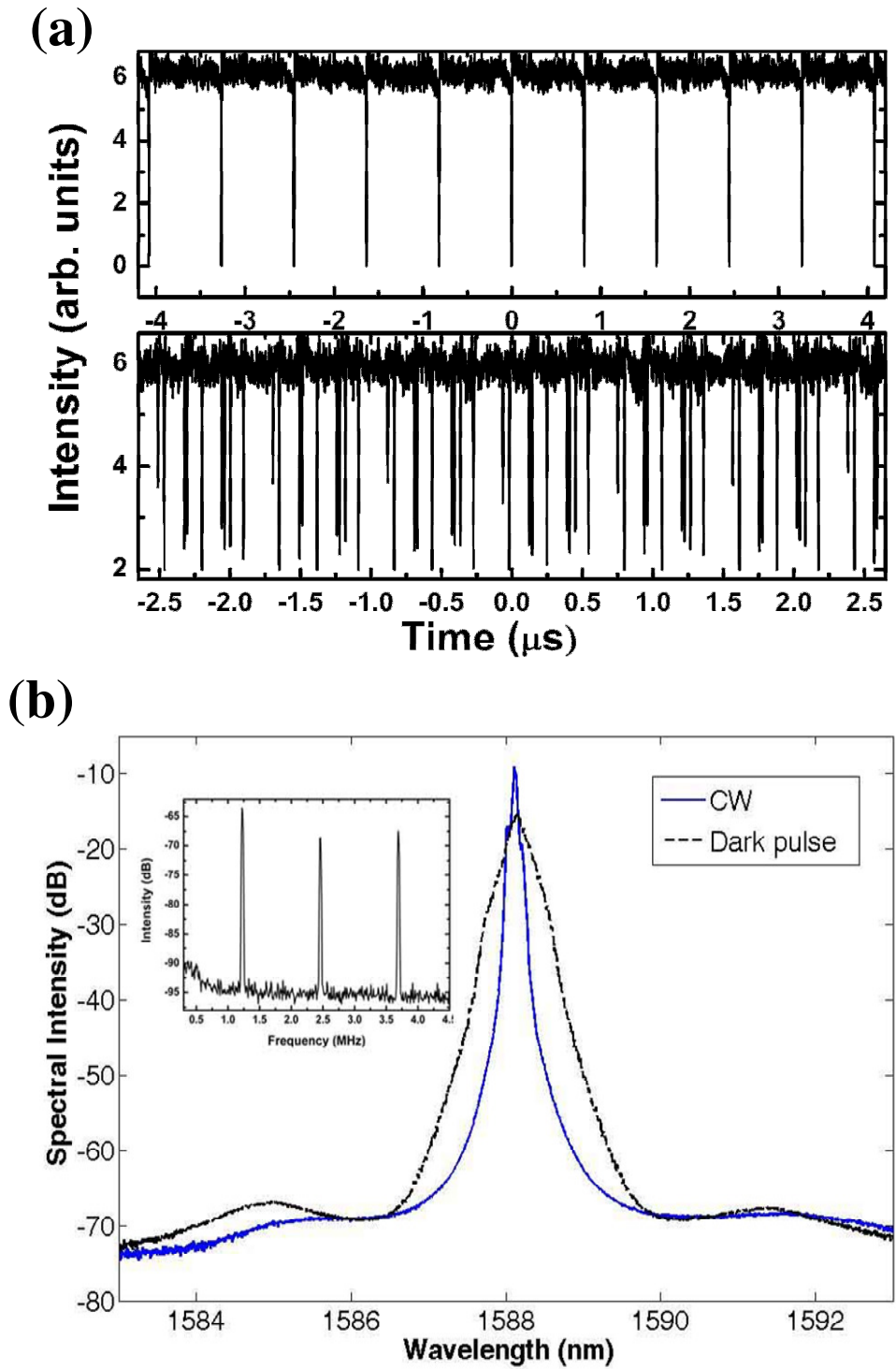

**Figure 6.2**: Dark pulse emission of the laser. (a) Oscilloscope traces, upper: single dark pulse emission; down: multiple dark pulse emission. (b) Optical spectra of the laser emissions. Inset: RF spectrum of the single dark pulse emission.





To obtain the above dark pulse emission, the LCPDB must be shifted away from the mode locking regime [115]. This was experimentally done by carefully adjusting the polarization controller. In the no mode locking regime strong stable CW laser emission was obtained, whose strength increased with the pump power. In our experiment as the launch pump power increased to ~2 W, the CW laser intensity suddenly became strongly fluctuated. Carefully checking the CW laser intensity fluctuation, it turned out that clusters of dark pulses were formed in the background of the CW intensity. The random movement and mutual interactions of the dark pulses caused the strong intensity fluctuation of the CW laser emission beam intensity. Experimentally, we found that through carefully adjusting the pump strength and orientations of the PC, the number of dark pulses could be significantly reduced. Eventually a state of stable single dark pulse emission as shown in **Figure 6.2**a (upper) was obtained.

Unlike the single bright pulse emission of the laser, the single dark pulse emission state was difficult to be maintained for long time. Probably due to the laser noise and/or weak environmental perturbations, new dark pulses always automatically appeared in the cavity, leading to states of multiple dark-pulse operation, an example is shown in **Figure 6.2**a (down trace). Different from the multiple gain-guided soliton operation of the laser, where all the solitons have the same pulse height and energy, known as the soliton energy quantization [115], obviously each of the multiple dark pulses has different shallowness, indicating that neither their energy nor their darkness is the same.

By removing 150 m DCF from the cavity, we had also experimentally studied dark pulse emission under short cavity length. Dark pulses could still be





obtained, however, their appearance required much higher pump intensity, and a stable single dark pulse state was difficult to achieve. In addition, by replacing the polarization dependent isolator with a polarization independent one, we could still obtain the dark pulse emission of the fiber laser. However, there is no NPR induced cavity feedback effect in the laser. Again in the laser it is hard to obtain a stable single dark pulse emission state. These experimental results suggest that the dark pulse formation could be an intrinsic feature of the all-normal dispersion cavity fiber lasers, and the NPR induced cavity feedback could have played an import role on the stability of the dark pulses in the laser.

To confirm our experimental observations, we have numerically simulated the dark pulse formation in our laser based on a round-trip model as described in [115]. To make the simulations comparable with our experiment, we used the following parameters: the orientation of the intra-cavity polarizer to the fiber fast birefringent axis $\Phi = 0.125 \, \pi$; nonlinear fiber coefficient $\gamma = 3 \, \text{W}^{-1}\text{km}^{-1}$; erbium fiber gain bandwidth $\Omega_g = 24$ nm; fiber dispersions $D''_{EDF} = -32$ (ps/nm) /km, $D''_{DCF} = -4$ (ps/nm)/km and $D''' = 0.1$ (ps$^2$/nm)/km; cavity length $L = 5.0$ m$_{EDF}$ + 157.6m$_{DCF}$ = 162.6 m; cavity birefringence $L/L_b = 0.1$ and the nonlinear gain saturation energy $P_{sat} = 50$ pJ.

We have always started our simulations with an arbitrary weak dip input. It was found that when the LCPDB was set in the laser mode locking regime, the self-started mode locking occurred and a bright pulse was always obtained. We have therefore focused our simulations on the LCPDB values in the non-mode-locking regime. With low laser gain, independent of the value of LCPDB no dark pulse could be obtained. However, under strong laser gain it was found





that if the LCPDB was so selected that the laser cavity provided a large negative cavity feedback (reverse saturable absorption), a dark pulse could automatically formed in the laser. **Figure 6.3**a shows a dark pulse numerically obtained under a gain coefficient of 900 km$^{-1}$ and a linear cavity phase delay bias of 1.0 π. **Figure 6.3**b shows the intensity and phase profiles of the dark pulse. The pulse intensity has a hyperbolic-tangent shape and a brutal phase jump close to π was observed at the minimum of the pulse, suggesting that it is a dark soliton. **Figure 6.3**c shows the spectrum of the calculated dark pulse. It closely resembles the spectra of the experimentally observed dark pulses, as shown in **Figure 6.2**b. The soliton feature of the pulse is also reflected by the appearance of spectral sidebands in its spectrum. Nevertheless, due to the unavoidable experimental noise these spectral sidebands could not be detected experimentally. Numerically further increasing the laser gain, the CW background level was boosted, while the absolute depth of the dark soliton kept the same, leading to a decrease on the soliton darkness. As G > 4000 km$^{-1}$, only a noisy CW background was observed.

Based on the numerical simulations we interpret the formation of dark pulses as a result of dark soliton shaping in the laser. Due to the normal dispersion of the cavity fibers, dark solitons are automatically formed under strong CW operation. As a dark soliton can be created by an arbitrary initial small dip on a CW background [146], and in the practical laser noise is unavoidable, many dark solitons are always initially formed in the laser. Formation of each individual dark soliton is uncorrelated. Depending on the initial "seed dips" the dark soliton have different darkness. Therefore, to obtain a stable single dark pulse





train of the laser emission, a certain competition mechanism among the solitons is required, which explains why a long cavity is beneficial for obtaining a stable single dark soliton in the laser. Finally, we point out that as the formed dark pulses have only a narrow spectral bandwidth no gain filter effects occur. Therefore, the effect of gain in the laser is purely to provide a strong CW and compensate the laser losses.

In conclusion, we have experimentally demonstrated a dark pulse emission fiber laser. It was shown that under strong CW operation, dark pulses could be automatically formed in a fiber laser made of all-normal dispersion fibers. Through introducing a strong negative cavity feedback mechanism, stable single pulse could be selected. Eventually a train of dark pulses at the fundamental cavity repetition rate was obtained. To our knowledge, this is the first demonstration of such a dark pulse emission laser.





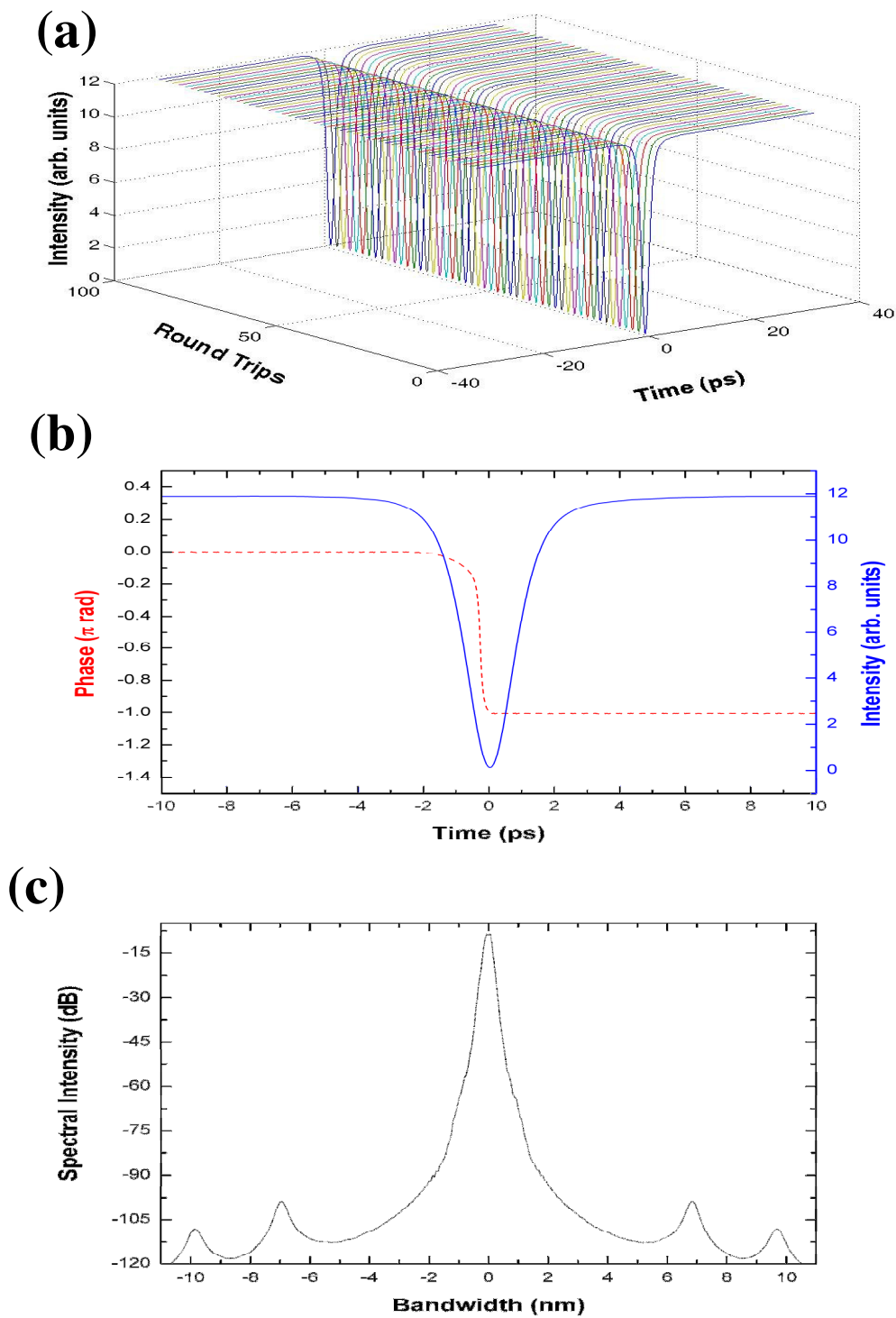

**Figure 6.3**: a dark pulse state numerically calculated. (a) Evolution with cavity roundtrips. (b) Intensity and phase profile. (c) Optical spectrum.





## 6.2    Dispersion managed dark solitons

A feature of fiber lasers is that their cavities can be easily dispersion managed (DM), meaning that the cavity is made of fibers with opposite sign of group velocity dispersion. Light circulation in a DM cavity fiber laser is analog to the light propagation in an endless DM fiber transmission line [157,158]. It has been theoretically shown that such a DM fiber transmission line supports both bright and dark solitons, known as the DM bright or dark solitons [159-161]. Experimentally, the DM bright solitons have been observed in DM cavity fiber lasers [162]. However, to the best of our knowledge, no experimental observation on the DM dark solitons has been reported. In this section, we report on the first experimental observation of DM dark solitons in a net normal dispersion erbium-doped fiber laser. We found experimentally that the formation of DM dark soliton has lower pump threshold as compared with the dark soliton formation in an all-normal-dispersion fiber laser. In addition, the DM dark soliton operation of a fiber laser is less sensitive to the environmental perturbations. Numerical simulations have also confirmed DM dark soliton formation in our fiber laser.

A dispersion-managed ring cavity fiber laser as shown in **Figure 6.4** was used. The cavity consists of a piece of ~5.0 m EDF with a GVD parameter of –32 (ps/nm)/km and ~8.0 m SMF with a GVD parameter of 18 (ps/nm)/km. The rest of fibers used are the DCF with a GVD parameter of –2 (ps/nm)/km. The total cavity length is ~18.2 m. A polarization dependent isolator was used to force the unidirectional operation of the cavity, and an in-line polarization controller (PC) was inserted in the cavity to fine-tune the birefringence of the cavity. A 50%





fiber coupler was used to output the signal. The laser was pumped by a high power Fiber Raman Laser source (KPS-BT2-RFL-1480-60-FA) of wavelength 1480 nm. The maximum pump power can reach to 5 W. An optical spectrum analyzer (Ando AQ-6315B) and a 350 MHz oscilloscope (Agilen 54641A) together with a 2 GHz photo-detector were simultaneously used to monitor the laser operation.

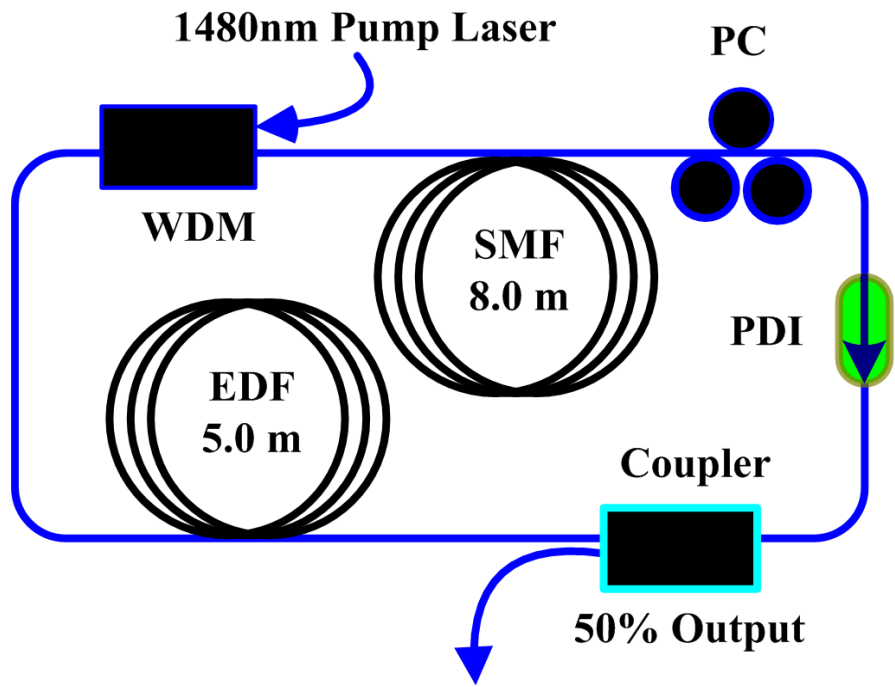

**Figure 6.4**: Schematic of the vector dark soliton fiber laser.

The total dispersion of the cavity is estimated ~0.3224 $ps^2$. We note that the laser has exactly the same cavity configuration as a nonlinear polarization rotation (NPR) mode locking fiber laser [163]. Indeed, depending on the linear cavity phase delay bias (LCPDB) setting, self-started mode-locking could be obtained in the laser. A typical mode locked state of the laser is shown in **Figure 6.5**. The mode locked pulses exhibit the DM dissipative soliton features.





The central wavelength of the solitons is at 1585.7 nm, and their spectrum has a 3 dB bandwidth of ~13.6 nm. The FWHM DM dissipative soliton pulse width is ~266 fs. Therefore, their time-bandwidth product (TBP) is ~0.452.

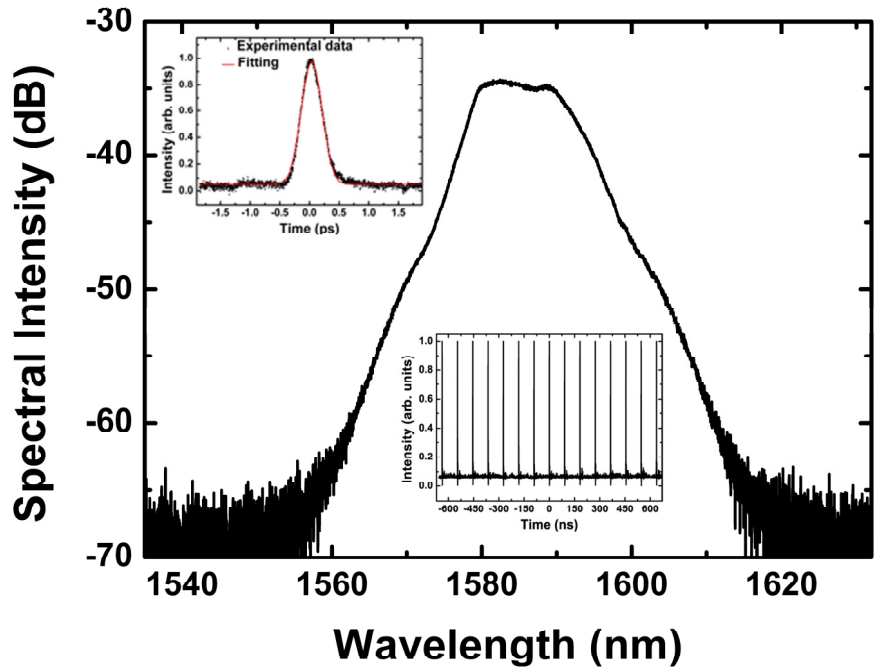

**Figure 6.5**: Optical spectrum of DM bright soliton. Insert: its autocorrelation trace and oscilloscope trace.

Starting from a bright dissipative soliton operation state, we then fixed all the other laser parameters but continuously tuned the orientation of the PC. It was observed that the laser emission changed from the bright soliton emission first to a CW operation state, and then to a dark soliton emission state, as illustrated in **Figure 6.6**. In **Figure 6.6**a, we have shown both the CW laser emission spectrum and the dark pulse emission spectrum. Under the dark pulse emission, the spectrum becomes obviously broader. We note that the dark pulse emission state is not a mode locked state of the laser, compared with **Figure 6.5**; the





LCPDB is now set in the non-mode-locking regime [163]. On the oscilloscope trace, a dark soliton pulse is identified by a narrow intensity dip on the strong CW background. Depending on the pump strength, either multiple or single dark solitons could be obtained. In the case of multiple dark solitons, the solitons have different darkness, indicating that different from the bright solitons, no soliton energy quantization effect exists. The FWHM width of the dark pulse is narrower than 500 ps, which is below the resolution of our detection system. Unfortunately, due to the low repetition rate of the dark pulses, the actual dark pulse width cannot be measured with the conventional autocorrelation technique either. Based on the 3 dB spectral bandwidth of ~0.8 nm, as shown in **Figure 6.6**a, we estimated that the pulse width is around ~3 ps, if a transform-limited hyperbolic-tangent profile is assumed.





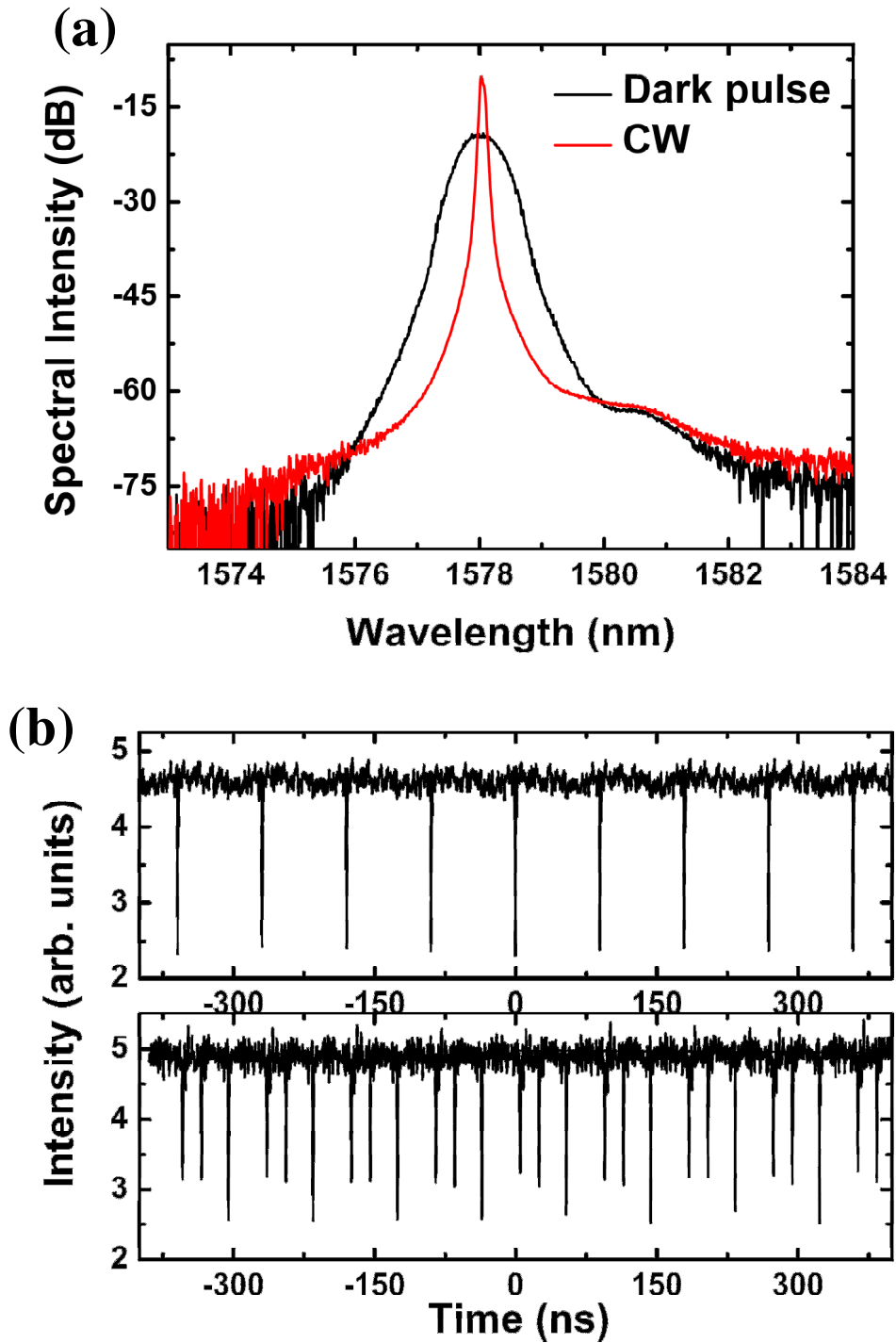

**Figure 6.6**: (a) Spectra of DM dark soliton and CW. (b) Oscilloscope traces, upper: single dark soliton; down: multiple dark solitons.

We have experimentally investigated the effect of the anomalous dispersion





SMF on the DM dark soliton by deliberately changing the SMF length. The net cavity dispersion then changed from all-normal to net-anomalous. When SMF was completely removed, and the laser has an all-normal dispersion short cavity, very high pump power, normally higher than 2.5 W, was required to obtain dark pulse emission. In this case it is difficult to obtain stable single dark soliton. New dark solitons were constantly automatically formed [163]. After SMF was spliced into the cavity and its length was increased, it was observed that the pump threshold for the dark soliton formation progressively decreased, and single dark soliton became less difficult to obtain. It seemed that the insertion of the anomalous dispersion SMF could stabilize the dark soliton propagation in the laser cavity. Under an appropriate cavity DM strength and pump strength, stable single dark soliton emission states have been obtained. We found experimentally that the 3 dB spectral bandwidth of the single dark soliton could not be broader than ~0.9 nm. Beyond it, the single dark soliton became unstable, and eventually a new dark soliton appeared. The laser displayed a clearly different multiple dark soliton formation scenario than an all-normal dispersion cavity fiber laser. When the anomalous dispersion SMF was elongated to ~10 m and thereby the net cavity dispersion shifted to the net-anomalous dispersion regime, no dark soliton was observed.

We further numerically simulated the DM dark soliton formation in our fiber laser. We used a model as described in [163]. To make the simulations comparable with our experiment, we used the following parameters: the orientation of the intra-cavity polarizer to the fiber fast birefringent axis $\Phi = 0.13 \ \pi$; the nonlinear fiber coefficient $\gamma = 3 \ \text{W}^{-1}\text{km}^{-1}$; the erbium fiber gain





bandwidth $\Omega_g = 24$ nm; fiber dispersions $D''_{EDF} = -32$ (ps/nm) /km, $D''_{SMF} = 18$ (ps/nm) /km, $D''_{DCF} = -2$ (ps/nm) /km and $D''' = 0.1$ (ps$^2$/nm)/km; cavity linear birefringence $L/L_b = 0.01$ and the nonlinear gain saturation energy $P_{sat} = 500$ pJ, $L_{EDF} = 5$ m and $L_{DCF} = 5.2$ m, and the length of SMF was changed from 0 m to 10 m, therefore, the total cavity dispersion was varied from 0.2215 ps$^2$ to $-$0.0125 ps$^2$.

In all our simulations an arbitrary weak dip input was used as the initial condition. It was found that when the LCPDB was biased in the non-mode-locking regime, stable dark soliton could always be obtained. In the case of an all-normal-dispersion cavity, we found that through progressively increasing the laser gain, a dark soliton can hardly break into multiple dark solitons. Increasing the pump strength increased the CW level, and an initially black soliton gradually became a gray soliton, and eventually vanished in the noisy CW background. However, if anomalous dispersion SMF was introduced, not only the formation threshold of the dark solitons became much lower, but also the formed DM dark solitons exhibited different features. As the pump strength is increased, the DM dark solitons split. **Figure 6.7** shows the numerically calculated parameter domain for stable single DM dark soliton of our laser. Depending on the net cavity dispersion, the range of gain variation for stable single DM dark soliton is limited, e.g. in a cavity with net-normal dispersion of ~0.0437 ps$^2$, where ~7.6 m SMF was used, gain could not exceed an upper threshold ~580 km$^{-1}$. Otherwise, the DM dark soliton would break automatically.





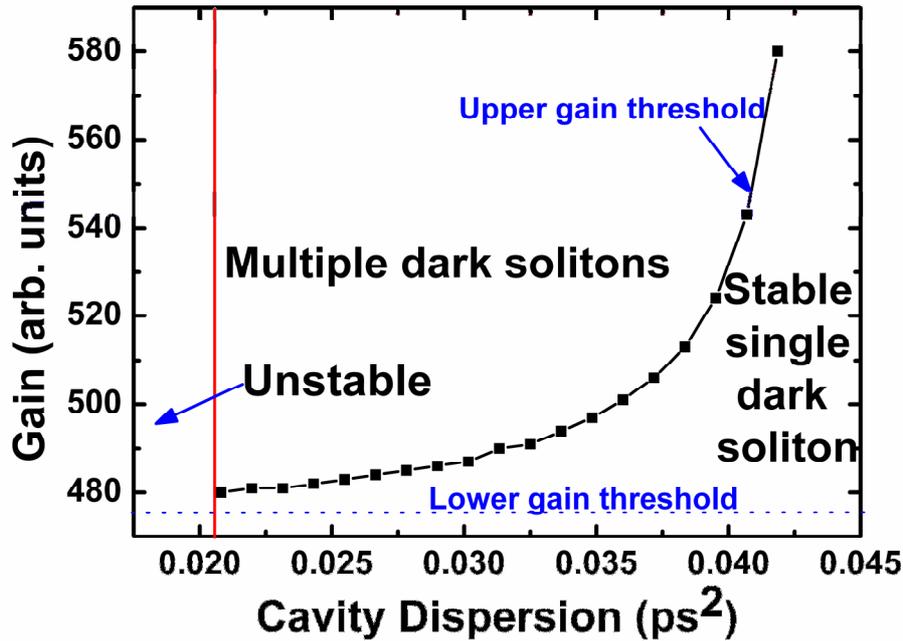

**Figure 6.7**: Region of existence of DM dark solitons in the (Dispersion, Gain) plane.

Bound states of dark solitons were also numerically obtained, as shown in **Figure 6.8**. In the simulation we have used net-dispersion 0.0343 ps$^2$. Under a low laser gain of 485 km$^{-1}$, very stable single DM dark soliton with pulse width of 3.7 ps and a modulation depth of 0.92, which indicates a nearly black DM soliton, was obtained. When the laser gain was raised to 510 km$^{-1}$, a new dark soliton was formed, as shown in **Figure 6.8**b. Like the experimental observations, the dark solitons have different darkness. The separation between the dark solitons is 3.8 ps, representing the formation of a DM black-grey soliton pair. It is to point out that although bound DM dark soliton pair was numerically obtained, it remains a challenge to experimentally observe it. The main difficulty is how to experimentally measure two closely spaced dark





pulses. A streak camera might be necessary to visualize it.

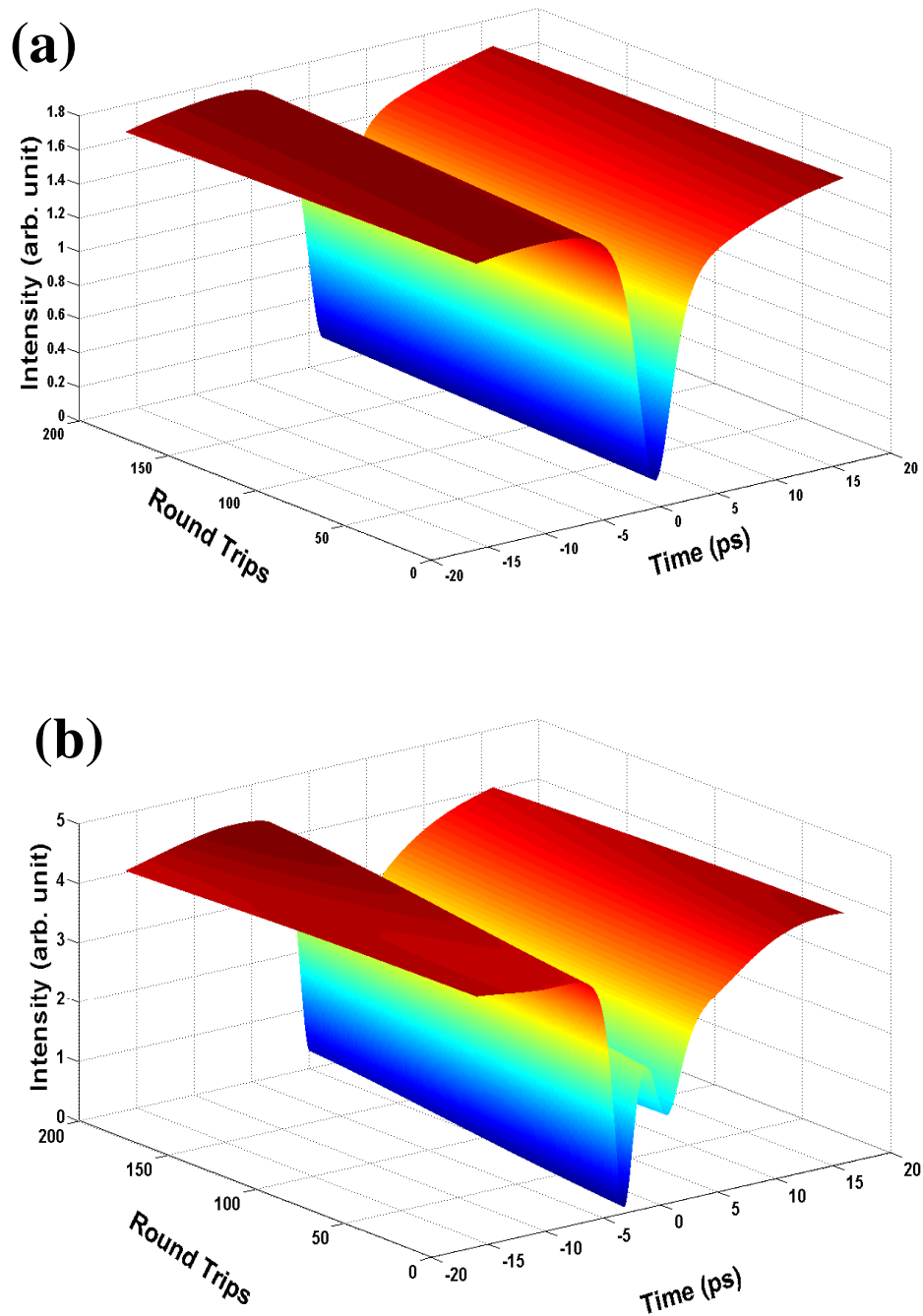

**Figure 6.8**: Evolution of DM dark solitons in time domain with the net-cavity dispersion 0.343 ps$^2$: (a) *Gain* = 485 km$^{-1}$; (b) *Gain* = 510 km$^{-1}$.





In conclusion, we have first experimentally observed DM dark solitons in a net-normal dispersion erbium-doped fiber laser. We found that in a DM cavity not only the formation threshold of a dark soliton is lower, as compared with that of an all-normal dispersion cavity, but also the state of stable single dark soliton could be easily obtained. DM dark soliton formation in our laser was also numerically simulated. Moreover, numerical simulations have revealed a bound state of black-grey solitons. Our study shows that dark soliton emission is an intrinsic feature of the normal dispersion fiber lasers.

## 6.3    Trapping of dark vector solitons

Optical soliton, in the form of a localized coherent structure, has been the object of intensive theoretical and experimental studies over the last decades [40]. It is now well-known that the NLSE could govern the formation of either bright or dark optical solitons depending on the sign of the fiber GVD [40]. For future photonic computing, optical logic gates and light control light devices, derived from soliton interaction by virtue of XPM, are attractive, since the general soliton features could be still sustained thanks to the conjunct interactions. It is physically and technologically meaningful to explore the bright and dark soliton interactions. Mutual bright soliton interaction as a result of XPM in optical fiber, also defined as soliton trapping, was first theoretically predicted by Curtis R. Menyuk [164, 165] and then experimentally validated [165-168]. Soliton trapping is denoted as two orthogonally polarized time-delayed optical solitons in birefringence fibers trapping each other and co-propagating jointly as long as their group velocity difference could be compensated through shifting their central frequencies in opposite directions by means of XPM. Kivshar





theoretically foretold a novel incoherently coupled dark vector soliton comprised by two grey solitons, belonging to two orthogonal polarization modes, strongly coupled with the help of XPM, in a highly birefringent fiber [169]. However, to the best of our knowledge, no incoherently coupled temporal dark vector soliton formation experiments have been reported.

Recently, we have experimentally achieved the formation of stable dark scalar solitons in an all-normal dispersion fiber laser [163]. It is to expect that through appropriately devising the cavity, dark vector soliton operation could be established if the polarization dependent component is replaced with a polarization insensitive device which profit to stabilize dark solitons. Triggered by facts that dark pulses could steadily exist in a cavity soliton laser in the presence of saturable absorber [170], a novel saturable absorber made of graphene, was specially introduced into an all normal dispersion fiber laser cavity. In this section, indeed, stable dark vector soliton could be ultimately observed. Depending on the pumping strength and cavity birefringence, both single and multiple dark vector solitons are realized. Further investigation demonstrates that the dark vector soliton is composed of two orthogonally polarized grey solitons incoherently coupled as a consequence of XPM, which is then numerically affirmed by our simulation model based on NLSEs. We deem that Kivshar's theory on dark vector soliton could be both experimentally and numerically justified.

Our fiber laser is schematically shown in **Figure 6.8**. The laser cavity is an all-normal dispersion fiber ring consisting of 5.0 m, 2880 ppm Erbium-doped fiber (EDF) with GVD of –32 (ps/nm)/km, and 157.6 m DCF with GVD of –2





(ps/nm)/km. A polarization insensitive isolator together with an in-line polarization controller (PC) was employed in the cavity to force the unidirectional operation of the ring and control the light polarization. A 50% fiber coupler was adopted to output the signal, and the laser was pumped by a high power Fiber Raman Laser source (KPS-BT2-RFL-1480-60-FA) of wavelength 1480 nm. The maximum pump power can reach as high as 5 W. All the passive components used (WDM, Coupler, Isolator) were made of the DCF. An optical spectrum analyzer (Ando AQ-6315B) and a 350 MHz oscilloscope (Agilen 54641A) together with two 2 GHz photo-detectors were utilized to simultaneously monitor the spectra and pulse train, respectively.

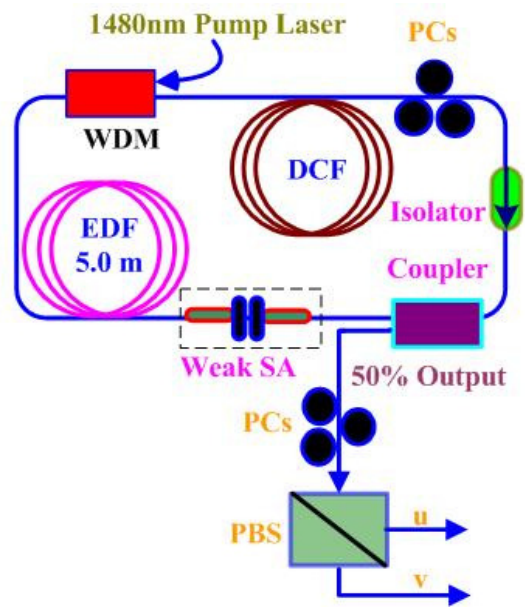

**Figure 6.9:** Schematic of the fiber laser.

In order to distinguish the laser emission along the two orthogonal principal polarization directions of the cavity, an in line PBS combined with an external cavity PCs which was operated to optimize the output polarization state, was spliced after the output coupler so as to split the output signals into two





orthogonal polarizations which were then simultaneously monitored by a multi-channeled oscilloscope associated with two identical 2 GHz photo-detectors. In particular, optical and electric path discrepancy between the two polarization-resolved signals should be eliminated as weak as possible. Recently, we have readily obtained scalar dark soliton in a fiber laser with a polarization dependent isolator under appropriate pump strength and negative cavity feedback [163]. But dark vector solitons were brittle and sensitive to environment perturbations if a polarization dependent isolator was used to replace the polarization dependent one. Opportunely, dark vector solitons could be stabilized by means of incorporating a weak saturable absorber, which represents a nonlinear loss term to suppress the dark soliton instabilities, inside the cavity. This weak saturable absorber is specially made of atomic layer graphene film attached onto fiber pigtails [97]. About 2~3 layers of graphene covered on the fibre core area is identified through a combination of Raman and contrast spectroscopy. It has a saturable absorption modulation depth of 64.4%, a saturation fluence of 0.71 µJ/cm$^2$ and a carrier relaxation time 1.67 ps (determined by a time-resolved pump-probe method). Due to the knowledge that G could be readily saturated, its mode locking ability is severely restrained, meaning that the formation of bright soliton turns to be formidable and thus chances of dark soliton formation become higher.

As soliton performances are strongly cavity birefringence and pump strength dependent, the output could be significantly modulated through rotating PCs and varying the pump. In the experiments, laser prefers to operate in the continuous wave (CW) regime, represented by a straight line above ground





level in the oscilloscope trace after optical-electrical conversion using a fast photodiode. Moreover, we noticed that when CW beam intensity kept sufficiently strong under a high pump power ~2 W, CW instability began to emerge and heavy intensity dips became visible and even could survive over a long period if PCs were precisely adjusted. Carefully examining the polarization resolved oscilloscope traces; it is to note that while along one principal polarization direction the laser emitted a train of intensity dips conjointly with a strong CW background (upper trace of **Figure 6.9**a), along the orthogonal polarization direction reciprocal dark soliton features was observed (lower trace of **Figure 6.9**a). Anyhow, tuning the external cavity PCs, which physically corresponds to affect the output fiber birefringence and eventually influence the dark soliton extinction ratio after passing through PBS, the two polarization resolved oscilloscope traces always maintained a single dark soliton pulse train despite the relative strength could be extensively adjusted. Further polarization resolved spectra as shown in **Figure 6.9**b illustrates that the two orthogonal polarization components have clearly different spectral distribution and ~0.1 nm central wavelength difference, which is comparably large enough in contrast with the 3 dB spectral bandwidth for each polarization ~0.5 nm, if the central spectra, as shown in the insert of **Figure 6.9**b, is deliberately inspected. Based on the above experimental observation, we could conclude that the observed dark soliton is not scalar (linearly polarized) but vector and each polarizations are incoherently coupled to form a group velocity locked dark vector soliton. Akin to the nature of dark scalar soliton [163], multiple dark vector solitons could be viably formed as shown in **Figure 6.10**.





If the pump strength is further increased, each of the dark solitons does not necessarily have the equivalent shallowness, i.e. the failure of "soliton energy quantization effect". In addition, juxtaposition of the zoom-in oscilloscope traces of dark vector solitons at different positions as shown in **Figure 6.10**. Obviously, it turns up that the relative shallowness ratio of each polarization could be unrelated, revealing that the darkness for each polarization is neither invariant nor "quantized" but fluctuant along the propagation. It could be traced back to the formation mechanism of dark soliton. Kivshar prognosticated that any dip on the CW background pulse of one polarization would instantly breed a similar dip in its orthogonal polarization mode incoherently coupled to the maternal pulse. Accordingly, as individual dark soliton arises from the weak perturbation of CW and thus pertains to their initial "seed dips", the uncorrelated formation renders the arbitrary darkness for each polarization. **Figure 6.10** also shows that the two polarizations are temporally synchronized and have almost the same pulse profiles and widths that are limited by the resolution of the electronic detection system. Due to the low pulse repetition rate, the pulse width of the dark solitons could not be measured with the autocorrelation.





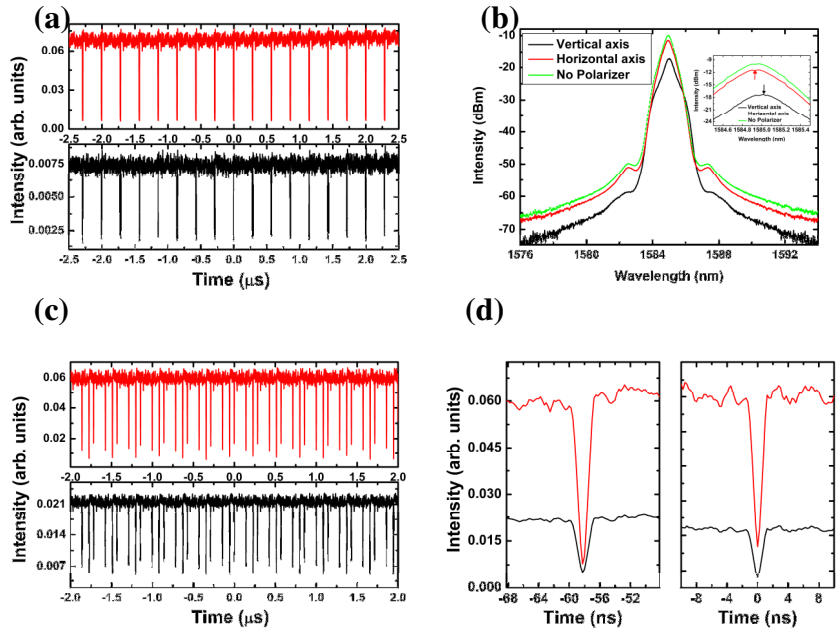

**Figure 6.10**: Polarization resolved (a): oscilloscope trace of single dark vector soliton in the cavity and (b) its corresponding optical spectra Insert: zoom in of (b) near the spectral center. (c) Polarization resolved oscilloscope trace of multiple dark vector soliton in the cavity. (d) Enlarge scale of (c) at different positions.

To confirm whether incoherently coupled dark solitons could be formed under the current cavity parameters; we numerically simulated the operation of the laser with the standard split-step Fourier technique to solve the equations and a so-called pulse tracing method to model the effects of laser oscillation [163]. We used the coupled Ginzburg-Landau equations to describe the pulse propagation in the weakly birefringent fibers:

To make the simulation possibly close to the experimental situation, we used the following parameters: $\gamma = 3$ W$^{-1}$km$^{-1}$, $\Omega_g = 16$ nm, $P_{sat} = 50$ pJ, $k''_{DCF} = 4.6$





ps$^2$/km, $k''_{EDF}$ = 41.6 ps$^2$/km, $k'''$ = –0.13 ps$^3$/km, $E_{sat}$ = 1 pJ, $l_0$ = 0.5, and $T_{rec}$ = 2 ps, Cavity length $L$ = 162.6 m.

To faithfully simulate the creation of incoherently coupled dark soliton, we always use an arbitrary weak pulse as the input initial condition and the weak saturable absorber functions in the transmission mode rather than reflection mode. Stable dark vector soliton could be conceivably obtained under diverse cavity birefringence regimes. **Figure 6.11** and **Figure 6.12** show the evolution of dark vector soliton within both time and frequency domain when the cavity beat length was chosen as $L_b = L$/40 and $L_b = L$/60, respectively. Numerically, we found that an antiphase type of periodic intensity variation between CW background inherent to the two orthogonally polarized dark solitons took place under weak cavity birefringence, for example, $L_b = 100L$. However, the stronger the linear cavity birefringence the weaker such intensity variation, eventually, coherent energy exchange vanished as seen in **Figure 6.11**a. Due to strong cavity birefringence where phase matching conditions for four wave mixing are challenging to fulfill [125], coherent coupling between the two polarization components of a dark vector soliton become vain while incoherent interaction arising from XPM still militates. This can explain why the two dark polarization components could be still synchronously entrapped in temporal domain and have little central frequency difference, as shown in **Figure 6.12**, even though the cavity birefringence $L_b = L$/40 is very strong.





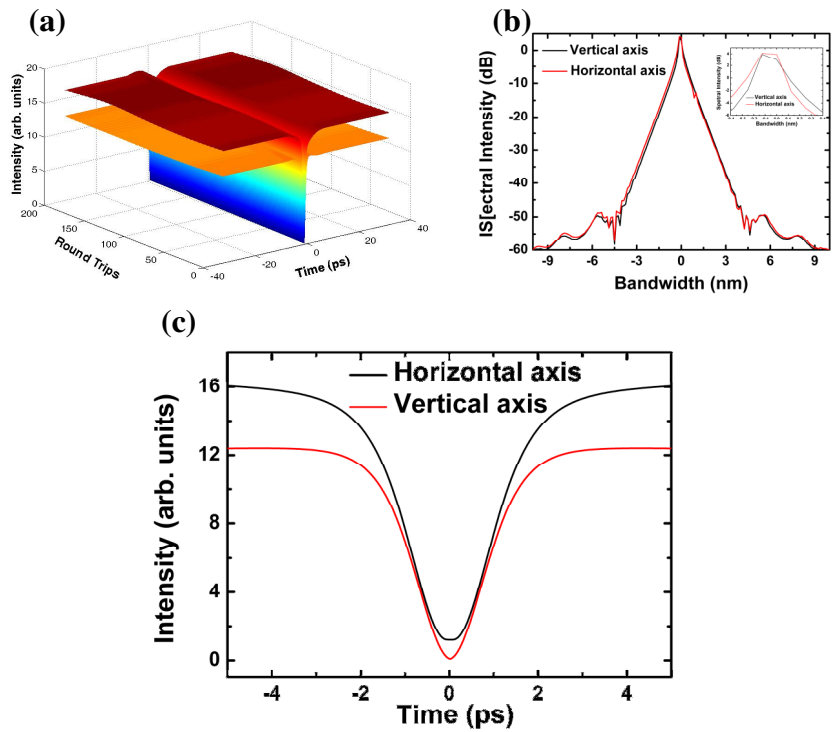

**Figure 6.11:** Stable dark vector soliton state numerically calculated. (a) Evolution of the dark vector soliton with cavity roundtrips. (b) Zoom in of (a). (c) The corresponding spectra and insert: zoom-in of (c). *Gain* = 1500. $L/L_b$ = 60.

Further strengthening the cavity birefringence, the two polarization components could be still concurrently bounded and propagates as one non-dispersive unit but they are temporally delayed, as shown in **Figure 6.11**b when $L_b = L/60$. Moreover, their central frequencies have 0.1 nm difference, matching well with our experimental observation. Whereas, under further higher cavity birefringence, the two dark soliton could no longer be trapped as one entity but break up and spread at their respective group velocities, indicating that a larger time delay correspond to a less effective attraction, and consequently uncoupled. Finally, we note that in the current simulation model, the dips of the





numerically calculated dark solitons could reach zero under weak cavity birefringence but are nonzero as shown in **Figure 6.11**b and **Figure 6.12**b, implying that they are grey solitons. CW backgrounds for each grey soliton are different, which confirms Kivshar's prediction that two grey solitons with different background intensities could be incoherently coupled by virtue of XPM. Furthermore, grey solitons in **Figure 6.12**b are slightly asymmetric near the central dips but converge to consistent asymptotic values provided that t→±∞. We conjecture that such asymmetry is a natural consequence of laser cavity effect, which does not exist in Kivshar's model, because apart from the balance of normal cavity dispersion and fiber nonlinear Kerr effect, a dark vector soliton is also subject to the cavity gain and losses, boundary condition, as well as the filtering effect caused by the strong cavity birefringence. Nevertheless, our numerical simulation model could well explain the formation of incoherently coupled dark vector soliton trapping in a fiber laser.





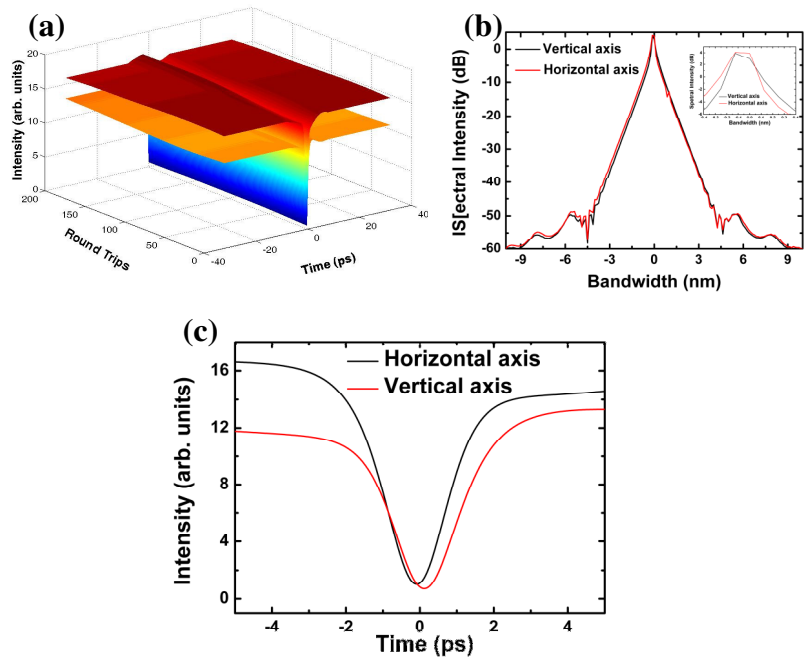

**Figure 6.12:** Stable dark vector soliton state numerically calculated. (a) Evolution of the dark vector soliton with cavity roundtrips. (b) Zoom in of (a). (c) The corresponding spectra and insert: zoom-in of (c). *Gain* = 1500. *L/L$_b$* = 40.

In summary, incoherently coupled dark vector soliton has been experimentally and numerically observed in an all normal dispersion fiber laser where a novel saturable absorber based on Graphene is purposely introduced to cooperatively stabilize the dark vector soliton. We believe our study could not only perfectly confirm Kivshar's theory but also verify the feasibility of dark soliton trapping technology. Although the dark soliton trapping is achieved in a fiber laser, due to the general applicability of NLSE, dark soliton trapping could be a universal hallmark of all coupled dark solitary waves, such as trapping of dark matter wave soliton in Bose-Einstein condensate.





# Chapter 7.  Dynamics of domain wall solitons

Domain walls, as topological structures separating different stable static arrangements in order to achieve minima of the Hamiltonian, are ubiquitous and generic in the entire field of nonlinear physics ranging from ferromagnetism theory to optics and Bose-Einstein condensate [9]. Zakharov and Mikhailov first theoretically predicted the existence of stable optical domain wall, independent of dispersion, purely through the interaction between the polarization states of two counter-propagating electromagnetic waves [171]. In nonlinear optical fibers, Wabnitz and Daino analytically foresaw the possibility of generating domain walls as a result of polarization switching [172]. Later, Haelterman, Sheppard and Malomed found that the dynamics of domain wall soliton could be well encapsulated by the incoherently coupled NLSE where both the nonlinear polarization coupling and normal dispersion were taken into account [173, 174]. In the subsequently work, S. Pitois *et al.* have experimentally studied polarization modulation instability (PMI) in a normally dispersive bimodal fiber by commingling two counter-propagating beams [175, 176]. Although a sort of domain wall structure caused by polarization switching was formed in their experiments, the soliton natures of domain wall soliton were still evidentially insufficient. In addition, the occurrence of domain walls that networks different types of stable eigenpolarization should be irrelevant to fiber dispersion, indicating that it should be possible to realize domain walls in both anomalous and normal dispersion regime.

As discussed in the Chapter 3, coherent and incoherent interactions among the





two polarization components of a bright-bright vector soliton strongly depends on the cavity birefringence. Particularly, if the cavity birefringence is strong enough, incoherent interactions exist; the coherent interactions occur on the condition that the cavity birefringence is sufficiently weak. Similar to bright-bright vector soliton, the dynamics of domain wall soliton is cavity birefringence sensitive as well. Under different birefringence realm, the structure of domain wall might be significantly different.

In Chapter 7, we try to investigate two types of domain wall soliton: coherently coupled and incoherently coupled. The Section 7.1 addressed that in weakly birefringent cavity erbium-doped fiber lasers, the phase of domain wall solitons could be locked: while a phase-locked dark-dark vector soliton was only observed in fiber lasers of positive dispersion, a phase-locked dark-bright vector soliton was obtained in fiber lasers of either positive or negative dispersion. Numerical simulations confirmed the experimental observations, and further showed that the observed vector solitons are the two types of phase-locked polarization domain-wall solitons theoretically predicted. The Section 7.2 further studied the general case of polarization domain wall under strong cavity birefringence and achieved the formation of vector dark polarization domain wall in anomalous dispersion cavity. In the section 7.3, the cavity birefringence was extremely large and dual wavelength domain wall soliton could be observed in all-normal dispersion cavity. Moreover, the coexistence of NLSEs type dark soliton and domain wall type dark soliton was presented in this section.





## 7.1   Coherently coupled domain wall solitons

Soliton formation is a fascinating phenomenon that has been observed in many diverse physical systems [40]. In the context of optics, soliton formation in the SMF has attracted the most attention. Light propagation in a SMF is governed by the NLSE, which admits formation of either bright or dark optical solitons depending on the sign of the fiber dispersion. When the vector nature of light is considered, light propagation in a SMF is then governed by the coupled NLSEs, which have much richer dynamics than the scalar NLSE. It has been theoretically shown that polarization coupling of light in a SMF could lead to formation of various types of vector solitons, including the bright-bright, dark-dark, and dark-bright [9] types of vector solitons. Formation of the incoherently coupled bright-bright and dark-dark types of vector solitons in birefringent SMFs has already been experimentally demonstrated in Chapter 4 and Chapter 7.

A fiber laser has its cavity mainly made of SMFs. Although light propagation in a fiber laser cavity is also subject to actions of other cavity components, such as the laser gain and cavity output coupler, it was found that optical solitons could still be formed in a fiber laser, and as far as the spectral bandwidth of the formed solitons is far narrower than the gain bandwidth, the dynamics of the formed solitons could be well described by the NLSE, or the coupled NLSEs when the cavity birefringence needs to be considered [58]. A fiber laser provides a unique nearly conservative system for the optical soliton studies.

Although majority of the theoretically predicted optical solitons in SMFs have





been experimentally confirmed, to the best of our knowledge, no experimental observation on the phase-locked polarization domain-wall solitons (PDWSs) has been reported. Formation of phase-locked PDWSs in diffractive or dispersive Kerr media was first theoretically predicted by Haelterman and Sheppard [173, 174]. They showed that Kerr media could sustain localized structures separating domains of different nonlinear polarization eigenstates. Two types of phase-locked PDWSs, one in the form of elliptically polarized dark-dark vector solitons, and the other one in the form of dark-bright vector solitons exist in fibers of normal dispersion. Although not named as PDWS, Christodoulides had also theoretically predicted a type of bright-dark vector soliton in weakly birefringent fibers [109]. He had given the pulse profiles of the dark and bright solitons, and pointed out that the solitons have the same pulse width, which is uniquely determined by the fiber group velocity dispersion (GVD) and strength of fiber birefringence. Based on our experimental studies, we suspect that the dark-bright vector soliton predicted by Christodoulides could be a special case of the dark-bright domain-wall solitons predicted by Haelterman and Sheppard. S. Pitois *et al* have experimentally investigated polarization modulation instability (PMI) in a birefringent fiber of normal dispersion [175, 176]. Although in their experiment a kind of PMI induced fast modulation structure was observed, no PDWSs were identified. The difficulties in observing the PDWSs in a SMF are that the fiber birefringence must be sufficiently small and maintained constant over a long distance, furthermore, coupling between the two polarization components must be strong enough.





In a previous experiment we have successfully demonstrated a high-order polarization locked bright-bright vector soliton in a mode-locked fiber laser [177]. To obtain the polarization locked bright-bright vector soliton the averaged cavity birefringence must also be kept sufficiently small. We note that previous experimental studies on the polarization dynamics of fiber lasers have shown that cavity feedback could significantly enhance the polarization coupling of light [178]. Worth of mentioning is that Williams and Roy had demonstrated a kind of anti-phase square-wave fast polarization dynamics in an erbium-doped fiber ring laser [179, 180]. The antiphase square-wave pulses observed by them are reminiscent of the polarization domain-wall structures. It is a well-known fact that a mode locked pulse in an anomalous dispersion fiber laser can be automatically shaped into a NLSE soliton. Considering the nature of PDWS formation in SMFs, it is to expect that under suitable conditions the antiphase square-wave pulses could be further shaped into PDWSs.

We designed two fiber lasers to investigate the formation of PDWSs. The overall laser setup is schematically shown in **Figure 7.1**. The laser cavity is a simple fiber ring comprising a segment of EDF as the laser gain medium, a wavelength-division-multiplexer for coupling in the pump light, and a 10% fiber output coupler for the laser output. A fiber-pigtailed polarization independent isolator and an in-line polarization controller are inserted in the cavity to force the unidirectional operation of the ring and to fine-tune the linear cavity birefringence. To study the effects of cavity dispersion on the PDWSs, we have made one laser with the purely anomalous dispersion fibers while the other the purely normal dispersion fibers. The laser emission along the two





orthogonal principal polarization directions of the cavity is resolved using a fiber polarization beam splitter (PBS).

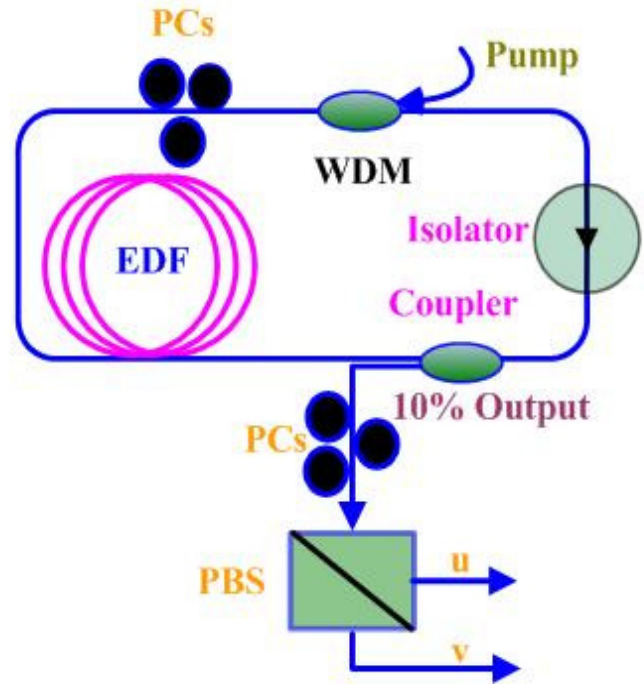

**Figure 7.1**: Setup of the fiber lasers. WDM: wavelength division multiplexer. EDF: erbium doped fiber. PC: Polarization controller. PBS: Polarization beam splitter.

The negative dispersion fiber laser has a cavity length of 14.7 m, with 6.4 m EDF of GVD of 10 (ps/nm)/km and 8.3 m SMF of GVD of 18 (ps/nm)/km. When pump power of the laser was increased to about 4~5 times of the laser threshold, the laser emission exhibited a periodic polarization switching between the two orthogonal principal polarization directions. Under the polarization resolved measurement it is represented as an anti-phase square pulse emission, which was explained previously as caused by the delayed cavity feedback and the polarization mode competition [179, 180]. Experimentally we found that associated with the polarization switching of the laser emission, a





PDWS was also formed, indicating that the laser emission along both of the two principal polarization directions were stable. At fixed pump strength, tuning the paddle orientation of the polarization controller, which corresponds to changing the average linear cavity birefringence, the square pulse width could be continuously changed. Multiple square-pulses could also be obtained under strong pumping. However, due to the intrinsic scalar modulation instability caused by the negative dispersion fibers, too strong pumping also resulted in formation of the NLSE type solitons within in the square pulses, which destabilizes the square pulses. Therefore, we have restricted us to the cases where no NLSE solitons appeared. Once the periodic square pulse emission of the laser was obtained, we then carefully decreased the linear cavity birefringence, which was experimentally detected by monitoring the lasing wavelength difference along the two polarization directions. In this way we could continuously decrease the square-pulse width, eventually a stable state as shown in **Figure 7.2** could be obtained. On the oscilloscope traces (**Figure 7.2**a) it is to see that while along one principal polarization direction the laser emitted a train of pulses (lower trace of **Figure 7.2**a), along the orthogonal polarization direction the laser emitted strong CW, and on the CW background there is a train of intensity dips (upper trace of **Figure 7.2**a). Each intensity dip corresponds temporally to an intensity pulse, and the dark-bright pulses repeat with the fundamental cavity repetition rate. **Figure 7.2**b shows a zoom-in of the bright and dark pulses. They have almost the same pulse profiles and widths that are limited by the resolution of the electronic detection system. Using a commercial autocorrelator we also measured the autocorrelation trace of the





bright pulses. A typical result is shown in the inset of **Figure 7.2**c. The autocorrelation trace has a good Sech$^2$-profile with a FWHM of about 4.8 ps, which gives that the pulse width is 3.1 ps. Due to the low pulse repetition rate, the pulse width of the dark pulses could not be measured with the autocorrelator.

**Figure 7.2**c shows the polarization resolved optical spectra of the laser emission. The soliton feature of the pulses is characterized by the appearance of the clear spectral sidebands on their spectra. We note that even the dark pulses have spectral sidebands, and especially whose spectral sidebands have the same locations as those of the bright pulses. The 3 dB spectral bandwidth of the bright pulses is 0.9 nm. Therefore, the bright solitons are near transform-limited. There is no obvious spectral difference between the dark and bright pulses, furthermore, no polarization evolution frequency [8] was observed on neither the bright nor the dark pulses, which clearly suggests that the phases of the pulses are locked. Therefore, the dark-bright pulses shown in **Figure 7.2**a, constitute a phase-locked dark-bright vector soliton (DBVS).





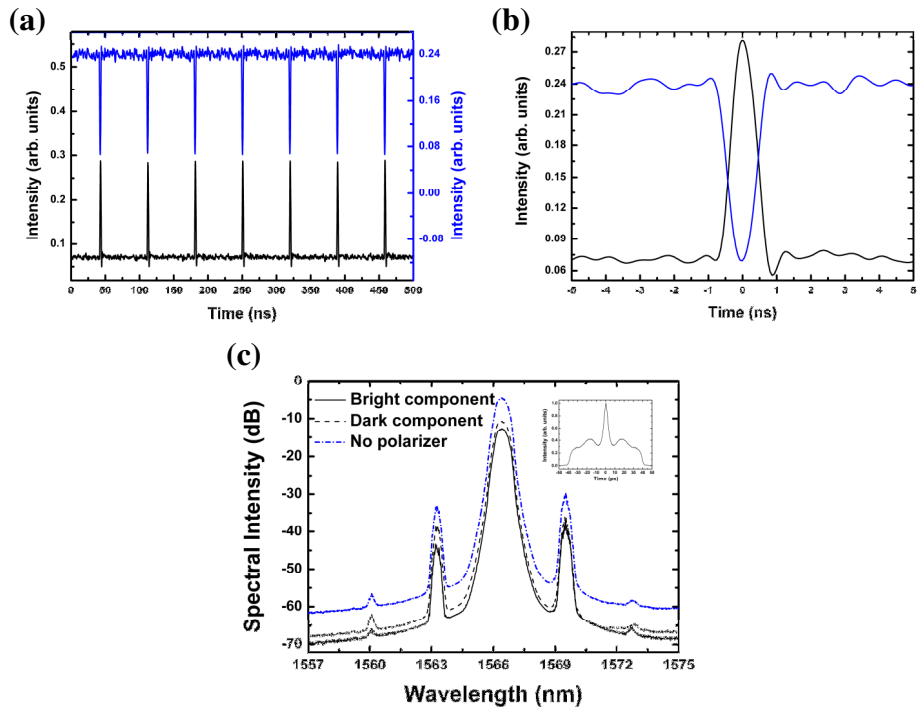

**Figure 7.2**: A typical dark-bright vector soliton emission of the lasers. (a) Oscilloscope traces. (b) Zoom-in of the dark-bright pulses. (c) Polarization resolved optical spectra; inset: Autocorrelation trace of the bright pulses.

Experimentally the DBVS could be obtained at various pump strengths. The higher the pump power, the easier is the state obtained. Under strong pumping occasionally multiple DBVSs have also been obtained. However, different from the cases of multiple bright vector solitons formed in a fiber laser, the multiple DBVSs could have different soliton energies, represented by their different pulse heights in the same oscilloscope trace. Energy of the bright and dark pulses in a vector soliton is always correlated. Corresponding to a bright soliton with a weak pulse, the dark soliton also has a shallow dip. In the case that the formed DBVSs have different soliton energies, slow relative movement between them was observed. Two vector solitons were observed to collide and





then merge together.

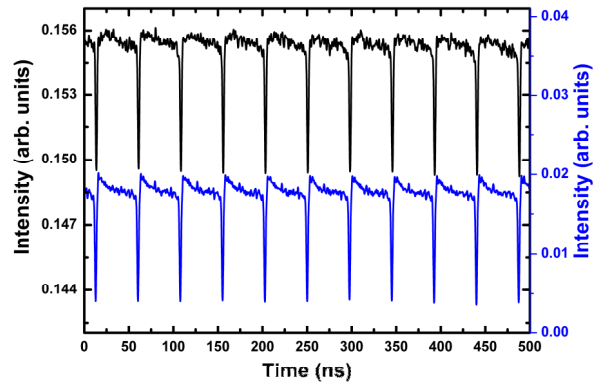

**Figure 7.3**: Oscilloscope traces of a typical dark-dark vector soliton emission of the positive dispersion fiber laser.

The positive dispersion fiber laser has a cavity length of 9.2 m. It comprises 5 m EDF of GVD of –32 (ps/nm)/km and 4.2 m DCF of GVD of –4 (ps/nm)/km. Periodic antiphase square-wave pulses have also been observed in the laser. However, because no intrinsic modulation instability exists, no fine structures were observed within the square pulses even under strong pumping. Exactly like those square-wave pulses observed in the negative dispersion fiber laser, decreasing the linear cavity birefringence the width of the square pulses could be reduced. With an appropriate selection of the cavity parameters, fundamental DBVSs have also been revealed in the laser, which confirms Christodoulides's prediction that formation of the DBVSs is independent of the fiber dispersion [181]. Apart from the DBVSs, in the laser in a narrow cavity birefringence regime we have further obtained a dark-dark type of vector soliton as shown in **Figure 7.3**. In the state the dark solitons were found static in the CW background, which is clearly different from the DBVSs observed. Based on the





polarization resolved soliton spectra and the evolution of the dark solitons with respect to the cavity roundtrips, we further confirmed that the phases of the dark solitons were locked. They constitute a fundamental phase-locked dark-dark vector soliton.

To confirm the dark-bright and dark-dark vector soliton formation in the fiber lasers, we further numerically simulated the operation of the lasers. To faithfully simulate the nonlinear light propagation in the weakly birefringent fibers and the laser cavity feedback, we used a model as described in [115]. Briefly, we circulated the light within a simulation window in the laser cavity. The light propagation in the cavity fibers was described by the coupled extended NLSEs. When the light propagated in the gain fiber, we also considered the gain effects. We had always started the calculations with an arbitrary weak pulse. After one cavity roundtrip of calculation, the result was then used as input for the next roundtrip of calculation, until a stead state of the light field was reached. To make the simulations possibly close to the experimental situations, we used the actual laser cavity lengths and the following parameters for the fibers: $\gamma = 3$ W$^{-1}$km$^{-1}$, $k''' = -0.13$ ps$^3$/km, $k''_{SMF} = -23$ ps$^2$/km, $k''_{EDF1} = -13$ ps$^2$/km, $k''_{DCF} = 5.2$ ps$^2$/km, $k''_{EDF2} = 41.6$ ps$^2$/km. For the gain fibers we used the gain bandwidth $\Omega_g = 16$ nm, and gain saturation energy $P_{sat,EDF1} = 50$ pJ, $P_{sat,EDF2} = 500$ pJ.

Independent of the sign of cavity dispersion, stable DBVSs could indeed be reproduced in our simulations. **Figure 7.4**a and **Figure 7.4**b shows one of the calculated DBVSs, obtained under a cavity linear birefringence of $\Delta n = n_u - n_v = 5.3 \times 10^{-9}$. Both solitons co-propagate in the cavity, and their phases are locked





along the propagation. Numerically it was noticed that in order to obtain a stable phase-locked DBVS, the cavity birefringence must be sufficiently small. Moreover, the pump power must be in an appropriate range. Spectral sidebands can also be identified on the calculated soliton spectra. Like the experimental observations, the sidebands of both solitons have the same locations on the soliton spectrum. Stable dark-dark vector solitons were also numerically obtained in the positive dispersion fiber laser, as shown in **Figure 7.4**c. It was confirmed that the dark-dark vector solitons are phase locked. We note that the dips of all the numerically calculated dark solitons have reached zero, which indicates that they are black solitons. However, the calculated bright soliton has a non-zero background. These features of the numerically calculated vector solitons well match those of the PDWSs theoretically predicted in [181], suggesting that the observed vector solitons are formed due to the coherent cross coupling between the two polarization components.





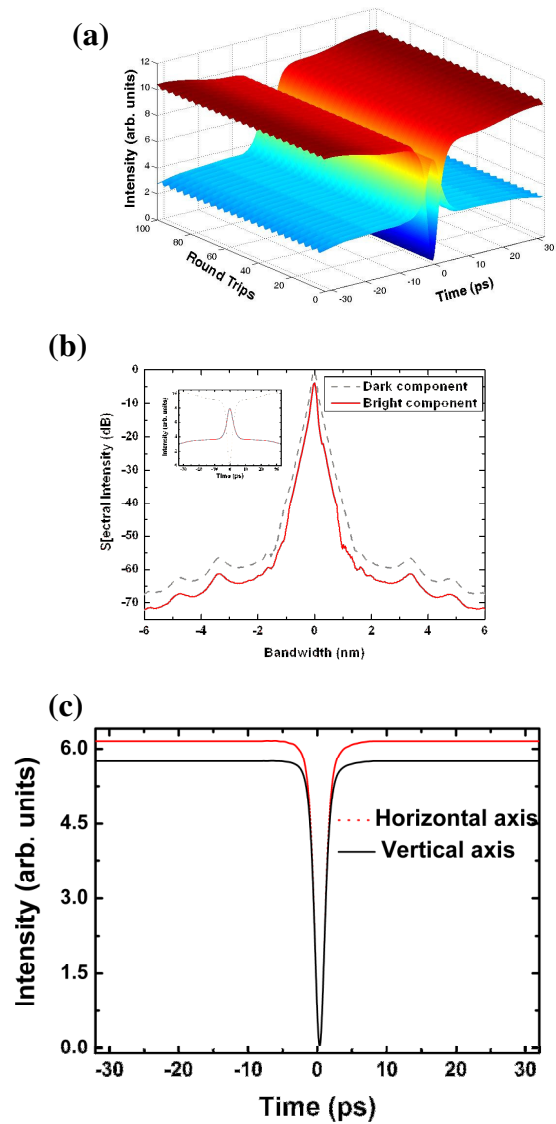

**Figure 7.4**: The dark-bright and dark-dark vector soliton states numerically calculated. (a) Evolution of the dark-bright vector soliton with the cavity roundtrips. (b) Optical spectra of the dark and bright solitons. Inset: soliton profiles of the dark and bright solitons. *Gain* = 120. (c) Soliton profiles of the dark-dark vector soliton numerically calculated. *Gain* = 200, *Δn* = $n_u$–$n_v$ = 1.7× $10^{-9}$.

Finally, we note that the spectral bandwidth of both the experimentally observed and the numerically calculated vector solitons are far narrower than





the gain bandwidth. We had also experimentally observed formation of the NLSE solitons in the negative dispersion fiber laser under strong pumping. The formed NLSE solitons are embedded in the square-wave pulses. Spectral bandwidth of the NLSE solitons was found much broader than those of the domain wall solitons. In addition, PMI was occasionally observed in our experiments. Its appearance had a higher threshold than that of the PDWSs. PMI introduced a high frequency intensity modulation on the square pulses of the lasers. Because of the much narrower spectral bandwidth of the observed PDWSs than the laser gain bandwidth, the effect of the laser gain is purely to balance the cavity loss.

In conclusion, we have first experimentally observed a fundamental phase-locked dark-bright and dark-dark vector soliton in weakly birefringent cavity fiber lasers, respectively. The vector solitons were formed due to the strong cross polarization coupling of light in the fiber lasers. They constitute the two types of the fundamental PDWS theoretical predicted for the coupled NLSEs. Since the coupled NLSEs describe a wide range of physical systems, such as the Bose-Einstein condensate, we believe our experimental observation could be of fundamental as well practical importance.

## 7.2    Vector dark domain wall solitons

Soliton operation of fiber lasers have been observed with anomalous cavity dispersion. It was found that the solitons formed could be described by the NLSE [40]. Formation of solitons in a fiber laser is a result of the mutual nonlinear interaction among the laser gain and losses, cavity dispersion and





fiber nonlinearity, as well as the cavity effects. Dynamics of the laser solitons should be described by the complex Ginzburg-Landau equation (CGLE) [9]. However, it was noticed that solitons formed in the anomalous dispersion cavity fiber lasers have normally a narrow spectral bandwidth, which is far narrower than the laser gain bandwidth. Consequently, no gain bandwidth filtering effect practically exists in the lasers, and the effect of laser gain is mainly to balance the cavity losses. It was confirmed experimentally when the effect of spectral filtering on the pulse shaping could no longer be ignored, the solitons formed in a fiber laser could not be described by the NLSE but the CGLE [9]. Solitons governed by the CGLE are particular example of the dissipative solitons. Recently, formation of dissipative solitons in fiber lasers has attracted considerable attention [16-27].

In addition to gain, the vector nature of light also needs to be considered for fiber lasers whose cavity consists of no polarizing components. In these fiber lasers the soliton dynamics is governed by the coupled NLSEs or CGLEs. Theoretical studies on the coupled NLSEs have shown that due to the cross polarization coupling, new types of solitons, such as the group velocity locked vector soliton [164], phase locked vector soliton [182], induced solitons [119], high order phase locked vector solitons [109], and incoherently coupled vector dark soliton [169] could be formed. Indeed, these solitons were experimentally observed in fiber lasers. Moreover, PDWs and a novel type of vector dark PDW soliton were also theoretically predicted using coupled NLSEs [183]. However, they have not been observed in fiber lasers. Phenomena related to PDW soliton have been observed in passive fibers using two counter-propagating laser beams





[175] and by polarization modulation instability [184]. In this section, we report the first experimental observation of the vector dark PDW solitons in a fiber ring laser.

Our experiment was conducted on a fiber laser schematically shown in **Figure 7.5**. The ring cavity is made of all-anomalous dispersion fiber, consisting of 6.4 m EDF with GVD of 10 (ps/nm)/km and 5.0 m SMF with GVD of 18 (ps/nm)/km. A polarization insensitive isolator was employed in the cavity to force the unidirectional operation of the ring, and an in-line PC was used to fine-tune the linear cavity birefringence. A 10% fiber coupler was used to output the signal. The laser was pumped by a high power Fiber Raman Laser source of wavelength 1480 nm. An in-line polarization PBS was used to separate the two orthogonal polarizations of the laser emission, and they were simultaneously measured with two identical 2 GHz photo-detectors and monitored in a multi-channeled oscilloscope.





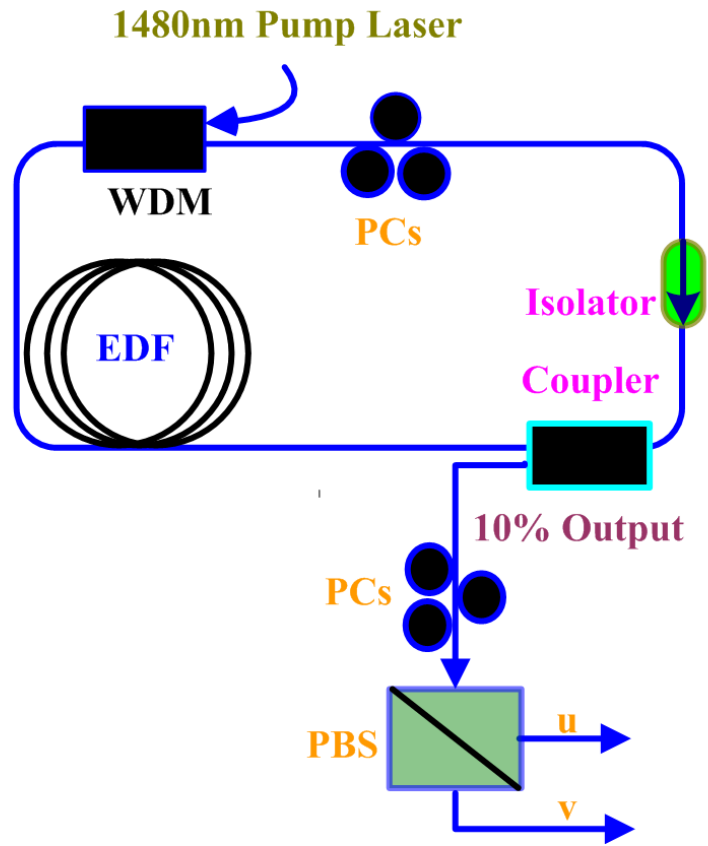

**Figure 7.5:** Schematic of the experimental setup.

A similar configuration fiber laser was previously investigated and a fast antiphase square-pulse polarization dynamics was observed [179, 180]. The observed fast antiphase polarization dynamics was interpreted as caused by the gain competition between the two cavity polarization modes and the cavity feedback. Feedback induced polarization switching in fiber laser was also previously reported [178]. Antiphase square pulse emission along the two orthogonal polarizations was observed in our laser. However, it was different from previous observations [178] in that the square pulse width varied with the cavity birefringence. **Figure 7.6** shows an experimentally measured square-pulse width variation with the orientation of one of the PC paddles. In a range of the paddle's orientation the laser emitted square pulses, and the square pulse





width could be continuously changed as the paddle's orientation was varied. At the two ends of the orientation range, the laser emitted stable CW, whose polarization is linear and orthogonal to each other, indicating that they are the two stable principal polarization states of the laser emission. We then fixed the orientation of the paddle at a position within the stable square pulse emission range and further studied the features of the laser polarization switching with the pump strength change. At a weak pumping, stable square pulses could still be obtained. It was found that the anti-phase intensity variation along the two orthogonal polarizations perfectly compensated each other. When the total laser emission intensity was measured, almost no signature of the polarization switching could be observed. However, as the laser emission intensity was increased, the anti-phase polarization switching was no longer compensated.





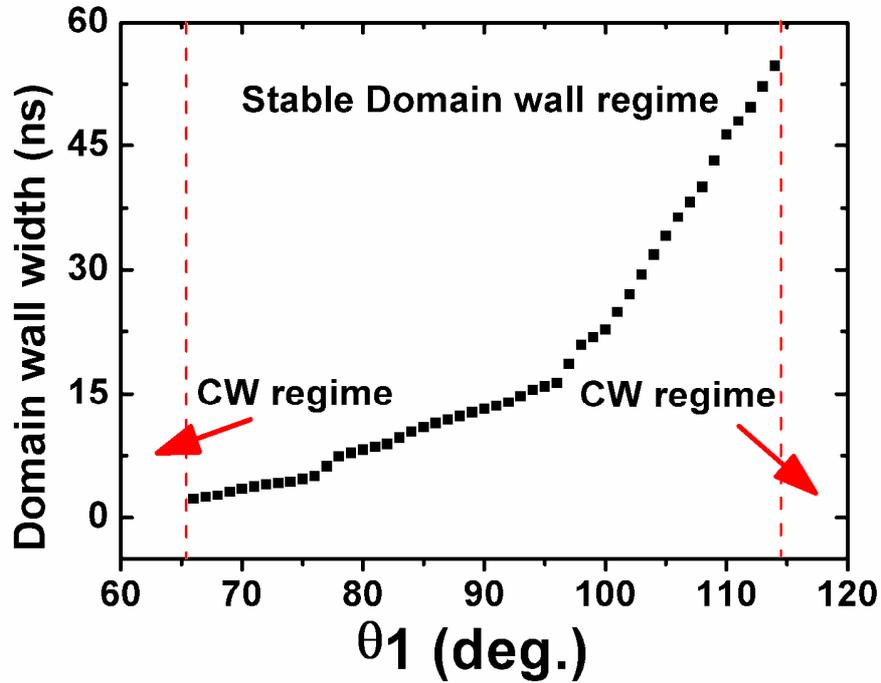

**Figure 7.6**: Duration variation of the square pulses versus the orientation angle of one of the paddles of the intra-cavity PC.

**Figure 7.7**a shows the oscilloscope trace of the total laser emission and one of the polarized emissions, respectively. Within one cavity roundtrip time there is one square-pulse along each polarization direction. Associated with one of the laser emission switching from one polarization to the other, an intensity dip appeared on the total laser emission. The profile of the intensity dip is stable with the cavity roundtrips, and each dip separates the two stable linear polarization states of the laser emission. **Figure 7.7**b shows the corresponding optical spectra of the laser emissions. Laser emissions along the two orthogonal polarization directions have obvious different wavelengths and spectral distributions, showing that the coupling between the two polarization





components is incoherent. In our experiments the cavity birefringence could be altered, by rotating the paddles of the PC or carefully bending the cavity fibers, eventually the wavelength separation between the two spectral peaks could be changed. Independent of the wavelength difference the intensity dip could always be obtained. Moreover, the width and depth of the dip varied with both the cavity birefringence and the pumping strength. At even higher pump strength, splitting of the square pulse could occur. Within one cavity roundtrip another square pulse could suddenly appear. The new square pulse was found unstable. It slowly moved in the cavity and eventually merged with the old one.

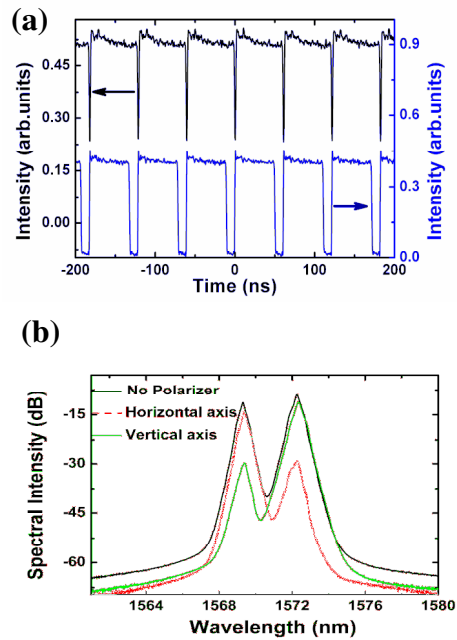

**Figure 7.7**: Vector dark polarization domain wall soliton emission of the laser. (a) Total laser emission (upper trace) and one of the polarized laser emissions (lower trace). (b) The corresponding optical spectra.

The intensity dips possess the characteristics of vector dark PDW soliton predicted by Haelterman and Shepperd [**174, 175**]; despite of the fact that the





two stable polarization domains are now orthogonal linear polarizations instead of circular polarizations. To confirm that such PDWs could exist in our laser, we further numerically simulated the operation of the laser, using the model as described in [115] but with no polarizer in cavity. To make the simulation possibly close to the experimental situation, we used the following parameters: $\gamma = 3$ W$^{-1}$km$^{-1}$, $\Omega_g = 16$ nm, $k''_{SMF} = -23$ ps$^2$/km, $k''_{EDF} = -13$ ps$^2$/km, $k''' = -0.13$ ps$^3$/km, $E_{sat} = 10$ pJ, cavity length $L = 11.4$ m, $L_b = L$ and $G = 120$ km$^{-1}$. To favor the creation of an incoherently coupled domain wall, a perfect polarization switching inside the simulation window was put in the initial condition. This corresponds to the initial existence of the linear polarization switching in our laser caused by the laser gain competition and cavity feedback.

A stable PDW soliton separating the two principal linear polarization states of the cavity could be numerically obtained. **Figure 7.8** shows the both polarization components of domain wall soliton could propagate without distortion and be conjointly trapped in the time domain. **Figure 7.8**b shows the domain walls along each of the polarizations and **Figure 7.8**c is the vector dark PDW soliton formed on the total laser emission and the ellipticity degree of the soliton. We adopted the definition of ellipticity degree $q = (\mu - \nu) / (\mu + \nu)$, where $q = \pm 1$ represents the two orthogonal linearly polarized states and $q = 0$ refers to a circularly polarized state [183]. The PDWs and soliton are stable and invariant with the cavity roundtrips. Numerically it was observed that even with very weak cavity birefringence, e.g. $L_b = 100L$, stable PDW soliton could still be obtained. However, if the cavity birefringence becomes too large, e.g. larger than $L_b = 0.5L$, no stable PDWs could be obtained.





Therefore, based on the numerical simulation and the features of the experimental phenomenon, we interpret the intensity dips shown in **Figure 7.7**a as a type of vector dark PDW soliton. To understand why the PDWs and vector dark soliton could be formed in our laser, we note that Malomed had once theoretically studied the interaction of two orthogonal linear polarizations in the twisted nonlinear fiber [185-187]. It was shown that PDWs between the two orthogonal linear polarizations of the fiber exist, and the fiber twist could give rise to an effective force driving the domain walls. Considering that both the gain competition and the cavity feedback could have the same role as the fiber twist, not only the domain walls but also the moving of the domain walls could be explained.





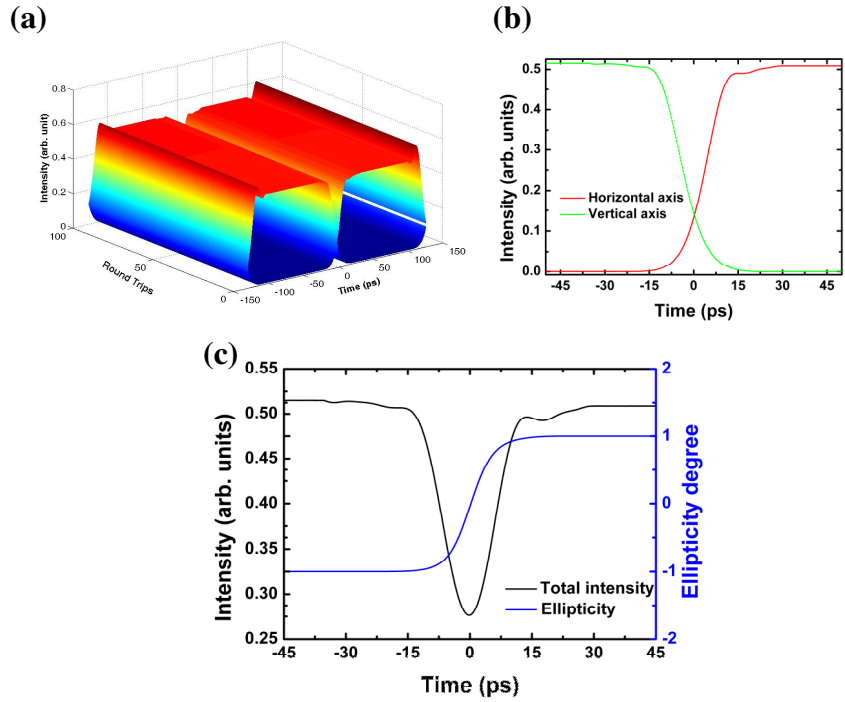

**Figure 7.8**: Polarization domain wall numerically calculated. (a) Evolution of the polarization domain wall with the cavity roundtrips. (b) Domain wall profiles at particular roundtrip. (c) The vector domain wall soliton and its ellipticity degree at particular roundtrip.

In conclusion, we have reported the experimental observation of PDWs and vector dark PDW soliton in a linear birefringence cavity fiber ring laser. The domain walls and solitons are found to separate the two stable orthogonal linear principal polarizations. We have further shown that the cavity feedback and the gain competition could have played an important role on the formation of such PDWs and the vector dark domain wall soliton.

### 7.3    Dual-wavelength domain wall solitons

In addition to the NLSE solitons, recently a novel new type of optical solitary





waves known as the polarization domain wall solitons (PDWSs) were also experimentally revealed in fiber lasers [188]. Formation of PDWSs was first theoretically predicted by Haelterman and Sheppard [183]. It was shown that the cross coupling between the two orthogonal polarization components of light propagating in a dispersive Kerr medium could lead to the formation of a stable localized structure that separates domains of the two orthogonal polarization fields. Cross coupling between waves is a common phenomenon in a wide range of nonlinear physical systems. The experimental confirmation of PDWSs suggests that similar domain wall solitons could also be observed in other nonlinear wave coupling systems [187]. In this section, we report on the experimental observation of a dual-wavelength optical domain wall soliton (DWS) in a fiber ring laser made of all-normal GVD fibers. We show both experimentally and numerically that strong coupling between two different wavelength beams in the fiber laser can result in the formation of DWSs, representing as a stable dark intensity pulse separating the two different wavelength laser emissions.

We used a fiber laser whose cavity is as shown in [163]. Briefly, the cavity is made of 5.0 m EDF with a GVD parameter of –32 (ps/nm)/km, 6.1 m DCF with a GVD parameter of –2 (ps/nm)/km. A polarization sensitive isolator was employed in the cavity to force the unidirectional operation of the ring cavity, and an in-line PC was used to fine-tune the linear cavity birefringence. A 10% fiber coupler was used to output the signal. The laser was pumped by a high power Fiber Raman Laser source of wavelength 1480 nm. The laser output was monitored with a 2 GHz photo-detector and displayed on a multi-channeled





oscilloscope. A major difference of the current laser to that of [163] is that in setting up the laser we have intentionally imposed large birefringence into the cavity. Consequently, the birefringence induced multi-pass filter effect becomes very strong [189] and can no longer be ignored for the laser. Under the combined action of the laser gain and the birefringent filter, the laser is forced to operate in a dual wavelength CW emission mode. **Figure 7.9**a shows a typical spectrum of the laser under dual-wavelength emission. In our experiment the strength of the laser emission increased with the pump power. As the intensity of the laser emission was continuously increased, it was observed that a dark intensity pulse as shown in **Figure 7.9**b could appear on the total laser output and the stronger the pump power, the narrower became the dark pulse, as shown in **Figure 7.9**c. Tuning the orientation of the intra cavity PC, the bandpass wavelengths of the birefringence filter can be shifted. Therefore, the relative strength of the two laser emissions could be altered. Eventually single wavelength emission could also be obtained. It was found that under single wavelength operation no above dark pulses could be observed, indicating that the appearance of the dark pulse was related to the coupling of the two different wavelength beams.





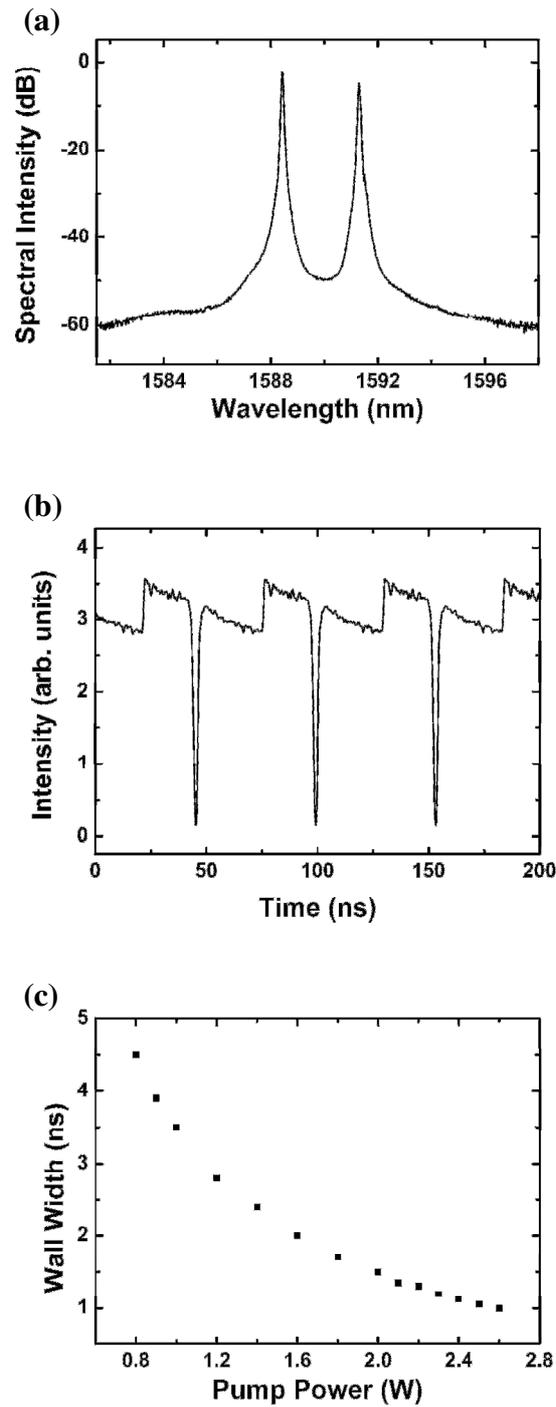

**Figure 7.9**: (a) Spectrum; (b) oscilloscope traces of the dual-wavelength optical

domain wall. (c): the wall duration as a function of the pump strength.

Careful examining the laser output trace shown in **Figure 7.9**b it came to our

attention that the laser emission exhibited two distinctive intensity levels. We





could experimentally identify that each of the intensity levels was related to the laser emission at one of the two wavelengths. Namely, the laser was not simultaneously emitting at the two wavelengths but alternating between them. The dark pulses always appeared at the position where the laser emission switched from one wavelength to the other. The dark pulses were very stable, showing that they are a DWS.

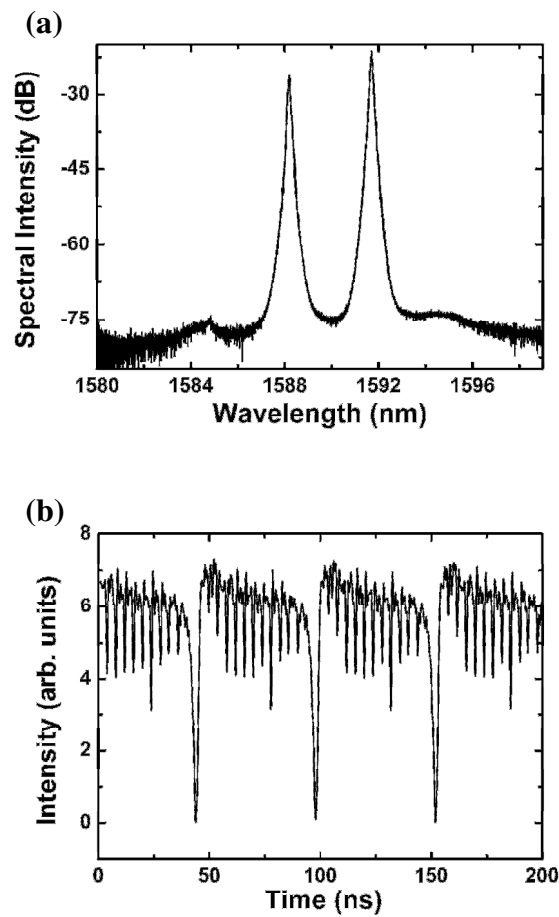

**Figure 7.10**: (a) Spectrum and (b) oscilloscope trace of the single dark soliton obtained through the paddles of PC. (For comparison)

Under even stronger pump power, a state as shown in **Figure 7.10** was also obtained, where a new type of dark pulses with much narrower pulse width





suddenly appeared. These new dark pulses moved with respect to the DWS. They not only have different darkness but also appeared randomly. We had studied previously the NLSE dark solitons in a fiber laser [163]. The observed features of the new dark pulses suggest that they are the NLSE dark solitons. In our experiment we could control the strength of the laser emission of either wavelength through shifting the filter frequencies. In this way we can suppress the appearance of the NLSE dark solitons on either of the two wavelength laser emissions. **Figure 7.11** shows a comparison of the laser emissions where one of the two-wavelength laser emissions is beyond the NLSE soliton threshold. **Figure 7.11** shows the laser emission spectra. The upper (lower) oscilloscope trace shown in **Figure 7.11**b corresponds to the spectrum whose longer (shorter) wavelength spectral line has become broadened. Associated with the appearance of the NLSE dark solitons the corresponding laser emission spectrum became further broadened. The experimental results shown in **Figure 7.10** and **Figure 7.11** clearly demonstrated that the appearance of the DWSs has much lower pump threshold than the NLSE dark solitons.





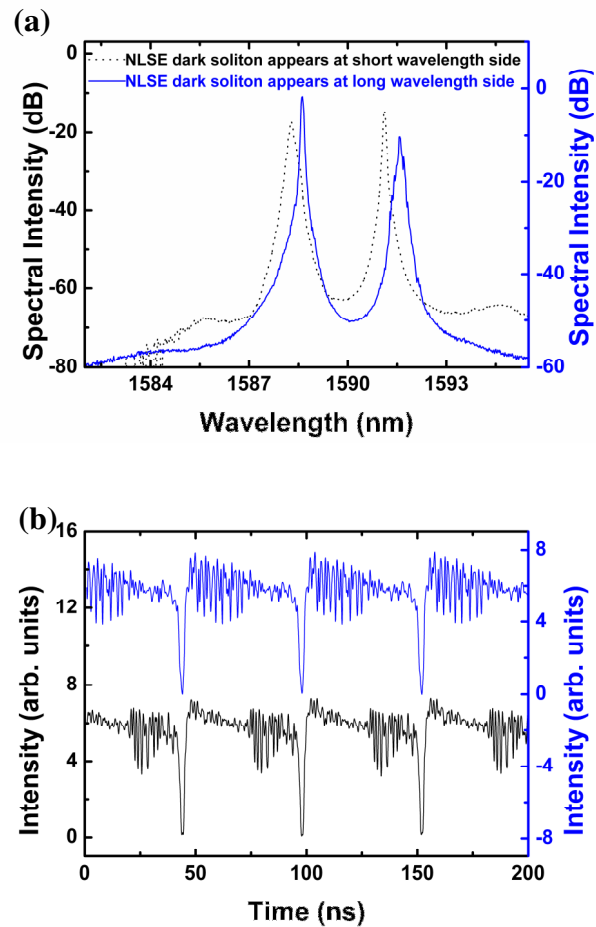

**Figure 7.11**: Generation of the multiple dark solitons at each individual wavelength of the optical domain wall. (a) Spectra and (b) oscilloscope traces, upper (lower) trace corresponds to spectral broadening at longer (shorter) wavelength.

To better understand the dual-wavelength DWS formation in our laser, we have further numerically simulated the operation of our laser under two wavelength emissions. The following coupled Ginzburg-Landau equations were used to describe the light propagation in the cavity fibers:





$$\frac{\partial u_1}{\partial z} = i\beta u_1 - \delta\frac{\partial u_1}{\partial t} - \frac{ik''}{2}\frac{\partial^2 u_1}{\partial t^2} + \frac{k'''}{6}\frac{\partial^3 u_1}{\partial t^3} + i\gamma\left(|u_1|^2 + 2|u_2|^2\right)u + \frac{g}{2}u_1 + \frac{g}{2\Omega_g^2}\frac{\partial^2 u_1}{\partial t^2}$$

$$\frac{\partial u_2}{\partial z} = -i\beta u_2 + \delta\frac{\partial u_2}{\partial t} - \frac{ik''}{2}\frac{\partial^2 u_2}{\partial t^2} + \frac{k'''}{6}\frac{\partial^3 u_2}{\partial t^3} + i\gamma\left(|u_2|^2 + 2|u_1|^2\right)v + \frac{g}{2}u_2 + \frac{g}{2\Omega_g^2}\frac{\partial^2 u_2}{\partial t^2}$$

where, $u_1$ and $u_2$ are the normalized envelopes of the optical pulses along the same polarization in the optical fiber but having different central wavelengths $\lambda_1$ and $\lambda_2$. $\beta = 2\pi\Delta n/(\lambda_1 + \lambda_2)$ is the wave-number difference between the two optical waves. $\delta = \beta(\lambda_1 + \lambda_2)/4\pi c$ is the inverse group velocity difference. $k''$ is the second order dispersion coefficient, $k'''$ is the third order dispersion coefficient and $\gamma$ represents the nonlinearity of the fiber. $g$ is the saturable gain coefficient of the fiber and $\Omega_g$ is the bandwidth of the laser gain. For undoped fibers $g = 0$; for erbium doped fiber, we considered its gain saturation as

$$g = G\exp\left[-\frac{\int(|u|^2 + |v|^2)dt}{P_{sat}}\right]$$

Moreover, we considered the cavity feedback effect by circulating the light in the cavity [11]. To make the simulations possibly close to the experimental situation, we have assumed that the two wavelength beams have the same group velocity and used the following parameters: $\gamma = 3$ W$^{-1}$km$^{-1}$, $\Omega_g = 16$ nm, $k''_{DCF} = 2.6$ ps$^2$/km, $k''_{EDF} = 41.6$ ps$^2$/km, $k''' = -0.13$ ps$^3$/km, $P_{sat} = 500$ pJ, cavity length $L = 11.1$ m, and $G = 120$ km$^{-1}$.

A weak dual-wavelength beam with 2.6 nm wavelength separation and an intensity switching between the two wavelengths was used as the initial condition. We let the light circulate in the cavity until a stable state is obtained.





The CGLEs were solved using the split-step method. We found numerically that a stable DWS separating the laser emissions of different wavelengths could indeed be formed in our laser, as shown in **Figure 7.12**. Evolution of the dual wavelength domain wall with the cavity roundtrips can be clearly seen in **Figure 7.12**a and **Figure 7.12**b. **Figure 7.12**c shows the domain walls and the corresponding dark DWS calculated. **Figure 7.12**d shows the optical spectrum of the laser emission.

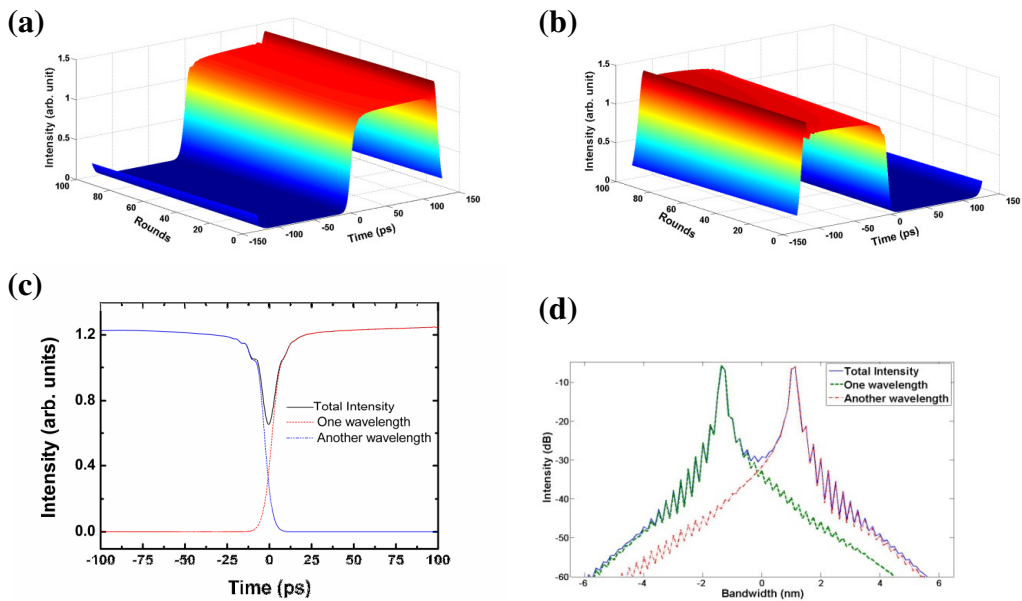

**Figure 7.12**: dual wavelength domain wall numerically calculated. Evolution of the dual wavelength domain wall with the cavity roundtrips: (a) one wavelength (shorter wavelength) (b) Another wavelength (longer wavelength). (c) Domain wall profiles at particular roundtrip (d) Its corresponding spectra.

We note that dual-wavelength DWS in a SMF was theoretically studied by Haelterman and Badolo [190]. To the best of our knowledge, no dual-wavelength optical DWSs have been experimentally confirmed. Based on our numerical simulations, we noticed that for the formation of the DWS an initial





intensity alternation between the two wavelengths is crucial. We believe that in our laser the gain competition between the two laser beams could have played an important role on forming such an initial condition. It had been shown experimentally previously that gain competition could cause antiphase dynamics between two different wavelength beams in a fiber laser [191].

In conclusion, we have experimentally observed a new type of dark soliton in an erbium-doped fiber laser made of all-normal GVD fibers. It was shown that the formation of the soliton was a result of the mutual coupling between two different wavelength beams and the formed soliton has the characteristic of separating the two different wavelength laser emissions. It is an optical DWS. In addition, our experimental result has shown that the appearance of DWS has a much lower pump threshold than the NLSE dark solitons, and under strong nonlinear coupling the dual wavelength emission fiber laser is not simultaneously emitting two wavelengths by alternating between the two wavelengths.





# Chapter 8. Graphene mode locked fiber lasers

Until now, passive mode-locking techniques are effective approaches to generate ultra-fast pulse and the dominant mode-locking technology is based on SESAMs, which use III-V semiconductor multiple quantum wells grown on distributed Bragg reflectors. However, there are a number of drawbacks associated with SESAMs. For example, SESAMs require complex and costly clean-room-based fabrication systems such as MOCVD or MBE, and an additional substrate removal process is needed in some cases; high-energy heavy-ion implantation is required to introduce defect sites in order to reduce the device recovery time (typically a few nanoseconds) to the picosecond regime required for short-pulse laser mode-locking applications; since the SESAM is a reflective device, its use is restricted to only certain types of linear cavity topologies. Other laser cavity topologies such as the ring-cavity design, which requires a transmission-mode device, offers advantages such as doubling the repetition rate for a given cavity length, and less sensitive to reflection-induced instability with the use of optical isolators, is not possible unless an optical circulator is employed, which increases cavity loss and laser complexity; SESAMs also suffer from a low optical damage threshold. But there had been no alternative saturable absorbing materials to compete with SESAMs for the passive mode-locking of fiber lasers.

Recently, by the virtue of the saturable absorption of SWCNTs in the near-infrared region with ultrafast saturation recovery times of 1 picosecond,





researchers have successfully produced a new type of effective saturable absorber quite different from SESAMs in structure and fabrication, and has, in fact, led to the demonstration of pico- or subpicosecond EDF lasers. In these lasers, solid SWCNT saturable absorbers have been formed by direct deposition of SWCNT films onto flat glass substrates, mirror substrates, or end facets of optical fibers. However, the non-uniform chiral properties of SWNTs present inherent problems for precise control of the properties of the saturable absorber. Furthermore, the presence of bundled and entangled SWNTs, catalyst particles, and the formation of bubbles cause high nonsaturable losses in the cavity, despite of the fact that the polymer host can circumvent some of these problems to some extent and afford ease of device integration. In addition, under large energy ultrashort pulses multi-photon effect induced oxidation occurs, which degrades the long term stability of the absorber [192].

In the following Chapter, we would like to introduce another novel type of saturable absorber: graphene which is a single two-dimensional (2D) atomic layer of carbon atom arranged in a hexagonal lattice. Although as an isolated film it is a zero bandgap semiconductor, we found that like SWCNTs, graphene also possesses saturable absorption. In particular, as it has no bandgap, its saturable absorption is wavelength independent. It is potentially possible to use graphene or graphene-polymer composite to make a wideband saturable absorber for laser mode locking. Furthermore, comparing with the SWCNTs, as graphene has a 2D structure it should have much smaller non-saturable loss and much higher damage threshold. Indeed, in our experiments with an erbium-doped fiber laser we have achieved self-started mode locking and stable soliton





pulse emission with high energy. Section 8.1 introduces the application of atomic layer pure graphene as a saturable absorber and Section 8.2 further report on the fabrication of graphene-polymer composite as a mode locker. In Section 8.3, the generation of vector soliton in a graphene mode locked fiber laser is discussed. In Section 8.4, we would briefly discuss the potential application of graphene as full-waveband saturable absorber in ultra-fast photonics. The prospect of graphene based ultrafast photonics will be briefly presented in Section 8.5.

## 8.1 Atomic Layer Graphene as Saturable Absorber

High power ultrashort optical pulses have widespread applications in industrial material processing, medical treatment and scientific researches. Recently, it has been demonstrated that such optical pulses can also been directly generated by the passively mode locked fiber laser oscillators. By using the all-normal dispersion dissipative soliton shaping technique, Ruehl *et al*. have demonstrated 10 nJ pulse generation in an EDF laser [193], and A. Chong *et al*. have generated 20 nJ pulses from a Yb-doped fiber laser [194]. The results have shown the promise of using the mode locked fiber lasers to replace the bulk ultrashort pulse solid-state lasers for many practical applications. However, the reported high power mode locked fiber lasers have a drawback. In order to enable the high power mode locking operation, an artificial saturable absorber formed based on the light interference was used as the mode locker. Therefore, the lasers are environmentally unstable. In order to sidestep this drawback, a real passive mode locker that can endure high optical power, particularly the SWCNTs based saturable absorbers should be used. Very recently, Y. W. Song





*et al.* reported a technique of using single walled carbon nanotubes as mode locker for high-energy pulse formation [83]. By employing the evanescent field interaction of light with the nanotubes, it was shown that the high optical power induced damage of the nanotube could be avoided. Consequently, stable mode locked pulses with 1.2 ps pulse width and 6.5 nJ pulse energy were generated in an erbium-doped fiber laser. In this section, we report on the large energy ultrashort pulse generation in an erbium-doped fiber laser passively mode locked with atomic layer graphene. We show that graphene can not only be used as a mode locker to mode lock fiber lasers, but also has high optical damage threshold. Stable mode locked pulses with pulse energy as high as 7.3 nJ and pulse width of 415 fs have been directly obtained from a dispersion-managed cavity fiber lasers.

Graphene is a single 2D atomic layer of carbon atoms arranged in a hexagonal lattice. An isolated graphene film is a zero bandgap semiconductor with a linear energy dispersion relation for both electrons and holes near Dirac point [195]. Saturable absorption in graphene is achieved due to the Pauli blocking of the electrons and holes for occupation of the energy levels in the conduction and valence bands that are resonant with the incident photons [97]. Recent advance in graphene research has shown that graphene saturable absorption has an ultrashort recovery time [97]. As graphene has a 2D structure, it has much smaller non-saturable loss and higher damage threshold. Therefore, it is expected that using graphene as a laser mode locker large energy ultrashort pulses could be generated.

Our fiber laser is schematically shown in **Figure 8.1**a. A piece of 5.0 m, 2880





ppm EDF with GVD of –32 (ps/nm)/km was used as the gain medium, and 24.2 m SMF with GVD 18 (ps/nm)/km was employed to compress the intra-cavity pulses and obtain ultrashort pulses. The net cavity fiber dispersion is estimated –0.3583 ps$^2$. A 30% fiber coupler was used to output the signal. The laser was pumped by a high power Fiber Raman Laser source (KPS-BT2-RFL-1480-60-FA) of wavelength 1480 nm, and the maximum pump power can be as high as 5 W. A polarization independent isolator was spliced in the cavity to force the unidirectional operation of the ring cavity, and an intra-cavity PC was used to fine tune the linear cavity birefringence. **Figure 8.1**b illustrates the atomic structure of graphene, which is a one-atom-thick planar sheet of sp$^2$-bonded carbon atoms that are densely packed in a honeycomb crystal lattice.

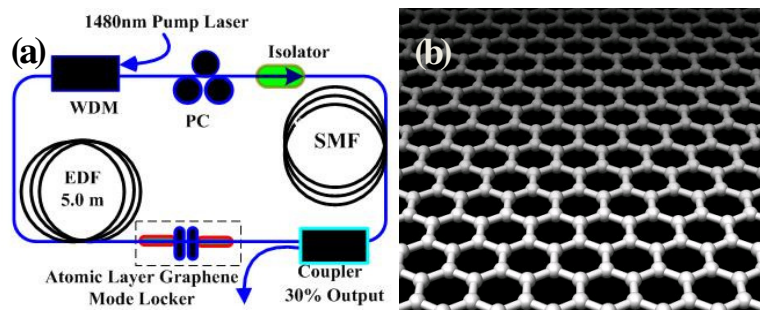

**Figure 8.1:** (a) Schematic of the fiber laser. (b) 2D atomic layer of carbon atoms arranged in a hexagonal lattice.

The graphene mode locker used in our experiment was made by transferring a piece of free standing graphene film onto a fiber pigtail by the mutual Van Der Waals forces. The few layers graphene thin film was synthesized on Ni/Si substrate by the CVD method [196]. The as-produced graphene was stripped off the substrate through an oxidizing etching treatment in an aqueous iron (III)





chloride (FeCl3) solution to obtain a freestanding atomic layer graphene film. **Figure 8.2**a shows a Raman spectrum of the graphene, from which the G band and 2D band are clearly resolved. The relative weak D band indicates a very high crystallinity of our samples. **Figure 8.2**b shows the Raman image plotted by the Raman peak of doped $SiO_2$, from which the fiber core area is located. And **Figure 8.2**c shows the Raman image plotted by G band of the graphene around the core area, which demonstrates that few layers graphene is coated on the fiber core. We have experimentally measured the normalized absorption modulation depth and saturation fluence of the graphene saturable absorber. They are ~66.5% and 0.71 MW/cm$^2$, respectively [97].





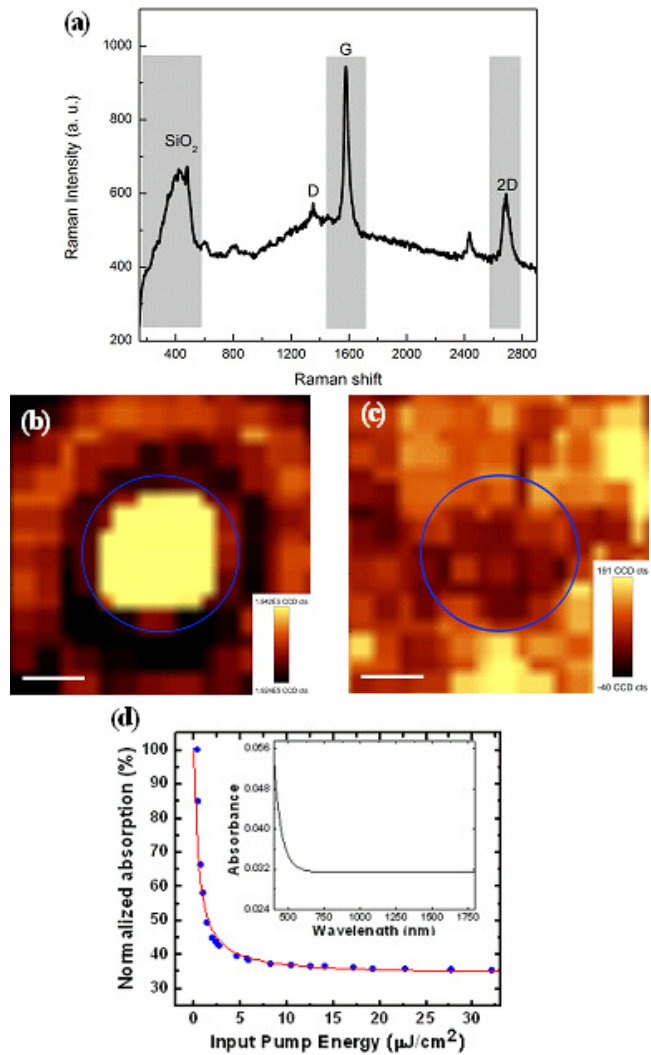

**Figure 8.2**: Characterization of graphene thin film covering on the fiber core. (a) Raman spectra of the grapheme film. (b) Raman image around the fiber core plotted by the intensity of the Raman peak of $SiO_2$. The scale bar is 3 µm. (c) Raman images around the fiber core plotted by the intensity of G band of graphene. The scale bar is 3 µm.

Self-started mode locking of the laser occurred at the incident pump power of about 130 mW. The mode locking state could then be maintained to the maximum accessible pump intensity of 3.5 W. **Figure 8.3** shows a typical mode locking state of the laser. **Figure 8.3**a is the optical spectrum of the mode





locked pulses. It is centered at 1576.3 nm and has a 3 dB bandwidth of ~10 nm. Although the net cavity fiber dispersion is anomalous, no Kelly sidebands are observed on the spectrum. **Figure 8.3**b is the measured autocorrelation trace of the mode locked pulses. It has a FWHM width of 590 fs. If a Gaussian-pulse profile is assumed, the pulse duration is 415 fs. The TBP of the pulses is ~0.518, indicating that the mode locked pulses are slightly chirped. **Figure 8.3**c shows the measured oscilloscope trace within nanosecond and millisecond (insert of **Figure 8.3**c) time scale. In all of the pump power range, the laser always emitted single pulse, no pulse breaking or multiple pulse operation was detected as confirmed with a high speed oscilloscope together with the autocorrelation trace measurement. The pulse circulated in the cavity with the fundamental cavity repetition time of 146 ns. Gradually increasing the pump strength, except the spectral bandwidth became broader, the autocorrelation profile and the spectral profile had always the same shapes. We also measured the RF spectrum of the mode locking state. **Figure 8.3**d shows a measurement made at a span of 10 kHz and a resolution bandwidth of 10 Hz. The fundamental peak located at the cavity repetition rate of 6.84 MHz has a signal-to-noise ratio of 65 dB. The insert of **Figure 8.3**d shows the wideband RF spectrum up to 100 MHz. The absence of spectral modulation in RF spectrum demonstrates that the laser operates well in the CW mode-locking regime.





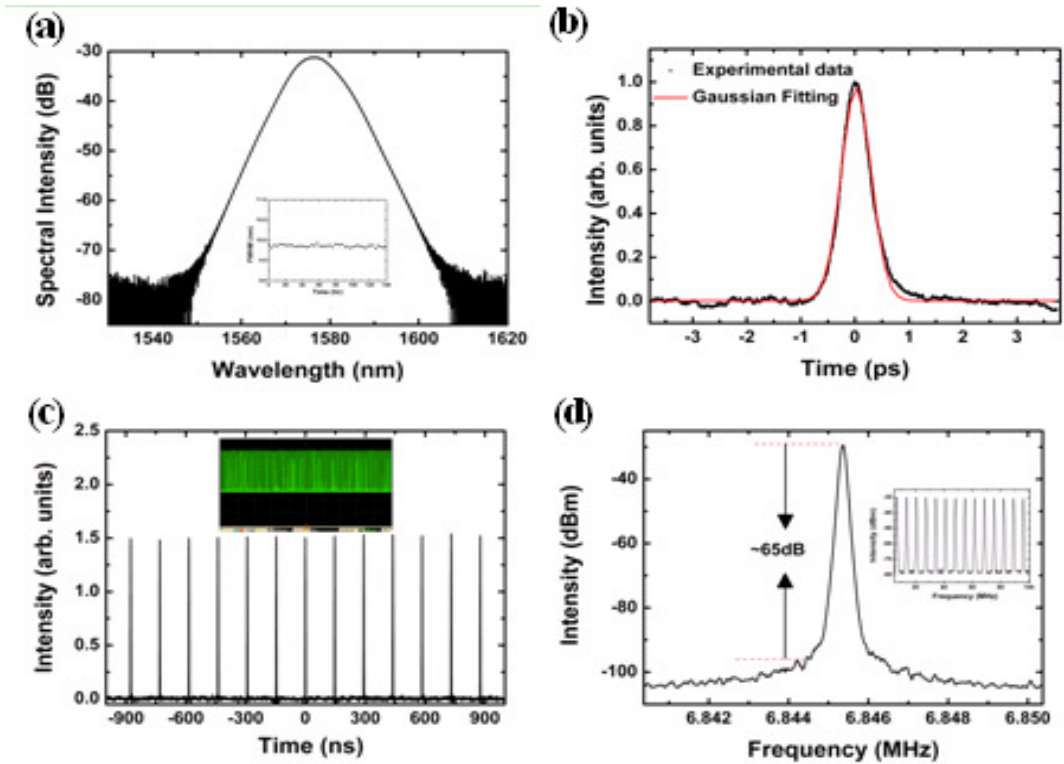

**Figure 8.3**: Pulse operation of the fiber laser. (a) Pulse spectra measured. Insert: long-term fluctuation of the FWHM. (b) Autocorrelation traces of the pulses. (c) An oscilloscope trace of the single pulse emission. Insert: pulse train of CW mode-locking in millisecond time scale. (d) The fundamental radio-frequency (RF) spectrum of the laser output. Insert: wideband RF spectrum up to 100 MHz.

**Figure 8.4** shows the single pulse energy change with the pump power. Starting from the laser mode locking threshold, the output power increased linearly with the pump power with a slope efficiency of 6.7%. The maximum achieved single pulse energy is as high as ~7.3 nJ. To the best of our knowledge, this is the highest pulse energy reported for ultrafast erbium-doped fiber laser mode locked with a real saturable absorber in cavity. At the maximum output power, the pulse width is ~415 fs, which gives the maximum peak power of 17.6 kW.





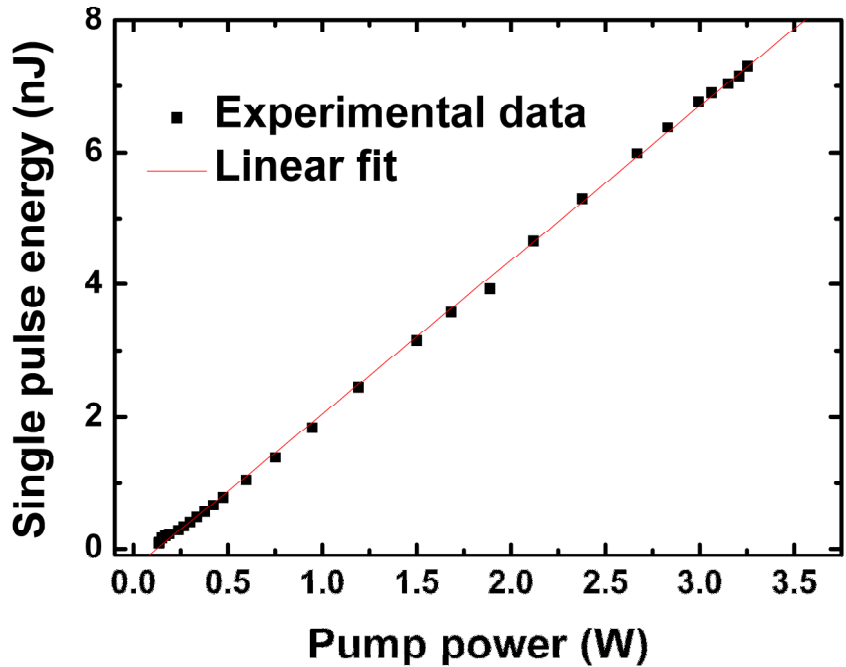

**Figure 8.4**: the single pulse energy in respect to the pump power.

To investigate the long-term stability of the atomic layer graphene mode locking, we monitored the mode locked laser operation for 140 hours. As shown in the insert of **Figure 8.3**a, the 3 dB spectral bandwidths of the output pulses were kept in a consistent value with little fluctuations. Despite of the fact that the graphene was radiated under an optical fluency of 52 mJ/cm$^2$ no obvious degradation on the mode locking performance was observed. Optical microscopy study on the atomic layer graphene film also confirmed that its morphology has kept intact.

In this section, we have demonstrated large energy mode locking of an erbium-doped fiber laser with atomic layer graphene as the saturable absorber. Stable mode locked pulses with single pulse energy as high as 7.5 nJ and 415 fs pulse width have been generated. Our experimental results showed that atomic layer





graphene could be a promising saturable absorber for high power laser mode locking.

## 8.2    Graphene-polymer composite as saturable absorber

To manipulate and engineer the saturable absorption properties of graphene, a series of physical or chemical treatments could be used towards graphene, including: tuning the optical modulation depth and saturable fluence through different layers of graphene or doping/intercalating with other materials. In Chapter 8, we demonstrated the application of graphene derivatives or graphene nano-composites (e. g. polymer-graphene, graphene gel) as saturable adsorber materials for the mode locking of fiber lasers. Due to the introduction of polymer host into this novel saturable absorber which helps to prevent graphene from oxidation, the damage threshold could be significantly enhanced. Furthermore, comparing with the SWCNTs, as graphene has a 2D structure it should have less surface tension, therefore, much higher damage threshold. Indeed, in our experiments with an erbium-doped fiber laser we have first demonstrated self-started mode locking of the laser with a grapheme-polymer membrane as the saturable absorber, and achieved stable soliton pulses with 700 fs pulse width and 3 nJ pulse energy.

Our fiber laser is similar to **Figure 8.1**a. A piece of 5.0 m EDF with GVD of –32 (ps/nm)/km was used as the gain medium, and 23.5 m SMF with GVD 18 (ps/nm)/km was employed in the cavity to compensate the normal dispersion of the EDF and ensure that the cavity has net anomalous GVD. The net cavity dispersion is estimated –0.3419 $ps^2$. A 30% fiber coupler was used to output the





signal, and the laser was pumped by a high power Fiber Raman Laser source (KPS-BT2-RFL-1480-60-FA) of wavelength 1480 nm. The maximum pump power can be as high as 5 W. A polarization independent isolator was spliced in the cavity to force the unidirectional operation of the ring cavity, and an intra-cavity polarization PC was used to fine-tune the linear cavity birefringence. An optical spectrum analyzer (Ando AQ-6315B) and a 350 MHz oscilloscope (Agilent 54641A) combined with a 2 GHz photo-detector was used to simultaneously monitor the spectra and pulse train, respectively. A graphene-polymer nanocomposite membrane with a thickness of ~ 10 µm inserted between two ferules was used as the saturable absorber for the laser mode locking. To make the graphene-polymer nanocomposite, graphene nanosheets were produced by chemically reducing the oxidized graphene exfoliated from graphite flakes [197]. The graphene nanosheets were then non-covalently functionalized with 1-pyrenebutanoic acid, succinimidyl ester (PBASE) to improve their solubility in organic solvents (i.e., ethanol and acetone) and compatibility with polymers (i.e., PVDF). The as-produced graphene (2 mg) was further dispersed in ethanol (3 mL) and mixed with PVDF solution comprising 1.5 g PVDF dissolved in 10 mL dimethylacetamide/acetone (2:3). The G-PVDF solution was then stirred for 24 hours at 60 ºC in sealed bottle to form wet paste for electrospinning. The electrospinning was carried out in a MECC NANON ELECTROSPINNING SETUP at a bias voltage of 30 kV and feeding rate of 0.5 mL/hour. **Figure 8.5**b is a scanning electron microscopy (SEM) image of the membrane. It shows that the membrane mainly comprises networks of the graphene-filled polymer nanofibers. The inset of **Figure 8.5**b is





a photo of a free-standing membrane. **Figure 8.5**c shows the transmission electron microscopy (TEM) image of a graphene-PVDF nanofiber. It reveals that the graphene nanosheets are well dispersed in the polymer matrix without obvious aggregation.

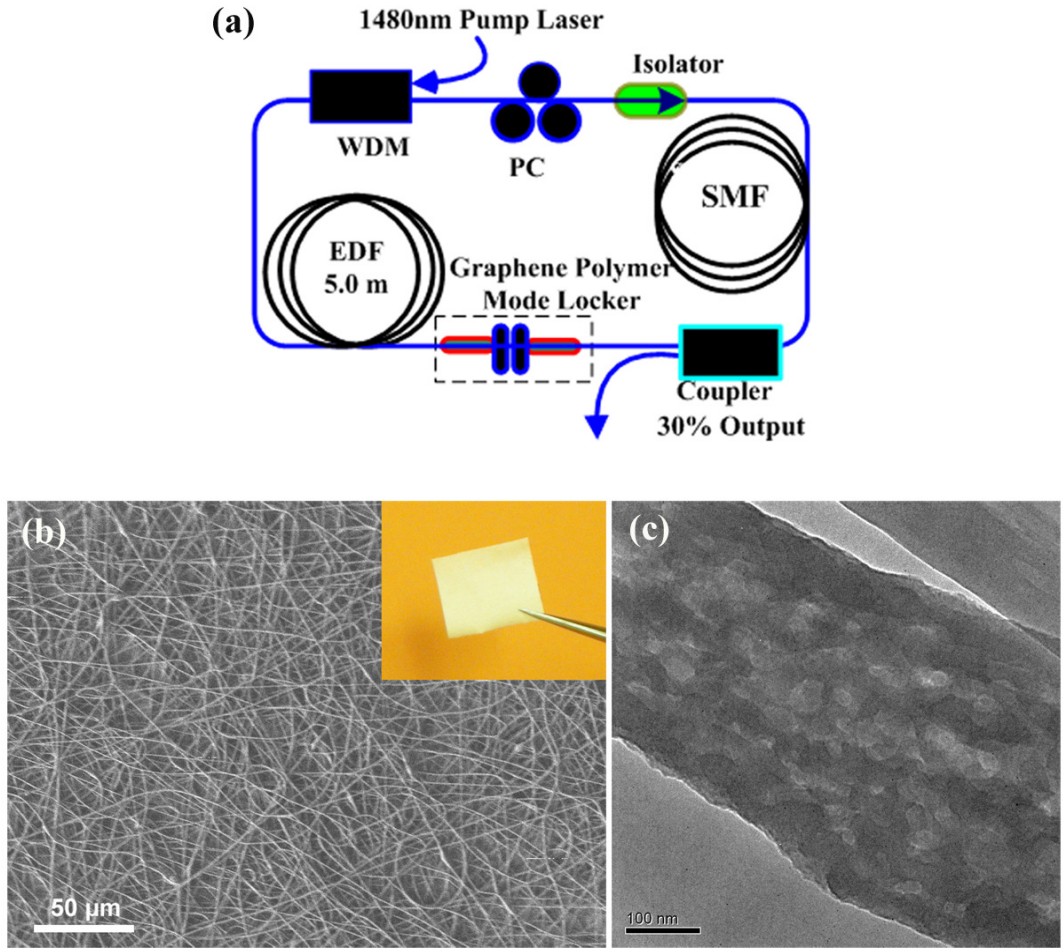

**Figure 8.5:** (a) SEM image of the graphene-polymer nanofiber networks. Inset: a photo of the free-standing graphene-polymer composite membrane. (b) Transmission electron microscopy (TEM) image of a graphene-PVDF nanofiber.





We have also measured the linear and nonlinear absorption of the graphene-polymer membrane. The linear absorption spectra of both graphene-based PVDF nanocomposite and pure PVDF were compared in **Figure 8.6**a , which shows that pure PVDF has a relatively low absorbance of ~35% in the L-band of the optical communication windows while graphene-based PVDF nano-composites has an enhanced absorbance of ~52%. The nonlinear absorption curve of **Figure 8.6**b measured at the wavelength of $\lambda = 1590$ nm shows that the graphene-based PVDF nanocomposite has a normalized modulation depth of ~28.3 % and a saturable fluence of 0.75 MW/cm$^2$, which is about an order of magnitude smaller than that of the SWCNTs based saturable absorber. Moreover, the insertion loss of the graphene-based PVDF nanocomposite was as low as ~1.5 dB. Previous studies have also shown that graphene and graphite thin films have both an ultrafast absorption recovery time constant of ~ 200 fs and a slower recovery constant of 2.5 ps to 5 ps [198].





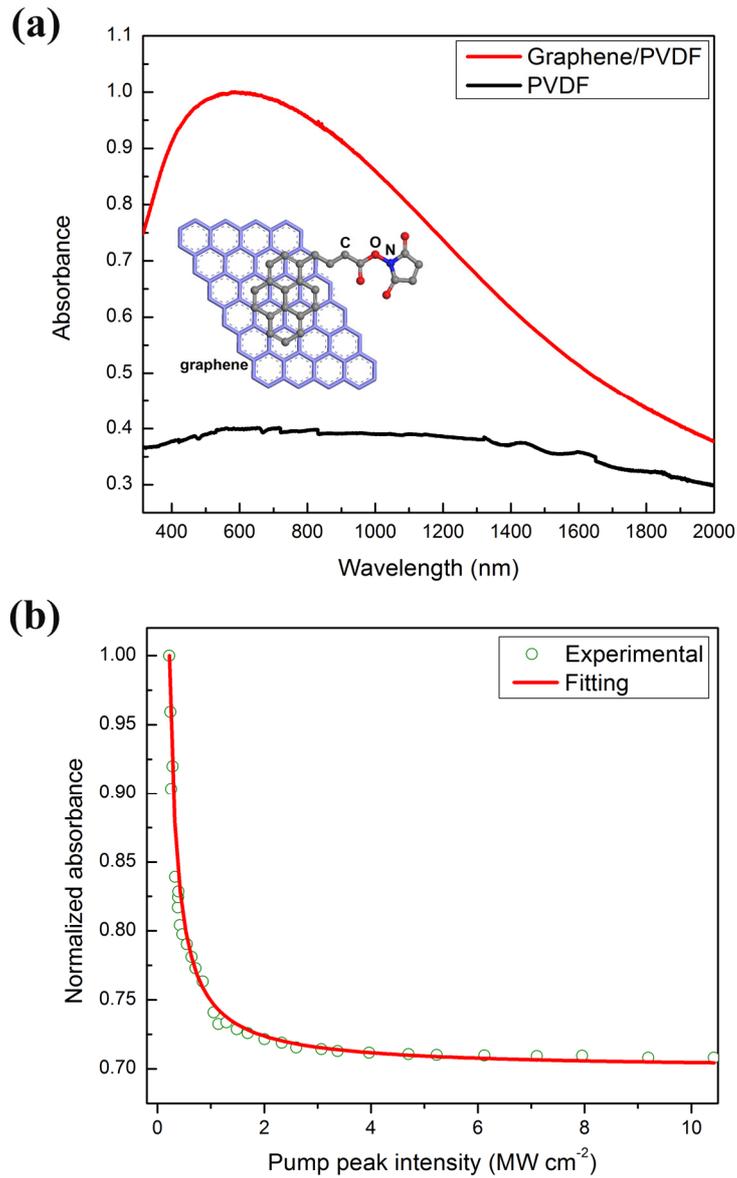

**Figure 8.6**: (a) UV-VIS-NIR absorption spectra of graphene-based PVDF nanocomposites and pure PVDF. The inset shows the chemical structure of the functionalized graphene. b) Power dependent nonlinear saturable absorption of graphene-based PVDF nanocomposites.

Mode locking of the laser self-started at a pump power of ~400 mW. **Figure 8.6**a shows the typical optical spectra of the mode locked laser emissions. The





spectra have a broad spectral bandwidth with obvious Kelly spectral sidebands, characterizing that the mode locked pulses are optical solitons. The central wavelength of the spectra is at 1589.68 nm, which is in the L-band of the optical communication windows. The 3 dB bandwidth of the spectra is about 5.0 nm. **Figure 8.7**b shows the measured autocorrelation trace of the solitons. It has a Sech$^2$-profile with a FWHM width of about 1.07 ps, which, divided by the decorrelation factor of 1.54, corresponds to a pulse width of 694 fs. The TBP of the pulses is 0.412, showing that the solitons are slightly chirped. We also measured the RF spectrum of the mode locking state. **Figure 8.7**d shows a measurement made at a span of 10 kHz and a resolution bandwidth of 10 Hz. The fundamental peak located at the cavity repetition rate of 6.95 MHz has a signal-to-noise ratio of 65 dB. The insert of **Figure 8.7**d shows the wideband RF spectrum up to 1 GHz. The absence of spectral modulation in RF spectrum indicates that the laser operates well in the CW mode-locking regime.

Different from the soliton operation of the erbium-doped fiber lasers mode locked with the NPR technique or a SESAM, no multiple solitons were formed in the cavity immediately after the mode locking. Instead only one soliton was always initially formed. This experimental result shows that the laser has a much lower effective mode locking threshold than those mode locked with SESAMs, which is traced back to the much smaller saturation fluence of the grapheme-polymer nanocomposite than the SESAMs. The single soliton operation could be maintained in the laser as the pump power was gradually increased to 2 W. Further increasing pump power, pulse breaking occurred. Eventually multiple solitons were formed in the laser. Under multiple soliton





operations occasionally harmonic mode locking was also observed. We have focused on the single soliton operation of the laser. The energy of the soliton increased with the pump power. A maximum output power of 13.1 dBm had been obtained under the pump power of 2 W, which indicates the single soliton energy as high as 3 nJ. Experimentally we found that the pump power could be increased to as high as 3.2 W and the laser output power could be as large as 17 dBm. Under the pump and laser operation condition the mode locking of the laser could still maintain for hours, which indicates that the graphene-polymer composite could endure at least an optical fluency as high as 21.4 mJ/cm$^2$. After the operation we had also checked the graphene-polymer composite film using the optical microscopy and found that its morphology was kept intact, which verified its strong thermal stability.

In order to investigate the long-term stability of the mode locking, we have recorded the soliton spectra of the laser in a 4-hour interval over 2 days, as shown in **Figure 8.7**a. It shows that the soliton parameters, these are the central wavelength, 3 dB spectral bandwidth, Kelly sideband positions and the spectral peak powers, have kept reasonably unchanged. Experimentally, it was also found that the soliton pulse width could be compressed to 524 fs after passing through a 10 m DCF of GVD –4 (ps/nm)/km. The result shows that the output solitons were negatively chirped. We note that the current experimental results were obtained by the first try of the mode locking technique. It is expected that through further careful design of the laser cavity and optimization on the saturable absorption parameters of the graphene-polymer composite, mode locked pulses with even larger pulse energy and narrower pulse width could be





generated.

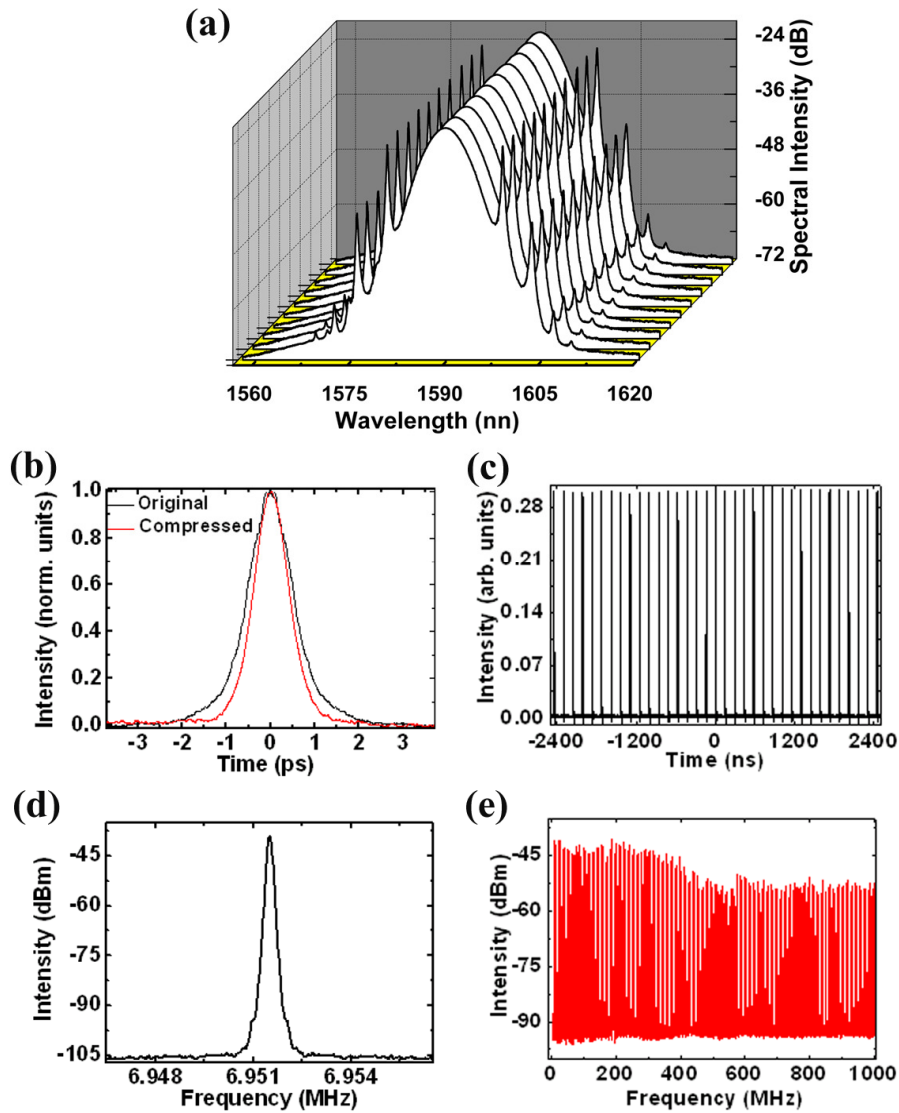

**Figure 8.7**: Soliton operation of the fiber laser. (a) Soliton spectra measured. (b) Autocorrelation traces of the solitons. (c) An oscilloscope trace of the laser emission. (d) The RF spectrum of the laser output. Insert: RF spectrum up to 1 GHz

In this section we have reported an erbium-doped soliton fiber laser with the graphene-polymer nanocomposite membrane as the mode locker. Self-started mode locking of the laser using a graphene-polymer composite was first





experimentally demonstrated, and stable soliton pulses at 1590 nm wavelength with 3 nJ pulse energy and 700 fs pulse width were directly generated from the laser. Our experimental results have clearly shown that a graphene-polymer composite membrane has not only the desired saturable absorption for laser mode locking, but also a large damage threshold. It could be a cost-effective saturable absorber for large energy fiber laser mode locking.

### 8.3     Graphene as a full-waveband saturable absorber

Graphene's unique electronic properties produce an unexpectedly optical conductance of monolayer graphene, with a startlingly simple value: it absorbs $\alpha = e^2/\hbar c \approx 2.3\%$ (where e is the electron charge, $\hbar$ is Dirac's constant and $c$ is the speed of light) of white light, where $\alpha$ is the fine-structure constant [199]. The absorbance has been predicted to be independent of frequency. In principle, the interband optical absorption in zero-gap graphene could be saturated readily under strong excitation due to Pauli blocking [97].

**Figure 8.8**a illustrates the excitation processes responsible for absorption of light in monolayer graphene, in which electrons from the valence band (orange) are excited into empty states in the conduction band (yellow). **Figure 8.8**b shows the subsequent thermalization of hot electrons to form a Fermi-Dirac distribution. As the excitation is increased to high intensity, the concentration of photogenerated carriers increases significantly over that of the intrinsic carrier concentrations and saturate the states near the edge of the conduction and valence bands (**Figure 8.8**c), this produces a Pauli blocking process which results in transparency to light at photon energies just above the band edge.





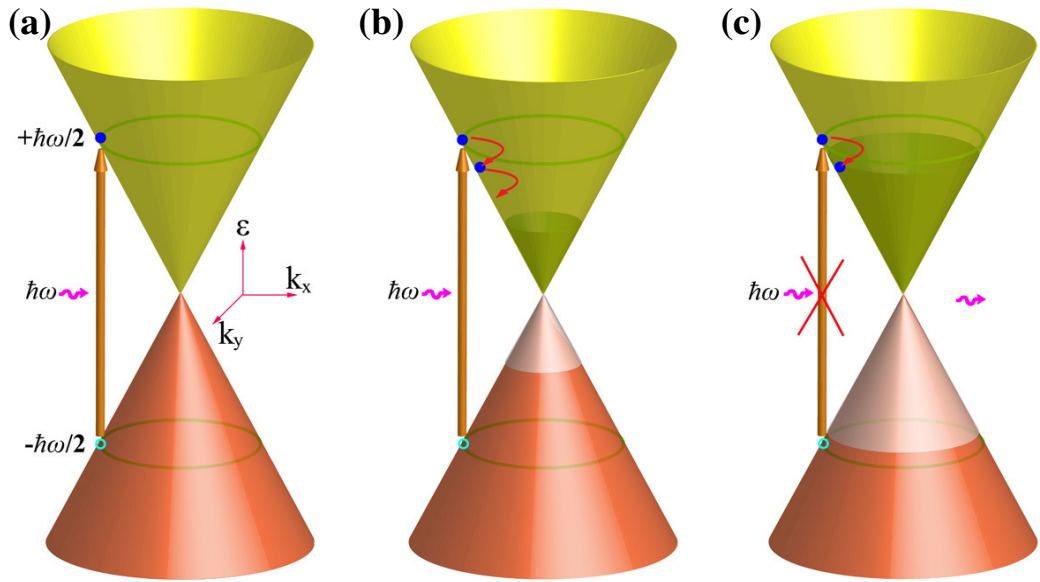

**Figure 8.8:** Absorption of light in graphene. (a) Schematic excitation processes responsible for absorption of light in graphene. The arrow indicates optical interband transition. (b) The photogenerated carriers thermalize and cool down within subpicosecond to form a hot Fermi-Dirac distribution, an equilibrium electron and hole distribution could be finally approached through intraband phonon scattering and electron-hole recombination. (c) At enough high excitation intensity, the photogenerated carriers cause the states near the edge of the conduction and valence bands to fill, blocking further absorption.

Since the optical absorption from graphene is frequency independent, the feature of saturable absorption is unrelated with the input wavelength, which means that theoretically graphene is a "full-waveband" saturable absorber operating at arbitrary wavelength. In order to confirm whether graphene could function as a wideband saturable absorber, we use the same experimental setup and parameters as those of **Figure 8.5**, we could observe a tuning range of 30 nm (1570-1600 nm) of the laser output wavelength, as shown in **Figure 8.9**. All





the output spectra at varied wavelength show two resonant sidebands due to continuum generation, indicating the signature of soliton operation [123]. This is an unambiguous evidence of stable mode locking status of our fiber laser. It was recently reported that wideband tuneability of laser output could be obtained from SWNTs by technically adapting a band-pass filter. The authors claimed that SWNTs with different chiralities and diameters in the mixture are fully made use of to harvest wideband absorption, but lack direct evidence.

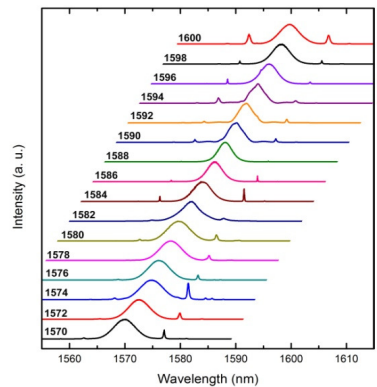

**Figure 8.9**: Wideband spectra tuning from 1570 to 1600 nm.

In our work, despite of the absence of intra-cavity filter, the slight residual polarization asymmetry of the components could still bring about the formation of an artificial "birefringent filter" in the cavity. It is well known that the transmission window of the "birefringent filter" could be continuously varied through slightly tuning the orientation of the polarization controllers, which corresponds to change the cavity birefringence [115]. If the transmission maximum of the "birefringent filter" coincides with the resonant nonlinear absorption of functionalized graphene polymer composites, lasing around this particular wavelength could be built. Theoretically, owing to the fact that the





"birefringent filter" could function as a full-waveband filter and graphene is a natural full-waveband saturable absorber, the current ultrafast Photonics setup could be potentially designed as a full-waveband ultra-fast laser source, i.e., pulse mode locked at arbitrary wavelengths. Unfortunately, due to the amplification bandwidth limitation from Erbium ion, we only obtain the tuneability from 1570 to 1600 nm. However, if Yttrium- , Bismuth- or Thulium-doped fiber, which has broader gain bandwidth, replace the Erbium- doped fiber, much wider tuneability could be achieved. To realize full-waveband ultra-fast lasing, Raman fiber instead of the conventional ion-based fiber must be used because amplification mechanism of the Raman fiber is based on the phonon-photon interaction, where bandwidth limitation effect is weak and insignificant, and that of the ion-based fiber arises from the energy band-gap excitation, where bandwidth limitation effect dominates. In summary, the wideband saturable absorption of graphene has successfully offered one compulsory element to achieve the full-waveband ultra-fast laser source and to sweep the remaining obstacle—gain bandwidth limitation deserves our future exploration.

## 8.4    Graphene mode locked vector dissipative solitons

Firstly, we experimentally investigated the polarization dependence of saturable absorption of the atomic layer graphene, as illustrated in **Figure 8.10**. A stable mode locking L-band fiber laser was used as the light source. The output of the laser was first amplified through a commercial erbium-doped fiber amplifier (EDFA), and then separated into two orthogonally polarized components with an in-line polarization beam splitter (PBS), and perpendicularly incident to the 2D atomic layer graphene. The saturable absorption feature of the graphene





sample under the illumination of each of the orthogonally polarized light was measured, respectively. To reduce any artificial discrepancy caused by the fluctuation of the input power, output powers (with or without passing through the graphene sample) were detected simultaneously through two set of power meters. **Figure 8.11** shows the measured polarization dependence of the graphene saturable absorption. The same normalized modulation depth of ~66 % for both polarizations was observed. The measured saturable fluencies for the two polarizations are 0.71 MW cm$^{-2}$ and 0.70 MW cm$^{-2}$, respectively, which is well within the measurement error range.

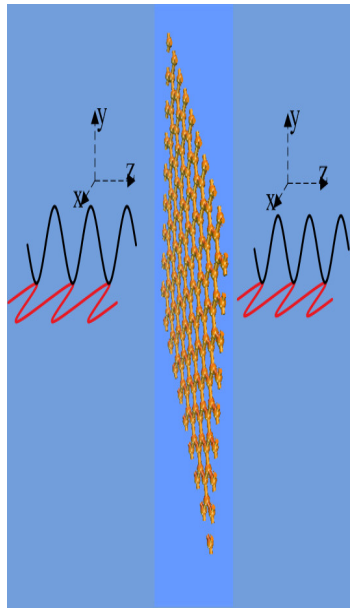

**Figure 8.10:** Illustration of optical conductivity of graphene: Incident light normal to the graphene layer (x-z plane).





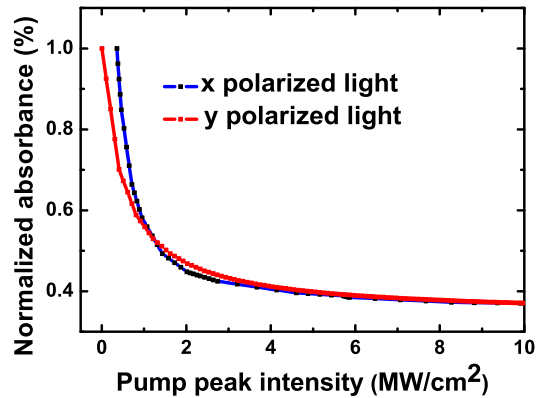

**Figure 8.11:** Polarization resolved saturable absorption curve of graphene-based mode locker.

Using the atomic layer graphene as a mode locker, we further investigated the passive mode locking of an erbium-doped fiber laser under normal cavity dispersions and the vector soliton formation in the laser. The laser has a ring cavity that consists of a segment of 5 m EDF with erbium concentration of 2880 ppm and GVD parameter of −32 (ps/nm)/km, a total length of 9.0 SMF with a GVD parameter of ~18 (ps/nm)/km, and ~118 m dispersion DCF with a GVD parameter of −2 (ps/nm)/km. The intra-cavity optical components such as the 10% fiber coupler, WDM and optical isolator were carefully selected. They all have polarization dependent losses less than 0.1 dB. A polarization independent isolator was used to force the unidirectional operation of the ring, and an intra-cavity PC was used to finely adjust the linear cavity birefringence. The graphene saturable absorber was inserted in the cavity through transferring a piece of free standing few layers graphene film onto the end facet of a fiber pigtail via Van Der Walls force. The laser was pumped by a high power Fiber Raman Laser source (KPS-BT2-RFL-1480-60-FA) of wavelength 1480 nm,





whose maximum pump power can reach as high as 5 W. The fiber laser has a typical dispersion-managed cavity configuration and its net cavity dispersion is ~0.3047 ps$^2$. Mode locking of the laser was achieved through slightly varying the orientation of the PC. Due to the nonlinear pulse propagation in the net normal dispersion cavity, the mode locked pulses were immediately automatically shaped into dissipative solitons. It exhibits the characteristic sharp steep spectral edges of the dissipative solitons formed in the normal dispersion fiber lasers, indicating that the mode locked pulses have been shaped into dissipative solitons, as shown in **Figure 8.12**. Using a 50 GHz high-speed oscilloscope (Tektronix CSA 8000) together with a 45 GHz photo-detector (New Focus 1014), we have measured the dissipative soliton pulse width and profile. The pulses have a width (FWHM) of ~111 ps. If the Sech$^2$ pulse shape is assumed, it gives the pulse width of ~71 ps. The 3 dB spectrum bandwidth of the pulses is ~7.18 nm, which gives a time-bandwidth product of ~63.7, indicating that the pulses are strongly chirped.

Experimentally we further identified that the formed dissipative solitons compose of two orthogonal polarization components, therefore, they are vector dissipative solitons. To experimentally resolve the two orthogonal polarization components of the solitons, an in-line PBS together with a PC were spliced to the output port of the fiber laser. The two orthogonally polarized outputs of the PBS were simultaneously measured with two identical 2 GHz photo-detectors and monitored with a multi-channel oscilloscope (Agilent 54641A) and an optical spectrum analyzer (Ando AQ-6315B). Depending on the net cavity birefringence, either the polarization rotating or the polarization locked





dissipative vector soliton was obtained in our experiments. **Figure 8.12**c shows the polarization resolved measurement of a polarization rotation dissipative vector soliton experimentally obtained. Unlike the dissipative solitons observed in fiber lasers mode locked with the NPR technique, where the polarization of the solitons emitted by the laser is fixed, the polarization of the solitons varies from pulse to pulse. The polarization rotation feature of the dissipative soliton is clearly reflected in **Figure 8.12**c. Without passing through a polarizer, the soliton pulse train has a uniform pulse height, while after passed through the PBS; the pulse height becomes periodically modulated. In particular, the periodic pulse height modulation for each of the orthogonal polarization directions is 90 degree out of phase. In **Figure 8.12**c, the pulse height returns back to its original value after every 9.9 µs, which indicates that the polarization state of the dissipative soliton rotates back to its original state after every 15 cavity roundtrips.





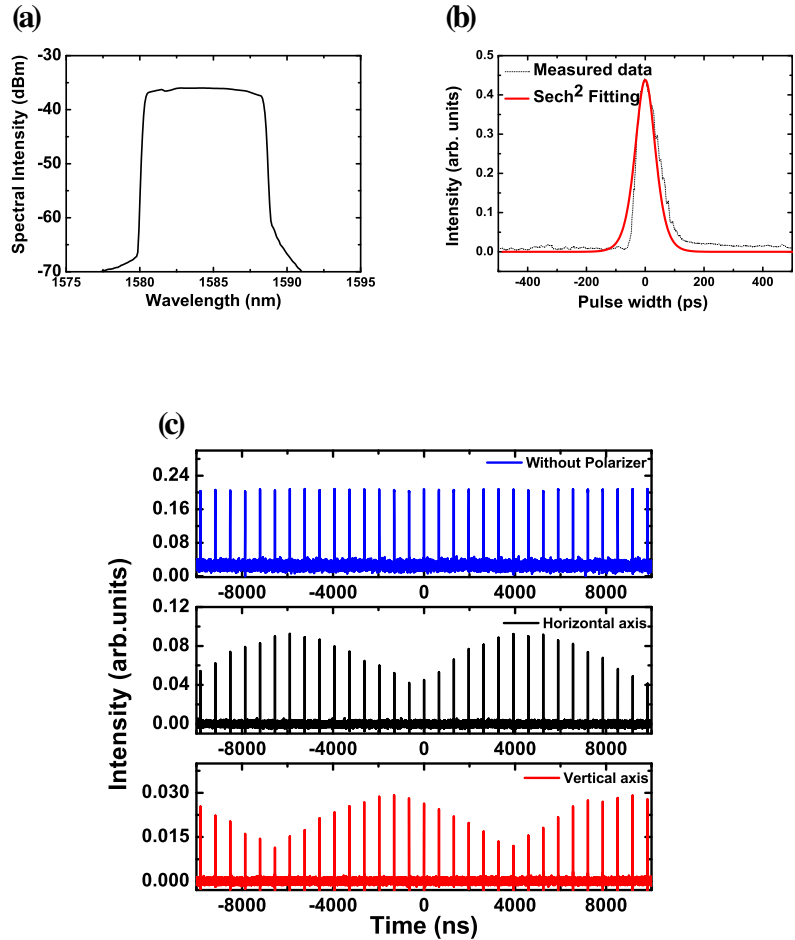

**Figure 8.12**: Dissipative vector soliton operation of the fiber laser. (a) Optical spectrum measured. (b) Pulse profile measured with a high speed oscilloscope. (c) Polarization resolved oscilloscope traces: the polarization of the soliton is rotating in the cavity.

Carefully adjusting the orientation of the intra-cavity PC, polarization locked dissipative vector solitons could also be obtained in our laser. **Figure 8.13** shows a state of the polarization locked DVS operation of the laser. In this case even measured after the extra cavity PBS, the two polarization-resolved pulse traces show uniform pulse trains. However, as the cavity birefringence was changed, the state then changed to a polarization rotation DVS state. For a





phase locked DVS operation state, it is possible to measure the soliton spectra along the long and short polarization ellipse axes. They are shown in **Figure 8.13**a. The two orthogonal polarization components have comparable spectral intensity and equal sharp edge to sharp edge spectral separation. However, carefully examining the fine structures of their spectra, it was found that they are obviously different. While the peak spectral position for one polarization component is at 1598.5 nm, the peak spectral position of the other is located at 1593.9 nm. The 3 dB spectral bandwidths of them are also different, one is 5.48 nm and the other is 5.58 nm. At first glance it is surprising that the peak spectral positions of the two orthogonal polarization components of a phase locked vector soliton could be different. However, considering that the vector soliton is a dissipative soliton, and the characteristics of the dissipative solitons are that they have large nonlinear frequency chirp and broad pulse width, this result might be possible.





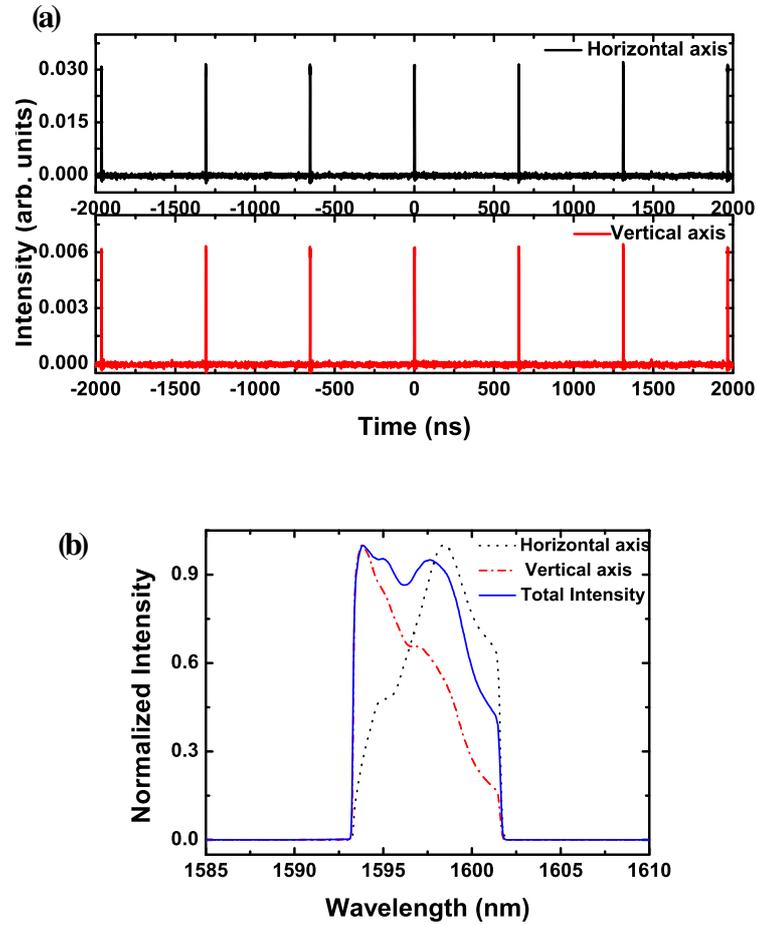

**Figure 8.13:** Polarization locked dissipative soliton operation of the fiber laser. (a) Polarization resolved oscilloscope trace; (b) the corresponding optical spectra.

In this section, we have experimentally investigated the vector soliton operation of erbium-doped fiber lasers mode locked with the atomic layer graphene. Our experimental results not only unambiguously confirmed the saturable absorption of the graphene and its application for passive fiber laser mode locking, but also showed that the saturable absorption of graphene is polarization insensitive when the light is incident perpendicular to the 2D





atomic layer. Due to that the saturable modulation depth of graphene saturable absorber could be easily changed, we expected that new dynamics of vector solitons could be further observed in the graphene mode locked vector soliton fiber lasers.

## 8.5    Prospect of graphene based ultrafast photonics

In Chapter 8, both atomic layer graphene (monolayer and few layers) and graphene polymer composite are promising candidates to function as saturable absorber to generate high energy ultrafast pulse. Owing to the wavelength-independent saturable absorption features of graphene, wideband tuneability of ultrafast pulses have been successfully achieved. Due to its polarization independent saturable absorption property, this novel type of saturable absorber is suitable for generating vector solitons. Although the current study only touched on the studies of graphene based ultra-fast saturable absorbers, Graphene, compared with today's widely investigated photonic materials, such as the silicon and gallium arsenide semiconductors, has the ponderous advantage: tunable photonic properties, achievable through chemical doping. By virtue of graphene's ultra-fast/polarization-insensitive/wide-band/ tunable photonic features, graphene-based ultra-fast photonic devices could have faster response, larger operation range and more flexible administration. Although we have only addressed on features and dynamics of graphene based ultra-fast photonics, due to the wide applicability of graphene which represents the first two dimensional material, results of this study could also be borrowed and applied into other low dimensional material systems. Graphene research is an exciting field that interfaces disciplines like chemistry, physics, materials





science and engineering. In the following years, researchers from the quantum electronics, ultra-fast photonics and nano-science research communities may find in graphene a new testing ground for the ideas and methods they have been researching on their own fields. This will perhaps lead to novel hybrid platform consisting of graphene, graphene-organics and electrically dope graphene, paving the way for the roadmap towards all-carbon photonic devices.





# Chapter 9.  Conclusions and Future Work

In this dissertation, dynamics of optical vector soliton operations in fiber laser have been extensively investigated. We have obtained a large amount of information on the vector soliton properties. Chapter 9 tends to summarize the accomplishments in the research and present some suggestions for the future work.

## 9.1    Summary of accomplishments

### 9.1.1    Vector bright-bright soliton

Firstly, we have studied the vector bright-bright soliton dynamics in fiber lasers passively mode locking through SESAM. A novel type of spectral sideband generation on the soliton spectra of the phase locked vector solitons in a passively mode-locked fiber ring laser has been experimentally and numerically confirmed. The polarization resolved study on the soliton spectrum revealed that the new sidebands were caused by the coherence energy exchange between the two orthogonal polarization components of the vector solitons. The formation of induced temporal solitons has been experimentally and numerically observed in a passively mode-locked fiber laser with birefringence cavity. It was found that the induced solitons were formed by the XPM between the two orthogonal polarization components of the birefringence laser, and the induced solitons could either have the same or different soliton frequency to the inducing soliton. As the induced solitons always have the same group velocity as that of the inducing soliton, they form vector solitons in the laser. To our





knowledge, this is the first experimental observation of temporal induced solitons. Moreover, we also study the effect of XPM on the vector solitons including the induced vector solitons, trapping of vector solitons both in purely negative dispersion cavity and dispersion managed cavity with net positive dispersion.

A novel type of high order phase locked vector soliton in a passively mode-locked fiber laser has been experimentally and numerically observed. The high order vector soliton is characterized by that its two orthogonal polarization components are phase locked, and the stronger polarization component is a single hump pulse, while the weaker component has a double-humped structure with 180° phase difference between the humps. Our experimental result firstly confirmed the theoretical predictions on the high order phase locked vector solitons in birefringent dispersive media.

Dissipative vector solitons have been experimentally demonstrated in a dispersion-managed fiber laser passively mode locked by a SESAM. It was found that despite of their large frequency chirp of the gain-guided solitons, polarization rotating and polarization locked dissipative vector solitons could still be formed in a fiber laser. In addition, formation of multiple dissipative vector solitons with identical soliton parameters and stable harmonic dissipative vector solitons mode-locking are also experimentally obtained.

Group interactions of vector solitons have been both experimentally and numerically observed; our experimental observation shows that two sets of vector soliton traveling at different group velocity as they have orthogonal





polarization states, leading to endlessly collision with each other. Further numerical investigation reveals that their polarization state could be varied significantly in the process of collision but would recover to their original sates once they are separated.

Polarization rotating and locking of DVSs have been both experimentally and numerically obtained in either a dispersion-managed or purely normal dispersion fiber laser cavity. The period of DVSs polarization rotation could be still locked to integer multiple of the cavity roundtrip. We have also experimentally shown than despite of the existence of the laser gain competition, the angle of polarization ellipse orientation between two sets of DVSs can be largely varied because larger chirp and the broader pulse separation could partially counteract the influence of laser gain competition. Numerical simulations have confirmed such polarization rotation.

Multi-wavelength DVSs in an all normal dispersion fiber laser passively mode-locked with SESAM have been firstly experimentally observed. Depending on the strength of the cavity birefringence, stable single-, dual- and triple-wavelength DVSs can be formed in the laser. The multi-wavelength soliton operation of the laser was experimentally investigated, and the formation mechanism of the multi-wavelength DVSs is discussed.

### 9.1.2    Scalar/vector dark soliton

In an all-normal dispersion fiber laser cavity, if SESAM is replaced with a polarization dependent isolator, scalar dark soliton emission could be obtained. It is believed that within a narrow operation regime, the mode locking behavior





would be suppressed and the fiber laser could operate in the non-mode-locking regime. Correspondingly, dark pulse rather than bright pulse emission was established. Through nonlinearity accumulation and further pulse shaping, eventually, dark soliton could be formed. Moreover, if a weak saturable absorber instead of the polarization dependent isolator was used, we could observe the vector dark-dark soliton and even the trapping of vector dark-dark soliton under strong cavity birefringence. Numerical simulations confirmed the above experimental observations, and further proved that the observed dark pulse/soliton could be a genetic feature of NLSE.

### 9.1.3   Domain wall soliton

Optical domain walls, characterized by topological structures separating different circular polarizations, have been experimentally observed in a fiber laser. It is found that the optical domain walls are irrelative to the cavity dispersion but dependent on the cavity birefringence. Particularly, similar to the bright-bright vector soliton, when the cavity is weakly birefringent, the two polarization components of the optical domain wall solitons are coherently coupled and phased locked. Moreover, they could be interpreted as dark-bright vector soliton which is a special case of domain wall solitons. When the cavity is largely birefringent, the two polarization components of the optical domain wall solitons are incoherently coupled and the wall widths are strongly related to the cavity birefringence. The numerical simulation could well reproduce the above experimental observations.





### 9.1.4   Existence domain of diverse vector solitons

Vector solitons, as multi-component solitons, are unique in that the mutual interaction among the soliton components can result in abundant nonlinear dynamics [9, 41, 109]. Determined by the strength of cavity birefringence, the two polarization components of a vector soliton formed in a fiber laser could have the same or different central frequencies, indicating that they could be coherently or incoherently coupled. Experimentally it was observed that under the coherent interaction, an extra spectral sideband caused by four wave mixing appeared while in the incoherent coupling, a strong soliton polarization component could sustain a weak soliton along its orthogonal state due to cross polarization coupling effect, forming the induced soliton. The cavity dispersion is another crucial ingredient impacting vector soliton behaviors. When fiber laser is operated in mode locking regime, NLSE type bright vector soliton can always be naturally formed in anomalous dispersion cavity owing to the balance between the anomalous dispersion and fiber Kerr nonlinearity, whereas in normal dispersion regime, dissipative bright vector solitons are generated as a result of an extra balance between gain and loss. However, it was further experimentally found that when a fiber laser was operated in non-mode-locking regime, NLSE type dark soliton could be generated only in the normal dispersion and the laser gain must be sufficiently high. In contrast to the NLSE type dark soliton, domain wall type dark soliton including: vector dark-bright and dark-dark soliton could be observed in both normal and anomalous dispersion cavity. This is because of the exclusive formation mechanism of domain wall soliton, which originated from the nonlinear coupling between the





two polarization components. Furthermore, laser gain effect also plays an important role on shaping the vector solitons. The formation of fundamental order vector soliton required relatively low pumping while high-order vector soliton needed higher pump. Correspondingly, dispersion, gain and birefringence together lead to colorful configurations of diverse vector solitons whose existence domain can be summarized in the following table.





|  | Dispersion | Gain | Birefringence |
|---|---|---|---|
| Coherent energy exchange | Anomalous to normal | Low | Weak |
| Polarization locked | Anomalous to normal | Moderate | Weak |
| Polarization rotating | Anomalous to normal | Moderate | Weak |
| Trapping of bright VSs | Anomalous to normal | Moderate to high | Strong |
| Induced solitons | Anomalous to normal | Moderate | Strong |
| High-order bright VSs | Anomalous to normal | Moderate | Weak |
| Dissipative bright VSs | Normal | Moderate to high | Moderate to strong |
| Dark VSs | Normal | Very high | moderate |
| Trapping of Dark solitons | Normal | Very high | Strong |
| Polarization DWSs | Anomalous to normal | Weak to high | Weak |
| Vector dark-bright DW | Anomalous to normal | Moderate | Very weak |
| Dual wavelength DW | Anomalous to normal | Moderate | Strong |

**Table 9.1** Existence domain of various vector solitons





### 9.1.5   Graphene based ultrafast saturable Absorbers

In contrast with graphite which nearly completely absorb the input light, graphene, a single 2D atomic layer of carbon atom arranged in a hexagonal lattice, exibits superior optical conductivity. Although it is a zero bandgap semiconductor, like the SWCNTs, graphene also holds the feature of saturable absorption due to Pauli blocking effect. Particularly, since it has no bandgap, the feature of saturable absorption was wavelength independent. Therefore, graphene could be potentially used as a wideband saturable absorber for ultrafast mode-locking fiber lasers. Furthermore, comparing with the SWCNTs, as graphene has a 2D structure, it should have much smaller non-saturable loss and much higher damage threshold. Indeed, in our experiments with an erbium-doped fiber laser we have achieved self-started mode locking and stable soliton pulse emission with high energy. The ultrafast recovery time of graphene also facilitates ultrashort pulse generation. The optical modulation depth can be tuned in a wide range by using single to multilayer graphene or doping/intercalating with other materials. By the virtue of graphene, single pulse energy up to 7.3 nJ in the erbium-doped fiber lasers, which to our knowledge is the highest pulse energy obtained in the erbium-doped fiber lasers with a real passive mode locker.

### 9.2   Recommendations for future work

The discovery of HOVS will induce a new wave on the study of vector solitons; it would be interesting to study the generation mechanism of HOVS both experimentally and numerically. Is it possible to observe other forms of HOVS such as: solitons along both polarizations are double humped, or even triple





humped? Our numerical simulations show a positive answer to it. What we need to do is to experimentally confirm them. The studies are especially essential in enriching the family of solitons.

It might be possible to replace the gain fiber (EDF) with SMF or high nonlinear fiber since such fiber could also provide gain through Stimulated Raman Scattering Effect. Therefore, vector Raman solitons could be observed and even such fiber laser cavity can be purely made of one kind of fiber. As the Raman conversion efficiency is rather high and the wavelength limitation is lift, such fiber laser has potential to provide high power output and ultra-fast pulses at arbitrary wavelength on condition that resonate wavelength coincides with the SESAM absorption wavelength. Furthermore, if SESAM is replaced with other mode locking component such as graphene whose operation wavelength is rather broad and damage threshold is rather high, such laser operation could be further boosted. If FBG is added in the cavity, Raman Gap (Bragg) soliton could be generated as well. Such special soliton is only theoretically propose but never experimentally confirmed.

Our numerical simulation also shows that for the polarization locked vector solitons, their polarization ellipse could rotate either right circularly or left circularly depending on the cavity parameters. Correspondingly, it would be possible to examine the handedness of output solitons with one kind of special fiber which only allows the transmission of right (or left) circularly polarized light. Therefore, handed vector soliton could be both observed experimentally and numerically. The handedness of optical solitons has not been studied till recently. Since left and right handed light exist, the soliton should also have





different handedness.

Bidirectional mode locking through two SESAMs and four ports Reciprocal Circulator could be achieved. In such cavity, the group interaction of vector soliton propagating in two directions could be investigated. As for our former laser operation, vector soliton could only propagate in one direction because of the unidirectional limitation originating from the three ports circulator. However, if the four ports Reciprocal Circulator are applied to replace the three ports Circulator, vector soliton could be generated and propagate in two directions: either clockwise or anticlockwise. Then two sets of vector solitons could be formed. Is there any difference or relationship between these two sets of vector soliton? Do they have strong group interaction? How are their polarization states? Could be one possible state: along one propagating direction, the polarization state of vector soliton is locked while the polarization state is rotating along the other propagating direction? Such interesting physical phenomena deserve our further researching.

Actually there are two types of domain wall soliton: Bloch type and Ising type. The Bloch domain wall is characterized as a nonzero intensity for each polarization component while for the Ising type domain wall its minimum intensity could reach zero. Experimentally, we have observed both types of domain wall solitons but their individual generation mechanisms remain unresolved, which deserves our future exploitation.

Through controlling the cavity birefringence and pumping strength, we could observe the transformation of dark-dark domain wall to dark-bright domain





wall. Moreover, the corresponding numerical simulation confirmed well such transformation. During our experimental studies on dark-bright vector soliton, it is possible to find that the minimum intensity of the dark component and bright component reach zero if a weak saturable absorber (graphene) is added into the cavity. Under appropriate adjustment including the saturable properties of graphene, the cavity birefringence and the pump strength, dark-bright vector soliton could be shaped into the ultimate special vector soliton: black-white vector soliton, which is never experimentally observed in any experimental system including fiber lasers.

Through chemical improvement on graphene, especially, the N-doping or P-doping, the electron density as well as the energy band spectrum of graphene might be varied widely. Whether such doped graphene could function as saturable absorbers? What are the saturable properties of doped graphene? Those issues need to be addressed in the near future.

In our experiments about graphene mode locking, the input laser is perpendicular to the plane of graphene atomic layer. However, could graphene guide optical waves in the direction parallel to the atomic layer? If it works, definitely, graphene would have more exciting applications in optoelectronics except being saturable absorbers.

# Author's Publications

**Journals:**

1. Dingyuan Tang, <u>**Han Zhang**</u>, Luming Zhao and Xuan Wu, "Observation of high-order polarization-locked vector solitons in a fiber laser," **Physical Review Letters**, 101, 153904 (2008).

2. <u>**Han Zhang**</u>, Dingyuan Tang, Luming Zhao, and Randall Knize, "Vector dark domain wall solitons in a fiber ring laser," **Optics Express**, 18, 4428 (2010).

3. <u>**Han Zhang**</u>, Dingyuan Tang, Luming Zhao, Qiaoliang Bao, and Kian Ping Loh, "Graphene mode locked, wavelength-tunable, dissipative soliton fiber laser," **Applied Physics Letters**, 96, 111112 (2010).

4. <u>**Han Zhang**</u>, Qiaoliang Bao, Dingyuan Tang, Luming Zhao, and Kianping Loh, "Large energy soliton erbium-doped fiber laser with a graphene-polymer composite mode locker," **Applied Physics Letters**, 95, 141103 (2009).

5. <u>**Han Zhang**</u>, Dingyuan Tang, Luming Zhao and Wu Xuan, "Dark pulse emission of a fiber laser,'' **Physical Review A**, 80, 045803 (2009).

6. <u>**Han Zhang**</u>, Dingyuan Tang, Luming Zhao, Qiaoliang Bao,and Kianping Loh,"Large energy mode locking of an erbium-doped fiber laser with atomic layer graphene, " **Optics Express**, 17,17630-17635 (2009).

Ping Loh, "Graphene-Polymer Nanofiber Membrane for Ultrafast Photonics," **Advanced Functional Materials**, 20, 782-791 (2010).

15. Qiaoliang Bao, **Han Zhang**, Yu Wang, Zhenhua Ni, Yongli Yan, Ze Xiang Shen, Kian Ping Loh ,and Ding Yuan Tang , "Atomic layer graphene as saturable absorber for ultrafast pulsed lasers, " **Advanced Functional materials**, 19, 3077–3083 (2009).

**Conferences:**

1. **H. Zhang**, D. Y. Tang, L. M. Zhao, Q. L. Bao, K. P. Loh, "Mode locking of fiber lasers with atomic layer graphene," Australian institute of physics Congress 2009.

2. **H. Zhang**, D. Y. Tang, L. M. Zhao, and X. Wu. "Dark Soliton Fiber Lasers", Australian institute of physics Congress 2009.

3. **H. Zhang**, D. Y. Tang, L. M. Zhao, and X. Wu. "Higher-order four wave mixing spectral sideband generation in fiber lasers", Australian institute of physics Congress 2008, Australian institute of physics Congress 2008.

4. **H. Zhang**, D. Y. Tang, L. M. Zhao, and X. Wu. "Polarization rotation of gain-guided vector solitons in a dispersion-managed fiber laser", Australian institute of physics Congress 2008.